\documentclass[preprint,aps,12pt,showpacs,nofootinbib,tightenlines,amsmath,amssymb]{revtex4}
\usepackage{amsmath}
\usepackage{graphicx}
\usepackage{amssymb}
\usepackage{color}

\newcommand{\be}{\begin{eqnarray}}
\newcommand{\ee}{\end{eqnarray}}

\newcommand{\p}{\partial}
\newcommand{\hp}{\hat{\partial}}

\newcommand{\nn}{\nonumber}

\newcommand{\diag}{\mathop{\rm diag}}
\newcommand{\cL}{{\mathcal L}}
\newcommand{\fkL}{\mathfrak{L}}

\newcommand{\cA}{{\mathcal A}}
\newcommand{\hcA}{ \hat{{\mathcal A}}}
\newcommand{\ckcA}{ \check{{\mathcal A}}}
\newcommand{\bvcA}{ \breve{{\mathcal A}}}
\newcommand{\tcA}{ \tilde{{\mathcal A}}}

\newcommand{\hcF}{ \hat{{\mathcal F}}}
\newcommand{\tcF}{ \tilde{{\mathcal F}}}

\newcommand{\ckcF}{ \check{{\mathcal F}}}

\newcommand{\mF}{\mathsf F}

\newcommand{\cF}{{\mathcal F}}
\newcommand{\cQ}{{\mathcal Q}}

\newcommand{\fkA}{\mathfrak{A}}
\newcommand{\fkF}{\mathfrak{F}}
\newcommand{\fkD}{\mathfrak{D}}

\newcommand{\ckcD}{\check{\cD}}

\newcommand{\cM}{{\mathcal M}}

\newcommand{\fF}{{\mathbf F}}
\newcommand{\hfF}{{\hat{\fF}}}
\newcommand{\hmF}{{\hat{\mF}}}

\newcommand{\fV}{{\mathbf V}}
\newcommand{\mV}{{\mathsf V}}
\newcommand{\cV}{{\mathcal V}}

\newcommand{\cD}{{\mathcal D}}

\newcommand{\hcD}{\hat{{\mathcal D}}}
\newcommand{\tcD}{\tilde{{\mathcal D}}}

\newcommand{\cG}{{\mathcal G}}
\newcommand{\fG}{{\mathbf G}}

\newcommand{\cJ}{{\mathcal J}}
\newcommand{\mJ}{{\mathsf J}}
\newcommand{\fJ}{{\mathbf J}}
\newcommand{\tfJ}{\tilde{\fJ}}

\newcommand{\hmJ}{\hat{\mJ}}
\newcommand{\hfJ}{\hat{\fJ}}

\newcommand{\hJ}{\hat{J}}
\newcommand{\hG}{\hat{G}}
\newcommand{\hT}{\hat{T}}
\newcommand{\htt}{\hat{t}}
\newcommand{\htk}{\hat{t}_{\kappa}}

\newcommand{\fI}{{\mathbf I}}

\newcommand{\fK}{{\mathbf K}}

\newcommand{\cW}{{\mathcal W}}
\newcommand{\cS}{{\mathcal S}}

\newcommand{\fGh}{{\mathbf G}_h}

\newcommand{\bchi}{\bar{\chi} }

\newcommand{\GeV}{\;\text{GeV}}

\newcommand{\wtjoin}{\,\widetilde{\Join}\, }

\newcommand{\1}{\mspace{1mu}}

\newcommand{\heth}{\hat{\eth}}

\newcommand{\fGa}{\mathbf \Gamma}
\newcommand{\bfGa}{\bar{\mathbf \Gamma}}
\newcommand{\vGa}{\varGamma}

\newcommand{\tvGa}{\tilde{\varGamma}}

\newcommand{\hfGa}{\hat{\fGa}}

\newcommand{\vka}{\varkappa}

\newcommand{\tvSi}{\tilde{\varSigma}}
\newcommand{\vSi}{\varSigma}

\newcommand{\fSi}{\mathbf \Sigma}
\newcommand{\hfSi}{\hat{\fSi}}
\newcommand{\bfSi}{\bar{\fSi}}

\newcommand{\tlam}{\tilde{\lambda}}

\newcommand{\cH}{\mathcal H }

\newcommand{\fH}{\mathbf H }
\newcommand{\mH}{\mathsf H }
\newcommand{\hmH}{\hat{\mH}}

\newcommand{\bhmH}{\bar{\hmH}}
\newcommand{\thmH}{\tilde{\hmH}}

\newcommand{\cR}{\mathcal R}
\newcommand{\fR}{\mathbf R}
\newcommand{\hfR}{\hat{\fR}}

\newcommand{\mT}{\mathsf T}
\newcommand{\fT}{\mathbf T}

\newcommand{\mR}{\mathsf R}

\newcommand{\ckS}{\check{S}}
\newcommand{\tS}{\tilde{S}}

\newcommand{\mM}{\mathsf M}

\newcommand{\mU}{\mathsf U}

\newcommand{\tmA}{\tilde{\mathsf A}}

\newcommand{\tA}{\tilde{A}}
\newcommand{\tB}{\tilde{B}}
\newcommand{\wtA}{\widetilde{A}}

\newcommand{\bA}{\bar{A}}
\newcommand{\bB}{\bar{B}}

\newcommand{\mA}{\mathsf A}
\newcommand{\mB}{\mathsf B}
\newcommand{\mC}{\mathsf C}
\newcommand{\mD}{\mathsf D}
\newcommand{\mE}{\mathsf E}

\newcommand{\fL}{\mathbf L}
\newcommand{\fM}{\mathbf M}
\newcommand{\fN}{\mathbf N}
\newcommand{\fP}{\mathbf P}
\newcommand{\fQ}{\mathbf Q}

\newcommand{\fA}{\mathbf A}
\newcommand{\fB}{\mathbf B}
\newcommand{\fC}{\mathbf C}
\newcommand{\fD}{\mathbf D}
\newcommand{\fE}{\mathbf E}
\newcommand{\fS}{\mathbf S}

\newcommand{\hbfA}{\hat{\bfA}}
\newcommand{\bhfA}{\bar{\hfA}}

\newcommand{\hfA}{\hat{\mathbf A}}
\newcommand{\bfA}{\bar{\mathbf A}}

\newcommand{\ckcC}{\check{\cC}}

\newcommand{\mW}{\mathsf W}

\newcommand{\la}{l_{\alpha}}
\newcommand{\Ma}{M_{\alpha}}

\newcommand{\bc}{\bar{c}}
\newcommand{\bC}{\bar{C}}

\newcommand{\bM}{\bar{M}}

\newcommand{\mOm}{\mathsf \Omega }
\newcommand{\fOm}{\mathbf \Omega }

\newcommand{\lamq}{\lambda_{q_c} }

\newcommand{\chih}{\hat{\chi}}

\newcommand{\bchih}{\hat{\bar{\chi}}}
\newcommand{\tchi}{\tilde{\chi}}

\newcommand{\tcH}{\tilde{\cH}}

\newcommand{\mk}{m_{\kappa}}
\newcommand{\hmk}{\hat{m}_{\kappa}}
\newcommand{\Mka}{M_{\kappa}}

\newcommand{\lka}{l_{\kappa}}
\newcommand{\hlka}{\hat{l}_{\kappa}}

\newcommand{\apk}{\alpha_{\kappa}}
\newcommand{\hHk}{\hat{H}_{\kappa}}
\newcommand{\Hka}{H_{\kappa}}
\newcommand{\Lk}{L_{\kappa}}
\newcommand{\hLk}{\hat{L}_{\kappa}}
\newcommand{\lak}{\lambda_{\kappa}}
\newcommand{\lachi}{\lambda_{\chi}}
\newcommand{\eka}{\eta_{\kappa}}
\newcommand{\cka}{c_{\kappa}}

\newcommand{\xia}{\xi_{\alpha}}
\newcommand{\dla}{d_{\lambda}}

\newcommand{\ta}{\tilde{a}}
\newcommand{\tb}{\tilde{b}}
\newcommand{\tc}{\tilde{c}}

\newcommand{\tU}{\tilde{U}}
\newcommand{\tD}{\tilde{D}}
\newcommand{\tQ}{\tilde{Q}}
\newcommand{\tM}{\tilde{M}}
\newcommand{\tN}{\tilde{N}}

\newcommand{\fPsi}{\mathbf \Psi}

\newcommand{\hx}{\hat{x}}

\newcommand{\htau}{\hat{\tau}}

\newcommand{\tC}{\tilde{C}}

\newcommand{\mG}{\mathsf{G}}

\newcommand{\cC}{\mathcal{C}}

\newcommand{\cT}{\mathcal{T}}

\newcommand{\hfT}{\hat{\fT}}

\newcommand{\hmT}{\hat{\mT}}

\newcommand{\CQc}{\mathcal{C}_{\cQ_c}}

\newcommand{\fQG}{\fQ_{\mG}}

\newcommand{\fQGf}{\fQ_{\mG_f}}
\newcommand{\fQGi}{\fQ_{\mG_i}}

\newcommand{\fQGpi}{\fQ_{\mG_i}^+}
\newcommand{\fQGni}{\fQ_{\mG_i}^-}

\newcommand{\fQGpf}{\fQ_{\mG_1}^+}
\newcommand{\fQGps}{\fQ_{\mG_2}^+}
\newcommand{\fQGpt}{\fQ_{\mG_3}^+}
\newcommand{\fQGpft}{\fQ_{\mG_4}^+}

\newcommand{\fQGnf}{\fQ_{\mG_1}^-}
\newcommand{\fQGns}{\fQ_{\mG_2}^-}
\newcommand{\fQGnt}{\fQ_{\mG_3}^-}
\newcommand{\fQGnft}{\fQ_{\mG_4}^-}
\newcommand{\fQGfn}{\fQ_{\mG_f}^-}
\newcommand{\fQGfbn}{\fQ_{\mG_{f'}}^-}
\newcommand{\fQGfbp}{\fQ_{\mG_{f'}}^+}
\newcommand{\fQGfp}{\fQ_{\mG_f}^+}
\newcommand{\fQGfpn}{\fQ_{\mG_f}^{\pm}}

\newcommand{\fQGft}{\fQ_{\mG_4}}

\newcommand{\fQU}{\fQ_{\mU}}

\newcommand{\fQUn}{\fQ_{\mU}^-}

\newcommand{\fQUp}{\fQ_{\mU}^+}

\newcommand{\fQE}{\fQ_{\mE}}

\newcommand{\fQH}{\fQ_{\mH}}

\newcommand{\hPsi}{\hat{\Psi}}
\newcommand{\tPsi}{\tilde{\Psi}}
\newcommand{\vPsi}{\varPsi}

\begin{document}
\def\intdk{\int\frac{d^4k}{(2\pi)^4}}
\def\sla{\hspace{-0.22cm}\slash}
\hfill


\title{ The foundation of the hyperunified field theory II \\ - fundamental interaction and evolving universe}  

\author{Yue-Liang Wu}\email{ylwu@itp.ac.cn}
\affiliation{$^1$Institute of Theoretical Physics, Chinese Academy of Sciences, Beijing 100190, China\\
$^2$International Centre for Theoretical Physics Asia-Pacific (ICTP-AP) \\ (Beijing/Hangzhou), UCAS, Beijing 100190, China \\
$^3$ Taiji Laboratory for Gravitational Wave Universe (Beijing/Hangzhou), University of Chinese Academy of Sciences (UCAS), Beijing 100049, China \\ 
$^4$ School of Fundamental Physics and Mathematical Sciences, Hangzhou Institute for Advanced Study, UCAS, Hangzhou 310024, China }


\begin{abstract}
In part I of the foundation of the hyperunified field theory, we have shown the presence of entangled hyperqubit-spinor field as fundamental building block with appearance of Minkowski hyper-spacetime as free-motion spacetime and emergence of inhomogeneous hyperspin symmetry as fundamental symmetry. In this paper as part II, we demonstrate by following along gauge invariance principle and scaling invariance hypothesis that the inhomogeneous hyperspin gauge symmetry and scaling gauge symmetry govern fundamental interactions and reveal the nature of gravity and spacetime. With the fiber bundle structure of biframe hyper-spacetime and emergence of non-commutative geometry, we explore methodically the gauge-geometry duality and genesis of gravitational interaction in locally flat gravigauge hyper-spacetime. A whole hyperunified field theory in 19-dimensional hyper-spacetime is established to unify not only all discovered leptons and quarks into hyperunified qubit-spinor field but also all known basic forces into hyperunified interaction governed by inhomogeneous hyperspin gauge symmetry. We present a systematic investigate on the hyperunified field theory with deriving the dynamics of fundamental fields as basic laws of nature and conservation laws relative to basic symmetries and showing Higgs-like bosons and three families of lepton-quark states. We provide a detailed analysis on inflationary early universe with evolving graviscalar field and a discussion on scaling gauge field as dark matter candidate and $\cQ_c$-spin scalar field as source of dark energy.
\end{abstract}

\pacs{12.10.-g, 04.50.+h, 04.50.-h,11.10.Kk \\
Keywords: gauge and scaling invariance principle, inhomogenous hyperspin gauge symmetry,
unification of all basic forces, gravigauge hyper-space-time, inflationary universe, dark matter and
dark energy.}

\maketitle

\begin{widetext}
\tableofcontents
\end{widetext}

\section{Introduction}

In the previous paper as part I of the foundation of the hyperunified field theory\cite{FHUFT-I}, starting from the motional property of functional field by proposing the maximum coherence motion principle and maximum entangled-qubits motion principle as two guiding principles, we have provided a detailed analysis and systematic investigation on the existence of entangled hyperqubit-spinor field and $\cQ_c$-spin scalar field as fundamental building blocks of nature and also the appearance of Minkowski hyper-spacetime and emergence of inhomogeneous hyperspin symmetry and scaling symmetry as fundamental symmetries of nature. In light of advantages of the action principle with path integral formulation, the two guiding principles presented in \cite{FHUFT-I} have been shown to lay the essential foundation of the hyperunified field theory\cite{HUFT,HUFTSB,HUFTTK} within the framework of gravitational quantum field theory\cite{GQFT,GQFTTK}. It has been demonstrated that a locally entangled state of qubits obeying maximum coherence motion principle brings about an entangled qubit-spinor field with automatic appearance of canonical anticommutative relation, and meanwhile both spacetime dimension and fundamental symmetry are determined by the motion-correlation $\cM_c$-spin charge $\cC_{\cM_c}=D_h$ of entangled qubit-spinor field. The $\cQ_c$-spin charge $\cC_{\cQ_c}=q_c$ has been shown to possess a periodic feature given by the mod 4 qubit number, which allows us to classify all entangled qubit-spinor fields and spacetime dimensions into four categories characterized by four $\cQ_c$-spin charges $\cC_{\cQ_c}=0,1,2,3$. In general, we come to a better understanding on the basic constituent of matter as locally entangled state of qubits and also the motional nature of matter, which includes the appearance of Minkowski spacetime with only one temporal dimension and the emergence of fundamental symmetry of matter and spacetime. To get a better understanding on the fundamental symmetry of nature, we have distinguished symmetry groups in Hilbert space from those in Minkowski spacetime, so that we arrive at an associated symmetry in which inhomogeneous hyperspin symmetry group and global scaling symmetry group in Hilbert space as internal space of basic fields are in association with inhomogeneous Lorentz-type symmetry group and scaling symmetry group in Minkowski spacetime as external spacetime of coordinates, where their symmetry group transformations must be coincidental so as to ensure the invariance of the action. In particular, it has been verified that the entangled decaqubit-spinor field with $\cM_c$-spin charge $\cC_{\cM_c}=19$ and $\cQ_c$-spin charge $\cC_{\cQ_c}=1$ brings on a hyperunified qubit-spinor field which unifies all known leptons and quarks into a single fundamental building block with prediction on the existence of mirror lepton-quark states, and its action possesses an associated symmetry in which the inhomogeneous hyperspin symmetry WS(1,18) is in association with inhomogeneous Lorentz-type/Poincar\'e-type symmetry PO(1,18). Such an associated symmetry provides a hyperunified symmetry as fundamental symmetry. So that we are able to provide reliable explanations on some longstanding open questions, such as why there exist leptons and quarks beyond one family in SM and why the currently observed universe is only four-dimensional spacetime.

In the present paper as part II of the foundation of the hyperunified field theory, we will further study the basic properties of entangled hyperqubit-spinor field and $\cQ_c$-spin scalar field as fundamental building blocks and also the essential features of inhomogeneous hyperspin symmetry and scaling symmetry as fundamental symmetries. As there remain some open questions to be issued, such as: what is acted as the fundamental interaction of nature? how does the fundamental symmetry govern basic forces? what is the nature of gravity? what is the basic structure of spacetime? how does early universe get inflationary expansion? what is a dark matter candidate? what is the nature of dark energy?  what is the nature of Higgs boson? how can we understand three families of chiral type leptons and quarks discovered by current experiments?  

To make issues on such open questions come to the main purpose of present paper as the part II of the foundation of the hyperunified field theory. We will begin with considering gauge invariance principle as guiding principle, which brings global symmetry in Hilbert space into local symmetry with the introduction of gauge field\cite{Weyl,YM} as basic field for characterizing fundamental interactions. When taking global inhomogeneous hyperspin symmetry and scaling symmetry in Hilbert space to be local symmetries based on the gauge invariance principle, we arrive at local inhomogeneous hyperspin gauge symmetry and scaling gauge symmetry as fundamental gauge symmetries which are shown to govern the fundamental interactions of nature. 

A biframe hyper-spacetime with fiber bundle structure is deduced from gauging inhomogeneous hyperspin symmetry with the introduction of hyper-gravigauge field, which leads locally flat gravigauge hyper-spacetime in Hilbert space to emerge as hyper-fiber bundle with globally flat Minkowski hyper-spacetime as base spacetime obeying special relativity\cite{SR}. We will demonstrate that the local scaling gauge symmetry of basic fields in Hilbert space together with global scaling symmetry of coordinates in Minkowski spacetime play an essential role for building a full hyperunified field theory. Therefore, the gauge invariance principle and scaling invariance hypothesis together with the maximum coherence motion principle and maximum entangled-qubits motion principle proposed in the previous paper\cite{FHUFT-I} will be shown to lay a whole foundation of the hyperunified field theory. In light of a fully constructed hyperunified field theory within the framework of gravitational quantum field theory, by corroborating both the gauge-gravity correspondence and gravity-geometry correspondence, we will come to a better understanding on the basic structure of spacetime and the nature of basic forces, which includes the genesis of gravity and the relation to general relativity\cite{GR,FGR}. The equations of motion of fundamental fields are derived to characterize the laws of nature and meanwhile various conservation laws are resulted from relevant symmetries. It will be shown that in locally flat gravigauge hyper-spacetime, the hyperspin gravigauge field behaves as an auxiliary field with emergence of non-commutative geometry characterized by the hyperspin gravigauge field. The gauge-geometry duality is verified and applied to obtain equivalent gauge-type and geometric Einstein-type gravitational equations of motion, which allows us to analyze electric-like and magnetic-like gravitational interactions. Furthermore, it makes us to achieve a better understanding on the evolution of universe with inflationary early universe and currently accelerated expanding universe from evolving graviscalar field and $\cQ_c$-spin scalar field, and meanwhile get a better comprehension on three families of chiral type leptons and quarks in the standard model (SM) of electroweak and strong interactions\cite{SM1,SM2a,SM2b,SM3a,SM3b,SM3c,SM4, SM5,SM6,SM7,SM8} in the framework of quantum field theory\cite{QFT1a,QFT1b,QFT2}. In particular, such a hyperunified field theory is shown to result in dark matter and dark energy candidates and bring on more Higgs-like bosons in four-dimensional spacetime. Anyway, we are going to make a systematic analysis and detailed investigation on the whole hyperunified field theory in this paper.

Our paper is organized as follows: a brief introduction is presented in Sect.1. In Sect. 2, we will make global $\cM_c$-spin/hyperspin symmetry and $\cQ_c$-spin symmetry as well as scaling symmetry as local gauge symmetries in Hilbert space by proposing gauge invariance principle as guiding principle, and the corresponding gauge fields as basic fields are introduced to characterize gauge interactions. The gauge covariant equation of motion for entangled hyperqubit-spinor field with scaling gauge invariance is derived to describe gravitational relativistic quantum theory of entangled hyperqubit-spinor field with the presence of hyper-gravigauge field and $\cQ_c$-spin scalar field. In Sect. 3, we will show that bicovariant vector field $\chih_{\fA}^{\;\; \fM}(x)$ and dual bicovariant vector field $\chi_{\fM}^{\; \fA}(x)$ as hyper-gravigauge field form locally flat gravigauge hyper-spacetime with emergence of non-commutative geometry, which brings on the introduction of biframe hyper-spacetime with hyper-fiber bundle structure. The fundamental interaction is shown to be governed by inhomogeneous hyperspin gauge symmetry WS(1,$D_h$-1). In particular, Abelian-type $\cW_e$-spin invariant-gauge field brings on the genesis of hyper-gravigauge field as bicovariant vector field $\chi_{\fM}^{\; \fA}(x)$ associated with graviscalar field and meanwhile leads to gravitational origin of gauge symmetry in hyper-spacetime, which demonstrates that gravitational interaction is truly characterized by gauge interaction. In Sect. 4, we are going to build a whole hyperunified field theory for describing both fundamental building blocks of nature and fundamental interactions of nature by following along gauge invariance principle and scaling invariance hypothesis. The hyperunified qubit-spinor field $\fPsi_{\fQH}(x)$ in 19-dimensional hyper-spacetime is regarded as fundamental building block of nature and the inhomogeneous hyperspin gauge interaction is viewed as fundamental interaction mediated through inhomogeneous hyperspin gauge field $\bvcA_{\fM}(x)$.  In Sect. 5, by reformulating such a hyperunified field theory in biframe hyper-spacetime to have hidden scaling gauge invariance, we are able to demonstrate the gauge-gravity correspondence and derive equations of motion of fundamental fields by applying for the least action principle. Meanwhile, various conserved currents are obtained with respect to gauge symmetries. In Sect. 6, we will apply Noether's theorem to such a whole hyperunified field theory in the presence of gravitational interaction and derive conservation laws and dynamic evolution equations from both global Poincar\'e-type group symmetry and conformal scaling symmetry of Minkowski hyper-spacetime. 

By defining hyper-spacetime gauge field from hyperspin gauge field in Sect. 7, we will reformulate an equivalent action for the whole hyperunified field theory in a hidden gauge formalism, where the hyper-gravigauge field plays an essential role as Goldstone-like boson. Such a hidden gauge formalism implies that dynamics of hyper-gravigauge field is equivalently characterized by that of Einstein-Hilbert type action which is governed by a hidden general linear group symmetry GL(1,$D_h$-1, R). So that we are able to demonstrate the gravity-geometry correspondence under flowing unitary gauge and also the profound correlation between gravitational interaction and emergent Riemann geometry in curved Riemannian hyper-spacetime. As the hyperunified field theory is in principle independent of the choice of coordinate systems, we will rewrite in Sect.8 the whole hyperunified field theory in a hidden coordinate formalism by utilizing the scaling gauge invariant hyper-gravigauge field as Goldstone-like boson. Such a hyperunified field theory reveals the geometry-gauge correspondence, so that the gravitational interaction is described by hyperspin gravigauge field. In locally flat gravigauge hyper-spacetime, such a hyperspin gravigauge field behaves as an auxiliary field and characterizes emergent non-commutative geometry. In Sect. 9, by taking entirety unitary gauge as gauge prescription to fix both hyperspin gauge symmetry and hidden general linear group symmetry, we will present a more radical investigation on the foundation of the hyperunified field theory within the framework of gravitational quantum field theory. By revealing the gauge-geometry duality in biframe hyper-spacetime, we are able to explore the dynamics of both hyper-gravigauge field and hyper-gravimetric gauge field with electric-like and magnetic-like gravitational interactions through gauge-type and geometric Einstein-type gravitational equations. In Sect. 10, we will analyze various gauge fixing conditions for scaling gauge symmetry in the whole hyperunified field theory under entirety unitary gauge. By choosing Einstein-type basis in the presence of fundamental mass scale $\Mka$, we will show how a conformally flat gravigauge hyper-spacetime as nonsingularity background enables us to describe the evolution of early universe with evolving graviscalar field in hyper-spacetime. In light of the so-called co-moving cosmic time,  the universe undergoes an inflationary expansion. We will show that the minimum potential of $\cQ_c$-spin scaling field not only prevents early universe from keeping inflationary expansion but also provides a tiny cosmic energy density as dark energy candidate characterized by fundamental cosmic mass scale $\mk$. In particular, a heavy scaling gauge field is found to be as dark matter candidate when inflationary universe comes to an end. In Sect. 11, we will analyze hyperunified symmetry structure and possible symmetry breaking mechanisms in the whole hyperunified field theory in 19D hyper-spacetime. To illustrate possible symmetry breaking mechanisms for hyperspin symmetry, we are going to focus on the action of entangled decaqubit-spinor field as hyperunified qubit-spinor field with showing the explicit group decomposition structure for hyperspin gauge field in locally flat gravigauge hyper-spacetime and examining a specific bulk structure of 19th spatial dimension in hyper-spacetime, which enables us to get a better comprehension on SM with Higgs-like bosons and three families of chiral type lepton-quark states in four-dimensional spacetime. Our conclusions and discussions will be presented in the final section.


\section{ Inhomogeneous hyperspin and $\cQ_c$-spin gauge symmetries and gravitational relativistic quantum theory with scaling gauge symmetry based on gauge invariance principle }

In the previous paper as the part I of the foundation of the hyperunified field theory\cite{FHUFT-I}, we have presented an extensive analysis on the categorization for qubit-spinor field and spacetime dimension and made a detailed discussion on the symmetry properties of entangled qubit-spinor field and Minkowski spacetime. In particular, it has been shown that the category-1 entangled decaqubit-spinor field $\Psi_{\fQE^{10}}(x)$ enables us to merge all known leptons and quarks in SM into a single hyperunified qubit-spinor field $\fPsi_{\fQH}(x)$ in 19-dimensional hyper-spacetime. Such a hyperunified qubit-spinor field possesses inhomogeneous hyperspin symmetry WS(1,18). Nevertheless, all previous discussions are mainly focused on a free motion of entangled qubit-spinor field without involving dynamics through interactions. In this section, we are going to demonstrate how the fundamental interaction can be introduced by making $\cM_C$-spin and $\cQ_c$-spin symmetries of entangled qubit-spinor field to be gauge symmetries in Hilbert space by proposing the {\it gauge invariance principle} as guiding principle. A gauge covariant equation of motion for entangled hyperqubit-spinor field with scaling gauge invariance will be derived to characterize gravitational relativistic quantum theory of entangled hyperqubit-spinor field in the presence of hyper-gravigauge field and $\cQ_c$-spin scalar field.


\subsection{Inhomogeneous hyperspin and $\cQ_c$-spin gauge symmetries with introduction of bicovariant vector field and gauge field by making gauge invariance principle}

It has been shown in ref.\cite{FHUFT-I} that the action of entangled hyperqubit-spinor field in category-$q_c$ possesses an associated symmetry which is presented by inhomogeneous hyperspin symmetry WS(1,$D_h$-1) and global scaling symmetry SG(1) in association with inhomogeneous Lorentz-type/Poincar\'e-type group symmetry PO(1,$D_h$-1) and conformal scaling symmetry SC(1) together with $\cQ_c$-spin symmetry SP($q_c$),
\be
G_S & = & SC(1)\ltimes P^{1,D_h-1}\ltimes SO(1,D_h-1)  \wtjoin SP(1,D_h-1)\rtimes W^{1,D_h-1} \times SG(1) \times SP(q_c) \nn \\
& = &  SC(1)\ltimes PO(1,D_h-1) \wtjoin WS(1,D_h-1) \times SG(1) \times SP(q_c), 
\ee 
where the symbol ``$ \wtjoin$" has been adopted to denote such an associated symmetry in which the group transformations of hyperspin group symmetry SP(1, $D_h$-1) and global scaling symmetry SG(1) in Hilbert space of basic fields must be coincidental with those of Lorentz-type group symmetry SO(1, $D_h$-1) and conformal scaling symmetry SC(1) in Minkowski spacetime of coordinates .

It is noticed that the inhomogeneous hyperspin symmetry and global scaling symmetry SG(1) as well as $\cQ_c$-spin symmetry operate on the entangled hyperqubit-spinor field and $\cQ_c$-spin scalar field in Hilbert space, while the inhomogeneous Lorentz-type/Poincar\'e-type group symmetry together with scaling symmetry SC(1) act on the coordinates in Minkowski hyper-spacetime. In general, they are distinguishable group symmetries. While when imposing symmetry invariance on the action under global transformations of hyperspin group symmetry SP(1, $D_h$-1) and scaling symmetry SG(1) for freely moving hyperqubit-spinor field in Hilbert space, such group symmetry transformations have to coincide with those of Lorentz-type group symmetry SO(1, $D_h$-1) and scaling symmetry SC(1) in Minkowski hyper-spacetime. Nevertheless, distinguishing group symmetries in Hilbert space of basic fields with those in Minkowski spacetime of coordinates becomes more essential when applying a gauge invariance principle to the action of entangled qubit-spinor field and scalar field as fundamental fields in Hilbert space. Moreover, such a distinguishment enables us to avoid automatically the longstanding obstacle on the unified field theory arising from the so-called Coleman-Mandula no-go theorem\cite{CM} about the impossibility of combining spacetime and internal symmetries in any except a trivial way.   

{\it Gauge invariance principle}: the laws of nature should be independent of the choice of local field configurations. So that all symmetries of entangled hyperqubit-spinor field and $\cQ_c$-spin scalar field as fundamental fields are postulated to be local gauge symmetries in Hilbert space with the introduction of relevant gauge fields. 

Based on such a gauge invariance principle as guiding principle, inhomogeneous hyperspin symmetry WS(1, $D_h$-1) and scaling symmetry SG(1) as well as $\cQ_c$-spin symmetry SP($q_c$) should be gauged to be local gauge symmetries which are referred to as {\it inhomogeneous hyperspin gauge symmetry} and {\it scaling gauge symmetry} as well as {\it $\cQ_c$-spin gauge symmetry}, respectively, and all their group operations act on entangled hyperqubit-spinor field and $\cQ_c$-spin scalar field in Hilbert space. Whereas inhomogeneous Lorentz-type/Poincar\'e-type group symmetry PO(1,$D_h$-1) and scaling symmetry SC(1) remain global symmetries in globally flat Minkowski hyper-spacetime, their group operations act on coordinate system. In such a consideration, the local inhomogeneous hyperspin gauge symmetry WS(1, $D_h$-1) and scaling gauge symmetry SG(1) in Hilbert space of basic fields do distinguish from the global inhomogeneous Lorentz-type/Poincar\'e-type group symmetry PO(1,$D_h$-1) and scaling symmetry in Minkowski hyper-spacetime of coordinates. 

Following along the gauge invariance principle, we should introduce gauge field $\fkA_{\fM}(x)$ relevant to inhomogeneous hyperspin gauge symmetry WS(1, $D_h$-1) and $\cQ_c$-spin gauge symmetry SP($q_c$) so as to preserve gauge invariance of the action. Concretely, it is realized by replacing the usual derivative operator of coordinates in globally flat Minkowski hyper-spacetime with the following {\it covariant derivative operator}: 
\be
& &  i\p_{\fM} \to i\fkD_{\fM} \equiv  i\p_{\fM} + \fkA_{\fM}(x) , \nn \\
& & \fkA_{\fM}(x) \equiv  \cA_{\fM}(x) +  \ckcA_{\fM}(x) + \tcA_{\fM}(x), \nn \\
& & \bvcA_{\fM}(x) \equiv \cA_{\fM}(x) +  \ckcA_{\fM}(x) \equiv \cA_{\fM}^{\; \fB\fC}(x)\, \frac{1}{2}\varSigma_{\fB\fC}  + \ckcA_{\fM}^{\fB}(x)\frac{1}{2}\varSigma_{- \fB} , \nn \\
& & \hcA_{\fM}(x) \equiv \cA_{\fM}(x) + \tcA_{\fM}(x) \equiv \cA_{\fM}^{\; \fB\fC}(x)\, \frac{1}{2}\varSigma_{\fB\fC} + \tcA_{\fM}^{\; pq}(x)\, \frac{1}{2}\tvSi_{pq} , 
\ee
with $\fM, \fB, \fC = 0,1,2,3,5,\ldots, D_h$ and $p,q=1,\cdots, q_c$. Where $\bvcA_{\fM}(x)$ is {\it inhomogeneous hyperspin gauge field} with $\cA_{\fM}^{\; \fB\fC}(x)$ denoting {\it hyperspin gauge field} of hyperspin gauge symmetry SP(1, $D_h$-1) and $\ckcA_{\fM}^{\fB}(x)$ representing $\cW_e$-spin gauge field relevant to translation-like $\cW_e$-spin Abelian gauge symmetry $W^{1,D_h-1}$. $\tcA_{\fM}^{\; pq}(x)$ is {\it $\cQ_c$-spin gauge field} of $\cQ_c$-spin gauge symmetry SP($q_c$). $\varSigma_{\fB\fC}$, $\varSigma_{- \fB}$ and $\tvSi_{pq}$ are relevant gauge group generators given by the commutators of $\vGa$-matrices\cite{FHUFT-I},
\be
& & \vSi_{\fB\fC} = \frac{i}{4} [ \vGa_{\fB}, \vGa_{\fC} ], \quad \vSi_{- \fB} = \frac{1}{2} \vGa_{\fB} \vGa_{-}, \quad \vGa_{-} = \frac{1}{2} ( 1 - \hat{\gamma}_{D_h + 1} ), \nn \\
& & \varSigma_{D_h+p\, D_h+q} =  \frac{i}{4} [ \vGa_{D_h + p}, \vGa_{D_h + q} ] = \frac{i}{4} [ \tvGa_{p}, \tvGa_{q} ] \equiv \tvSi_{pq}. 
\ee
Where the hyperspin gauge field $\cA_{\fM}(x)$ belongs to adjoint representation of hyperspin gauge group symmetry SP(1, $D_h$-1). 

To ensure both local inhomogeneous hyperspin gauge symmetry WS(1, $D_h$-1) and global inhomogeneous Lorentz-type/Poincar\'e-type group symmetry PO(1,$D_h$-1), we should introduce a {\it bicovaraint vector field} $\chih_{\fA}^{\;\, \fM}(x)$ in addition to the gauge field $\fkA_{\fM}(x)$. So that the kinematic term of the action is extended to be as follows:
\be
\vGa^{\fA} \delta_{\fA}^{\;\; \fM} i\p_{\fM} \to \vGa^{\fA}\chih_{\fA}^{\;\, \fM}(x) i\fkD_{\fM} , 
\ee
with $\fA, \fM = 0,1,2,3,5,\ldots, D_h$. Where the constant bicovariant vector $\delta_{\fA}^{\;\; \fM}$ is replaced by the bicovaraint vector field $\chih_{\fA}^{\;\, \fM}(x)$.  

With the above considerations, the action for a freely moving entangled hyperqubit-spinor field with global inhomogeneous hyperspin symmetry and scaling symmetry in associating with global inhomogeneous Lorentz-type/Poincar\'e-type group symmetry and scaling symmetry should be generalized to be gauge invariant action, which possesses local inhomogeneous hyperspin gauge symmetry WS(1, $D_h$-1) and scaling gauge symmetry SG(1) jointly together with global inhomogeneous Lorentz-type/Poincar\'e-type group symmetry PO(1,$D_h$-1) and scaling symmetry SC(1). 

Let us begin with the following gauge invariant action:
\be  \label{actionfQEGS}
\cS_{\fQE}^{(q_c)} & \equiv & \int [d^{D_h^{(q_c,k)}}x] \,  \{ \bar{\Psi}_{\fQE}^{(q_c,k)}(x)  \vSi_{-}^{\fA}\chih_{\fA}^{\;\, \fM}(x) i\fkD_{\fM}  \Psi_{\fQE}^{(q_c,k)}(x) \nn \\
& + & \lamq\phi_{p}(x)  \bar{\Psi}_{\fQE}^{(q_c,k)}(x) \vSi_{-}^{D_h+p} \Psi_{\fQE}^{(q_c,k)}(x) \}, 
\ee
with $\fA, \fM= 0,1,2,3,5, \cdots, D_h^{(q_c,k)}$ and $p=0,\cdots, q_c$. As discussed in the previous paper as the part I\cite{FHUFT-I}, $q_c$ represents $\cQ_c$-spin charge $\CQc=q_c$ with $q_c=0,1,2,3$ as four categoric charges, and $\lamq$ is a coupling constant with $\lambda_0=\lambda_3=0$ since there exist no such coupling terms for $q_c=0,3$. $D_h^{(q_c,k)}$ denotes the categoric dimension of hyper-spacetime given by the relation $D_h^{(q_c,k)} = D_{q_c} + 8k$ with $D_{q_c}= 2,3,4,6$ as four basic dimensions and $k=0,1,\cdots$ as periodic number. $\Psi_{\fQE}^{(q_c,k)}(x)$ denotes the categoric entangled hyperqubit-spinor field in category-$q_c$ and $k$-th period, which has dimensions $2^{Q_N^{(q_c,k)}+1}$ in qubit-spinor representation of Hilbert space with $Q_N^{(q_c,k)}$ labeling the categoric qubit number given by the relation $Q_N^{(q_c,k)}=q_c + 4k$. 

The matrices $\vSi_{-}^{\fA}$ and $\vSi_{-}^{D_h+p} $ are defined via the anticommuting $\vGa$-matrices $\vGa^{\hfA}= (\vGa^{\fA}, \vGa^{D_h + p}) $ ($\hfA= 0,1,2,3,5,\cdots, D_h+q_c$) as follows:
\be
& & \vSi_{-}^{\fA} = \frac{1}{2} \vGa^{\fA}\vGa_{-}, \quad \vGa_{-} = \frac{1}{2} ( 1 - \hat{\gamma}_{D_h + 1} ), 
 \nn \\
& & \vSi_{-}^{D_h+p} = \frac{1}{2} \vGa^{D_h + p}\vGa_{-} =  \frac{1}{2} \tvGa^{p}\vGa_{-} \equiv \tvSi_{-}^p. 
\ee

The action in Eq.(\ref{actionfQEGS}) is invariant under global inhomogeneous Lorentz-type/Poincar\'e-type group symmetry PO(1,$D_h$-1) with the following Lorentz-type group transformation:
\be
& & x^{\fM} \to x^{'\fM} = L^{\fM}_{\; \; \; \fN}\; x^{\fN}, \quad L^{\fM}_{\; \;\; \fN} \in \mbox{SO}(1, D_h\mbox{-}1), \nn \\
& & \chih_{\fA}^{\;\, \fM}(x) \to \chih_{\fA}^{'\;\, \fM}(x') = L^{\fM}_{\; \; \; \fN} \chih_{\fA}^{\;\, \fN} (x), \nn \\
& & \fkA_{\fM}(x) \to \fkA'_{\fM}(x') = L_{\fM}^{\; \; \; \fN}  \fkA_{\fN}(x) , \nn \\
& &  \Psi_{\fQE}^{(q_c,k)}(x) \to  \Psi_{\fQE}^{' (q_c,k)}(x') =  \Psi_{\fQE}^{(q_c,k)}(x) , 
\ee
and translation group transformation:
\begin{eqnarray}
 x^{\fM} \to x^{'\fM} = x^{\fM} + a^{\fM}\, , \qquad  a^{\fM} \in P^{1, D_h-1}\, , 
 \end{eqnarray}
with $a^{\fM}$ a constant vector in globally flat Minkowski hyper-spacetime.

The action in Eq.(\ref{actionfQEGS}) possesses the local inhomogeneous hyperspin gauge symmetry WS(1, $D_h$-1) which is separated to the global inhomogeneous Lorentz-type group symmetry. The local hyperspin gauge transformation is made as follows:
\be
& & \bvcA'_{\fM}(x) =   S(\Lambda) i \partial_{\fM} S^{-1}(\Lambda) +  S(\Lambda) \bvcA_{\fM}(x) S^{-1}(\Lambda)
, \nn \\
& & \Psi_{\fQE}^{' (q_c,k)}(x) =  S(\Lambda) \Psi_{\fQE}^{(q_c,k)}(x) ,  \quad S(\Lambda) = e^{i\varpi_{\fA\fB}(x) \varSigma^{\fA\fB}/2} \in \mbox{SP}(1, D_h\mbox{-}1), 
\ee
with the following explicit transformation property:
\be
& & S^{-1}(\Lambda) \vSi_{-}^{\fA} S(\Lambda) = \Lambda^{\fA}_{\;\;\; \fB}(x)\; \vSi_{-}^{\fB}, \quad \chih_{\fA}^{\;\, \fM}(x) \to \chih_{\fA}^{'\;\, \fM}(x) = \Lambda_{\fA}^{\; \; \fB}(x) \chih_{\fB}^{\;\, \fM} (x), \nn \\
& & \cA_{\fM}^{' \fA\fB}(x) =  \frac{1}{2} \left(\Lambda^{\fA}_{\;\;\; \fC}\p_{\fM}  \Lambda^{\fC\fB} -  \Lambda^{\fB}_{\;\;\; \fC} \p_{\fM}  \Lambda^{\fC\fA} \right) + \Lambda^{\fA}_{\;\;\; \fC} \Lambda^{\fB}_{\;\;\; \fD} \cA_{\fM}^{ \fC\fD}(x) , \nn \\
& &  \ckcA_{\fM}^{' \fA}(x) =  \Lambda^{\fA}_{\;\;\; \fB}(x) \ckcA_{\fM}^{\fB}(x), \quad \Lambda^{\fA}_{\;\;\; \fB}(x)  \in \mbox{SP}(1, D_h\mbox{-}1) .
\ee
The local $\cW_e$-spin gauge transformation is given as follows:
\be
& & \bvcA'_{\fM}(x) =   \ckS(x) i \partial_{\fM} \ckS^{-1}(x) +  \ckS(x)\bvcA_{\fM}(x) \ckS^{-1}(x), \nn \\
& &  \Psi_{\fQE}^{' (q_c,k)}(x) =  \ckS(x) \Psi_{\fQE}^{(q_c,k)}(x) , \quad \ckS(x) = e^{i \varpi_{\fA}(x) \vSi_{-}^{\fA}/2 }  \in W^{1,D_h-1}, 
\ee
with the following explicit form:
\be \label{TESGS}
& & \ckS(x) = 1 + i \varpi_{\fA}(x) \vSi_{-}^{\fA}/2, \quad \cA_{\fM}^{' \fA}(x) \equiv   \cA_{\fM}^{\fA}(x) , \nn \\
& & \Psi_{\fQE}^{' (q_c,k)}(x) = \Psi_{\fQE}^{(q_c,k)}(x) + \check{\Psi}_{\fQE}^{(q_c,k)}(x), \quad \check{\Psi}_{\fQE}^{(q_c,k)}(x) \equiv i \frac{1}{2}\varpi_{\fA}(x) \vSi_{-}^{\fA} \Psi_{\fQE}^{(q_c,k)}(x),  \nn \\
& & \ckcA_{\fM}^{' \fA}(x) =   \ckcA_{\fM}^{\fA}(x)  + \cD_{\fM}  \varpi^{\fA}(x), \nn \\
& & \cD_{\fM}  \varpi^{\fA}(x) \equiv \p_{\fM}  \varpi^{\fA}(x) + \cA_{\fM \fB}^{\; \fA}(x)  \varpi^{\fB}(x) .
\ee   
The intrinsic $\cQ_c$-spin gauge invariance of the action in Eq.(\ref{actionfQEGS}) is easily checked under the following gauge transformation: 
\be
& &\tcA'_{\fM}(x) =  \tilde{S}(\Lambda)i \p_{\fM} \tilde{S}^{-1}(\Lambda) +  \tilde{S}(\Lambda) \tcA_{\fM}(x) \tilde{S}^{-1}(\Lambda) ,\nn \\
& & \Psi_{\fQE}^{' (q_c,k)}(x) =  \tilde{S}(\Lambda) \Psi_{\fQE}^{(q_c,k)}(x) ,  \quad  \tilde{S}(\Lambda) = e^{i\varpi_{pq}(x) \tvSi^{pq}/2}  \in SP(q_c), 
\ee
with the explicit forms given by, 
\be
& & \tilde{S}^{-1}(\Lambda) \tvSi_{-}^{p} \tilde{S}(\Lambda) = \Lambda^{p}_{\;\;\; q}(x)\; \tvSi_{-}^{q} , \quad  \Lambda^{p}_{\; \;  q}(x)  \in SP(q_c), \nn \\
& & \cA_{\fM}^{' pq}(x) =  \frac{1}{2} \left(\Lambda^{p}_{\;\;p'} \p_{\fM}  \Lambda^{p' q} -  \Lambda^{q}_{\;\;p'} \p_{\fM}  \Lambda^{p' p} \right) +  \Lambda^{p}_{\;\; p'}  \Lambda^{q}_{\;\;q'} \cA_{\fM}^{ p' q'}(x), \nn \\
& &  \phi_{p}(x) \to \phi'_{p}(x) = \Lambda_{p}^{\; \;  q}(x) \phi_{q}(x) . 
\ee
In obtaining the above transformation properties, we have used the following identities:
\be \label{TP}
& & S^{-1}(\Lambda) \vGa^{\fA} S(\Lambda)= \Lambda^{\fA}_{\;\; \fB}(x) \vGa^{\fB}, \quad \vGa^{\fC} \vSi_{\fA\fB}  \vGa_{\fC} = (D_h-4) \vSi_{\fA\fB}, \nn \\
& & S(\Lambda) i\p_{\fM} S^{-1}(\Lambda) =  \frac{1}{4} \left(\Lambda^{\fA}_{\;\;\; \fC}\p_{\fM}  \Lambda^{\fC\fB} -  \Lambda^{\fB}_{\;\;\; \fC} \p_{\fM}  \Lambda^{\fC\fA} \right) \vSi_{\fA\fB} , \nn \\
& & \tS^{-1}(\Lambda) \tvGa^{p} \tS(\Lambda)= \Lambda^{p}_{\;\; q}(x) \tvGa^{q}, \quad \tvGa^{p'} \tvSi_{p q}  \vGa_{p'} = (D_h-4) \tvSi_{p q}, \nn \\
& & \tS(\Lambda) i\p_{\fM} \tS^{-1}(\Lambda) =  \frac{1}{4} \left(\Lambda^{p}_{\;\;\; p'}\p_{\fM}  \Lambda^{p' q} -  \Lambda^{q}_{\;\; p'} \p_{\fM}  \Lambda^{p' p} \right) \tvSi_{p q} ,
\ee
where the matrices $S(\Lambda)$ and $\tilde{S}(\Lambda)$ as well as $\Lambda^{\fA}_{\;\;\; \fB}(x)$ and $\Lambda^{p}_{\;\;\; q}(x)$ correspond to local group elements in the qubit-spinor representations as well as adjoin representations of gauge groups SP(1,$D_h$-1) and SP($q_c$), which should be manifest as the categoric entangled hyperqubit-spinor field is the presented in qubit-spinor representation and both hyperspin gauge field and $\cQ_c$-spin gauge field belong to the adjoint representations of gauge symmetry groups SP(1, $D_h$-1) and SP($q_c$), respectively. The bicovariant vector field and scalar field are in the vector representations of gauge symmetry groups SP(1, $D_h$-1) and SP($q_c$), respectively.

From the general covariant derivative $\fkD_{\fM}$, the field strength of gauge field $\fkA_{\fM}(x)$ is defined as follows: 
\be \label{GFS}
& & \fkF_{\fM\fN} \equiv i [\fkD_{\fM}, \fkD_{\fN} ] = \cF_{\fM\fN} + \ckcF_{\fM\fN} + \tcF_{\fM\fN}, \nn \\
& & \cF_{\fM\fN}(x) \equiv  \cF_{\fM\fN}^{\fA\fB}(x)\frac{1}{2}\vSi_{\fA\fB} = \p_{\fM} \cA_{\fN}(x) - \p_{\fN} \cA_{\fM}(x) - i [ \cA_{\fM}(x) ,  \cA_{\fN}(x) ], \nn \\
& & \ckcF_{\fM\fN}(x) \equiv \ckcF_{\fM\fN}^{\fA}(x)\frac{1}{2}\vSi_{- \fA}=\p_{\fM} \ckcA_{\fN}(x) - \p_{\fN} \ckcA_{\fM}(x) - i (\cA_{\fM}(x)\ckcA_{\fN}(x) - \cA_{\fN}(x)\ckcA_{\fM}(x) ), \nn \\
& & \tcF_{\fM\fN}(x) \equiv \tcF_{\fM\fN}^{p q}(x)\frac{1}{2}\tvSi_{p q} = \p_{\fM} \tcA_{\fN}(x) - \p_{\fN} \tcA_{\fM}(x) - i [ \tcA_{\fM}(x) ,  \tcA_{\fN}(x) ],
\ee 

In general, the action in Eq.(\ref{actionfQEGS}) possesses the following {\it joint symmetry}, 
\be \label{JGLS}
G_S & \equiv &  PO(1, D_h-1)\Join WS(1, D_h-1) \times SP(q_c) ,
\ee
where PO(1,$D_h$-1) remains global inhomogeneous Lorentz-type/Poincar\'e-type group symmetry,  
\be
 PO(1, D_h-1) \equiv P^{1, D_h-1} \ltimes SO(1, D_h-1) , 
 \ee
which characterizes symmetry property of globally flat Minkowski hyper-spacetime. 

We would like to address that the global inhomogeneous Lorentz-type/Poincar\'e-type group PO(1, $D_h$-1) and local inhomogeneous hyperspin gauge group WS(1, $D_h$-1) do not form a direct product group symmetry as the former is a global symmetry and the later is a local symmetry, they are presented as a {\it joint symmetry} denoted by the symbol ``$\Join$.


\subsection{ Scaling gauge symmetry based on gauge invariance principle and the gauge covariant equation of motion of entangled hyperqubit-spinor field as gravitational relativistic quantum theory } 

It can be verified that the action in Eq.(\ref{actionfQEGS}) is invariant under the following global scaling transformation:
\be
& & x^{\fM} \to x^{'\fM} = \lambda^{-1} \1 x^{\fM}, \nn \\
& & \fkA_{\fM}(x) \to \fkA'_{\fM}(x') =  \lambda\1 \fkA_{\fM}(x), \nn \\
& & \Psi_{\fQE}^{(q_c,k)}(x) \to \Psi_{\fQE}^{' (q_c,k)}(x') = \lambda^{\frac{D_h-1}{2}}\1 \Psi_{\fQE}^{(q_c,k)}(x) , 
\ee
with $\lambda$ a constant scaling factor. 

To distinguish scaling symmetry of coordinate system in Minkowski hyper-spacetime from that of entangled hyperqubit-spinor field in Hilbert space, we should extend the global scaling symmetry of entangled hyperqubit-spinor field to be local scaling gauge symmetry based on the gauge invariance principle, which has the following scaling gauge transformation:
\be
& & \Psi_{\fQE}^{(q_c,k)}(x) \to \Psi_{\fQE}^{' (q_c,k)}(x) = \xi^{\frac{D_h-1}{2}}(x) \Psi_{\fQE}^{(q_c,k)}(x), \nn \\
& & \hat{\chi}_{\fA}^{\;\; \fM}(x) \to  \hat{\chi}_{\fA}^{'\; \fM}(x) = \xi(x) \hat{\chi}_{\fA}^{\;\; \fM}(x), \nn \\
& & \phi_{p}(x) \to  \phi_{p}(x) = \xi(x) \phi_{p}(x) ,
\ee
where $\xi(x)$ is a functional scaling factor. Following along the gauge invariance principle, we should introduce a {\it scaling gauge field} $\mW_{\fM}(x)$ with the following gauge transformation:
\be
\mW_{\fM}(x) \to \mW'_{\fM}(x) = \mW_{\fM}(x) - \p_{\fM}\ln \xi(x) ,
\ee
which was initiated by Weyl\cite{Weyl} to characterize an electromagnetic field though it was turned out to be incorrect.

To preserve the scaling gauge symmetry, the action in Eq.(\ref{actionfQEGS}) is extended in formal to be as follows:
\be  \label{actionfQESG}
\cS_{\fQE}^{(q_c)} & \equiv & \int [d^{D_h^{(q_c,k)}}x] \, \chi(x)\,  \{ \bar{\Psi}_{\fQE}^{(q_c,k)}(x)  \vSi_{-}^{\fA}\chih_{\fA}^{\;\, \fM}(x) i\fkD_{\fM}  \Psi_{\fQE}^{(q_c,k)}(x) \nn \\
& + & \lamq  \phi_p(x)\bar{\Psi}_{\fQE}^{(q_c,k)}(x) \tvSi_{-}^{p} \Psi_{\fQE}^{(q_c,k)}(x) \},
\ee
with the generalized covariant derivative,
\be
i\fkD_{\fM} \equiv  i\p_{\fM} + \cA_{\fM}(x) +  \ckcA_{\fM}(x) + \tcA_{\fM}(x) + \frac{D_h-1}{2}\mW_{\fM}(x) . 
\ee
It is noted that the multiplier factor $\chi(x)$ as the inverse of the determinant of bicovaraint vector field $\chih_{\fA}^{\;\, \fM}(x)$ has been introduced by regarding $\chih_{\fA}^{\;\, \fM}(x)$ as a matrix field, i.e.:
\be
& &  \chi(x) \equiv 1/\chih(x), \quad \chih(x) = \det \chih_{\fA}^{\; \; \fM}(x) \neq 0 ,
\ee
which is multiplied to the action so as to ensure the scaling gauge invariance with the following scaling gauge transformation: 
\be
& &  \chi(x) \to \chi'(x) = \xi^{-D_h}(x) \chi(x) . 
\ee

From the matrix property,  $\chi(x)$ can be expressed as the determinant of {\it dual bicovaraint vector field} $\chi_{\fM}^{\;\; \fA} (x)$ which is viewed as an inverse matrix field of the matrix field $\chih_{\fA}^{\;\, \fM}(x)$. Such a dual bicovaraint vector field is defined as follows:
\be
& & \chi_{\fM}^{\;\; \fA} (x)\, \chih^{\;\; \fM}_{\fB}(x)   =  \chi_{\fM}^{\;\; \fA} (x) \chih_{\fB\, \fN}(x) \eta^{\fM\fN} = \eta^{\;\; \fA}_{\fB} , \nn \\
& &  \chi_{\fM}^{\;\; \fA}(x) \chih^{\;\;\fN}_{\fA}(x) = \chi_{\fM\, \fA} (x) \chih_{\fB}^{\;\; \fN}(x)  \eta^{\fA\fB} = \eta_{\fM}^{\;\;\fN} , \nn \\
& &  \chi(x) = \det \chi_{\fM}^{\;\; \fA}(x) = \chih^{-1}(x), 
\ee
where the matrix fields $\chi_{\fM}^{\;\; \fA} (x)$ and $\chih^{\;\; \fM}_{\fB}(x)$ satisfy orthonormality conditions.   

From the self-conjugated chiral property of entangled hyperqubit-spinor field in category-$q_c$ and the transpose property of $\vGa$-matrices under complex charge conjugation, i.e.:
\be
& & \Psi_{\fQE}^{(q_c,k)\, c}(x)= C_{\fQE}^{(q_c,k)} \bar{\Psi}_{\fQE}^{(q_c,k)\, T}(x) = \Psi_{\fQE}^{(q_c,k)}(x) , \nn \\
& & C_{\fQE}^{(q_c,k)}\vGa^{\fA}\left(C_{\fQE}^{(q_c,k)}\right)^{-1} = - \left(\vGa^{\fA}\right)^T, \nn \\
& & C_{\fQE}^{(q_c,k)}\vGa^{D_h + p}\left(C_{\fQE}^{(q_c,k)}\right)^{-1} = \left(\vGa^{D_h + p}\right)^T , 
\ee 
together with the chirality property of entangled hyperqubit-spinor field, it can be verified that the gauge interactions of entangled hyperqubit-spinor field with both $\cW_e$-spin gauge field $\ckcA_{\fM}(x)$ and scaling gauge field $\mW_{\fM}(x)$ should be absent in the hermitian action presented in Eq.(\ref{actionfQEGS}).

The gauge invariant hermitian action in Eq.(\ref{actionfQEGS}) can actually be rewritten as follows:
\be  \label{actionfQESGI}
\cS_{\fQE}^{(q_c)} & \equiv & \int [d^{D_h^{(q_c,k)}}x] \, \chi(x)\,  \{\, \bar{\Psi}_{\fQE}^{(q_c,k)}(x)  \vSi_{-}^{\fA}\chih_{\fA}^{\;\, \fM}(x) i\hcD_{\fM}  \Psi_{\fQE}^{(q_c,k)}(x) \nn \\
& + & \lamq  \phi_p(x)\bar{\Psi}_{\fQE}^{(q_c,k)}(x) \tvSi_{-}^{p} \Psi_{\fQE}^{(q_c,k)}(x) \} .
\ee
with the definitions,
\be
& & i\hcD_{\fM} \equiv   i\p_{\fM} + \hcA_{\fM}(x), \quad  \hcA_{\fM}(x) \equiv \cA_{\fM}(x) + \tcA_{\fM}(x) . 
\ee

Such an action possesses the following {\it joint symmetry}:
\be \label{GLSS}
G_S= SC(1)\ltimes PO(1,D_h-1)\Join WS(1,D_h-1)\rtimes SG(1)\times SP(q_c),
\ee
where SC(1) and SG(1) denote {\it global scaling symmetry} and {\it local scaling gauge symmetry}, respectively.

From the gauge invariant action in Eq.(\ref{actionfQESGI}), we can derive the following equation of motion for categoric entangled hyperqubit-spinor field:
\be \label{EMfQE}
& & \vSi_{-}^{\fA} \chih_{\fA}^{\;\; \fM}(x) \left(i\hcD_{\fM} - i\mV_{\fM}(x) \right) \Psi_{\fQE}^{(q_c,k)}(x) 
 -\lamq \phi_p(x) \tvSi_{-}^{p} \Psi_{\fQE}^{(q_c,k)}(x)  =   0 , 
\ee
where $\mV_{\fM}(x)$ represents an {\it induced gauge field} arising from the bi-covariant vector field and inhomogeneous hyperspin gauge field via a hidden gauge formalism, which has the following explicit form:
\be
\mV_{\fM}(x) & \equiv &  \frac{1}{2}\left(\chi(x) \chih_{\fB}^{\;\; \fN}(x)\right)\cD_{\fN}\left(\chih(x)\chi_{\fM}^{\;\; \fB}(x) \right),
\ee
where $\cD_{\fN}\left(\chih(x)\chi_{\fM}^{\;\; \fB}(x)\right)$ defines the covariant derivative of dual bicovaraint vector field $\chi_{\fM}^{\;\; \fB} (x)$,
\be \label{CD}
& & \cD_{\fN}\left(\chih(x)\chi_{\fM}^{\;\; \fB}(x)\right) \equiv \p_{\fN} \left(\chih(x)\chi_{\fM}^{\;\; \fB}(x)\right) + \cA_{\fN\, \fC}^{\fB}(x) \left(\chih(x)\chi_{\fM}^{\;\; \fC}(x)\right) . 
\ee
It can be checked that $\mV_{\fM}(x)$ preserves local scaling gauge invariance in the above equation of motion with the following transformation property:
\be
\mV_{\fM}(x) & \to & \mV'_{\fM} (x) = \mV_{\fM}(x) + \frac{1}{2}(D_h-1)\p_{\fM}\xi(x) .
\ee

It can be verified that the gauge covariant equation of motion Eq.(\ref{EMfQE}) is local scaling gauge invariant. Let us now derive the quadratic form for the equation of motion Eq.(\ref{EMfQE}), which is found to have the following form:
\be \label{EMfQE2}
& & \chih^{\fM\fN} (\hat{\nabla}_{\fM} - \mV_{\fM} ) ( \hcD_{\fN} - \mV_{\fN} ) \Psi_{\fQE}^{(q_c,k)}(x) =   \varSigma^{\fA\fB}\chih_{\fA}^{\;\; \fM} \chih_{\fB}^{\;\; \fN} \nn \\
& & \qquad \quad \cdot  \left( \cF_{\fM\fN}+ \tcF_{\fM\fN} - {\cal G}_{\fM\fN}^{\fC} \chih_{\fC}^{\;\; \fP} i ( \hcD _{\fP} - \mV_{\fP} ) - i \mV_{\fM\fN}\, \right) \Psi_{\fQE}^{(q_c,k)}(x)  \nn \\
& & \qquad \quad - \left( \lamq^2 \phi^2_p  +  \lamq \vGa^{\fA} \tvGa^{p} \chih_{\fA}^{\;\; \fM} (i\tcD_{\fM}\phi_p) \,  \right) \Psi_{\fQE}^{(q_c,k)}(x), 
 \ee
where we have introduced the following definition:
 \be \label{STF}
 & & \chih^{\fM\fN}(x) = \chih_{\fA}^{\;\; \fM}(x) \chih_{\fB}^{\;\; \fN}(x) \eta^{\fA\fB} ,
 \ee
 for the symmetric tensor field, and
 \be \label{VFCD}
 & & \hat{\nabla}_{\fM} ( \hcD_{\fN}  - \mV_{\fN} ) \equiv \hcD_{\fM} ( \hcD_{\fN} - \mV_{\fN} ) - {\mathit A}_{\fM\fN}^{\fP} ( \hcD_{\fP} - \mV_{\fP} ) ,\nn \\
 & & {\mathit A}_{\fM\fN}^{\fP}  \equiv   \chih_{\fA}^{\;\; \fP}  \cD_{\fM}\chi_{\fN}^{\;\;\fA} =  \chih_{\fA}^{\;\; \fP}(\p_{\fM} \chi_{\fN}^{\;\; \fA} + \cA_{\fM\, \fB}^{\fA}  \chi_{\fN}^{\;\;\fB}\, ) , \nn \\
 & & \tcD_{\fM}\phi_p \equiv \p_{\fM} \phi_p - \tcA_{\fM\, p} ^{\; q} \phi_q, 
 \ee
 for the covariant derivatives $\hat{\nabla}_{\fM}$ and $\tcD_{\fM}$. On the right-hand side of Eq.(\ref{EMfQE2}), $\cF_{\fM\fN}(x)$ and $\tcF_{\fM\fN}(x)$ define the field strengths of {\it hyperspin gauge field} $\cA_{\fM}(x)$ and $\cQ_c$-spin gauge field $\tcA_{\fM}(x)$, respectively, with the following explicit forms:
\be
& & \cF_{\fM\fN}(x) \equiv  {\cal F}_{\fM\fN}^{\fA\fB}(x)\, \frac{1}{2}\varSigma_{\fA\fB}  = \partial_{\fM} {\cal A}_{\fN}(x) - \partial_{\fN} {\cal A}_{\fM}(x) - i [{\cal A}_{\fM}(x) ,  {\cal A}_{\fN}(x) ], \nn \\
& & \tcF_{\fM \fN}(x) \equiv  {\cal F}_{\fM\fN}^{p q}(x)\, \frac{1}{2}\tvSi_{p q}  = \p_{\fM} \tcA_{\fN}(x) - \p_{\fN} \tcA_{\fM}(x) - i [\tcA_{\fM}(x) ,  \tcA_{\fN}(x) ], 
\ee
and $\mV_{\fM\fN}$ is the {\it induced gauge field strength} defined as follows:
 \be
 \mV_{\fM\fN} \equiv  \p_{\fM}\mV_{\fN} -  \p_{\fN}\mV_{\fM} .
 \ee

The tensor field ${\cal G}_{\fM\fN}^{\fA}$ is defined as follows:
 \be \label{HGFFS}
 \cG_{\fM\fN}^{\fA}  & \equiv & \cD_{\fM} \chi_{\fN}^{\;\; \fA} - \cD_{\fN} \chi_{\fM}^{\;\; \fA} \nn \\
 & = &  \p_{\fM} \chi_{\fN}^{\;\; \fA}(x) - \p_{\fN} \chi_{\fM}^{\;\; \fA}(x) + \cA_{\fM\, \fB}^{\fA}(x)  \chi_{\fN}^{\;\;\fB}(x)  - \cA_{\fN\, \fB}^{\fA}(x)  \chi_{\fM}^{\;\;\fB}(x),
 \ee
which provides the gauge covariant field strength of dual bicovaraint vector field $\chi_{\fM}^{\; \fA}(x)$ with $\cD_{\fM} \chi_{\fN}^{\;\; \fA}$ representing the covariant derivative of $\chi_{\fM}^{\; \fA}(x)$. It is noticed that the dual bicovaraint vector field $\chi_{\fM}^{\; \fA}(x)$ behaves as a gauge-type vector field and its inverse as bicovaraint vector field $\chih_{\fA}^{\;\; \fM}(x)$ couples to the motion of entangled hyperqubit-spinor field, so that $\chi_{\fM}^{\; \fA}(x)$ is expected to characterize gravitational interaction in hyper-spacetime. For convenience, $\chi_{\fM}^{\;\; \fA}$ is referred to as {\it hyper-gravigauge field} in hyper-spacetime, which will be demonstrated below for its genesis as gauge field. 

The equations of motion presented in Eqs.(\ref{EMfQE}) and (\ref{EMfQE2}) for categoric entangled hyperqubit-spinor field $\Psi_{\fQE}^{(q_c,k)}(x)$ possess global Poincar\'e-type group symmetry and scaling symmetry jointly with local inhomogeneous hyperspin symmetry and $\cQ_c$-spin gauge symmetry as well as local scaling gauge symmetry. The gauge symmetries are described correspondingly by the hyperspin gauge field $\cA_{\fM}^{\fA\fB}(x)$ and $\cQ_c$-spin gauge field $\tcA_{\fM}^{\fA\fB}(x)$ as well as hyper-gravigauge field $\chi_{\fM}^{\; \fA}(x)$ and scaling gauge field $\mW_{\fM}$. Such equations characterize the dynamics of entangled hyperqubit-spinor field, which brings on {\it gravitational relativistic quantum theory} in hyper-spacetime and is regarded as a generalization to Dirac relativistic quantum theory\cite{DE}. The dynamics of entangled hyperqubit-spinor field is generally governed by the hyperspin gauge field strength $\cF_{\fM\fN}^{\fA\fB}$ and hyper-gravigauge field strength ${\cal G}_{\fM\fN}^{\fA}$ as well as $\cQ_c$-spin gauge field strength $\tcF_{\fM\fN}^{\fA\fB}$. 

It is noticed that the induced gauge field $\mV_{\fM}(x)$ and its field strength $\mV_{\fM\fN}$ distinguish from other gauge fields and field strengths due to the presence of an associated imaginary factor $``i"$, which appears naturally so as to keep the local scaling gauge invariance in the equation of motion of entangled hyperqubit-spinor field. For convenience of mention, we may refer to $\mV_{\fM}(x)$ as {\it graviscaling induced gauge field} and $\mV_{\fM\fN}$ as {\it graviscaling induced gauge field strength}. It is clear that the appearance of graviscaling induced gauge field and its field strength will lead distribution amplitudes of entangled hyperqubit-spinor field to be boosted by the induced scaling gauge factor. 

Therefore, the gauge invariance principle as guiding principle brings all symmetries of basic fields in Hilbert space to be local gauge symmetries with the introduction of gauge fields, which enables us to reveal the dynamics of entangled hyperqubit-spinor field as fundamental building block.


\section{ Hyper-fiber bundle structure of biframe hyper-spacetime with gravitational origin of gauge symmetry and $\cW_e$-spin invariant-gauge field as genesis of hyper-gravigauge field with graviscalar and gravivector fields }

The gauge invariance of the action of categoric entangled hyperqubit-spinor field brings on the introduction of bicovariant vector field $\chih_{\fA}^{\;\; \fM}(x)$. The dual bicovariant vector field $\chi_{\fM}^{\; \fA}(x)$ leads to a gauge-type field strength, which makes $\chi_{\fM}^{\; \fA}(x)$ to behave as a hyper-gravigauge field in hyper-spacetime. We are going to further demonstrate that such bicovariant vector fields form locally flat gravigauge hyper-spacetime and bring about the concept of biframe hyper-spacetime. Mathematically, such a biframe hyper-spacetime sets up a hyper-fiber bundle structure with Minkowski hyper-spacetime as base spacetime. Such a locally flat gravigauge hyper-spacetime is shown to be characterized by non-commutative geometry with Lie algebra structure factor determined by the field strength of hyper-gravigauge field. We will show that it is the hyper-gravigauge field $\chi_{\fM}^{\; \fA}(x)$ that brings on the gravitational origin of gauge symmetry in hyper-spacetime. Furthermore, we will verify that the hyper-gravigauge field $\chi_{\fM}^{\; \fA}(x)$ is identified to be the $\cW_e$-spin invariant-gauge field which is introduced via the translation-like $\cW_e$-spin Abelian-type gauge symmetry as subgroup gauge symmetry of inhomogeneous hyperspin gauge symmetry. Therefore, we get better understanding on the nature of gravity that the gravitational interaction is truly a gauge interaction and also better comprehension on the nature of spacetime that the observable spacetime is actually a biframe spacetime.


\subsection{ Locally flat gravigauge hyper-spacetime with emergence of non-commutative geometry and biframe hyper-spacetime with hyper-fiber bundle structure }

The Poincar\'e-type group symmetry PO(1, $D_h$-1) = P$^{1, D_h\mbox{-}1}$ $\ltimes$ SO(1, $D_h$-1) characterizes globally flat Minkowski hyper-spacetime $M_h$ which is known to be an {\it affine spacetime}. The ordinary  derivative vector operator $\partial_{\fM} \equiv \partial/\partial x^{\fM}$ at point $x$ of $M_h$ defines a {\it tangent basis } $\{\partial_{\fM}\}\equiv \{\partial/\partial x^{\fM}\} $ for {\it tangent hyper-spacetime} $T_h$ over $M_h$. A displacement vector $dx^{\fM}$ at point $x$ of $M_h$ provides a {\it dual tangent basis} $\{dx^{\fM}\}$ for {\it dual tangent hyper-spacetime} $T_h^{\ast}$ over $M_h$. The tangent basis and dual tangent basis satisfy the following dual condition:
\be
 \langle dx^{\fM},\, \partial/\partial x^{\fN}  \rangle = \frac{\partial x^{\fM}}{\partial x^{\fN}} = \eta_{\fN}^{\; \fM} .
\ee

In Hilbert space, the introduction of bicovaraint vector field $\chih_{\fA}^{\;\, \fM}(x)$ and its dual bicovariant vector field $\chi_{\fM}^{\;\; \fA}(x)$ not only preserves inhomogeneous hyperspin gauge symmetry for characterizing the dynamics of entangled hyperqubit-spinor field, but also allows us to define {\it field vector operator} $\eth_{\fA}$ and its {\it dual field vector operator} $ \delta\zeta^{\fA}$ at point $x$ of $M_h$ with respective to ordinary coordinate derivative vector operator $\partial_{\fM}$ and displacement vector operator $dx^{\fM}$ as follows: 
\be \label{eth}
& & \eth_{\fA} \equiv  \chih_{\fA}^{\;\, \fM}(x) \partial_{\fM} , \nn \\
& & \delta\zeta^{\fA} \equiv \chi_{\fM}^{\;\; \fA} (x) dx^{\fM} ,
\ee
which forms the field basis $\{\eth_{\fA} \}$ and dual field basis $ \{\delta\zeta^{\fA}\}$ for locally flat hyper-spacetime over globally flat Minkowski hyper-spacetime $M_h$. 

The dual field bases satisfy the following dual condition:
 \be
 & & \langle \delta\zeta^{\fA},  \eth_{\fB}\rangle  \equiv \chi_{\fM}^{\; \fA}(x)  \chih_{\fB}^{\;\fN} (x)  \langle dx^{\fM} , \partial_{\fN} \rangle = \chi_{\fM}^{\;\; \fA}(x)  \chih_{\fB}^{\;\; \fN} (x)  \eta_{\fN}^{\; \fM} = \eta_{\fB}^{\;\, \fA} \, .
\ee
Such a pair of dual field bases $\{\eth_{\fA} \}$ and $ \{\delta\zeta^{\fA}\} $ form a pair of dual locally flat hyper-spacetimes over globally flat Minkowski hyper-spacetime $M_h$. It is easy to verify that the field basis $\{\eth_{\fA}\}$ does not commute and satisfies the following non-commutation relation: 
\be \label{NCR}
& &   [ i \eth_{\fA} ,\; i\eth_{\fB}] = i f_{\fA\fB}^{\; \; \; \fC}(x)\,  i\eth_{\fC}, \nn \\
& & f_{\fA\fB}^{\; \; \; \fC}(x) \equiv - \chih_{\fA}^{\;\; \fM}(x) \chih_{\fB}^{\;\; \fN}(x) \mG_{\fM\fN}^{\; \fC}(x) , \quad \mG_{\fM\fN}^{\; \fC}(x) \equiv\p_{\fM}\chi_{\fN}^{\;\; \fC}(x) - \p_{\fN}\chi_{\fM}^{\;\; \fC}(x).
\ee 
which shows that such a non-commutation relation brings on a Lie algebra with non-constant group structure factor $f_{\fA\fB}^{\; \fC}(x)$ characterized by the tensor field $\mG_{\fM\fN}^{\; \fC}(x)$. Such a tensor field is presented as an Abelian-type field strength of hyper-gravigauge field $\chi_{\fM}^{\;\; \fC}(x)$. Therefore, the above non-commutation relation brings about a {\it non-commutative geometry} in locally flat hyper-spacetime, which is described by {\it Abelian-type hyper-gravigauge field strength} $\mG_{\fM\fN}^{\; \fC}(x)$. 

For the convenience of mention, the dual field bases $ \{\eth_{\fA} \} $ and $ \{\delta\zeta^{\fA}\} $ formed by the hyper-gravigauge field are called as a pair of dual {\it hyper-gravigauge field bases}. Correspondingly, the dual locally flat hyper-spacetimes formed by the dual hyper-gravigauge field bases are referred to as locally flat dual {\it gravigauge hyper-spacetimes}. Specifically, the gravigauge hyper-spacetime formed from hyper-gravigauge field basis $\{\eth_{\fA}\}$ is called as tangent-like hyper-spacetime denoted by $G_h$, and the dual gravigauge hyper-spacetime formed from dual hyper-gravigauge field basis $ \{\delta\zeta^{\fA}\} $ is called as dual tangent-like hyper-spacetime denoted by $G_h^{\ast}$. 

In general, the hyper-gravigauge field $\chi_{\fM}^{\;\; \fA}(x)$ is considered to be sided on the dual tangent Minkowski hyper-spacetime $T_h^{\ast}$ and valued on the dual gravigauge hyper-spacetime $G_h^{\ast}$. The dual hyper-gravigauge field $\hat{\chi}_{\fA}^{\;\;\fM}(x)$ is defined on the gravigauge hyper-spacetime $G_h$ and valued on the tangent Minkowski hyper-spacetime $T_h$. Namely, $\hat{\chi}_{\fA}^{\;\;\fM}(x)$ transforms as bicovariant vector field under transformations of both hyperspin gauge symmetry SP(1, $D_h$-1) and global Lorentz-type group symmetry SO(1, $D_h$-1). The non-commutation relation for hyper-gravigauge field basis $\{\eth_{\fA}\}$ displays that locally flat gravigauge hyper-spacetime $G_h$ is correlated to {\it non-commutative geometry}. 

The gravigauge hyper-spacetime $G_h$ and tangent Minkowski hyper-spacetime $T_h$ form a {\it biframe hyper-spacetime} $T_h\times G_h$ associated with a dual {\it biframe hyper-spacetime} $T_h^{\ast}\times G_h^{\ast}$ over coordinate hyper-spacetime $M_h$. 

Mathematically,  globally and locally flat vector spacetimes allow for a canonical identification of vectors in tangent Minkowski hyper-spacetime $T_h$ at points with vectors in Minkowski hyper-spacetime itself $M_h$, and also for a canonical identification of vectors at points with its dual vectors at the same points. For the meaning in physics, either tangent or dual tangent Minkowski hyper-spacetime over globally flat Minkowski hyper-spacetime is regarded as {\it free-motion hyper-spacetime} $\fM_h$, while locally flat gravigauge hyper-spacetime is viewed as {\it emergent hyper-spacetime} $\fGh$.  The canonical identification for the vector spacetimes brings about a simple structure of biframe hyper-spacetime,
\be
& & \fB_{h} =  \fM_h \times \bf{G}_h , \nn \\
 & & \fM_h \equiv T_h \cong T_h^{\ast} \cong  M_h , \quad \fGh \equiv G_h \cong G_h^{\ast} .
\ee
Such a biframe hyper-spacetime structure forms {\it gravigauge hyper-fiber bundle} $\fE_h$ with locally flat emergent gravigauge hyper-spacetime as a hyper-fiber $\fGh$ and globally flat free-motion Minkowksi hyper-spacetime as a base spacetime $\fM_h$. 

In mathematics, the hyper-fiber bundle $\fE_h$ is related to the product hyper-spacetime $\fM_h\times \fG_h$ as biframe hyper-spacetime via a continuous surjective map $\Pi_{\chi}$, which projects the bundle $\fE_h$ to the base spacetime $\fM_h$. Geometrically, the {\it gravigauge hyper-fiber bundle structure} of biframe hyper-spacetime is expressed as  $(\fE_h, \fM_h, \Pi_{\chi}, \fG_h) $. In general, the hyper-fiber bundle $\fE_h$ with the surjective map $\Pi_{\chi}$ is expressed as follows with the trivial case: 
\be
& & \Pi_{\chi}: \; \; \fE_h \to \fM_h , \nn \\
& & \fE_h \sim  \fB_h = \fM_h \times \fG_h .
\ee


\subsection{Gravitational origin of gauge symmetry and the hyperspin gravigauge and covariant-gauge fields with gauge covariant field strengths }

In the usual gauge theory, the internal gauge symmetry is attributed in formal to a pure gauge field, one can always choose a configuration via a gauge transformation to eliminate unphysical part arising from gauge symmetry. By making such a gauge prescription in unitary gauge, the gauge field keeps only independent degrees of freedom. Nevertheless, for hyperspin gauge symmetry SP(1,$D_h$-1), the situation appears to be distinguishable as it is resulted from gauging global hyperspin symmetry which is initiated from a motion-correlation $\cM_c$-spin symmetry. In fact, when building the gauge invariant action of hyperqubit-spinor field, we are led to introduce not only the hyperspin gauge field $\cA_{\fM}(x)$ but also the bicovariant vector field $\chih_{\fA}^{\;\; \fM}(x)$. It is conceivable that such a bicovariant vector field must be in correlation to hyperspin gauge symmetry. To corroborate such an idea,  let us decompose the hyperspin gauge field $\cA_{\fM}(x)$ into two parts $\mOm_{\fM}(x)$ and $\mA_{\fM}(x)$ as follows:
\be \label{HSGFDC}
\cA_{\fM}(x) = \mOm_{\fM}(x) + \mA_{\fM}(x) \equiv \cA_{\fM}^{\fA\fB}\frac{1}{2}\vSi_{\fA\fB} , 
\ee 
so that $\mOm_{\fM}(x)$ obeys an inhomogeneous transformation of hyperspin gauge symmetry and $\mA_{\fM}(x)$ transforms homogeneously under hyperspin gauge transformation, i.e.:
\be \label{HSGFGT}
& & \mOm_{\fM}(x) \to \mOm'_{\fM}(x) = S(\Lambda) i\p_{\fM} S^{-1}(\Lambda)  + S(\Lambda) \mOm_{\fM}(x) S^{-1}(\Lambda) , \nn \\
& & \mA_{\fM}(x) \to \mA'_{\fM}(x) = S(\Lambda) \mA_{\fM}(x) S^{-1}(\Lambda), \quad S(\Lambda) \in SP(1,D_h-1).
\ee
For a usual internal gauge field, $\mOm_{\fM}$ is generally taken as a pure gauge field with vanishing field strength. In the present case, the gauge field $\mOm_{\fM}$ with inhomogeneous gauge transformation is presumed to correlate with the hyper-gravigauge field $\chi_{\fM}^{\; \fA}$. This is because when turning hyperspin gauge symmetry to be global symmetry with group transformation being coincidental to that of global Lorentz-type group symmetry, the hyperspin gauge field $\cA_{\fM}$ disappears and hyper-gravigauge field $\chi_{\fM}^{\; \fA}$ becomes the constant bicovariant vector $\delta_{\fM}^{\; \fA}$.

It can be verified that $\mOm_{\fM}$ does correlate to the hyper-gravigauge field $\chi_{\fM}^{\; \fA}$ and its explicit form is uniquely given as follows:
\be \label{HSGGF}
\mOm_{\fM}^{\fA\fB} & = & \frac{1}{2}\left( \hat{\chi}^{\fA\fN} \mG_{\fM\fN}^{\fB} - \hat{\chi}^{\fB\fN} \mG_{\fM\fN}^{\fA} -  \hat{\chi}^{\fA\fP}  \hat{\chi}^{\fB\fQ}  \mG_{\fP\fQ}^{\fC} \chi_{\fM \fC } \right) ,
\ee
where the antisymmetric tensor $\mG_{\fM\fN}^{\fA}$ is defined as Abelian-type field strength of hyper-gravigauge field $\chi_{\fM}^{\; \fA}$ shown in Eq.(\ref{NCR}). The gauge transformation of hyper-gravigauge field $\chi_{\fM}^{\; \fA}$ in the vector representation of hyperspin gauge symmetry SP(1,$D_h$-1) does lead $\mOm_{\fM}^{\fA\fB}$ to get the following proper gauge transformation:
\be \label{HSGFGT1}
& & \chi_{\fM}^{'\; \fA} =  \Lambda^{\fA}_{\;\; \fC} \chi_{\fM}^{\; \fC} , \quad  \chih_{\fA}^{' \; \fM}=  \Lambda_{\fA}^{\;\; \fC} \chih_{\fC}^{\; \fM} ,\quad  \Lambda^{\fA}_{\; \fC}  \in \mbox{SP(1,$D_h$-1)} , \nn \\
& & \mOm_{\fM}^{'\fA\fB} =  \frac{1}{2} ( \Lambda^{\fA}_{\; \fC}  \p_{\fM}  \Lambda^{\fB\fC}- \Lambda^{\fB}_{\; \fC}  \p_{\fM}  \Lambda^{\fA\fC} )+ \Lambda^{\fA}_{\; \fC}  \Lambda^{\fB}_{\; \fD} \mOm_{\fM}^{\fC\fD} ,
\ee
where $\mOm_{\fM}^{\fA\fB}$ displays an inhomogeneous gauge transformation in the adjoint representation of hyperspin gauge symmetry SP(1,$D_h$-1). 

As $\mOm_{\fM}^{\fA\fB}(x)$ is completely determined by the hyper-gravigauge field $\chi_{\fM}^{\; \fA}(x)$, it involves no additional independent degrees of freedom in the decomposition of hyperspin gauge field $\cA_{\fM}(x)$. For convenience, we may refer to $\mOm_{\fM}^{\fA\fB}(x)$ as {\it hyperspin gravigauge field} and $\mA_{\fM}^{\fA\fB}(x)$ as {\it hyperspin covariant-gauge field}. 

It is not possible to make a gauge transformation to eliminate the hyperspin gravigauge field $\mOm_{\fM}^{\fA\fB}(x)$, which is completely different from the usual internal gauge field. To check that explicitly, let us decompose the hyperspin gauge field strength into the following two parts:
\be
\cF_{\fM\fN}^{\fA\fB} &\equiv & \mR_{\fM\fN}^{\fA\fB} + \mF_{\fM\fN}^{\fA\fB} \, , 
\ee
with the explicit forms,
\be \label{FS1}
& & \mR_{\fM\fN}^{\fA\fB} = \partial_{\fM} \mOm_{\fN}^{\fA\fB} - \partial_{\fN} \mOm_{\fM}^{\fA\fB} + \mOm_{\fM \fC}^{\fA} \mOm_{\fN}^{\fC \fB} -  \mOm_{\fN \fC}^{\fA} \mOm_{\fM}^{\fC \fB},  
\nn \\  
& & \mF_{\fM\fN}^{\fA\fB} =  \mD_{\fM} \mA_{\fN}^{\fA\fB} - \mD_{\fN} \mA_{\fM}^{\fA\fB} +  \mA_{\fM \fC}^{\fA} \mA_{\fN}^{\fC \fB} -  \mA_{\fN \fC}^{\fA} \mA_{\fM}^{\fC \fB}  , \nn \\
& & \mD_{\fM} \mA_{\fN}^{\fA\fB} = \p_{\fM}  \mA_{\fN}^{\fA\fB}  +  \mOm_{\fM \fC}^{\fA} \mA_{\fN}^{\fC \fB} + \mOm_{\fM \fC}^{\fB} \mA_{\fN}^{\fA \fC}  ,
\ee
where $\mR_{\fM\fN}^{\fA\fB}$ is purely the field strength of hyperspin gravigauge field $\mOm_{\fM}^{\fA\fB}(x)$, which is not vanishing.

The decomposition of hyperspin gauge field enables us to obtain two gauge covariant field strengths $\mR_{\fM\fN}^{\fA\fB}$ and $\mF_{\fM\fN}^{\fA\fB}$ in correspondence to the hyperspin gravigauge field $\mOm_{\fM}^{\fA\fB}(x)$ and hyperspin covariant-gauge field $\mA_{\fM}^{\fA\fB}(x)$. The hyperspin gravigauge field $\mOm_{\fM}^{\fA\fB}(x)$ and its gauge covariant field strength $\mR_{\fM\fN}^{\fA\fB}$ characterize the gravitational interaction originated from hyper-gravigauge field.

From the above analysis, we come to the statement about the {\it gravitational origin of gauge symmetry} that the hyperspin gauge symmetry SP(1,$D_h$-1) of entangled qubit-spinor field is essentially governed by the hyper-gravigauge field $\chi_{\fM}^{\; \fA}(x)$ with gauge transformation in the vector representation of SP(1,$D_h$-1).


\subsection{ $\cW_e$-spin invariant-gauge field as genesis of hyper-gravigauge field in the presence of graviscalar field }

The hyper-gravigauge field $\chi_{\fM}^{\; \fA}(x)$ is defined as the dual field of bicovariant vector field $\chih_{\fA}^{\; \fM}(x)$ which is introduced when gauging inhomogeneous hyperspin symmetry based on the gauge invariance principle as guiding principle. Although $\chi_{\fM}^{\; \fA}(x)$ behaves as a gauge-type vector field indicated from its field strength shown either in the equation of motion (Eq.(\ref{HGFFS})) or in the non-commutation relation (Eq.(\ref{NCR})), while it remains inapparent whether $\chi_{\fM}^{\; \fA}(x)$ can be regarded as a gauge field introduced directly from a gauge symmetry by just following along the gauge invariance principle. For this purpose, it is interesting to notice that the field strength in Eq.(\ref{HGFFS}) for $\chi_{\fM}^{\; \fA}(x)$ has an analogous formalism with the field strength in Eq.(\ref{GFS}) for $\cW_e$-spin gauge field $\ckcA_{\fM}^{\fA}(x)$. As $\ckcA_{\fM}^{\fA}(x)$ is introduced via the $\cW_e$-spin gauge symmetry W$^{1,D_h-1}$ which is the subgroup symmetry of inhomogeneous hyperspin gauge symmetry WS(1,$D_h$-1), we are motivated to explore the correlation between the hyper-gravigauge field $\chi_{\fM}^{\; \fA}(x)$ and $\cW_e$-spin gauge field $\ckcA_{\fM}^{\fA}(x)$. 

In analogous to the decomposition of hyperspin gauge field $\cA_{\fM}^{\fA\fB}(x)$, let us decompose the $\cW_e$-spin gauge field $\ckcA_{\fM}^{\fA}(x)$ into the following two parts: 
\be
\ckcA_{\fM}(x) \equiv \ckcA_{\fM}^{\fA}(x) \frac{1}{2}\vSi_{-\fA} =  (\fOm_{\fM}^{\fA}(x) +  \fA_{\fM}^{\fA}(x) ) \frac{1}{2}\vSi_{-\fA} , 
\ee
where $\fOm_{\fM}^{\fA}(x)$ is supposed to have inhomogeneous gauge transformation and $\fA_{\fM}^{\fA}(x)$ becomes gauge invariant under the $\cW_e$-spin gauge transformation shown in Eq.(\ref{TESGS}), while they all transform as covariant vector fields under the hyperspin gauge transformation. Their explicit transformations are presented as follows:
\be
& & \fOm_{\fM}^{\fA}(x) \to  \fOm_{\fM}^{' \fA}(x) =  \fOm_{\fM}^{\fA}(x) + \cD_{\fM}\varpi^{\fA}(x), \nn \\
& & \fA_{\fM}^{\fA}(x) \to  \fA_{\fM}^{' \fA}(x) =  \fA_{\fM}^{\fA}(x) , 
\ee
under the $\cW_e$-spin gauge transformation, and
\be
& &  \fOm_{\fM}^{\fA}(x) \to  \fOm_{\fM}^{' \fA}(x) = \Lambda^{\fA}_{\;\; \fB} \fOm_{\fM}^{\fB}(x), \nn \\
& & \fA_{\fM}^{\fA}(x) \to  \fA_{\fM}^{' \fA}(x) = \Lambda^{\fA}_{\;\; \fB} \fA_{\fM}^{\fB}(x),  
\ee
under the hyperspin gauge transformation. 

For convenience of mention, $\fOm_{\fM}^{\fA}(x)$ is referred to as {\it $\cW_e$-spin gravigauge field} and $\fA_{\fM}^{\fA}(x)$ as {\it $\cW_e$-spin invariant-gauge field}. In general, the combination of $\fA_{\fM}^{\fA\fB}(x)$ and $\fA_{\fM}^{\fA}(x)$, i.e., $\breve{\fA}_{\fM}(x)\equiv \fA_{\fM}^{\fA\fB}(x)\frac{1}{2}\vSi_{\fA\fB} + \fA_{\fM}^{\fA}(x)\frac{1}{2}\vSi_{-\fA}$, is regarded as {\it inhomogeneous hyperspin covariant-gauge field}. 

In order to keep the initial degrees of freedom for the inhomogeneous hyperspin gauge field, there must exist certain relations among $\cW_e$-spin gravigauge field $\fOm_{\fM}^{\fA}(x)$, $\cW_e$-spin invarian-gauge field $\fA_{\fM}^{\fA}(x)$, hyper-gravigauge field $\chi_{\fM}^{\; \fA}(x)$ and hyperspin gauge field $\cA_{\fM}^{\fA\fB}(x)$. Indeed, it is natural to propose the following relations:
\be \label{TESGGF}
& & \fOm_{\fM}^{\fA}(x) \equiv  \cD_{\fM}\vka^{\fA}(x) = \p_{\fM} \vka^{\fA}(x) + \cA_{\fM \fB}^{\fA}(x) \vka^{\fB}(x), \nn \\
& & \fA_{\fM}^{\fA}(x) \equiv \phi(x) \chi_{\fM}^{\; \fA}(x), 
\ee
where $\vka^{\fA}(x)$ is introduced as a vector field in locally flat gravigauge hyper-spacetime and $\phi(x)$ is taken as a real {\it scalar field} to keep a canonical mass dimension. To ensure the $\cW_e$-spin gauge transformation, the vector field $\vka^{\fA}(x)$ must transform as follows:
\be \label{TCST}
\vka^{\fA}(x) \to \vka^{' \fA}(x) = \vka^{\fA}(x) + \varpi^{\fA}(x) .
\ee
which is regarded as a local translation of vector field $\vka^{\fA}(x)$ in locally flat gravigauge hyper-spacetime. As the above transformation reflects the symmetry property of translation-like $\cW_e$-spin Abelian-type gauge group, $\vka^{\fA}(x)$ is viewed as a {\it $\cW_e$-spin vector field} in locally flat gravigauge hyper-spacetime.  

From the above considerations, it is the {\it $\cW_e$-spin invariant-gauge field} $\fA_{\fM}^{\fA}(x)$ that brings on the hyper-gravigauge field $\chi_{\fM}^{\; \fA}(x)$ in association with a scalar field $\phi(x)$. Therefore, the hyper-gravigauge field $\chi_{\fM}^{\; \fA}(x)$ arises from the $\cW_e$-spin invariant-gauge field associated with scalar field $\phi(x)$. Such a scalar field is considered to be a gravitational scalar field and referred to as {\it graviscalar field} for short. 

It can be checked that the $\cW_e$-spin vector field $\vka^{\fA}(x)$ has zero scaling charges for both global and local scaling symmetries SC(1) and SG(1). Under the transformations of SC(1) and SG(1) symmetries, the graviscalar field $\phi(x)$ transforms as follows:
\be
& & \phi(x) \to \phi'(x') = \lambda\, \phi(x), \quad x' = \lambda^{-1} x, \quad \lambda \in SC(1), \nn \\
& & \phi(x) \to \phi'(x') = \xi(x)\phi(x), \quad \xi(x) \in SG(1) .
\ee

The field strength of $\cW_e$-spin gauge field can also be decomposed into the following two parts:
\be \label{TCSFS}
& & \ckcF_{\fM\fN}^{\fA}(x) \equiv \fR_{\fM\fN}^{\fA}(x) + \fF_{\fM\fN}^{\fA}(x), \nn \\
& & \fR_{\fM\fN}^{\fA}(x) = \cD_{\fM}\fOm_{\fN}^{\fA}(x) -  \cD_{\fN}\fOm_{\fM}^{\fA}(x) =  \cF_{\fM\fN}^{\fA\fB}(x) \vka_{\fB}(x) , \nn \\
& & \fF_{\fM\fN}^{\fA}(x) = \cD_{\fM}\fA_{\fN}^{\fA}(x) -  \cD_{\fN}\fA_{\fM}^{\fA}(x) \equiv \phi(x) \cG_{\fM\fN}^{\fA}(x), 
\ee
where $\cG_{\fM\fN}^{\fA}(x)$ defines a hyperspin gauge and scaling gauge covariant field strength of hyper-gravigauge field $\chi_{\fM}^{\fA}(x)$,  
\be \label{HGGFS2}
\cG_{\fM\fN}^{\fA}(x) & \equiv & \cD_{\fM}\chi_{\fN}^{\fA}(x) -  \cD_{\fN}\chi_{\fM}^{\fA}(x) = d_{\fM}\chi_{\fN}^{\fA}(x) -  d_{\fN}\chi_{\fM}^{\fA}(x)  +  \cA_{\fM\fB}^{\fA}\chi_{\fM}^{\fB}(x) -  \cA_{\fN\fB}^{\fA}\chi_{\fM}^{\fB}(x)  \nn \\
& \equiv & \fG_{\fM\fN}^{\fA}(x) + \cA_{\fM\fB}^{\fA}\chi_{\fM}^{\fB}(x) -  \cA_{\fN\fB}^{\fA}\chi_{\fM}^{\fB}(x) , 
\ee
with $d_{\fM}$ defined as {\it scaling gauge covariant derivative}. Where we have introduced the gauge-type field strength of hyper-gravigauge field as follows: 
\be \label{HGGFS0}
& & \fG_{\fM\fN}^{\fA}(x) \equiv d_{\fM}\chi_{\fN}^{\fA}(x) -  d_{\fN}\chi_{\fM}^{\fA}(x) , \nn \\
& & d_{\fM}\equiv \p_{\fM} + \fS_{\fM}, \quad \fS_{\fM} \equiv \frac{1}{2}\p_{\fM}\ln \phi^2(x), 
\ee
which becomes scaling gauge covariance but not hyperspin gauge covariance. Where $\fS_{\fM}$ is regarded as a {\it pure graviscaling gauge field} given by the graviscalar field $\phi(x)$.

Therefore, we arrive at the statement that it is the $\cW_e$-spin gauge symmetry as subgroup gauge symmetry of inhomogeneous hyperspin gauge symmetry that brings on the genesis of dual bicovariant vector field $\chi_{\fM}^{\; \fA}(x)$ as hyper-gravigauge field, which provides a natural explanation on the gauge property of hyper-gravigauge field appearing either in the equation of motion shown in Eq.(\ref{HGFFS}) or in the non-commutative relation presented in Eq.(\ref{NCR}).


 \subsection{ Biframe displacement correspondence in biframe hyper-spacetime and the dimensionless hyper-gravicoordinate displacement and derivative with respect to hyper-gravivector field in locally flat gravigauge hyper-spacetime}

The $\cW_e$-spin gauge transformation is simply presented by a local translation of $\cW_e$-spin vector field $\vka^{\fA}(x)$ in locally flat gravigauge hyper-spacetime as shown in Eq.(\ref{TCST}). It is manifest that under the $\cW_e$-spin gauge transformation, $\fF_{\fM\fN}^{\fA}(x)$ becomes $\cW_e$-spin invariant-gauge field strength and   the field strength $\fR_{\fM\fN}^{\fA}(x)$ transforms as follows,
\be
\fR_{\fM\fN}^{\fA}(x) \to \fR_{\fM\fN}^{' \fA}(x) = \fR_{\fM\fN}^{\fA}(x) + \cF_{\fM\fN}^{\fA\fB}(x) \varpi_{\fB}(x) .
\ee

Let us regard $\cW_e$-spin vector field $\vka^{\fA}(x)$ as {\it dimensionless hyper-gravivector field} in locally flat gravigauge hyper-spacetime, so that the local translation can be represented via a local translation operator defined in formal by the following coordinate-like field derivative:
\be \label{HGCD}
\hat{\eth}_{\fA} \equiv \frac{\delta }{\delta \vka^{\fA}}, 
\ee
which is in analogous to the ordinary translation operator `$\p_{\fM}$' of coordinates in globally flat Minkowski hyper-spacetime. Similarly, we can introduce the coordinate-like field displacement $\delta\vka^{\fA}$ in locally flat gravigauge hyper-spacetime. For convenience of mention, the displacement $\delta\vka^{\fA}$ is referred to as {\it dimensionless hyper-gravicoordinate displacement} and the derivative $\hat{\eth}_{\fA}$ is called as {\it dimensionless hyper-gravicoordinate derivative}.

As the basic structure of biframe hyper-spacetime is described solely by the hyper-gravigauge field originated from the $\cW_e$-spin invariant-gauge field, it is natural to consider {\it biframe displacement correspondence} which states that the hyper-gravicoordinate displacement of hyper-gravivector field $\vka^{\fA}(x)$ in locally flat gravigauge hyper-spacetime is directly associated to the ordinary coordinate displacement in globally flat Minkowski hyper-spacetime through the $\cW_e$-spin invariant-gauge field. Namely, when the locally flat gravigauge hyper-spacetime becomes a globally flat one, the hyper-gravicoordinate displacement $\delta\vka^{\fA}$ should be coincidental to the ordinary coordinate displacement. Based on such a biframe displacement correspondence and biframe hyper-spacetime structure, we are able to express the dimensionless hyper-gravicoordinate displacement $\delta\vka^{\fA}$ and derivative $\hat{\eth}_{\fA}$ as follows:
\be \label{HGCDP}
\delta\vka^{\fA} &\equiv&  \fA_{\fM}^{\;\;\fA}(x)dx^{\fM} = \phi(x)\chi_{\fM}^{\;\;\fA}(x)dx^{\fM} \equiv  \phi(x) \delta\zeta^{\fA} , \nn \\
\hat{\eth}_{\fA} & \equiv & \hat{\fA}_{\fA}^{\;\;\fM}(x)\p_{\fM} = \phi^{-1}(x)\chih_{\fA}^{\;\;\fM}(x) \p_{\fM} = \phi^{-1}(x)\eth_{\fA} ,
\ee 
with $\phi(x)$ the graviscalar field. Where the dimensionless hyper-gravicoordinate displacement $\delta\vka^{\fA}$ and derivative $\hat{\eth}_{\fA}$ with respect to hyper-gravivector field are related to the corresponding dimensionful vector field displacement $\delta\zeta^{\fA}$ and derivative $\eth_{\fA}$ via the graviscalar field. They are all characterized by the hyper-gravigauge field which is regarded as bicovariant vector field defined in biframe hyper-spacetime. 

Once locally flat gravigauge hyper-spacetime approaches to a globally flat one, we are led to the following special case:
\be \label{FMS}
& & \chih_{\fM}^{\;\;\fA}(x) \to \eta_{\fM}^{\;\; \fA}, \quad \phi(x) \to \Mka, \quad \delta\zeta^{\fA} \to \eta_{\fM}^{\;\; \fA} dx^{\fM} = dx^{\fA}, \nn \\
& & \delta\vka^{\fA} \to \Mka \delta\zeta^{\fA} = \Mka \eta_{\fM}^{\;\; \fA} dx^{\fM}, 
\ee
where $\Mka$ is supposed to be a {\it fundamental mass scale}. Correspondingly,  $\delta\zeta^{\fA}$ and $\eth_{\fA}$ are considered as dimensionful {\it hyper-gravicoordinate displacement and derivative}, respectively, in locally flat gravigauge hyper-spacetime with the following forms:
\be \label{DHDCDD}
& & \delta\zeta^{\fA} \equiv  \phi^{-1}(x) \fA_{\fM}^{\;\;\fA}(x)dx^{\fM} = \chi_{\fM}^{\;\;\fA}(x)dx^{\fM}, \nn \\
& & \eth_{\fA} \equiv \frac{\delta}{\delta \zeta^{\fA}} \equiv \phi(x) \hat{\fA}_{\fA}^{\;\;\fM}(x)\frac{\p}{\p x^{\fM}} = \chih_{\fA}^{\;\;\fM}(x)\p_{\fM}. \nn
\ee


\section{Hyperunified field theory for fundamental building block and fundamental interaction based on gauge invariance principle and scaling invariance hypothesis }

Following along the maximum entangled qubit-spinor principle as guiding principle, we have shown in the previous paper as part I of the foundation of the hyperunified field theory\cite{FHUFT-I} that the hyperunified qubit-spinor field is obtained as entangled hyperqubit-spinor field with least $q_c$-spin charge. The category-0 entangled enneaqubit-spinor field with $q_c$-spin charge $q_c=0$ is regarded as the minimal hyperunified qubit-spinor field, and the category-1 entangled decaqubit-spinor field with $q_c$-spin charge $q_c=1$ is demonstrated to be the hyperunified qubit-spinor field that can unify all known leptons and quarks\cite{QK1,QK2} in SM. Based on the gauge invariance principle as guiding principle, we present in previous sections a general analysis that the fundamental interaction is governed by inhomogeneous hyperspin gauge symmetry WS(1,$D_h$-1). In particular, the gravitational interaction is shown to be governed by the translation-like $\cW_e$-spin Abelian-type gauge symmetry which is the subgroup gauge symmetry of inhomogeneous hyperspin gauge symmetry. The hyper-gravigauge field is identified to be the $\cW_e$-spin invariant-gauge field in association with the graviscalar field. Therefore, we get a better understanding on the gravitational origin of gauge symmetry and genesis of gravitational interaction. In this section, we are going to build a whole hyperunified field theory for characterizing both fundamental building block and fundamental interaction based on the gauge invariance principle and scaling invariance hypothesis. Specifically, we will take the entangled decaqubit-spinor field as hyperunified qubit-spinor field $\fPsi_{\fQH}(x)$ to be the fundamental building block of nature and meanwhile the inhomogeneous hyperspin gauge interaction as hyperunified gauge interaction to be the fundamental interaction mediated via the inhomogeneous hyperspin gauge field $\bvcA_{\fM}(x)$ in 19-dimensional hyper-spacetime. In addition, we will consider scaling gauge interaction and $\cQ_c$-spin scalar interaction as fundamental interactions mediated through the scaling gauge field $\mW_{\fM}(x)$ and $\cQ_c$-spin scalar field $\phi_1(x)$.


\subsection{ Hyperunified qubit-spinor structure of entangled decaqubit-spinor field as fundamental building block and the inhomogeneous hyperspin symmetry WS(1,18) as hyperunified symmetry in gravigauge hyper-spacetime }

Let us begin with returning to the entangled decaqubit-spinor field in category-1 with $\cM_c$-spin charge $\cC_{\cM_c}=19$ and $\cQ_c$-spin charge $\cC_{\cQ_c}=1$, which has been shown in the previous paper\cite{FHUFT-I} to be the minimal entangled hyperqubit-spinor field that unifies all known leptons and quarks into a single hyperunified qubit-spinor field $\fPsi_{\fQH}(x)$. To explicitly reflect the basic properties of chiral type lepton-quark states and mirror lepton-quark states, it is useful to represent it in an alternative qubit-spinor structure as follows:
\be \label{ffQH}
\fPsi_{\fQH}(x) & \equiv & \binom{ \fPsi_{\fQUn}(x)}{ \fPsi_{\fQUp}(x) } = \begin{pmatrix}
\Psi_{\fQUn}(x) \\ \hPsi_{\fQUn}(x) \\ \Psi_{\fQUp}(x) \\ \hPsi_{\fQUp}(x)  ,
\end{pmatrix} = \begin{pmatrix}
\Psi_{\fQGnf}(x) \\ \Psi_{\fQGns}(x) \\ \Psi_{\fQGnt}(x) \\ \Psi_{\fQGnft}(x)  \\ 
\Psi_{\fQGpf}(x) \\ \Psi_{\fQGps}(x) \\ \Psi_{\fQGpt}(x) \\  \Psi_{\fQGpft}(x) 
\end{pmatrix} ,
\ee
where $\fPsi_{\fQUn}(x)$ is composed of two ultra-grand unified qubit-spinor fields $\Psi_{\fQUn}(x)$ and $\hPsi_{\fQUn}(x)$ with negative U-parity, which corresponds to four families of chiral type lepton-quark states,  and $\fPsi_{\fQUp}(x)$ consists of two ultra-grand unified qubit-spinor fields $\Psi_{\fQUp}(x)$ and $\hPsi_{\fQUp}(x)$ with positive U-parity, which provides four families of mirror lepton-quark states, i.e.:
\be \label{ffQU}
& & \fPsi_{\fQUn}(x) = \binom{\Psi_{\fQUn}(x)}{\hPsi_{\fQUn}(x)} = \begin{pmatrix}
\Psi_{\fQGnf}(x) \\ \Psi_{\fQGns}(x) \\ \Psi_{\fQGnt}(x) \\ \Psi_{\fQGnft}(x)  
\end{pmatrix} , \nn \\
& & \fPsi_{\fQUp}(x) = \binom{\Psi_{\fQUp}(x)}{\hPsi_{\fQUp}(x)} = \begin{pmatrix}
\Psi_{\fQGpf}(x) \\ \Psi_{\fQGps}(x) \\ \Psi_{\fQGpt}(x) \\ \Psi_{\fQGpft}(x) 
\end{pmatrix} 
\ee
where both $\fPsi_{\fQUn}(x)$ and $\fPsi_{\fQUp}(x)$ are regarded as {\it paired ultra-grand unified qubit-spinor field} with negative U-parity and positive U-parity, respectively. 

$\Psi_{\fQGfn}(x)$ and $\Psi_{\fQGfp}(x)$ ($f=1,2,3,4$) represent four families of grand unified qubit-spinor fields with respective to chiral type lepton-quark states and mirror lepton-quark states, they are explicitly given as follows:
\be \label{fQGpn}
& & \Psi_{\fQGfn}(x) \equiv 
\begin{pmatrix}
\Psi_{\fQ_{\mC_f}^4}^{u\, L}(x) \\  \Psi_{\fQ_{\mC_f}^4}^{d\, R\, \bc}(x)  \\ \Psi_{\fQ_{\mC_f}^4}^{d\, L}(x) \\ - \Psi_{\fQ_{\mC_f}^4}^{u\, R\, \bc}(x)  \\
\Psi_{\fQ_{\mC_f}^4}^{u\, R}(x) \\  \Psi_{\fQ_{\mC_f}^4}^{d\, L\, \bc}(x)  \\ \Psi_{\fQ_{\mC_f}^4}^{d\, R}(x) \\ - \Psi_{\fQ_{\mC_f}^4}^{u\, L\, \bc}(x) 
\end{pmatrix}, \quad
\Psi_{\fQGfp}(x) \equiv
\begin{pmatrix}
\tPsi_{\fQ_{\mC_f}^4}^{u\, R}(x) \\  \tPsi_{\fQ_{\mC_f}^4}^{d\, L\, \bc}(x)  \\ \tPsi_{\fQ_{\mC_f}^4}^{d\, R}(x) \\ - \tPsi_{\fQ_{\mC_f}^4}^{u\, L\, \bc}(x)  \\
\tPsi_{\fQ_{\mC_f}^4}^{u\, L}(x) \\  \tPsi_{\fQ_{\mC_f}^4}^{d\, R\, \bc}(x)  \\ \tPsi_{\fQ_{\mC_f}^4}^{d\, L}(x) \\ - \tPsi_{\fQ_{\mC_f}^4}^{u\, R\, \bc}(x) 
\end{pmatrix} ,
\ee
where $\Psi_{\fQ_{\mC_f}^4}^{q\, L, R}(x)$ and $\tPsi_{\fQ_{\mC_f}^4}^{q\, L, R}(x)$ ($f=1,2,3,4$) are four kinds of {\it mirror pairs of chiral complex tetraqubit-spinor fields} with complex charge-conjugated ones $\Psi_{\fQ_{\mC_f}^4}^{q\, L, R\, \bc}(x)$ and $\tPsi_{\fQ_{\mC_f}^4}^{q\, L, R\, \bc}(x)$. They have the following explicit forms:
\be \label{fQC4}
& &  \Psi_{\fQ_{\mC_f}^4}^{q\, L, R}(x) \equiv \begin{pmatrix}
Q_f^{r}(x) \\ Q_f^{b}(x) \\ Q_f^{g}(x) \\ Q_f^{w}(x) 
\end{pmatrix}^q_{L, R}, \quad  
\tPsi_{\fQ_{\mC_f}^4}^{q\, L, R}(x) \equiv \begin{pmatrix}
\tQ_f^{r}(x) \\ \tQ_f^{b}(x) \\ \tQ_f^{g}(x) \\ \tQ_f^{w}(x) 
\end{pmatrix}^q_{L, R}, \nn \\
& & \Psi_{\fQ_{\mC_f}^4}^{q\, \bc\, L, R}(x) \equiv \begin{pmatrix}
Q^{r\, \bc}_{f}(x) \\ Q^{b\, \bc}_{f}(x) \\ Q^{g\, \bc}_{f}(x) \\ Q^{w\, \bc}_{f}(x) 
\end{pmatrix}^q_{L, R} ,  \quad  \tPsi_{\fQ_{\mC_f}^4}^{q\, \bc\, L, R}(x) \equiv \begin{pmatrix}
\tQ^{r\, \bc}_{f}(x) \\ \tQ^{b\, \bc}_{f}(x) \\ \tQ^{g\, \bc}_{f}(x) \\ \tQ^{w\, \bc}_{f}(x) 
\end{pmatrix}^q_{L, R} , \nn \\
& & Q^{\alpha\, \bc}_{f}(x) = C_D \bar{Q}^{\alpha\, T}_{f}(x), \quad \tQ^{\alpha\, \bc}_{f}(x)= C_D \bar{\tQ}^{\alpha\, T}_{f}(x), \quad C_D = - i\sigma_3\otimes \sigma_2 .
\ee
The superscripts $q=u$ and $q=d$ represent the up-type and down-type lepton-quark states $Q^{\alpha\, u}_{f\, L, R}(x) = U_{f\, L, R}^{\alpha}(x)$ and $Q^{\alpha\, d}_{f\, L, R}(x) = D_{f\, L, R}^{\alpha}(x)$ with the corresponding mirror lepton-quark states $\tQ^{\alpha\, u}_{f\, L, R}(x) = \tU_{f\, L, R}^{\alpha}(x)$ and $\tQ^{\alpha\, d}_{f\, L, R}(x) = \tD_{f\, L, R}^{\alpha}(x)$. The superscript $\alpha$ labels four color-spin charges $\alpha = r, b, g, w$ with respective to `red', `blue', `green' and `white', and the subscript $f=1,2, 3,4$ denotes four families of chiral type lepton-quark states and mirror lepton-quark states.  The letters `L' and `R' stand for the left-handed and right-handed lepton-quark states and mirror lepton-quark states. 

Explicitly, each family of chiral type lepton-quark states in SM and mirror lepton-quark states beyond SM can be expressed by the so-called westward and eastward entangled hyperqubit-spinor fields\cite{HUFT} with the following explicit forms:
\be \label{WEQH}
& & \Psi_{W_f}^{T}(x) \equiv  \Psi_{\fQ_{\mG_f}^{-}}^{T}(x)  \nn \\
& & \; =[ (U_f^{r}, U_f^{b}, U_f^{g}, U_f^{w}, D^{r}_{fc}, D^{b}_{fc}, D^{g}_{fc}, D^{w}_{fc}, D_f^{r}, D_f^{b}, D_f^{g}, D_f^{w}, -U^{r}_{fc}, -U^{b}_{fc}, -U^{g}_{fc}, -U^{w}_{fc})_L ,   \nn \\
& & \;\; (U_f^{r}, U_f^{b}, U_f^{g}, U_f^{w}, D^{r}_{fc}, D^{b}_{fc}, D^{g}_{fc}, D^{w}_{fc}, D_f^{r},  D_f^{b}, D_f^{g}, D_f^{w}, -U^{r}_{fc}, -U^{b}_{fc}, -U^{g}_{fc}, -U^{w}_{fc})_R ]^T , \nn \\
& & \Psi_{E_f}^{T}(x) \equiv  \Psi_{\fQ_{\mG_f}^{+}}^{T}(x)  \nn \\
& & \; =[ (\tU_f^{r}, \tU_f^{b}, \tU_f^{g}, \tU_f^{w}, \tD^{r}_{fc}, \tD^{b}_{fc}, \tD^{g}_{fc}, \tD^{w}_{fc}, \tD_f^{r}, \tD_f^{b}, \tD_f^{g}, \tD_f^{w}, -\tU^{r}_{fc}, -\tU^{b}_{fc}, -\tU^{g}_{fc}, -\tU^{w}_{fc})_R ,   \nn \\
& & \;\; (\tU_f^{r}, \tU_f^{b}, \tU_f^{g}, \tU_f^{w}, \tD^{r}_{fc}, \tD^{b}_{fc}, \tD^{g}_{fc}, \tD^{w}_{fc}, \tD_f^{r},  \tD_f^{b}, \tD_f^{g}, \tD_f^{w}, -\tU^{r}_{fc}, -\tU^{b}_{fc}, -\tU^{g}_{fc}, -\tU^{w}_{fc})_L ]^T ,
\ee
with $f=1,2,3,4$. The superscript $T$ denotes the transposition of column matrix.  

In such a hyperqubit-spinor structure, the kinetic action of hyperunified qubit-spinor field $\Psi_{\fQH}(x)$ in globally flat Minkowski hyper-spacetime can be represented as follows:
\be  \label{actionfQHK}
\cS_{\fQH} & \equiv & \int d^{19}x \,  \{ \bar{\Psi}_{\fQH}(x)  \delta_{\fA}^{\;\fM}  \vSi_{-}^{\fA} i\p_{\fM} \Psi_{\fQH}(x) - \phi_1(x)  \bar{\Psi}_{\fQH}(x)   \tvSi_{-} \Psi_{\fQH}(x) \},
\ee
with $\fA, \fM= 0,1,2,3,5, \cdots, 19$. The matrices $\vSi_{-}^{\fA}$ and $\tvSi_{-}$ are defined via $\vGa$-matrices $\vGa^{\hfA}\equiv (\vGa^{\fA}, \vGa^{20})\equiv (\vGa^a, \vGa^A, \tvGa) $ ($\hfA=0,1,2,3,5,\cdots, 20$) with the following explicit structure:
\be \label{GMffQH1}
& & \varSigma_{-}^{\fA} = \frac{1}{2} \vGa^{\fA}\vGa_{-}, \quad \fA = 0,1,2,3,5, \cdots 19, \nn \\
& & \varSigma_{-}^{20} = \frac{1}{2} \vGa^{20}\vGa_{-} \equiv \tvSi_{-} = \frac{1}{2} \tvGa \vGa_{-}, \quad \vGa_{-} = \frac{1}{2} ( 1 - \tilde{\gamma}_{21} ) ,  
\ee
and 
\be \label{GMffQH2}
& & \vGa^0 =\;\; \; \sigma_0 \otimes \sigma_0 \otimes \sigma_0 \otimes \sigma_0 \otimes \sigma_0 \otimes \sigma_0 \otimes \sigma_0 \otimes\sigma_0 \otimes \sigma_1 \otimes \sigma_0, \nn \\
& &  \vGa^1 =\;\;  i \sigma_0 \otimes \sigma_0 \otimes \sigma_0 \otimes \sigma_0 \otimes \sigma_0 \otimes \sigma_0 \otimes \sigma_0 \otimes \sigma_0\otimes \sigma_2\otimes \sigma_1, \nn \\
& & \vGa^2 = \;\;  i \sigma_0 \otimes \sigma_0 \otimes \sigma_0 \otimes \sigma_0 \otimes \sigma_0 \otimes \sigma_0 \otimes  \sigma_0 \otimes\sigma_0\otimes  \sigma_2\otimes \sigma_2, \nn \\
& & \vGa^3 = \;\; i \sigma_0 \otimes \sigma_0 \otimes \sigma_0 \otimes \sigma_0 \otimes \sigma_0 \otimes \sigma_0 \otimes \sigma_0 \otimes\sigma_0\otimes  \sigma_2\otimes \sigma_3, \nn \\
& &  \vGa^5 = -i \sigma_0 \otimes \sigma_0 \otimes \sigma_2 \otimes \sigma_1 \otimes \sigma_0 \otimes \sigma_1 \otimes \sigma_0 \otimes \sigma_2\otimes  \sigma_3\otimes \sigma_0, \nn \\
& & \vGa^6 = -i \sigma_0 \otimes \sigma_0 \otimes \sigma_2 \otimes \sigma_1 \otimes \sigma_0 \otimes \sigma_2 \otimes  \sigma_3 \otimes\sigma_2\otimes  \sigma_3\otimes \sigma_0, \nn \\
& & \vGa^7 = -i \sigma_0 \otimes \sigma_0 \otimes \sigma_2 \otimes \sigma_1 \otimes \sigma_0 \otimes \sigma_1 \otimes \sigma_2 \otimes\sigma_3\otimes  \sigma_3\otimes \sigma_0 , \nn \\
& &  \vGa^8 = -i \sigma_0 \otimes \sigma_0 \otimes \sigma_2 \otimes \sigma_1 \otimes \sigma_0 \otimes \sigma_2 \otimes \sigma_2 \otimes  \sigma_0\otimes \sigma_3\otimes \sigma_0, \nn \\
& &  \vGa^9 = -i \sigma_0 \otimes \sigma_0 \otimes \sigma_2 \otimes \sigma_1 \otimes \sigma_0 \otimes \sigma_1 \otimes  \sigma_2 \otimes \sigma_1\otimes \sigma_3\otimes \sigma_0 , \nn \\
& &  \vGa^{10} =  -i \sigma_0 \otimes \sigma_0 \otimes \sigma_2 \otimes \sigma_1 \otimes \sigma_0 \otimes \sigma_2 \otimes \sigma_1 \otimes \sigma_2\otimes \sigma_3\otimes \sigma_0 , \nn \\
& & \vGa^{11} =  -i \sigma_0 \otimes \sigma_0 \otimes \sigma_2 \otimes\sigma_2 \otimes \sigma_0 \otimes \sigma_0 \otimes  \sigma_0 \otimes \sigma_0\otimes \sigma_3\otimes \sigma_0 , \nn \\
& & \vGa^{12} =  -i \sigma_0 \otimes \sigma_0 \otimes \sigma_2 \otimes\sigma_1 \otimes \sigma_1 \otimes \sigma_3 \otimes  \sigma_0 \otimes \sigma_0\otimes \sigma_3\otimes \sigma_0 , \nn \\
& & \vGa^{13} =  -i \sigma_0 \otimes \sigma_0 \otimes \sigma_2 \otimes\sigma_1 \otimes \sigma_2 \otimes \sigma_3 \otimes  \sigma_0 \otimes \sigma_0\otimes \sigma_3\otimes \sigma_0 , \nn \\
& & \vGa^{14} =  -i \sigma_0 \otimes \sigma_0 \otimes \sigma_2 \otimes \sigma_1 \otimes \sigma_3 \otimes  \sigma_3 \otimes  \sigma_0 \otimes \sigma_0\otimes \sigma_3\otimes \sigma_0 , \nn \\ 
& &  \vGa^{15} =\; i\sigma_1 \otimes \sigma_2 \otimes \sigma_1\otimes \sigma_0 \otimes \sigma_0 \otimes  \sigma_0 \otimes  \sigma_0 \otimes \sigma_0\otimes \sigma_3\otimes \sigma_0 , \nn \\
& &  \vGa^{16} =\; i\sigma_1 \otimes \sigma_2 \otimes \sigma_2 \otimes \sigma_3 \otimes \sigma_0 \otimes  \sigma_0 \otimes  \sigma_0 \otimes \sigma_0\otimes \sigma_3\otimes \sigma_0, \nn \\
& &  \vGa^{17} =\; i\sigma_1 \otimes \sigma_2 \otimes \sigma_3\otimes \sigma_0 \otimes \sigma_0 \otimes  \sigma_0 \otimes  \sigma_0 \otimes \sigma_0\otimes \sigma_3\otimes \sigma_0 , \nn \\
& & \vGa^{18} = \; i \sigma_1 \otimes \sigma_1 \otimes \sigma_0 \otimes \sigma_3 \otimes \sigma_0 \otimes  \sigma_0 \otimes  \sigma_0 \otimes \sigma_0\otimes \sigma_3\otimes \sigma_0 , \nn \\
& & \vGa^{19} = \; i \sigma_1 \otimes \sigma_3 \otimes \sigma_0 \otimes \sigma_3 \otimes \sigma_0 \otimes  \sigma_0 \otimes  \sigma_0 \otimes \sigma_0\otimes \sigma_3\otimes \sigma_0 , \nn \\
& & \vGa^{20} = \; i \sigma_2 \otimes \sigma_0 \otimes \sigma_0 \otimes \sigma_3 \otimes \sigma_0 \otimes  \sigma_0 \otimes  \sigma_0 \otimes \sigma_0\otimes \sigma_3\otimes \sigma_0 \equiv \tvGa, \nn \\
& &  \tilde{\gamma}_{21} =\;\; \sigma_3\otimes \sigma_0\otimes \sigma_0\otimes \sigma_3 \otimes \sigma_0 \otimes  \sigma_0 \otimes  \sigma_0 \otimes \sigma_0\otimes \sigma_3\otimes \sigma_0 \nn \\
& & \quad \;\;  = \;\; \sigma_3\otimes\sigma_0\otimes  \hat{\gamma}_{17}=  \sigma_3\otimes\sigma_0\otimes \sigma_0\otimes \hat{\gamma}_{15} . 
\ee
Such a hyperunified qubit-spinor field $\fPsi_{\fQH}(x)$ satisfies the following self-conjugated chiral conditions:
\be
& &  \fPsi_{\fQH}^{c}(x) = \fC_{\fQH}\bar{\fPsi}_{\fQH}^{T}(x) =  \fPsi_{\fQH}(x) ,\quad \tilde{\gamma}_{21} \fPsi_{\fQH}(x) = - \fPsi_{\fQH}(x), \nn \\
& & \fC_{\fQH} = \sigma_0\otimes  \sigma_0\otimes C_{\fQU} = \sigma_0\otimes  \sigma_0\otimes \sigma_0\otimes C_{\fQG} \nn \\ 
& & =  \sigma_0\otimes  \sigma_0\otimes \sigma_0\otimes \sigma_1 \otimes \sigma_2 \otimes  \sigma_2 \otimes  \sigma_0 \otimes \sigma_0\otimes C_D , 
\ee
with $C_D= - i\sigma_3\times \sigma_2$ defined as the usual charge conjugation matrix in four-dimensional spacetime.


\subsection{Scaling properties of biframe hyper-spacetime and fundamental fields with scaling gauge invariant hyperspin gravigauge field and field strength }

To characterize the basic properties of fundamental fields and biframe hyper-spacetime formed from globally flat Minkowski hyper-spacetime and locally flat gravigauge hyper-spacetime, we are going to assign {\it global and local scaling charges} for all fundamental fields and biframe hyper-spacetime bases. 

The {\it global scaling charge} $\cC_s$ is defined as the power of scaling factor $\lambda$ under the following global conformal scaling transformation: 
\be
& & x^{\fM} \to x^{'\fM} = \lambda^{-1} x^{\fM}, \quad \cC_s = -1, \nn \\
& & \fPsi_{\fQH}(x) \to \fPsi'_{\fQH}(x') = \lambda^{\frac{D_h-1}{2}} \fPsi_{\fQH}(x), \quad \cC_s = \frac{D_h-1}{2} ,\nn \\
& & \cA_{\fM}(x) \to  \cA'_{\fM}(x') = \lambda \cA_{\fM}(x), \quad \mW_{\fM}(x) \to  \mW'_{\fM}(x') = \lambda \mW_{\fM}(x),  \quad \cC_s = 1, \nn \\
& & [d^{D_h}x] \to [d^{D_h}x'] = \lambda^{-D_h} [d^{D_h}x], \quad \cC_s = -D_h .
\ee
Similarly, the {\it local scaling charge} $\ckcC_s$ is defined as the power of scaling factor $\xi(x)$ under the following local scaling gauge transformation:
\be
& & \chi_{\fM}^{\;\;\fA}(x) \to  \chi_{\fM}^{'\;\;\fA}(x) =  \xi^{-1}(x) \chi_{\fM}^{\;\;\fA}(x) , \quad \ckcC_s = -1, \nn \\
& & \fPsi_{\fQH}(x) \to \fPsi'_{\fQH}(x) = \xi^{\frac{D_h-1}{2}}(x) \fPsi_{\fQH}(x), \quad \ckcC_s = \frac{D_h-1}{2} ,\nn \\
& & \cA_{\fM}(x) \to  \cA'_{\fM}(x) =  \cA_{\fM}(x),  \quad  \mW_{\fM}(x) \to  \mW'_{\fM}(x) =  \mW_{\fM}(x), \quad \ckcC_s = 0 , \nn \\
& & \chih_{\fA}^{\;\;\fM}(x) \to  \chih_{\fA}^{'\;\;\fM}(x) =  \xi(x) \chih_{\fA}^{\;\;\fM}(x) , \quad \ckcC_s = 1, 
\nn \\
& & \chi(x)\equiv \det  \chi_{\fM}^{\;\;\fA}(x) \to  \chi'(x) =  \xi^{-D_h}(x) \chi(x) , \quad \ckcC_s = -D_h .
\ee
The fields and operators with both global and local scaling charges are presented as follows: 
\be
& & \phi_p(x) \to \phi'_p(x') =  \lambda \phi_p(x), \quad \phi_p(x) \to \phi'_p(x) =  \xi(x) \phi_p(x), \quad \cC_s = 1, \; \ckcC_s = 1, \nn \\
& & \phi(x) \to  \phi'(x') =  \lambda\, \phi(x), \quad \phi(x) \to  \phi'(x) =  \xi(x) \phi(x), \quad \cC_s = 1,\; \ckcC_s = 1, \nn \\
& & \delta\zeta^{\fA}\equiv \chi_{\fM}^{\;\; \fA}(x) dx^{\fM} \to \delta\zeta^{' \fA} = \xi^{-1}(x)\lambda^{-1}  \delta\zeta^{\fA} , \quad \cC_s= - 1, \;\; \ckcC_s = -1 , \nn \\
& & \eth_{\fA} \equiv \chih_{\fA}^{\;\; \fM}(x)\p_{\fM} \to \eth'_{\fA} = \xi(x)\lambda\,  \eth_{\fA} , \quad \cC_s=1, \;\; \ckcC_s = 1  , \nn \\
& & [\delta^{D_h}\zeta]  \to [\delta^{D_h}\zeta'] =  \xi^{-D_h}(x)  \lambda^{-D_h} [\delta^{D_h}\zeta], \quad \cC_s = -D_h, \; \ckcC_s = -D_h, 
\ee
where $[\delta^{D_h}\zeta]\equiv [d^{D_h}x]\, \chi(x)$ is regarded as the hyper-gravicoordinate integral measure in locally flat gravigauge hyper-spacetime.

It is noticed that the inhomogeneous hyperspin gauge field $\bvcA_{\fM}(x)=\cA_{\fM}(x)+\ckcA_{\fM}(x)$ and scaling gauge field $\mW_{\fM}(x)$ are in general invariant under the local scaling transformation, which has zero local scaling charge $\ckcC_s=0$. Nevertheless, when the hyperspin gauge field is decomposed into the hyperspin gravigauge field and covariant-gauge field as shown in Eq.(\ref{HSGFDC}), the hyperspin gravigauge field $\mOm_{\fM}^{\fA\fB}(x)$ in Eq.(\ref{HSGGF}) is no longer local scaling gauge invariant, so that the hyperspin covariant-gauge field $\mA_{\fM}^{\fA\fB}(x)$ is also not local scaling gauge invariant. It is appropriate to decompose the hyperspin gauge field into local scaling gauge invariant hyperspin gravigauge field $\fOm_{\fM}^{\fA\fB}(x)$ and covariant-gauge field $\fA_{\fM}^{\fA\fB}(x)$ as follows:
\be \label{HSGGF1}
& & \cA_{\fM}^{\fA\fB}(x) = \fOm_{\fM}^{\fA\fB}(x) + \fA_{\fM}^{\fA\fB}(x), \nn \\
& & \fOm_{\fM}^{\fA\fB} = \frac{1}{2}\left( \hat{\chi}^{\fA\fN} \fG_{\fM\fN}^{\fB} - \hat{\chi}^{\fB\fN} \fG_{\fM\fN}^{\fA} -  \hat{\chi}^{\fA\fP}  \hat{\chi}^{\fB\fQ}  \fG_{\fP\fQ}^{\fC} \chi_{\fM \fC } \right)  \nn \\
& & \qquad \;\, \equiv \frac{1}{2}\left( \hat{\fA}^{\fA\fN} \mF_{\fM\fN}^{\fB} - \hat{\fA}^{\fB\fN} \mF_{\fM\fN}^{\fA} -  \hat{\fA}^{\fA\fP}  \hat{\fA}^{\fB\fQ}  \mF_{\fP\fQ}^{\fC} \fA_{\fM \fC } \right) ,
\ee
where $\fG_{\fM\fN}^{\fA}(x)$ is local scaling gauge covariant field strength defined in Eq.(\ref{HGGFS0}) for the hyper-gravigauge field $\chi_{\fM}^{\;\;\fA}(x)$, which is initiated from the local scaling invariant and $\cW_e$-spin invariant-gauge field $\fA_{\fM}^{\;\;\fA}(x)$ with the following relations:
\be \label{HSGGF2}
& & \fA_{\fM}^{\;\;\fA}(x) \equiv  \phi(x) \chi_{\fM}^{\;\;\fA}(x) , \quad \hat{\fA}_{\fA}^{\;\;\fM}(x) \equiv  \phi^{-1}(x) \chih_{\fA}^{\;\;\fM}(x), \nn \\
& & \mF_{\fM\fN}^{\fA}(x)=  \p_{\fM}\fA_{\fN}^{\fA}(x) -  \p_{\fN}\fA_{\fM}^{\fA}(x) \equiv \phi(x) \fG_{\fM\fN}^{\fA}(x) ,
\ee
where $\hat{\fA}_{\fA}^{\;\;\fM}(x)$ is viewed as {\it dual $\cW_e$-spin invariant-gauge field}. It becomes manifest that the field strength $\mF_{\fM\fN}^{\fA}(x)$ is both local scaling invariant and $\cW_e$-spin gauge invariant but no longer hyperspin gauge covariant. We have used the different letter style to distinguish $\mF_{\fM\fN}^{\fA}(x)$ from inhomogeneous hyperspin gauge covariant and local scaling invariant field strength $\fF_{\fM\fN}^{\fA}(x)$ defined in Eq.(\ref{TCSFS}). 

In terms of the local scaling invariant hyperspin gravigauge field $\fOm_{\fM}^{\fA\fB}(x)$ and covariant-gauge field $\fA_{\fM}^{\fA\fB}(x)$, their corresponding local scaling invariant field strengths are defined as follows:
\be
& & \cF_{\fM\fN}^{\fA\fB}(x) \equiv  \fR_{\fM\fN}^{\fA\fB}(x) + \fF_{\fM\fN}^{\fA\fB}(x) , \nn \\
& & \fR_{\fM\fN}^{\fA\fB} = \p_{\fM} \fOm_{\fN}^{\fA\fB} - \p_{\fN} \fOm_{\fM}^{\fA\fB} + \fOm_{\fM \fC}^{\fA} \fOm_{\fN}^{\fC \fB} -  \fOm_{\fN \fC}^{\fA} \fOm_{\fM}^{\fC \fB},  
\nn \\  
& & \fF_{\fM\fN}^{\fA\fB} =  \fD_{\fM} \fA_{\fN}^{\fA\fB} - \fD_{\fN} \fA_{\fM}^{\fA\fB} +  \fA_{\fM \fC}^{\fA} \fA_{\fN}^{\fC \fB} -  \fA_{\fN \fC}^{\fA} \fA_{\fM}^{\fC \fB}  , \nn \\
& & \fD_{\fM} \fA_{\fN}^{\fA\fB} = \p_{\fM}  \fA_{\fN}^{\fA\fB}  +  \fOm_{\fM \fC}^{\fA} \fA_{\fN}^{\fC \fB} + \fOm_{\fM \fC}^{\fB} \fA_{\fN}^{\fA \fC}  ,
\ee
It is easy to check that both $\fR_{\fM\fN}^{\fA\fB}$ and $\fF_{\fM\fN}^{\fA\fB}(x)$ become local scaling gauge invariant field strengths.


\subsection{ Hyperunified field theory for fundamental building block and fundamental interaction based on gauge invariance principle and scaling invariance hypothesis }

We now turn to build a whole hyperunified field theory for the category-1 entangled decaqubit-spinor field as hyperunified qubit-spinor field with inhomogeneous hyperspin gauge symmetry WS(1,$D_h$-1) and global inhomogeneous Lorentz-type/Poincar\'e-type group symmetry PO(1,$D_h$-1) in $D_h=19$ dimensional biframe hyper-spacetime. 

Considering the hyperunified qubit-spinor field $\fPsi_{\fQH}(x)$ in 19-dimensional hyper-spacetime as fundamental building block which involves 512 independent degrees of freedom in the qubit-spinor representation of Hilbert space, and taking the inhomogeneous hyperspin symmetry WS(1,18) as gauge symmetry by following along the gauge invariance principle, we are led to introduce the inhomogeneous hyperspin gauge field $\bvcA_{\fM}(x)$ which mediates the fundamental interaction. Such a gauge field concerns both hyperspin gauge field $\cA_{\fM}^{\fA\fB}(x)$ and $\cW_e$-spin invariant-gauge field $\fA_{\fM}^{\fA}(x)$, which belong to the adjoint representation and vector representation of hyperspin gauge symmetry SP(1,18), respectively. The global and local scaling charges corresponding to global scaling symmetry SC(1) and scaling gauge symmetry SG(1) allow us to identify the graviscalar field $\phi(x)$ and hyper-gravigauge field $\chi_{\fM}^{\fA}(x)$ from the $\cW_e$-spin invariant-gauge field $\fA_{\fM}^{\fA}(x) \equiv \phi(x) \chi_{\fM}^{\fA}(x)$ as well as $\cQ_c$-spin scalar field $\phi_1(x)$ relevant to the $\cQ_c$-spin charge $\cQ_c=q_c=1$. In addition, the scaling gauge symmetry SG(1) brings on the scaling gauge field $\mW_{\fM}(x)$. For convenience, we should list all relevant fields as follows:
\be
\fPsi_{\fQH}(x), \; \cA_{\fM}^{\fA\fB}(x), \; \chi_{\fM}^{\fA}(x), \; \phi(x),  \; \phi_1(x), \; \mW_{\fM}(x), \nn
\ee
which are all basic fields concerned in establishing a whole hyperunified field theory. The gravitational origin of gauge symmetry indicates that the inhomogeneous hyperspin gauge symmetry is governed by hyper-gravigauge field $\chi_{\fM}^{\fA}(x)$ via the hyperspin gravigauge field $\fOm_{\fM}^{\fA\fB}(x)$ defined in Eq.(\ref{HSGGF}) and also by $\cW_e$-spin gravigauge field  $\fOm_{\fM}^{\fA}(x)$ given in Eq.(\ref{TESGGF}).

The gauge invariant and scaling invariant action for hyperunified qubit-spinor field $\fPsi_{\fQH}(x)$ in category-1 can be read from Eq.(\ref{actionfQESGI}) in 19-dimensional biframe hyper-spacetime. In constructing the action in Eq.(\ref{actionfQESGI}), the scaling invariance for both global and local scaling symmetries plays a crucial role in addition to the gauge invariance principle. To construct the basic action for all fundamental fields, the scaling symmetry is still supposed to play an essential role. For that, we are motivated to make a {\it scaling invariance hypothesis} which postulates that the action is built to keep scaling invariance for both global and local scaling symmetries in addition to the inhomogeneous hyperspin gauge symmetry by following along gauge invariance principle. Furthermore, it is supposed that the basic gauge field and scalar field also obey the simplest motion postulate that they all involve just the lowest quadratic derivative in the gauge invariant and scaling invariant action.

Following along the scaling invariance hypothesis and gauge invariance principle together with the maximum coherence motion principle and maximum entangled quibit-spinor principle, we obtain definitively the following basic action for a whole hyperunified field theory:
\be  \label{actionHUFT}
\cS_{\mH\mU} & \equiv & \int [d^{D_h}x] \, \chi(x)\,  \{\, \bar{\fPsi}_{\fQH} \vSi_{-}^{\fA}\chih_{\fA}^{\;\, \fM} i\cD_{\fM}  \fPsi_{\fQH} - \phi_1 \bar{\fPsi}_{\fQH} \tvSi_{-}\fPsi_{\fQH} \nn \\
& + & \phi^{D_h-4} [\, - \chih_{\fC}^{\;\fM}\chih_{\fD}^{\;\fN} \chih_{\fC'}^{\;\fM'}\chih_{\fD'}^{\;\fN'}   (\,  \hat{\eta}^{\fC\fD\fC'\fD'}_{\fA\fB\fA'\fB'} g_H^{-2}  \frac{1}{4}  \cF_{\fM\fN}^{\fA\fB} \cF_{\fM'\fN'}^{\fA'\fB'}  \nn \\
&  + & \tilde{\eta}^{\fC\fD\fC'\fD'}_{\fA\fA'} g_H^{-2} \frac{1}{4}  \mF_{\fM\fN}^{\fA}\mF_{\fM'\fN'}^{\fA'}   - \hat{\eta}^{\fC\fD\fC'\fD'}  g_W^{-2}  \frac{1}{4} \mF_{\fM\fN} \mF_{\fM'\fN'}\, )  \nn \\
& + &  \frac{1}{2}  \chih_{\fC}^{\;\fM}\chih_{\fD}^{\;\fN} \eta^{\fC\fD} g_H^{-2}\beta_G^2 \phi_1^2  (\cA_{\fM}^{\fA\fB} - \fOm_{\fM}^{\fA\fB})(\cA_{\fN\fA\fB} - \fOm_{\fN\fA\fB}) \nn \\
& + &  \frac{1}{2} \chih_{\fC}^{\;\fM}\chih_{\fD}^{\;\fN} \eta^{\fC\fD}  (\lambda_S^2 D_{\fM} \phi D_{\fN}\phi +  \lambda_Q^2 D_{\fM} \phi_1 D_{\fN}\phi_1 ) - \lambda_D^2 \phi_1^{4} \mF(\frac{\phi_1}{\phi}) \, ] \, \} ,
\ee
with $\fA, \fM= 0,1,2,3,5, \cdots, D_h$ and $D_h=19$. Where $g_H$ and $g_W$ are two gauge coupling constants, $\beta_G$, $\lambda_S$, $\lambda_Q$ and $\lambda_D$ are scalar coupling parameters and $\mF(\frac{\phi_1}{\phi})$ represents a general scaling invariant potential relevant to two scalar fields. $\vSi_{-}^{\fA}$ and $\tvSi_{-}$ matrices are defined via the $\vGa$-matrices $\vGa^{\hfA}= (\vGa^a, \vGa^A, \vGa^{20}=\tvGa ) $ ($\hfA=0,1,2,3,5,\cdots, 20$) presented in Eq.(\ref{GMffQH2}). All definitions in the basic action are already given in previous relevant sections, we represent them as follows just for convenience:
\be
& & i\cD_{\fM} \equiv   i\p_{\fM} + \cA_{\fM}(x), \quad 
\cA_{\fM}(x) = \cA_{\fM}^{\fB\fC}(x) \frac{1}{2}\vSi_{\fB\fC} , \nn \\
& & \fOm_{\fM}^{\fA\fB} = \frac{1}{2}\left( \hat{\chi}^{\fA\fN} \fG_{\fM\fN}^{\fB} - \hat{\chi}^{\fB\fN} \fG_{\fM\fN}^{\fA} -  \hat{\chi}^{\fA\fP}  \hat{\chi}^{\fB\fQ}  \fG_{\fP\fQ}^{\fC} \chi_{\fM \fC } \right)  \nn \\
& & D_{\fM}\phi = (\p_{\fM} + \mW_{\fM})\phi, \quad D_{\fM}\phi_1 = (\p_{\fM} + \mW_{\fM})\phi_1
\ee
for the covariant derivatives and gauge fields, and 
\be \label{GCFS}
& & \cF_{\fM\fN}^{\fA\fB} =  \p_{\fM} \cA_{\fN}^{\fA\fB} - \p_{\fN} \cA_{\fM}^{\fA\fB} +  \cA_{\fM \fC}^{\fA} \cA_{\fN}^{\fC \fB} -  \cA_{\fN \fC}^{\fA} \cA_{\fM}^{\fC \fB}  , \nn \\
& & \mF_{\fM\fN}^{\fA} =  \p_{\fM}\fA_{\fN}^{\;\fA} -  \p_{\fN}\fA_{\fM}^{\; \fA} = \phi^2\fG_{\fM\fN}^{\fA}, \nn \\
& & \fG_{\fM\fN}^{\fA} =  d_{\fM}\chi_{\fN}^{\;\fA} -  d_{\fN}\chi_{\fM}^{\; \fA} , \quad d_{\fM} \equiv \p_{\fM} + \p_{\fM}\ln \phi , \nn \\
& &  \mF_{\fM\fN} =  \p_{\fM}\mW_{\fN} -  \p_{\fN}\mW_{\fM},  
\ee
for the gauge covariant field strengths. 

The constant tensor factors involved in the action are given by   
\be \label{GItensor}
& & \hat{\eta}^{\fC\fD\fC'\fD'}_{\fA\fB\fA'\fB'} \equiv  \eta^{\fC\fC'} \eta^{\fD\fD'} \eta_{\fA\fA'} \eta_{\fB\fB'}   , \quad 
\hat{\eta}^{\fC\fD\fC'\fD'} \equiv \eta^{\fC\fC'} \eta^{\fD\fD'} , \nn \\
& & \tilde{\eta}^{\fC\fD\fC'\fD'}_{\fA\fA'}  \equiv  \hat{\eta}^{\fC\fD\fC'\fD'}_{\fA\fA'}    - \hat{\eta}^{\fC\fD\1 \fC'\fD'}_{(\fA\fA')}  + 3 \hat{\eta}^{\fC\fD\1 \fC'\fD'}_{[\fA\fA']},
\ee
with the definitions,
\be \label{SAtensor}
& & \hat{\eta}^{\fC\fD\1 \fC'\fD'}_{(\fA\fA')} \equiv \eta^{\fC\fC'} \frac{1}{2} (  \eta_{\fA}^{\; \fD'} \eta_{\fA'}^{\; \fD} + \eta_{\fA}^{\; \fD} \eta_{\fA'}^{\; \fD'} )  + \frac{1}{2} ( \eta_{\fA}^{\; \fC'} \eta_{\fA'}^{\; \fC} + \eta_{\fA}^{\; \fC} \eta_{\fA'}^{\; \fC'} ) \eta^{\fD\fD'} , \nn \\
& &  \hat{\eta}^{\fC\fD\1 \fC'\fD'}_{[\fA\fA']} \equiv \eta^{\fC\fC'} \frac{1}{2}( \eta_{\fA}^{\; \fD'} \eta_{\fA'}^{\; \fD} - \eta_{\fA}^{\; \fD} \eta_{\fA'}^{\; \fD'} ) +  \frac{1}{2} (\eta_{\fA}^{\; \fC'} \eta_{\fA'}^{\; \fC} - \eta_{\fA}^{\; \fC} \eta_{\fA'}^{\; \fC'}  ) \eta^{\fD\fD'} , \nn \\
& & \hat{\eta}^{\fC\fD\fC'\fD'}_{\fA\fA'}  \equiv \eta^{\fC\fC'} \eta^{\fD\fD'} \eta_{\fA\fA'} .
\ee

The tensor factor $\tilde{\eta}^{\fC\fD\fC'\fD'}_{\fA\fA'} $ may be called as {\it hypespin gauge invariance tensor}\cite{HUFT} as it brings the dynamic term of Abelian-type hyper-gravigauge field strength 
\be
\tilde{\eta}^{\fC\fD\fC'\fD'}_{\fA\fA'} \mF_{\fM\fN}^{\fA}\mF_{\fM'\fN'}^{\fA'},
\ee
to be gauge invariant under hyperspin gauge transformation although the Abelian-type hyper-gravigauge field strength $\mF_{\fM\fN}^{\fA}$ is not a gauge covariant one. The quadratic interaction term between gauge field and $\cQ_c$-spin scalar field $\phi_1$ is manifestly gauge invariant as it concerns the difference between hyperspin gauge field and hyperspin gravigauge field, i.e., $(\cA_{\fM}^{\fA\fB} - \fOm_{\fM}^{\fA\fB})$, so that their inhomogeneous gauge transformations cancel each other. Note that such a quadratic term appears as a mass-like term of gauge field.

The action in Eq.(\ref{actionHUFT}) possesses the following {\it joint symmetry}:
\be \label{GLSS}
G_S= SC(1)\ltimes PO(1,18)\Join WS(1,18)\rtimes SG(1) ,
\ee
where SC(1) and SG(1) represent the {\it global scaling symmetry} and {\it local scaling gauge symmetry}, respectively. $\mF_{\fM\fN}$ in Eq.(\ref{GCFS}) defines the field strength of scaling gauge field $\mW_{\fM}$. PO(1,18) remains global inhomogeneous Lorentz-type/Poincar\'e-type group symmetry and WS(1,18)  represents inhomogeneous hyperspin gauge symmetry, which brings about a joint global and local symmetry in biframe hyper-spacetime. Only when locally flat gravigauge hyper-spacetime goes to be globally flat one, then the motion-correlation inhomogeneous hyperspin symmetry WS(1,18) becomes to be associated with inhomogeneous Lorentz-type/Poincar\'e-type group symmetry PO(1,18), so that the group transformations of SP(1,18) and SO(1,18) have to be coincidental, which is realized from their isomorphic property SP(1,18)$\cong$SO(1,18).


\section{Hyperunified field theory in hidden scaling gauge formalism with gauge-gravity correspondence and the dynamics of fundamental fields with conserved currents }

The action of hyperunified field theory presented in Eq.(\ref{actionHUFT}) is fully built based on the gauge invariance principle and scaling invariance hypothesis, which enables us to demonstrate the gauge-gravity correspondence. As such a hyperunified field theory is formulated in biframe hyper-spacetime, we are able to derive equations of motion for all fundamental fields by following along the least action principle and obtain various conserved currents with respect to gauge symmetries. 


\subsection{Hyperunified field theory in hidden scaling gauge formalism with gauge-gravity correspondence and the kinetic motion of fundamental fields as the source of gravitational interaction }

It is noticed from the action in Eq.(\ref{actionHUFT}) that the $\cW_e$-spin gauge field $\ckcA_{\fM}^{\;\;\fA}$ does not interact with the hyperunified qubit-spinor field $\fPsi_{\fQH}$ via the gauge covariant derivative as like the usual case just following along the gauge invariance principle, this is because the group generators of $\cW_e$-spin gauge symmetry get the nilpotent condition $(\vSi_{-}^{\fA})^2=0$. In another word, the $\cW_e$-spin gauge interaction via the gauge covariant derivative becomes vanishing, the reason is that such an interaction term brings on the change of chirality for the hyperunified qubit-spinor field $\fPsi_{\fQH}$ which is an entangled hyperqubit-spinor field with self-conjugated chiral structure. Instead, the $\cW_e$-spin gauge field gets to interact with the hyperunified qubit-spinor field $\fPsi_{\fQH}$ through coupling the dual $\cW_e$-spin invariant-gauge field $\hat{\fA}_{\fA}^{\;\;\fM}\equiv \phi^{-1}\chih_{\fA}^{\;\;\fM}$ to the covariant derivative operator $\cD_{\fM}$ as kinetic term of hyperunified qubit-spinor field $\fPsi_{\fQH}$, i.e.: 
\be
\vSi_{-}^{\fA} \hat{\fA}_{\fA}^{\;\;\fM}\cD_{\fM}\equiv  \hmH^{\fM\fN} \fA_{\fM}^{\;\;\fA} \vSi_{- \fA} \cD_{\fN} , 
\ee
where we have introduced the scaling gauge invariant dual tensors,
\be \label{DTF}
& & \hfA_{\fA}^{\;\fM}  \fA_{\fN}^{\; \fB}  \eta_{\fM}^{\; \fN} = \eta_{\fA}^{\; \fB} , \quad 
 \hfA_{\fA}^{\;\fM}  \fA_{\fN}^{\; \fB}  \eta_{\fB}^{\; \fA} = \eta_{\fN}^{\; \fM} ,  \nn \\
 & &\hmH^{\fM\fN} \equiv \hfA_{\fC}^{\;\fM}\hfA_{\fD}^{\;\fN} \eta^{\fC\fD} \equiv \phi^{-2} \chih^{\fM\fN} ,  \nn \\
& & \hmH^{\fM\fP} \mH_{\fP\fN} = \eta^{\fM}_{\; \; \fN}, \quad \mH_{\fM\fP}  \hmH^{\fP\fN} = \eta_{\fM}^{\; \; \fN} , \nn \\
& & \mH_{\fM\fN} \equiv \fA_{\fM}^{\;\fC}\fA_{\fN}^{\;\fD} \eta_{\fC\fD} \equiv \phi^{2} \chi_{\fM\fN} . 
\ee
Geometrically, the dual tensors $\mH_{\fM\fN}$ ($\chi_{\fM\fN}$) and $\hmH^{\fM\fN}$ ($\chih^{\fM\fN}$) are referred to as dual {\it hyper-gravimetric tensor fields}. 

In the whole hyperunified field theory presented in Eq.(\ref{actionHUFT}), the bicovariant vector field $\chih_{\fA}^{\;\;\fM}$ corresponding to the dual $\cW_e$-spin invariant-gauge field $\chih_{\fA}^{\;\;\fM}\equiv \phi \hat{\fA}_{\fA}^{\;\;\fM}$ couples to the covariant derivative operator $\cD_{\fM}$, which arises not only from supplying both the local hyperspin gauge symmetry and global Lorentz-type group symmetry, but also from preserving the local scaling gauge symmetry. Therefore, the local scaling gauge invariance plays a crucial role in the building of a whole hyperunified field theory.

The gauge symmetry often involves redundant degrees of freedom, which demands to make a gauge prescription with setting up gauge fixing condition. In order to eliminate the redundant degree of freedom arising from local scaling gauge symmetry, let us rewrite the action given in Eq.(\ref{actionHUFT}) into the following formalism:
\be  \label{actionHUFTHSG}
\cS_{\mH\mU} & \equiv & \int [d^{D_h}x] \, \fkA(x)\, \fkL  \nn \\
& = & \int [d^{D_h}x] \, \fkA(x)\,  \{\,  \hmH^{\fM\fN} \bar{\fPsi}_{\fQH} \fA_{\fM\fA}  \vSi_{-}^{\fA} i\cD_{\fN}  \fPsi_{\fQH}  - \beta_Q \sinh\chi_s\,  \bar{\fPsi}_{\fQH} \tvSi_{-}\fPsi_{\fQH} \nn \\
& - &  \frac{1}{4} \hfA_{\fC}^{\;\fM}\hfA_{\fD}^{\;\fN} \hfA_{\fC'}^{\;\fM'}\hfA_{\fD'}^{\;\fN'} [ \hat{\eta}^{\fC\fD\fC'\fD'}_{\fA\fB\fA'\fB'} g_H^{-2} \cF_{\fM\fN}^{\fA\fB}  \cF_{\fM'\fN'}^{\fA'\fB'} - \tilde{\eta}^{\fC\fD\fC'\fD'}_{\fA\fA'} g_H^{-2} \mF_{\fM\fN}^{\fA}\mF_{\fM'\fN'}^{\fA'} \nn \\ 
& - & \bar{\eta}^{\fC\fD\fC'\fD'}_{\fA\fA'} g_H^{-2}\beta_G^2 \beta_Q^2 \sinh^2\chi_s\, \fF_{\fM\fN}^{\fA}\fF_{\fM'\fN'}^{\fA'}  +  \hat{\eta}^{\fC\fD\fC'\fD'} g_W^{-2}\mF_{\fM\fN} \mF_{\fM'\fN'} ] \nn \\
& + &  \frac{1}{2} \lambda_S^2( 1 + \sinh^2\chi_s ) \, \hmH^{\fM\fN} \mW_{\fM} \mW_{\fN}  +  \frac{1}{2} \hmH^{\fM\fN} \lambda_S^2 \p_{\fM} \chi_s \p_{\fN}\chi_s  - \lambda_D^2 \cF(\chi_s)  \, \} ,
\ee
with the definitions,
\be \label{DTM}
& & \fkA(x) \equiv \det \fA_{\fM}^{\;\;\fA} = \phi^{D_h} \det \chi_{\fM}^{\;\;\fA} = \phi^{D_h}(x) \chi(x), \nn \\& & \hat{\fkA}(x) \equiv \det \hfA_{\fA}^{\;\;\fM} = \phi^{-D_h} \det \chih_{\fA}^{\;\;\fM} = \phi^{-D_h}(x) \chih(x) = 1/\fkA(x) , 
\ee
and 
\be \label{GGF}
& & \mF_{\fM\fN}^{\fA} =  \p_{\fM}\fA_{\fN}^{\fA} -  \p_{\fN}\fA_{\fM}^{\fA} \equiv \phi(x) \fG_{\fM\fN}^{\fA}, \nn \\
 & & \fF_{\fM\fN}^{\fA} = \cD_{\fM}\fA_{\fN}^{\fA} -  \cD_{\fN}\fA_{\fM}^{\fA} = \mF_{\fM\fN}^{\fA} + \cA_{\fM\fB}^{\fA} \fA_{\fN}^{\fB} - \cA_{\fN\fB}^{\fA} \fA_{\fM}^{\fB} ,
 \ee
where $\fF_{\fM\fN}^{\fA}$ defines gauge covariant field strength under hyperspin gauge transformations, which is also local scaling gauge invariant field strength.
 
 We have also made the following definitions:
 \be
 & &  \fPsi_{\fQH} \to \phi^{\frac{D_h-1}{2}} \fPsi_{\fQH}, \nn \\
 & & \mW_{\fM} \to \mW_{\fM} - \frac{1}{2}\p_{\fM}\ln (\lambda_S^2 \phi^2 + \lambda_Q^2\phi_1^2) , \nn \\
 & & \frac{\lambda_S^2}{1 + \frac{\lambda_Q^2\phi_1^2}{\lambda_S^2\phi^2} } \p_{\fM}(\frac{\lambda_Q\phi_1}{\lambda_S\phi}) \p_{\fM}(\frac{\lambda_Q\phi_1}{\lambda_S\phi})  \equiv \lambda_S^2 \p_{\fM}\chi_s(x)\p_{\fN}\chi_s(x) ,
 \ee
where the explicit relation between the field $\chi_s(x)$ and the ratio $\phi_1(x)/\phi(x)$ is obtained as follows:
 \be 
 \chi_s(x) & \equiv & \sinh^{-1}\frac{\lambda_Q\phi_1(x)}{\lambda_S \phi(x)},  \nn \\
 \frac{\phi_1(x)}{\phi(x)}  & \equiv &  \beta_Q \sinh\chi_s(x) , 
 \quad  \beta_Q \equiv \frac{\lambda_S}{\lambda_Q}.  
 \ee
The dimensionless field $\chi_s(x)$ may be referred to as scaling invariant {\it $\cQ_c$-spin scaling field} which has zero global and local scaling charges. $\cF(\chi_s)$ is considered to be a general potential of $\cQ_c$-spin scaling field. 

The tensor factor $\bar{\eta}^{\fC\fD\1 \fC'\fD'}_{\fA\fA'}$ is defined as follows:  
\be \label{GMtensor}
\bar{\eta}^{\fC\fD\1 \fC'\fD'}_{\fA\fA'}   \equiv  \hat{\eta}^{\fC\fD\fC'\fD'}_{\fA\fA'}    + \frac{1}{2} ( \hat{\eta}^{\fC\fD\fC'\fD'}_{\fA\fA'} + \hat{\eta}^{\fC\fD\1 \fC'\fD'}_{(\fA\fA')}  + \hat{\eta}^{\fC\fD\1 \fC'\fD'}_{[\fA\fA']} ) , 
\ee
with the tensors $\hat{\eta}^{\fC\fD\fC'\fD'}_{\fA\fA'}$, $\hat{\eta}^{\fC\fD\1 \fC'\fD'}_{(\fA\fA')}$ and $\hat{\eta}^{\fC\fD\1 \fC'\fD'}_{[\fA\fA']}$ given in Eqs.(\ref{GItensor}) and (\ref{SAtensor}). Such a constant tensor $\bar{\eta}^{\fC\fD\1 \fC'\fD'}_{\fA\fA'}$ couples to the gauge covariant dynamic term of hyper-gravigauge field to reproduce the gauge invariant quadratic interaction term concerning the difference between the hyperspin gauge field $\cA_{\fM}^{\fA\fB}$ and hyperspin gravigauge field $\fOm_{\fM}^{\fA\fB}$, i.e.:
\be
\frac{1}{4}\hfA_{\fC}^{\;\fM}\hfA_{\fD}^{\;\fN} \hfA_{\fC'}^{\;\fM'}\hfA_{\fD'}^{\;\fN'} \bar{\eta}^{\fC\fD\fC'\fD'}_{\fA\fA'} \fF_{\fM\fN}^{\fA}\fF_{\fM'\fN'}^{\fA'} = \frac{1}{2}\hmH^{\fM\fN} (\cA_{\fM}^{\fA\fB} - \fOm_{\fM}^{\fA\fB}) (\cA_{\fN \fA\fB} - \fOm_{\fN\fA\fB}). \nn
\ee
It will be seen below that the gauge covariant dynamic formalism of hyper-gravigauge field becomes more convenient in deriving the equation of motion of hyper-gravigauge field. 

In the action Eq.(\ref{actionHUFTHSG}), the scaling gauge symmetry becomes a hidden symmetry. Namely, all basic fields have zero local scaling charge $\ckcC_s=0$. In addition, the redefined hyperunified qubit-spinor field $\fPsi_{\fQH}$ and $\cQ_c$-spin scaling field $\chi_s(x)$ get zero global scaling charge $\cC_s=0$. We may refer to the action in Eq.(\ref{actionHUFTHSG}) as {\it hyperunified field theory in hidden scaling gauge formalism}.

It is noticed that the graviscalar field $\phi(x)$ does not appear explicitly in the action presented in Eq.(\ref{actionHUFTHSG}). We arrive at five fundamental fields which include the hyperunified qubit-spinor field $\fPsi_{\fQH}$, the hyperspin gauge field $\cA_{\fM}^{\fA\fB}$, the $\cW_e$-spin invariant-gauge field $\fA_{\fM}^{\;\;\fA}$, the scaling gauge field $\mW_{\fM}$ and the $\cQ_c$-spin scaling field $\chi_s(x)$. Nevertheless, the independent degrees of freedom should be unchanged as the graviscalar field is considered to be absorbed into the scaling gauge field $\mW_{\fM}$, which appears to be a {\it massive-like gauge field}. 

In such a hidden scaling gauge formalism of hyperunified field theory, the dual $\cW_e$-spin invariant-gauge field $\hat{\fA}_{\fA}^{\;\;\fM}$ couples to either the coordinate derivative operator $\p_{\fM}$ that reflects the motion of fundamental fields or the fundamental gauge field that characterizes the interaction of fundamental fields. So that the $\cW_e$-spin invariant-gauge field $\fA_{\fM}^{\;\;\fA}$ as hyper-gravigauge field with the field strength $\mF_{\fM\fN}^{\fA}$ describes the fundamental gravitational interaction for all motional fields and vector fields in Minkowski hyper-spacetime.  

Therefore, we come to the statement that the kinetic motion of fundamental fields emerges as the source of gravitational interaction which is truly characterized by the gauge interaction. As a consequence, the fundamental gravitational interaction exists in all motional fields and a whole hyperunified field theory in hidden scaling gauge formalism reveals the {\it gauge-gravity correspondence}.


\subsection{Equations of motion and dynamics of fundamental fields  }

The action in Eq.(\ref{actionHUFTHSG}) for the hyperunified field theory enables us to derive equations of motion for all fundamental fields from the least action principle, which can be applied to describe the dynamics of fundamental fields and study gravitational relativistic quantum theory in hyper-spacetime. 

Let us first present the equation of motion for the hyperunified qubit-spinor field $\fPsi_{\fQH}$,
\be \label{EMfQH}
& & \vSi_{-}^{\fA} \hfA_{\fA}^{\;\; \fM}i \left(\cD _{\fM} -\fV_{\fM} \right) \fPsi_{\fQH} 
 -\beta_Q \sinh\chi_s \tvSi_{-} \fPsi_{\fQH} =   0 , 
\ee
where $\fV_{\fM}$ is regarded as {\it graviscaling induced gauge field} defined as follows: 
\be
\fV_{\fM} & \equiv & \frac{1}{2} \fkA\, \hfA_{\fB}^{\;\; \fN}\cD_{\fN}(\hat{\fkA} \fA_{\fM}^{\;\; \fB}) , 
\ee
which is local scaling gauge invariant. Where $\cD_{\fN}(\hat{\fkA} \fA_{\fM}^{\;\; \fB})$ concerns the covariant derivative for the $\cW_e$-spin invariant-gauge field $\fA_{\fM}^{\;\;\fB}(x)$,
\be \label{CD1}
& & \cD_{\fN}\fA_{\fM}^{\;\; \fB}(x)  \equiv \p_{\fN} \fA_{\fM}^{\;\; \fB}(x) 
+ \cA_{\fN\, \fC}^{\fB}(x)  \fA_{\fM}^{\;\;\fC}(x) . 
\ee

The quadratic form for the above equation of motion is found to be.
\be \label{EMfQH2}
& &\hmH^{\fM\fN} (\nabla_{\fM} - \fV_{\fM} ) ( \cD _{\fN} - \fV_{\fN} ) \Psi_{\fQH} 
\nn \\
& & \;\; = - \left( \beta_Q^2 \sinh^2\chi_s +  \vGa^{\fA} \tvGa\hfA_{\fA}^{\;\; \fM}\beta_Q \cosh\chi_s \, (i\p_{\fM}\chi_s) \,  \right) \Psi_{\fQH} \nn \\
& & \; \; + \varSigma^{\fA\fB}\hfA_{\fA}^{\;\; \fM} \hfA_{\fB}^{\;\; \fN}  \left( \cF_{\fM\fN} - \fF_{\fM\fN}^{\fC} \hfA_{\fC}^{\;\; \fP} i ( \cD _{\fP} - \fV_{\fP} ) - i \fV_{\fM\fN}\, \right) \Psi_{\fQH}, 
 \ee
with the definition for the covariant derivative,
 \be \label{CD2}
 & & \nabla_{\fM} ( \cD _{\fN}  - \fV_{\fN} ) \equiv \cD _{\fM} ( \cD _{\fN} -\fV_{\fN} ) - \cA_{\fM\fN}^{\fP} ( \cD _{\fP} -\fV_{\fP} ) ,\nn \\
 & & \cA_{\fM\fN}^{\fP}  \equiv   \hfA_{\fA}^{\;\; \fP}  \cD_{\fM}\fA_{\fN}^{\;\;\fA} =  \hfA_{\fA}^{\;\; \fP}(\p_{\fM} \fA_{\fN}^{\;\; \fA} + \cA_{\fM\, \fB}^{\fA}  \fA_{\fN}^{\;\;\fB}\, ) . 
 \ee
Where $\cF_{\fM\fN}(x)$ is the field strength of hyperspin gauge field $\cA_{\fM}(x)$ and $\fF_{\fM\fN}^{\fC}$ the field strength of $\cW_e$-spin invariant-gauge field $\fA_{\fM}^{\; \fC}(x)$ presented in Eq.(\ref{TCSFS}).  The graviscaling induced gauge field strength $\fV_{\fM\fN}$ is defined as follows: 
 \be
 \fV_{\fM\fN}  =  \p_{\fM}\fV_{\fN}-  \p_{\fN}\fV_{\fM} .
 \ee

We now turn to the equation of motion for the hyperspin gauge field $\cA_{\fM}^{\fA\fB}$, which is found to be, 
\be   \label{EMHSGF}
\cD_{\fN} \left( \fkA\,\hmH^{\fM\fM'}\hmH^{\fN\fN'}g_H^{-2}\cF_{\fM'\fN' \fA\fB}\right) =  \cJ_{\fA\fB}^{\; \fM} , 
\ee
where $\cJ_{\fA\fB}^{\;\fM}$ is bicovariant tensor current with the following explicit form:
\be \label{BCTC}
\cJ_{\fA\fB}^{\;\fM} =  \fkA\,  \frac{1}{2} \hfA_{\fC}^{\;\; \fM} 
\bar{\fPsi}_{\fQH} \{ \vSi_{-}^{\fC} \; \;  \varSigma_{\fA\fB} \} \fPsi_{\fQH} +  \fkA\, \frac{1}{2} \bhmH^{\fM\fM'\fN'}_{[\fA\fB]\fA'} g_H^{-2}\beta_G^2\beta_Q^2 \sinh^2\chi_s\, \fF_{\fM'\fN'}^{\fA'}  .
\ee
We have introduced the tensor $\bhmH^{\fM\fM'\fN'}_{[\fA\fB]\fA'}$ defined as follows:
\be \label{Tbar}
& & \bhmH^{\fM\fM'\fN'}_{[\fA\fB]\fA'} \equiv \bhmH^{[\fM\fN]\fM'\fN'}_{\fA\fA'}  \fA_{\fN\fB} - \bhmH^{[\fM\fN]\fM'\fN'}_{\fB\fA'} \fA_{\fN\fA} , \nn \\
& & \bhmH^{[\fM\fN]\fM'\fN'}_{\fA\fA'}  \equiv  \frac{1}{2} (\,  \bhmH^{\fM\fN\1 \fM'\fN'}_{\fA\fA'}  -  \bhmH^{\fN\fM\1 \fM'\fN'}_{\fA\fA'}  \, ), \nn \\
& & \bhmH^{\fM\fN\fM'\fN'}_{\fA\fA'}  \equiv  \hfA_{\fC}^{\;\fM}\hfA_{\fD}^{\;\fN} \hfA_{\fC'}^{\;\fM'}\hfA_{\fD'}^{\;\fN'} \bar{\eta}^{\fC\fD\fC'\fD'}_{\fA\fA'} .
\ee

Let us now derive the equation of motion for the $\cW_e$-spin invariant-gauge field $\fA_{\fM}^{\;\; \fA}$ as hyper-gravigauge field $\fA_{\fM}^{\;\; \fA}= \phi \chi_{\fM}^{\;\; \fA}$. The explicit form is found to be as follows:
\be  \label{EMfA}
\p_{\fN} \left( \fkA\,\thmH^{[\fM\fN]\fM'\fN'}_{\fA\fA'} g_H^{-2}\mF_{\fM'\fN' }^{\fA'} \right) + \cD_{\fN} ( \fkA\, \bhmH^{[\fM\fN]\fM'\fN'}_{\fA\fA'} g_H^{-2}\beta_G^2  \beta_Q^2 \sinh^2\chi_s\, \fF_{\fM'\fN' }^{\fA'} )= \fJ_{\fA}^{\;\; \fM}  , 
\ee
with $\fJ_{\fA}^{\;\fM}$ the bicovariant vector current given by,
\be \label{BVCfA}
 \fJ_{\fA}^{\;\; \fM} & = &  - \fkA\, \hfA_{\fA}^{\;\;\fM} \fkL +  \fkA\, \hfA_{\fA}^{\; \; \fP}  \{\,  \hfA_{\fC}^{\;\; \fM}  \bar{\fPsi}_{\fQH} \vSi_{-}^{\fC} i \cD_{\fP} \fPsi_{\fQH}     \nn \\
& - &\hmH^{[\fM\fN]\fM'\fN'}_{\fA''\fB\; \fA'\fB'} g_H^{-2} \cF_{\fP\fN}^{\fA''\fB} \cF_{\fM'\fN'}^{\fA'\fB'} + \thmH^{[\fM\fN]\fM'\fN'}_{\fA''\fA'} g_H^{-2} \mF_{\fP\fN}^{\fA''} \mF_{\fM'\fN'}^{\fA'}   \nn \\
& + &  \bhmH^{[\fM\fN]\fM'\fN'}_{\fA''\fA'}g_H^{-2}\beta_G^2 \beta_Q^2 \sinh^2\chi_s\, \fF_{\fP\fN}^{\fA''} \fF_{\fM'\fN'}^{\fA'}  +  \hmH^{\fM\fN} \lambda_S^2\p_{\fP}\chi_s  \p_{\fN} \chi_s   \nn \\
& - & g_W^{-2}\hmH^{\fM\fM'} \hmH^{\fN\fN'}\mF_{\fP\fN} \mF_{\fM'\fN'} +  \lambda_S^2(1+ \sinh^2\chi_s ) \hmH^{\fM\fN} \mW_{\fP} \mW_{\fN} \, \} ,
\ee
where the tensors $\hmH^{[\fM\fN]\fM'\fN'}_{\fA''\fB\; \fA'\fB'}$ and $\thmH^{[\fM\fN]\fM'\fN'}_{\fA\fA'}$ are defined as follows:
\be \label{Tensors}
& &\hmH^{[\fM\fN]\fM'\fN'}_{\fA''\fB\; \fA'\fB'} =  \frac{1}{2} (\,\hmH^{\fM\fN\fM'\fN'}_{\fA''\fB\; \fA'\fB'} -\hmH^{\fN\fM\fM'\fN'}_{\fA''\fB\; \fA'\fB'} ), \nn \\
& &\hmH^{\fM\fN\fM'\fN'}_{\fA''\fB\; \fA'\fB'} \equiv \hfA_{\fC}^{\;\fM}\hfA_{\fD}^{\;\fN} \hfA_{\fC'}^{\;\fM'}\hfA_{\fD'}^{\;\fN'} \hat{\eta}^{\fC\fD\fC'\fD'}_{\fA''\fB \fA'\fB'} , \nn \\
& &\thmH^{[\fM\fN]\fM'\fN'}_{\fA\fA'}  \equiv  \frac{1}{2} (\, \thmH^{\fM\fN\1 \fM'\fN'}_{\fA\fA'}  - \thmH^{\fN\fM\1 \fM'\fN'}_{\fA\fA'}  \, ), \nn \\
& &\thmH^{\fM\fN\fM'\fN'}_{\fA\fA'}  \equiv \hfA_{\fC}^{\;\fM}\hfA_{\fD}^{\;\fN} \hfA_{\fC'}^{\;\fM'}\hfA_{\fD'}^{\;\fN'} \tilde{\eta}^{\fC\fD\fC'\fD'}_{\fA\fA'} ,
\ee
and the tensor $\bhmH^{[\fM\fN]\fM'\fN'}_{\fA\fA'}$ is given in Eq.(\ref{Tbar}).

It is straightforward to obtain the equation of motion for the scaling gauge field $\mW_{\fM}$
\be   \label{EMSGF}
 \p_{\fN} \left( \fkA \,\hmH^{\fM\fM'}\hmH^{\fN\fN'} g_W^{-2} \mF_{\fM'\fN'} \right)  =   \mJ^{\fM} , 
\ee
with the vector current
\be
 \mJ^{\fM} = - \fkA\, \lambda_S^2 (1 + \sinh^2\chi_s) \,  \hmH^{\fM\fN} \mW_{\fN}  .
\ee

It is easy to write down the equation of motion for the {\it $\cQ_c$-spin scaling field},  
\be   \label{EMCSF}
\p_{\fM}  ( \fkA\,\hmH^{\fM\fN}\lambda_S^2 \p_{\fN}\chi_s ) =  \mJ_s , 
\ee
with the scalar current,
\be
\mJ_s & = & - \fkA\,  \beta_Q\cosh\chi_s\, \bar{\fPsi}_{\fQH} \tvSi_{-}\fPsi_{\fQH}   + \fkA\, \beta_Q^2 \sinh\chi_s\cosh\chi_s \,  \hmH^{\fM\fN} \mW_{\fM} \mW_{\fN} \nn \\
& + & \fkA\, \frac{1}{2} \hfA_{\fC}^{\;\fM}\hfA_{\fD}^{\;\fN} \hfA_{\fC'}^{\;\fM'}\hfA_{\fD'}^{\;\fN'}  \bar{\eta}^{\fC\fD\fC'\fD'}_{\fA\fA'} g_H^{-2}\beta_G^2 \beta_Q^2\sinh\chi_s\cosh\chi_s\,  \fF_{\fM\fN}^{\fA}\fF_{\fM'\fN'}^{\fA'}  \nn \\ 
& & - \fkA\, \lambda_D^2 \cF'(\chi_s) .
\ee
with
\be
\cF'(\chi_s) \equiv \frac{\p}{\p \chi_s}  \cF(\chi_s) . \nn
\ee

It is noticed that all equations of motion for five fundamental fields are global scaling invariant.

\subsection{ Conserved currents in correspondence to gauge symmetries in hyperunified field theory}

From Noether's theorem\cite{NT} that for every differentiable symmetry generated by the local action, there exists the corresponding conserved current. We are going to show the conserved currents generated by the inhomogeneous hyperspin gauge symmetry WS(1,18) and the scaling gauge symmetry SG(1). 

The bicovariant tensor current $\cJ^{\;\fM}_{ \fA\fB}$ generated from the hyperspin gauge symmetry SP(1,$D_h$-1) is a conserved current with respect to the covariant derivative, 
\be   \label{CBCTC}
\cD_{\fM}\cJ^{\;\fM}_{ \fA\fB} = 0 ,
\ee
which can be verified from the equation of motion of hyperspin gauge field,
\be
& & \cD_{\fM} \cD_{\fN} (\fkA\,\hmH^{\fM\fM'}\hmH^{\fN\fN'}\cF_{\fM'\fN' }^{\; \fA\fB} ) \nn \\
& &  \quad =\p_{\fM} \p_{\fN} ( \fkA\, \hmH^{\fM\fM'}\hmH^{\fN\fN'}) \cF_{\fM'\fN' }^{\; \fA\fB}  + \p_{\fM} ( \fkA\, \hmH^{\fM\fM'}\hmH^{\fN\fN'} ) \cD_{\fN}  \cF_{\fM'\fN' }^{\; \fA\fB} \nn \\
& &\quad +  \p_{\fN} (  \fkA\,\hmH^{\fM\fM'}\hmH^{\fN\fN'} ) \cD_{\fM} \cF_{\fM'\fN' }^{\;\; \fA\fB}   + \fkA\, \hmH^{\fM\fM'}\hmH^{\fN\fN'} \cD_{\fM} \cD_{\fN} \cF_{\fM'\fN' }^{\; \fA\fB} \nn \\
 & & \quad =  \fkA\, \hmH^{\fM\fM'}\hmH^{\fN\fN'} (  \cF_{\fM\fN}^{\;\; \fA\fC}  \cF_{\fM'\fN' }^{\;\; \fC'\fB}  - \cF_{\fM'\fN' }^{\;\; \fA\fC} \cF_{\fM\fN}^{\;\; \fC'\fB} ) \eta_{\fC\fC'} = 0 . \nn
\ee
In obtaining the above identity, the symmetric and antisymmetric properties of the tensor and gauge field strength have been used to bring the cancellation. 

From the conserved current of bicovariant tensor current $\cJ^{\fM}_{\;\; \fA\fB}$ presented in Eq.(\ref{CBCTC}), we are able to derive the following dynamical equation: 
\be \label{DESAMT}
 \cD_{\fM} \cS^{\fM}_{\;\; \fA\fB} =  - \cJ_{[\fA\fB]}, 
\ee
for the bicovariant tensor current $\cS^{\fM}_{\;\; \fA\fB}$ and antisymmetric tensor current $\cJ_{[\fA\fB]}$ with the following definitions:
\be  \label{HSEMT}
& & \cS^{\fM}_{\;\; \fA\fB} =  \fkA\,  \frac{1}{2} \hfA_{\fC}^{\;\; \fM} 
\bar{\fPsi}_{\fQH} \{ \vSi_{-}^{\fC} \; \;  \varSigma_{\fA\fB} \} \fPsi_{\fQH}, \nn \\
& &   \cJ_{[\fA\fB]} \equiv \tfJ_{[\fA\fB]}  -  \fkA\, \frac{1}{4}g_H^{-2}\beta_G^2 \beta_Q^2 \sinh^2\chi_s \, \eta_{[\fA\fB]}^{\fA'\fB'}  \bhmH^{\fM\fN\fM'\fN'}_{\fA'\fC} \fF_{\fM\fN\fB'}\fF_{\fM'\fN'}^{\fC} , \nn \\
& & \tfJ_{[\fA\fB]} = \tfJ_{\fA}^{\; \fM} \fA_{\fM \fB} - \tfJ_{\fB}^{\; \fM}\fA_{\fM \fA} , \quad \eta_{[\fA\fB]}^{\fA'\fB'} = \eta_{\fA}^{\; \fA'} \eta_{\fB}^{\; \fB'} - \eta_{\fB}^{\; \fA'} \eta_{\fA}^{\; \fB'}, \nn \\
& & \tfJ_{\fA}^{\; \fM} \equiv \fJ_{\fA}^{\; \fM} - \p_{\fN} \left( \fkA\,\thmH^{[\fM\fN]\fM'\fN'}_{\fA\fA'} g_H^{-2}\mF_{\fM'\fN' }^{\fA'} \right) . 
\ee
where $\cS^{\fM}_{\;\; \fA\fB}$ is regarded as {\it bicovariant hyperspin angular momentum tensor}. 

It is noticed that the dynamics of bicovariant hyperspin angular momentum tensor $\cS^{\fM}_{\;\; \fA\fB}$ is governed by the antisymmetric tensor current $\cJ_{[\fA\fB]}$. Such a tensor current is correlated to the hyper-gravigauge field strength and bicovariant vector current $\fJ_{\fA}^{\; \fM}$ which determines the dynamics of hyper-gravigauge field $\fA_{\fM}^{\;\; \fA}$.

The dynamical equation presented in Eq.(\ref{DESAMT}) may be referred to as {\it dynamic evolution equation} for the hyperspin angular momentum tensor. The bicovariant vector current $\tfJ_{\fA}^{\; \fM}$ provides the source for the equation of motion of hyper-gravigauge field $\fA_{\fM}^{\;\; \fA}$ as follows:
\be \label{EMfAb}
\cD_{\fN} ( \fkA\, \bhmH^{[\fM\fN]\fM'\fN'}_{\fA\fA'} g_H^{-2}\beta_G^2 \beta_Q^2 \sinh^2\chi_s \, \fF_{\fM'\fN' }^{\fA'} ) = \tfJ_{\fA}^{\; \fM} .
\ee
Note that neither $\fJ_{\fA}^{\;\;\fM}$ or $\tfJ_{\fA}^{\;\; \fM}$ is conserved. 

It is useful to define the following alternative bicovariant vector current:
\be \label{mJ}
\mJ_{\fA}^{\;\;\fM} \equiv \fJ_{\fA}^{\;\;\fM} - \cD_{\fN} ( \fkA\, \bhmH^{[\fM\fN]\fM'\fN'}_{\fA\fA'} g_H^{-2}\beta_G^2 \beta_Q^2 \sinh^2\chi_s \, \fF_{\fM'\fN' }^{\fA'} ) , 
\ee
so that the equation of motion presented in Eq.(\ref{EMfA}) for the hyper-gravigauge field $\fA_{\fM}^{\;\; \fA}$ can be rewritten into the following simple form:
\be \label{EMfAt}
\p_{\fN} \left( \fkA\,\thmH^{[\fM\fN]\fM'\fN'}_{\fA\fA'} g_H^{-2}\mF_{\fM'\fN' }^{\fA'} \right) = \mJ_{\fA}^{\;\;\fM}.
\ee
It can be verified that the bicovariant vector current $\mJ_{\fA}^{\;\;\fM}$ is conserved due to the following identity:
\be
 \p_{\fM} \mJ_{\fA}^{\;\;\fM} =  \p_{\fM} \p_{\fN} \left( \fkA\,\thmH^{[\fM\fN]\fM'\fN'}_{\fA\fA'} g_H^{-2}\mF_{\fM'\fN' }^{\fA'} \right) = 0 ,
\ee
which is attributed to the antisymmetric property of the tensor factor $\thmH^{[\fM\fN]\fM'\fN'}_{\fA\fA'}$. Such a conserved bicovariant vector current reflects the translation-like $\cW_e$-spin Abelian-type gauge symmetry. 

The equations of motion shown in Eqs.(\ref{EMfAb}) and (\ref{EMfAt}) enable us to obtain the following dynamical equations for both bicovariant vector currents $\tfJ_{\fA}^{\;\;\fM}$ and $\fJ_{\fA}^{\;\;\fM}$,
\be
& &  \cD_{\fM}\tfJ_{\fA}^{\;\;\fM} = - \frac{1}{2}  \fkA\, \bhmH^{[\fM\fN]\fM'\fN'}_{\fA''\fA'} g_H^{-2}\beta_G^2 \beta_Q^2 \sinh^2\chi_s \,  \cF_{\fM\fN\fA}^{\fA''}\fF_{\fM'\fN' }^{\fA'} , \nn \\
& & \p_{\fM}\fJ_{\fA}^{\;\;\fM} = - \p_{\fM} (  \fkA\,\bhmH^{[\fM\fN]\fM'\fN'}_{\fA''\fA'} g_H^{-2}\beta_G^2 \beta_Q^2 \sinh^2\chi_s \, \cA_{\fN\fA}^{\fA''} \fF_{\fM'\fN' }^{\fA'} ) .
\ee

For convenience of mention, $\mJ_{\fA}^{\;\;\fM}$ is regarded as {\it conserved bicovariant vector current} of hyper-gravigauge field, $\fJ_{\fA}^{\;\;\fM}$ as {\it total bicovariant vector current} and $\tfJ_{\fA}^{\;\;\fM}$ as {\it twisting bicovariant vector current} which governs the bicovariant hyperspin angular momentum tensor.

The local scaling gauge symmetry leads to the conserved vector current as follows:
\be   \label{CVCSGF}
  \partial_{\fM} \mJ^{\fM}  = \partial_{\fM} \partial_{\fN} ( \fkA \hmH^{\fM\fM'}\hmH^{\fN\fN'}  \mF_{\fM'\fN'}) \equiv 0 , 
\ee
which provides the following condition for the scaling gauge field when applying it to the explicit form of the current in Eq.(\ref{EMSGF}),
\be
 \p_{\fM} \left( \fkA\, (1 + \sinh^2\chi_s) \hmH^{\fM\fN} \mW_{\fN} \right) =0  .
\ee


\section{ Conservation law and dynamic evolution equation from global symmetry of Minkowski hyper-spacetime in hyperunified field theory }

The hyperunified field theory is formulated in globally flat Minkowski hyper-spacetime which is regarded as {\it free-motion hyper-spacetime} $\fM_h$, the action in Eq.(\ref{actionHUFTHSG}) possesses the global Poincar\'e-type group symmetry PO(1,18) and global scaling symmetry SC(1). In the absence of gravitational interaction, it is well known that the global Poincar\'e symmetry leads to the conservation laws for both energy momentum tensor and total angular momentum tensor, which follows Noether's theorem\cite{NT} that every differentiable symmetry of the action has the corresponding conservation law. We should apply Noether's theorem to hyperunified field theory in the presence of gravitational interaction and derive the conservation laws and dynamic evolution equations arising from global Poincar\'e-type group symmetry and scaling symmetry of Minkowski hyper-spacetime.

\subsection{ Conservation law of hyper-stress energy-momentum tensor from translation group symmetry of Minkowski hyper-spacetime in hyperunified field theory }

Let us first derive the conservation law for global translation group symmetry. Consider the global translation group transformation of coordinates in globally flat Minkowski hyper-spacetime as follows:
\be
x^{\fM} \to x^{'\fM} = x^{\fM} + a^{\fM} , \nn 
\ee
with $a^{\fM}$ an arbitrary constant vector. When applying for the variational principle, the invariance of the action under such a translation leads to the following relation:
\be 
\Delta \cS_{\mH\mU}= \int [d^{D_h}x] \, \p_{\fP} (\cT_{\fM}^{\;\, \fP}) a^{\fM} = 0 . 
\ee
which results in the following conservation law:
\be \label{EMCL}
\p_{\fP} \cT_{\fM}^{\;\, \fP}= 0 ,
\ee
where $\cT_{\fM}^{\;\, \fP}$ is referred to as {\it bicovariant hyper-stress energy-momentum tensor} in hyper-spacetime. From the action of hyperunified field theory given in Eq.(\ref{actionHUFTHSG}), we obtain the following explicit form for the bicovariant hyper-stress energy-momentum tensor:
\be \label{EMT}
\cT_{\fM}^{\;\, \fP} & = &  - \fkA\, \eta_{\fM}^{\; \fP} \fkL +  \fkA\, \{ \hfA_{\fA}^{\; \fP} \bar{\fPsi}_{\fQH} \vSi_{-}^{\fA} i \cD_{\fM} \fPsi_{\fQH}     \nn \\
& - &\hmH^{[\fP\fN]\fM'\fN'}_{\fA\fB\; \fA'\fB'} g_H^{-2} \cF_{\fM\fN}^{\fA\fB} \cF_{\fM'\fN'}^{\fA'\fB'} + \thmH^{[\fP\fN]\fM'\fN'}_{\fA\fA'} g_H^{-2} \mF_{\fM\fN}^{\fA} \mF_{\fM'\fN'}^{\fA'} \nn \\
& + & \bhmH^{[\fP\fN]\fM'\fN'}_{\fA\fA'}g_H^{-2}\beta_G^2 \beta_Q^2 \sinh^2\chi_s \, \fF_{\fM\fN}^{\fA} \fF_{\fM'\fN'}^{\fA'} +   \lambda_S^2 \p_{\fM}\chi_s  \p_{\fN} \chi_s  \hmH^{\fN\fP} \nn \\
& - & g_W^{-2}\hmH^{\fP\fM'} \hmH^{\fN\fN'} \mF_{\fM\fN} \mF_{\fM'\fN'}+ \lambda_S^2(1 + \sinh^2\chi_s )\, \mW_{\fM} \mW_{\fN}  \hmH^{\fN\fP}    \, \} ,
\ee
which is gauge invariant. In deriving such a gauge-invariant formalism, the equations of motion for the fundamental fields $\cA_{\fM}^{\fA\fB}$, $\fA_{\fM}^{\;\fA}$, $\mW_{\fM}$ and $\chi_s$ have been used and the surface terms have been ignored. 

It is interesting to notice the following relation between the bicovariant hyper-stress energy-momentum tensor and bicovariant vector current:
\be
& & \cT_{\fM}^{\;\;\fP} \equiv   \fA_{\fM}^{\;\;\fA}\fJ_{\fA}^{\; \;\fP},
\ee
which indicates that the conserved bicovariant hyper-stress energy-momentum tensor $\cT_{\fM}^{\;\;\fP}$ generated from the global translation group symmetry of coordinates in globally flat Minkowski hyper-spacetime correlates to the bicovariant vector current $\fJ_{\fA}^{\; \;\fP}$ which governs the dynamics of $\cW_e$-spin invariant-gauge field $\fA_{\fM}^{\;\;\fA}$ resulted from the local translation-like $\cW_e$-spin group symmetry of entangled hyperqubit-spinor field in locally flat gravigauge hyper-spacetime.

Therefore, we arrive at the statement that for the hyperunified field theory built based on the gauge invariance principle and scaling invariance hypothesis in light of biframe hyper-spacetime, the gauge-invariant bicovariant hyper-stress energy-momentum tensor $\cT_{\fM}^{\;\, \fN}$ derived from the translation group symmetry of Minkowski hyper-spacetime is a conserved tensor even in the presence of gravitational gauge interaction.


\subsection{ Dynamic evolution equations of hyper-rotation angular momentum tensor from Lorentz-type group symmetry and hyper-conformal scaling momentum tensor from conformal scaling symmetry in hyperunified field theory }

We now consider the global Lorentz-type group symmetry in Minkowski hyper-spacetime with the following infinitesimal group transformation of coordinates:
\be
x'^{\fM} = x^{\fM} + \delta L^{\fM}_{\; \fN} \, x^{\fN} . \nn
\ee
Based on the principle of least action, the invariance of the action under such a transformation leads to the following {\it dynamic evolution equation} : 
\be \label{DERAMT}
\p_{\fP} \cL^{\fP,\fM\fN} =  \cT^{[\fM\fN]} ,
\ee
with $\cL^{\fP,\fM\fN}$ and $\cT^{[\fM\fN]}$ given explicitly as follows,
 \be \label{RAMT}
& & \cL^{\fP,\fM\fN} \equiv \cT^{\;\,\fP}_{\fM'}\eta^{\fM'\fM} x^{\fN} -  \cT^{\;\, \fP}_{\fM'}  \eta^{\fM'\fN}x^{\fM} , \nn \\
& & \cT^{[\fM\fN]} \equiv \eta^{\fM\fM'}\cT_{\fM'}^{\; \, \fN}  - \eta^{\fN\fM'}\cT_{\fM'}^{\; \, \fM} ,
\ee
where $\cL^{\fP,\fM\fN}$ defines {\it bicovariant hyper-rotation angular momentum tensor} in association with the bicovariant hyper-stress energy momentum tensor in Minkowski hyper-spacetime. As the antisymmetric part of bicovariant hyper-stress energy-momentum tensor is not necessary to vanish in the hyperunified field theory due to the presence of antisymmetric hyper-gravigauge field, the covariant hyper-rotation angular momentum tensor obeys a dynamic evolution equation presented in Eq.(\ref{DERAMT}) rather than a homogeneously conserved law. 
 
It is useful to express the antisymmetric bicovariant hyper-stress energy-momentum tensor $\cT^{[\fM\fN]}$ into the following two parts:
\be \label{ASEMT}
& & \cT^{[\fM\fN]}  \equiv  \cT_a^{[\fM\fN]} + \cT_s^{[\fM\fN]}, \nn \\
& &  \cT_a^{[\fM\fN]}  \equiv \mH_{\fM'\fN'}^{\; [\fM\fN]} \hfT^{[\fM'\fN']}, \quad \cT_s^{[\fM\fN]} \equiv \mH_{\fM'\fN'}^{\; [\fM\fN]} \hfT^{(\fM'\fN')}, \nn \\
& & \hfT^{[\fM\fN]} \equiv \hfT^{\fM\fN} - \hfT^{\fN\fM}, \quad \hfT^{(\fM\fN)} \equiv \hfT^{\fM\fN} + \hfT^{\fN\fM} , \nn \\
& & \mH_{\fM'\fN'}^{\; [\fM\fN]} \equiv \frac{1}{2} ( \mH_{\fM'}^{\;\; \fM}\eta_{\fN'}^{\;\; \fN} - \mH_{\fM'}^{\;\; \fN}\eta_{\fN'}^{\;\; \fM} ), 
\ee
with
\be \label{CEMT}
& & \hfT^{\fM\fN} \equiv\hmH^{\fM\fP}\cT_{\fP}^{\;\;\fN} \equiv \fJ_{\fA}^{\; \;\fN} \hfA^{\fA\fM}, 
\ee 
which presents an alternative tensor $\hfT^{\fM\fN}$ defined either from the bicovariant hyper-stress energy-momentum tensor $\cT_{\fP}^{\;\;\fN}$ raised with the tensor $\hmH^{\fM\fP}$ or from the total bicovariant vector current $\fJ_{\fA}^{\; \;\fN}$ projected by the dual $\cW_e$-spin invariant-gauge field $\hfA_{\fA}^{\;\;\fM}$. 

To distinguish the tensor $\hfT^{\fM\fN}$ from the bicovariant hyper-stress energy-momentum tensor $\cT_{\fP}^{\;\;\fN}$, we may refer to $\hfT^{\fM\fN}$ as {\it covariant hyper-stress energy-momentum tensor}, which can be discriminated from their global scaling charges as $\cT_{\fP}^{\;\;\fN}$ and $\hfT^{\fM\fN}$ have global scaling charges $\cC_s=D_h$ and $\cC_s=D_h-2$, respectively.  It is useful to decompose the covariant hyper-stress energy-momentum tensor $\hfT^{\fM\fN}$ into antisymmetric part $\hfT^{[\fM'\fN']}$ and symmetric part $\hfT^{(\fM'\fN')}$, which brings the antisymmetric bicovariant hyper-stress energy-momentum tensor $\cT^{[\fM\fN]}$ into two parts $\cT_a^{[\fM\fN]}$ and $\cT_s^{[\fM\fN]}$. $\cT_a^{[\fM\fN]}$ is the antisymmetric bicovariant hyper-stress energy-momentum tensor associated with antisymmetric covariant hyper-stress energy-momentum tensor $\hfT^{[\fM\fN]}$, and $\cT_s^{[\fM\fN]}$ is the antisymmetric bicovariant hyper-stress energy-momentum tensor associated with symmetric covariant hyper-stress energy-momentum tensor $\hfT^{(\fM\fN)}$.

In general, the dynamic evolution equation for the bicovariant hyper-rotation angular momentum tensor $\cL^{\fP,\fM\fN}$ can be rewritten as follows,
\be
& & \p_{\fP} \cL^{\fP,\fM\fN} =  \cT_a^{[\fM\fN]} + \cT_s^{[\fM\fN]},
\ee 
which is governed by two kinds of sources $\cT_a^{[\fM\fN]}$ and $\cT_s^{[\fM\fN]}$. It is straightforward to
show that when turning locally flat gravigauge hyper-spacetime into globally flat Minkowski hyper-spacetime, the source part $\cT_s^{[\fM\fN]}$ goes to be vanishing, i.e.:
\be
 & & \mH_{\fM'}^{\;\; \fM}\propto \eta_{\fM'}^{\;\; \fM} , \quad \mH_{\fM'\fN'}^{\; [\fM\fN]} \propto  \eta_{\fM'\fN'}^{[\fM\fN]}  = - \eta_{\fN'\fM'}^{[\fM\fN]}, \nn \\
 & & \cT_s^{[\fM\fN]} \propto \eta_{\fM'\fN'}^{[\fM\fN]} \hfT^{(\fM'\fN')} = - \cT_s^{[\fM\fN]} \to 0 .
\ee

Let us turn to the global conformal scaling symmetry. The invariance of the action leads to the following dynamical equation: 
\be  \label{DECSS}
\left(x^{\fM} \frac{\p}{\p x^{\fM}} + D_h\right) (\fkA\, \fkL ) + \p_{\fM} \cT^{\fM}  - \cT = 0 , 
\ee
with $\cT^{\fM}$ and $\cT$ defined as follows:
 \be \label{TVS}
  \cT^{\fM} \equiv \cT^{\;\, \fM}_{\fN}\, x^{\fN}  \, , \qquad \cT \equiv \cT_{\fM}^{\;\, \fN} \eta_{\fN}^{\;\, \fM} ,
 \ee
where $\cT^{\fM}$ is regarded as {\it hyper-conformal scaling momentum tensor} and $\cT= \cT_{\fM}^{\; \fM}$ is the trace of bicovariant hyper-stress energy-momentum tensor. 

From the property that the integral $\int [d\hx] \lambda^{D_h}\,  \chi(\lambda x)\, \fkL (\lambda x) $ is actually independent of $\lambda$, and by taking the differentiation with respect to $\lambda$ at $\lambda =1$, we obtain the following identity:
 \be
 \int [d\hx] (x^{\fM} \frac{\partial}{\partial x^{\fM}} + D_h) (\fkA\, \fkL ) = 0 , \nn
 \ee  
 which leads the dynamical equation in Eq.(\ref{DECSS}) to be simplified as follows:
 \be
 \p_{\fM} \cT^{\fM} = \cT .
 \ee
Such an equation brings on {\it dynamic evolution equation} for the global conformal scaling symmetry in Minkowski hyper-spacetime. Only when the hyper-stress energy-momentum tensor becomes traceless $\cT= 0$, the tensor vector becomes conserved, i.e., $\p_{\fM} \cT^{\fM} = 0$. In general, the existence of $\cQ_c$-spin scaling field $\chi_s(x)$ and its coupling to the hyperunified qubit-spinor field $\fPsi_{\fQH}(x)$ gives a nonzero tensor scalar $\cT\neq 0$ in hyperunified field theory.

Therefore, the dynamic evolution equation for the bicovariant hyper-rotation angular momentum tensor $\cL^{\fP,\fM\fN}$ is in general governed by the antisymmetric bicovariant hyper-stress energy-momentum tensor  $\cT^{[\fM\fN]}$, and the dynamic evolution equation for the hyper-conformal scaling momentum tensor $\cT^{\fM}$ is determined by the trace of bicovariant hyper-stress energy-momentum tensor $\cT$.
 
 
 \subsection{Dynamic evolution equation for the total bicovariant hyper-angular momentum tensor in hyperunified field theory }
 
In globally flat four-dimensional Minkowski spacetime with the absence of gravitational interaction, the total angular momentum tensor as the sum of rotation and spin angular momentum tensors is known to be a conserved tensor. In the hyperunified field theory, the bicovariant hyper-rotation angular momentum tensor and bicovariant hyperspin angular momentum tensor are shown to obey two independent dynamic evolution equations with respective to global and local symmetries in globally flat Minkowski hyper-spacetime and locally flat gravigauge hyper-spacetime, respectively. To verify the total hyper-angular momentum conservation law when turning locally flat gravigauge hyper-spacetime to be globally flat one, it is useful to introduce the total bicovariant hyper-angular momentum tensor in a hidden locally flat gravigauge hyper-spacetime.

Let us first make a redefinition for the bicovariant hyperspin angular momentum tensor $\cS^{\fP}_{\;\; \fA\fB}$. It is appropriate to project $\cS^{\fP}_{\;\; \fA\fB}$ into the following tensor form via the dual $\cW_e$-spin invariant-gauge field $\hfA_{\fA}^{\;\;\fM}$: 
\be \label{HSAMT}
\cS^{\fP,\fM\fN}  & = & \cS^{\fP}_{\;\; \fA\fB}\hfA^{\fA\fM'}\hfA^{\fB\fN'} \mH_{\fM'\fN'}^{\; [\fM\fN]}
\nn \\
& = & \fkA\,  \frac{1}{2} \hfA_{\fC}^{\;\; \fP} \bar{\fPsi}_{\fQH} \{ \vSi_{-}^{\fC} \; \;  \varSigma_{\fA\fB} \} \fPsi_{\fQH}\hfA^{\fA\fM'}\hfA^{\fB\fN'} \mH_{\fM'\fN'}^{\; [\fM\fN]}, 
 \ee
 with $\mH_{\fM'\fN'}^{\; [\fM\fN]}$ defined in Eq.(\ref{ASEMT}). In terms of the above tensor form, we are able from the dynamical equation in Eq.(\ref{DESAMT}) to derive the following dynamic evolution equation for the bicovariant hyperspin angular momentum tensor:
\be \label{DESAMT2}
 \p_{\fP} \cS^{\fP,\fM\fN}  & = &  - \cT_a^{[\fM\fN]}  - \cT_{\fS}^{[\fM\fN]} + \cT_{\chi_s}^{[\fM\fN]}   + \cT_{c}^{[\fM\fN]} ,
 \ee
with $\cT_a^{[\fM\fN]}$ defined in Eq.(\ref{ASEMT}). Where we have introduced the following definitions: 
\be
& & \cT_{\fS}^{[\fM\fN]} \equiv \mH_{\; \,\fM'\fN'\fN''}^{[\fM\fN]\fM''}\cA_{\fP\fM''}^{\fN''} \hat{\fS}^{\fP,\fM'\fN'} , \quad \hat{\fS}^{\fP,\fM'\fN'} \equiv  \cS_{\fA\fB}^{\fP}\hfA^{\fA\fM'}\hfA^{\fB\fN'} , \nn \\
& & \mH_{\; \, \fM'\fN'\fN''}^{[\fM\fN]\fM''} \equiv \frac{1}{2} [  \mH_{\fM'\fN''} (\eta_{\; \fN'}^{\fM} \eta^{\fN\fM''} -  \eta_{\; \fN'}^{\fN} \eta^{\fM\fM''})   -  (\mH_{\; \fN'}^{\fM} \eta_{\; \fN''}^{\fN} -  \mH_{\; \fN'}^{\fN} \eta_{\; \fN''}^{\fM}) \eta_{\fM'}^{\; \fM''}  ], \nn \\
& & \cA_{\fP\fM''}^{\fN''} \equiv \hfA_{\fA}^{\;\; \fN''}\cD_{\fP}\fA_{\fM''}^{\;\fA} = \hfA_{\fA}^{\;\; \fN''}(\p_{\fP}\fA_{\fM''}^{\;\fA} + \cA_{\fP\fB}^{\fA}\fA_{\fM''}^{\;\fB} ) , 
\ee
for the antisymmetric bicovariant tensor $\cT_{\fS}^{[\fM\fN]}$ which arises from the inhomogeneous hyperspin gauge interaction of hyperunified qubit-spinor field, and 
\be
& & \cT_{\chi_s}^{[\fM\fN]} \equiv \bar{\eta}^{[\fM\fN]\fA\fB} \bhmH^{\fM''\fN''\fM'\fN'}_{\fA\fA'} g_H^{-2}\beta_G^2 \beta_Q^2 \sinh^2\chi_s \, \frac{1}{4}\fF_{\fM''\fN''\fB}\fF_{\fM'\fN'}^{\fA'} , \nn \\
& &\bar{\eta}^{[\fM\fN]\fA\fB} \equiv  \mH_{\fM'\fN'}^{\; [\fM\fN]}\hfA^{\fC\fM'}\hfA^{\fD\fN'} \eta_{[\fC\fD]}^{\fA\fB},
\ee
for the antisymmetric bicovariant tensor $\cT_{\chi_s}^{[\fM\fN]}$ which comes from the $\cQ_c$-spin scaling field interacting with the hyperspin gauge field, as well as
\be  \label{CASEMT}
& & \cT_{c}^{[\fM\fN]} \equiv \mH_{\fM'\fN'}^{\; [\fM\fN]} \hmT^{[\fM'\fN']}, \quad \hmT^{[\fM\fN]} =   \mJ_{[\fA\fB]}\hfA^{\fA\fM}\hfA^{\fB\fN} \equiv \hmT^{\fM\fN} - \hmT^{\fN\fM} , \nn \\
& & \hmT^{\fM\fN} \equiv \hfA^{\fA\fM} \mJ_{\fA}^{\; \fN}  = \hfA^{\fA\fM}\p_{\fP} ( \fkA\,\thmH^{[\fN\fP]\fM'\fN'}_{\fA\fA'} g_H^{-2}\mF_{\fM'\fN' }^{\fA'} ) , 
\ee
for the antisymmetric bicovariant tensor $\cT_{c}^{[\fM\fN]}$ which results from the antisymmetric covariant tensor $\hmT^{[\fM\fN]}$. Where $\hmT^{[\fM\fN]}$ is defined by the conserved bicovariant vector current $\mJ_{\fA}^{\; \fM}$ with $\hmT^{\fM\fN} \equiv \hfA^{\fA\fM} \mJ_{\fA}^{\; \fN}$.

In terms of the bicovariant hyper-rotation and hyperspin angular momentum tensors $\cL^{\fP,\fM\fN}$  and $\cS^{\fP,\fM\fN}$, we are able to define the {\it total bicovariant hyper-angular momentum tensor} as follows:
\be
 \cJ^{\fP,\fM\fN}  & \equiv & \cL^{\fP,\fM\fN}  + \cS^{\fP,\fM\fN} , 
\ee
which satisfies the following dynamic evolution equation: 
\be 
 \p_{\fP} \cJ^{\fP,\fM\fN}  & = & \cT_s^{[\fM\fN]}  -  \cT_{\fS}^{[\fM\fN]} + \cT_{\chi_s}^{[\fM\fN]}   + \cT_c^{[\fM\fN]} ,
\ee
which shows that the total hyper-angular momentum tensor is no longer homogeneously conserved due to the presence of gauge invariant gravitational interaction in hidden locally flat gravigauge hyper-spacetime. It is noticed that the source term of antisymmetric bicovariant hyper-stress energy-momentum tensor $\cT_{a}^{[\fM\fN]}$ in both $\cL^{\fP,\fM\fN}$ and $\cS^{\fP,\fM\fN}$ cancels each other due to the sign flip in their dynamic evolution equations.  

To verify the conservation law of total bicovariant hyper-angular momentum tensor in globally flat Mincowski hyper-spacetime with the absence of gravitational interaction, let us turn the local inhomogeneous hyperspin gauge symmetry into global symmetry by taking gauge fields to approach the following limiting case:
\be
 & & \cA_{\fM}^{\fA\fB} \to 0, \quad \hfA_{\fA}^{\;\;\fM} \to \Mka\eta_{\fA}^{\;\; \fM}, 
 \ee
 which leads to the following relations:
 \be
 & &  \cT_s^{[\fM\fN]},\; \cT_{\fS}^{[\fM\fN]} , \; \cT_{\chi_s}^{[\fM\fN]} ,\;  \cT_c^{[\fM\fN]} \to 0 , \nn \\
 & &  \p_{\fP} \cS^{\fP,\fM\fN}  = -  \cT_a^{[\fM\fN]}, \nn \\
 & &  \p_{\fP} \cL^{\fP,\fM\fN}  =  + \cT_a^{[\fM\fN]} .
 \ee
As a consequence, we arrive at the following equation:
\be
\p_{\fP} {\cal J}^{\fP,\fM\fN}  & = & \p_{\fP} ( \cL^{\fP.\fM\fN}  + \cS^{\fP,\fM\fN}  )  =0 ,
\ee
which does verify the conservation law of total bicovariant hyper-angular momentum tensor in globally flat Minkowski hyper-spacetime.

Therefore, we come to the observation that for the hyperunified qubit-spinor field as fundamental building block with inhomogeneous hyperspin gauge interaction as fundamental interaction, its covariant hyper-stress energy-momentum tensor is generally asymmetric. Neither the bicovariant hyper-rotation angular momentum tensor nor the bicovariant hyperspin angular momentum tensor is homogeneously conserved due to the antisymmetric property of the covariant hyper-stress energy-momentum tensor in hyperunified field theory, they all get dynamic evolution equations which are governed by the antisymmetric bicovariant hyper-stress energy-momentum tensor. Only in the absence of inhomogeneous hyperspin gauge interaction with turning locally flat gravigauge hyper-spacetime to globally flat Minkowski hyper-spacetime, the total hyper-angular momentum tensor becomes homogeneously conserved due to the appearance of cancellation, where the bicovariant hyper-rotation angular momentum tensor and hyperspin angular momentum tensor are governed by the same antisymmetric hyper-stress energy-momentum tensor but with an opposite sign.


\section{ Hyperunified field theory in hidden gauge formalism with gravity-geometry correspondence and Einstein-Hilbert type action with appearance of Riemann geometry and emergent group symmetry GL(D$_h$,R) as generalization of gauge invarance principle under flowing unitary gauge }

The basic action of hyperunified field theory presented in Eq.(\ref{actionHUFTHSG}) is built based on the gauge invariance principle and scaling invariance hypothesis with inhomogeneous hyperspin gauge symmetry and global Poincar\'e-type group symmetry. The $\cW_e$-spin invariant-gauge field as hyper-gravigauge field is taken as a projection operator to define the bicovariant hyperspin angular momentum tensor given in Eq. (\ref{HSAMT}). In this section, we are going to define a hyper-spacetime gauge field from the hyperspin gauge field and present an equivalent action of hyperunified field theory in a hidden gauge formalism. The hyper-gravigauge field as projection operator is shown to play an essential role as Goldstone-like boson. Such a hidden gauge formalism of the action enables us to investigate more profound correlations between gravitational interaction and Riemann geometry of hyper-spacetime. We will demonstrate how the dynamics of hyper-gravigauge field can equivalently be characterized by the dynamics of emergent Riemann geometry in light of Einstein-Hilbert type action, which is governed by an emergent general linear group symmetry GL($D_h$, R) and regarded as the generalization of gauge invariance principle. So that we are able to corroborate the gravity-geometry correspondence in curved Riemannian hyper-spacetime.


\subsection{ Hyper-gravigauge field as Goldstone-like boson and the hyper-spacetime gauge field with gauge covariant field strength in hidden gauge formalism }

The $\cW_e$-spin invariant-gauge field $\fA_{\fM}^{\;\;\fA}$ as hyper-gravigauge field is a bicovariant vector field spanned in biframe hyper-spacetime, which allows us to project the tensors in the representation of locally flat gravigauge hyper-spacetime into those in the representation of globally flat Minkowski hyper-spacetime. 

Let us firstly define a hidden gauge invariant field from the hyperspin gauge field by applying for the hyper-gravigauge field as projection operator, i.e.:
\be \label{HSTGF}
 & & \cA_{\fM\fQ}^{\fP}   \equiv  \hfA_{\fA}^{\;\; \fP} \p_{\fM} \fA_{\fQ}^{\;\;\fA} + \hfA_{\fA}^{\;\; \fP}   \cA_{\fM\1 \fB}^{\fA} \fA_{\fQ}^{\;\;\fB} , 
\ee
which already appears in the covariant derivative when deriving the quadratic form of equation of motion given in Eq.(\ref{CD2}) for the hyperunified qubit-spinor field, we may refer to such a gauge field $\cA_{\fM\fQ}^{\fP}$ as {\it hyper-spacetime gauge field}.

As the hyperspin gauge field $ \cA_{\fM}^{\fA\fB}$ exhibits gravitational origin of gauge symmetry by decomposing it into the sum of hyperspin gravigauge field $\fOm_{\fM}^{\fA\fB}$ and covariant-gauge field $\fA_{\fM}^{\fA\fB}$ presented in Eqs.(\ref{HSGGF1})-(\ref{HSGGF2}), we should also decompose the hyper-spacetime gauge field $\cA_{\fM\fQ}^{\fP}$ into the following two parts in an analogous way:
\be \label{HSTGF}
 \cA_{\fM\fQ}^{\fP}   \equiv   \fGa_{\fM\fQ}^{\fP}  +  \fA_{\fM\fQ}^{\fP} ,
\ee
with the explicit forms, 
\be \label{HSGMF}
\fGa_{\fM\fQ}^{\fP}  & \equiv &  \hfA_{\fA}^{\;\; \fP} \p_{\fM} \fA_{\fQ}^{\;\;\fA} +  \hfA_{\fA}^{\;\; \fP}   \fOm_{\fM\1 \fB}^{\fA} \fA_{\fQ}^{\;\;\fB} , \nn \\
& = & \frac{1}{2}\hmH^{\fP\fL} (\p_{\fM} \mH_{\fL\fQ} + \p_{\fQ} \mH_{\fL\fM} - \p_{\fL}\mH_{\fM\fQ} ) =\fGa_{\fQ\fM}^{\fP}, \nn \\
 \fA_{\fM\fQ}^{\fP} & \equiv &  \hfA_{\fA}^{\;\; \fP} \fA_{\fM\fB}^{\fA} \fA_{\fQ}^{\;\; \fB} ,
 \ee
which are all local scaling gauge invariant.  

The hyper-gravigauge field $\fA_{\fM}^{\;\;\fA}$ may be regarded as {\it Goldstone-like boson field} which transforms as bicovariant vector field under hyperspin gauge group and Lorentz-type group transformations. In light of the hyper-spacetime gauge field and field strength, the hyperspin gauge symmetry SP(1,$D_h$-1) is transmuted into a hidden gauge symmetry via the Goldstone-like hyper-gravigauge field $\fA_{\fM}^{\; \fA}$. 

It is noticed that the symmetric tensor fields $\mH_{\fM\fN}$ and $\hmH^{\fM\fN}$ defined in Eq.(\ref{DTF}) are presented as the product of two hyper-gravigauge fields by contracting the vector indices in locally flat gravigauge hyper-spacetime. Such symmetric tensor fields are considered as composite fields with both hidden hyperspin gauge symmetry and scaling gauge symmetry, which are referred to as Goldstone-like dual {\it hyper-gravimetric fields} geometrically. 

The gauge field $\fGa_{\fM\fQ}^{\fP}$ is a symmetric gauge field $\fGa_{\fM\fQ}^{\fP} = \fGa_{\fQ\fM}^{\fP}$ which is completely determined by the dual Goldstone-like hyper-gravimetric fields $\mH_{\fM\fN}$ and $\hmH^{\fM\fN}$. We may refer to such a gauge field as scaling gauge invariant {\it hyper-spacetime gravimetric-gauge field}. For convenience of mention, the gauge field $\fA_{\fM\fQ}^{\fP} $ is called as {\it hyper-spacetime covariant-gauge field} .

From the hyper-spacetime gauge field, we are able to define the corresponding field strength as follows: 
\be
 & & \cF_{\fM\fN\fQ}^{\fP}  = \nabla_{\fM} \cA_{\fN\fQ}^{\fP} - \nabla_{\fN} \cA_{\fM\fQ}^{\fP} + \cA_{\fM\fL}^{\fP} \cA_{\fN\fQ}^{\fL}   - \cA_{\fN\fL}^{\fP} \cA_{\fM\fQ}^{\fL} , \nn \\
& & \nabla_{\fM} \cA_{\fN\fQ}^{\fP} = \p_{\fM} \cA_{\fN\fQ}^{\fP} - \fGa_{\fM\fQ}^{\fL} \cA_{\fN\fL}^{\fP} + \fGa_{\fM\fL}^{\fP}  \cA_{\fN\fQ}^{\fL} ,
\ee
which can also be decomposed into the following two parts:
\be \label{HSGFS}
& &  \cF_{\fM\fN\fQ}^{\fP} \equiv  \fR_{\fM\fN\fQ}^{\fP} + \fF_{\fM\fN\fQ}^{\fP} , \nn \\
& & \fR_{\fM\fN\fQ}^{\;\fP}  = \p_{\fM} \fGa_{\fN\fQ}^{\fP} - \p_{\fN} \fGa_{\fM\fQ}^{\fP}  + \fGa_{\fM\fL}^{\fP} \fGa_{\fN\fQ}^{\fL}  - \fGa_{\fN\fL}^{\fP} \fGa_{\fM\fQ}^{\fL}, \nn \\
& & \fF_{\fM\fN\fQ}^{\;\fP}  = \nabla_{\fM} \fA_{\fN\fQ}^{\fP} - \nabla_{\fN} \fA_{\fM\fQ}^{\fP} + \fA_{\fM\fL}^{\fP} \fA_{\fN\fQ}^{\fL}   - \fA_{\fN\fL}^{\fP} \fA_{\fM\fQ}^{\fL}  ,\nn  \\
& & \nabla_{\fM} \fA_{\fN\fQ}^{\fP} = \p_{\fM} \fA_{\fN\fQ}^{\fP} - \fGa_{\fM\fQ}^{\fL} \fA_{\fN\fL}^{\fP} + \fGa_{\fM\fL}^{\fP}  \fA_{\fN\fQ}^{\fL} .
\ee
Where the tensor fields $\fR_{\fM\fN\fQ}^{\;\fP}$ and $\fF_{\fM\fN\fQ}^{\;\fP}$ defined in the hidden gauge formalism are referred to as scaling gauge invariant {\it hyper-spacetime gravimetric-gauge field strength} and {\it hyper-spacetime covariant-gauge field strength}, respectively.

In obtaining the above field strengths in a hidden gauge formalism, we have used the following identities: 
\be \label{IDS}
& & \nabla_{\fM}\fA_{\fN}^{\; \fA} = \p_{\fM} \fA_{\fN}^{\; \fA} + \fOm_{\fM \fB}^{\fA} \fA_{\fN}^{\; \fB}  
- \fGa_{\fM\fN}^{\fP} \fA_{\fP}^{\; \fA} =0 , \nn \\
& & \nabla_{\fM} \hfA_{\fB}^{\;\, \fP} =  \p_{\fM} \hfA_{\fB}^{ \;\, \fP} 
- \fOm_{\fM \fB}^{\fA} \hfA_{\fA}^{\;\,\fP} + \fGa_{\fM\fN}^{\fP} \hfA_{\fA}^{\; \fN}=0 , \nn \\
& & \nabla_{\fM} \mH_{\fP\fQ} = \p_{\fM} \mH_{\fP\fQ}  - \fGa_{\fM\fP}^{\fN} \mH_{\fN\fQ} - \fGa_{\fM\fQ}^{\fN} \mH_{\fP\fN} = 0 , \nn \\
& & \nabla_{\fM} \hmH^{\fP\fQ} = \p_{\fM} \hmH^{\fP\fQ}  + \fGa_{\fM\fN}^{\fP} \hmH^{\fN\fQ} + \fGa_{\fM\fN}^{\fQ}\hmH^{\fP\fN} = 0 ,
\ee
which enable us to verify the following relations:
\be \label{GGeR}
& & \hcF_{\fM\fN}^{\fP\fQ} = \cF_{\fM\fN\fQ'}^{\fP}\hmH^{\fQ'\fQ} = \cF_{\fM\fN\fP'\fQ'}\hmH^{\fP'\fP}\hmH^{\fQ'\fQ} =  \cF_{\fM\fN}^{\fA\fB} \hfA_{\fA}^{\; \fP} \hfA_{\fB}^{\; \fQ} =  - \hcF_{\fM\fN}^{\fQ\fP},  \nn \\
& & \hfR_{\fM\fN}^{\fP\fQ} = \fR_{\fM\fN\fQ'}^{\fP}\hmH^{\fQ'\fQ} = \fR_{\fM\fN\fP'\fQ'}\hmH^{\fP'\fP}\hmH^{\fQ'\fQ} =  \fR_{\fM\fN}^{\fA\fB} \hfA_{\fA}^{\; \fP} \hfA_{\fB}^{\; \fQ} =  - \hfR_{\fM\fN}^{\fQ\fP}, \nn \\
& & \hfF_{\fM\fN}^{\fP\fQ} = \fF_{\fM\fN\fQ'}^{\fP}\hmH^{\fQ'\fQ} = \fF_{\fM\fN\fP'\fQ'}\hmH^{\fP'\fP}\hmH^{\fQ'\fQ} =  \fF_{\fM\fN}^{\fA\fB} \hfA_{\fA}^{\; \fP} \hfA_{\fB}^{\; \fQ} =  - \hfF_{\fM\fN}^{\fQ\fP}, \nn \\
& & \cF_{\fM\fN}^{\fA\fB} =  \hcF_{\fM\fN}^{\fP\fQ} \fA_{\fP}^{\; \fA} \fA_{\fQ}^{\; \fB}= \hfA^{\fA\fP} \hfA^{\fB\fQ}  \cF_{\fM\fN\fP\fQ} . 
\ee

It is seen that the scaling gauge invariant hyper-spacetime gravimetric-gauge field strength $\fR_{\fM\fN\fQ}^{\fP}$ is described by the scaling gauge invariant Goldstone-like hyper-gravimetric field $\mH_{\fM\fN}$, which defines scaling gauge invariant {\it Riemann-type curvature tensor}. $\mH_{\fM\fN}$ is regarded as a basic field for hyper-spacetime gravimetric-gauge interaction in the hidden gauge formalism.


\subsection{Hyperunified field theory in hidden gauge formalism and Einstein-Hilbert type action on gravitational interaction of hyper-spacetime gravimetric-gauge field}

To formulate the basic action of hyperunified field theory in the hidden gauge formalism, it is useful to corroborate the following identity:
\be \label{RSCI}
 \frac{1}{4} \fkA\, \thmH^{\fM\fN\fM'\fN'}_{\fA\fA'} \mF_{\fM\fN}^{\fA}\mF_{\fM'\fN'}^{\fA'}  + 2 \p_{\fM} (\fkA \hmH^{\fM\fP} \hfA_{\fA}^{\;\fQ} \mF_{\fP\fQ}^{\fA} ) = \fkA\, \fR , 
 \ee
where the second term on the left-hand side is a total derivative. The scalar tensor $\fR$ on the right-hand side defines scaling gauge invariant {\it Ricci-type scalar curvature} in hyper-spacetime. It is useful to notice the following general relations for the hyper-spacetime gravimetric-gauge field:
\be
& & \fGa_{\fM\fP}^{\fP} = \eta_{\; \fP}^{\fQ}\fGa_{\fM\fQ}^{\fP} = \p_{\fM}\ln \fkA , \quad  \fGa_{\fM}^{\fP\fM} \equiv  \hmH^{\fM\fQ}\fGa_{\fM\fQ}^{\fP} = - \hat{\fkA} \p_{\fM} (\fkA\, \hmH^{\fM\fP}) .
\ee

In general, the scaling gauge invariant {\it Ricci-type curvature tensor} can uniquely be resulted from contracting the scaling gauge invariant Riemann-type curvature tensor, 
 \be
\fR_{\fM\fN} \equiv \fR_{\fM\fQ\fN}^{\fP} \eta^{\fQ}_{\;\;\fP} = - \fR_{\fQ\fM\fN}^{\fP} \eta^{\fQ}_{\;\;\fP} = - \fR_{\fP\fM\fN}^{\fP} , 
\ee 
which is a symmetric tensor with the following explicit form:
\be \label{RCT}
 \fR_{\fM\fN} = \fR_{\fN\fM} = \nabla_{\fM} \p_{\fN}\ln \fkA  - \p_{\fP} \fGa_{\fM\fN}^{\fP}  + \fGa_{\fM\fL}^{\fP} \fGa_{\fP\fN}^{\fL} .
\ee
The scaling gauge invariant Ricci-type scalar curvature in hyper-spacetime is obtained from contracting the Ricci-type curvature tensor, 
\be \label{RSC}
\fR = \hmH^{\fM\fN} \fR_{\fM\fN} = \hmH^{\fM\fN}  \fR_{\fM\fQ\fP\fN}\hmH^{\fQ\fP} = -  \hmH^{\fM\fN} \fR_{\fM\fQ\fN\fP} \hmH^{\fQ\fP} .
\ee

It is interesting to notice the following identity:
\be \label{QI}
\frac{1}{4} \bhmH^{\fM\fN\fM'\fN'}_{\fA\fA'} \fF_{\fM\fN}^{\fA}\fF_{\fM'\fN'}^{\fA'} & = & - \frac{1}{2} \hmH^{\fM\fM'} ( \cA_{\fM\fQ}^{\fP} - \fGa_{\fM\fQ}^{\fP} )(\cA_{\fM'\fP}^{\fQ} - \fGa_{\fM'\fP}^{\fQ}) ,
\ee
which brings on the quadratic interaction of hyper-spacetime gauge field.

Applying the identities and relations in Eqs.(\ref{GGeR})-(\ref{QI}), we are able to rewrite the basic action in Eq.(\ref{actionHUFTHSG}) into the following form formulated in the hidden gauge formalism:
\be  \label{actionHUFTHGF}
\cS_{\mH\mU} & \equiv & \int [d^{D_h}x] \, \fkA(x)\, \fkL = \int [d^{D_h}x] \, \fkA(x)\, \nn \\
& \cdot & \{\, \bar{\fPsi}_{\fQH} \hfSi_{-}^{\fM}  i\cD_{\fM}  \fPsi_{\fQH} -  \beta_Q \sinh\chi_s \, \bar{\fPsi}_{\fQH} \tvSi_{-}\fPsi_{\fQH}  \nn \\
& + &  \hmH^{\fM\fM'} \hmH^{\fN\fN'} \frac{1}{4} g_H^{-2} \cF_{\fM\fN\fQ}^{\fP}  \cF_{\fM'\fN'\fP}^{\fQ}  +  g_H^{-2} \fR \nn \\
& - & \frac{1}{2} g_H^{-2}\beta_G^2\beta_Q^2 \sinh^2\chi_s \,  \hmH^{\fM\fM'} ( \cA_{\fM\fQ}^{\fP} - \fGa_{\fM\fQ}^{\fP} )(\cA_{\fM'\fP}^{\fQ} - \fGa_{\fM'\fP}^{\fQ})  \nn \\
& - & \hmH^{\fM\fM'} \hmH^{\fN\fN'} \frac{1}{4} g_W^{-2} \mF_{\fM\fN} \mF_{\fM'\fN'} + \frac{1}{2} \lambda_S^2 (1 +\sinh^2\chi_s ) \hmH^{\fM\fM'}  \mW_{\fM} \mW_{\fM'}  \nn \\
&  + &  \frac{1}{2} \hmH^{\fM\fN} \lambda_S^2 \p_{\fM} \chi_s \p_{\fN}\chi_s  - \lambda_D^2 \cF(\chi_s)  \, \} ,
\ee
where we have ignored the surface term $2 \p_{\fM} (\fkA \hmH^{\fM\fP} \hfA_{\fA}^{\;\fQ} \mF_{\fP\fQ}^{\fA} )$. The following definitions in the hidden gauge formalism have implicitly adopted:
\be
& & i\cD_{\fM} \equiv i\p_{\fM} + (\fOm_{[\fM\fP\fQ]} +  \cA_{[\fM\fP\fQ]})\frac{1}{2}\hfSi^{\fP\fQ}, \nn \\
& & \cA_{[\fM\fP\fQ]} \equiv \frac{1}{3} (\cA_{\fM\fP\fQ} + \cA_{\fP\fQ\fM}  + \cA_{\fQ\fM\fP} ) , \nn \\
& & \fOm_{[\fM\fP\fQ]} \equiv  \frac{1}{3} (\fOm_{\fM\fP\fQ} + \fOm_{\fP\fQ\fM}  + \fOm_{\fQ\fM\fP} ) \nn \\
& & \qquad \quad \;\; \equiv \frac{1}{3} ( \p_{\fM} \fA_{\fP}^{\;\;\fC}\, \fA_{\fQ\fC} + \p_{\fP} \fA_{\fQ}^{\;\;\fC}\, \fA_{\fM\fC}  + \p_{\fQ} \fA_{\fM}^{\;\;\fC}\, \fA_{\fP\fC}) , \nn \\
& & \fOm_{\fM\fP\fQ} \equiv  \fOm_{\fM\fA\fB} \fA_{\fP}^{\;\; \fA} \fA_{\fQ}^{\;\; \fB} ,\quad \cA_{\fM\fP\fQ}  \equiv \mH_{\fP\fP'}\cA_{\fM\fQ}^{\fP'},  \nn \\
& & \hfSi_{-}^{\fM} \equiv \fGa^{\fM} \vGa_{-} , \quad \hfGa^{\fM} \equiv \hfA_{\fA}^{\;\;\fM} \vGa^{\fA}, \nn \\
& &  \hfSi^{\fP\fQ} \equiv \hfA_{\fA}^{\;\;\fP} \hfA_{\fB}^{\;\;\fQ} \varSigma^{\fA\fB} = \frac{i}{4} [\hfGa^{\fP}, \hfGa^{\fQ} ] ,  
\ee
where $\fOm_{[\fM\fP\fQ]}$ and $\cA_{[\fM\fP\fQ]}$ are total antisymmetric fields since the hyperunified qubit-spinor field couples only to totally antisymmetric gauge fields due to its self-conjugated chiral property. $\fOm_{\fM\fP\fQ}$ is referred to as {\it hyper-spacetime gravigauge field} which characterizes the gravitational origin of gauge symmetry. It is noticed that the above defined $\vGa$-matrices are no longer constant matrices, the matrices $\hfGa^{\fM}$ are regarded as {\it hyper-gravigauge $\Gamma$-matrices} and $\hfSi^{\fP\fQ}$ together with $\hfSi_{-}^{\fM}$ are viewed as {\it hyper-gravigauge inhomogeneous hyperspin group generators}. 
  
In the action shown in Eq.(\ref{actionHUFTHGF}), the Ricci-type scalar curvature $\fR$ brings about the scaling gauge invariant {\it Einstein-Hilbert type action} which describes the gravitational interaction as dynamics of scaling gauge invariant hyper-gravimetric field $\mH_{\fM\fN}$.


\subsection{ Hyperunified field theory in hidden gauge formalism with emergent general linear group symmetry GL($D_h$,R) as generalization of gauge invariance principle and the appearance of Riemann geometry in curved hyper-spacetime }

The action in hidden gauge formalism shown in Eq.(\ref{actionHUFTHGF}) is obtained from the action in Eq.(\ref{actionHUFTHSG}) via projecting locally flat gravigauge hyper-spacetime into globally flat free-motion hyper-spacetime $\fM_h$ by utilizing the hyper-gravigauge field $\fA_{\fM}^{\;\fA}(x)$ ($ \hfA_{\fA}^{\;\fM}(x)$) as Goldstone-like boson. The action in Eq.(\ref{actionHUFTHSG}) is built based on the gauge invariance principle and scaling invariance hypothesis within the framework of quantum field theory, so that it possesses global Poincar\'e-type group symmetry. In general, the laws of nature should be independent of the choice of coordinate systems, which indicates that the action should hold in any frame of coordinates. To explicitly verify that, let us turn to the geometric analysis.

Geometrically, when considering the hyperspin gauge and scaling gauge invariant tensor field $\mH_{\fM\fN}(x)$ as hyper-gravimetric field of hyper-spacetime, the hyper-spacetime gravimetric-gauge field $\fGa_{\fM\fQ}^{\fP}(x)$ defined in Eq.(\ref{HSGMF}) is regarded as scaling gauge invariant {\it Levi-Civita hyper-connection} or scaling gauge invariant {\it Christoffel symbols} of the second kind in hyper-spacetime. By lowering the index of the second kind Christoffel symbols $\fGa_{\fM\fQ}^{\fP}(x)$ via the hyper-gravimetric field $\mH_{\fM\fN}(x)$, we obtain scaling gauge invariant Christoffel symbols of the first kind as follows:
\be \label{1CS}
\fGa_{\fM\fP\fQ} & = & \mH_{\fP\fL}\fGa_{\fM\fQ}^{\fL}  = \frac{1}{2} (\, \p_{\fM} \mH_{\fQ \fP} + \p_{\fQ} \mH_{\fM \fP} - \p_{\fP}\mH_{\fM\fQ} \, ) \nn \\
& = & \fGa_{\fQ\fP\fM} \equiv \fGa_{\fM[\fP\fQ]}  + \frac{1}{2} \p_{\fM} \mH_{\fP \fQ} .
\ee
Similarly, by raising the index of the second kind Christoffel symbols $\fGa_{\fM\fQ}^{\fP}(x)$ via the dual hyper-gravimetric field $\hmH_{\fM\fN}(x)$, we obtain the following Christoffel symbols:
\be \label{3CS}
\fGa^{\fM\fP\fQ} & = & \hmH^{\fM\fM'}\fGa_{\fM'\fQ'}^{\fP} \hmH^{\fQ'\fQ} = \frac{1}{2} (\, \hat{\p}^{\fP} \hmH^{\fQ \fM} - \hat{\p}^{\fQ} \hmH^{\fP \fM} - \hat{\p}^{\fM} \hmH^{\fP \fQ} \, ) \nn \\
& = & \fGa^{\fQ\fP\fM}  \equiv \hfGa^{\fM[\fP\fQ]}  - \frac{1}{2} \hat{\p}^{\fM} \hmH^{\fQ \fP} .
\ee
Where the antisymmetric part of hyper-spacetime gravimetric-gauge field can simply be expressed as follows:
\be
& & \fGa_{\fM[\fP\fQ]} = - \frac{1}{2} ( \p_{\fP} \mH_{\fQ\fM} - \p_{\fQ} \mH_{\fP\fM} ) , \nn \\
& & \hfGa^{\fM[\fP\fQ]}= \frac{1}{2} ( \hat{\p}^{\fP} \hmH^{\fQ\fM} - \hat{\p}^{\fQ} \hmH^{\fP\fM} ),  \quad \hat{\p}^{\fP}\equiv \hmH^{\fP\fM} \p_{\fM} . 
\ee

Christoffel symbols in hyper-spacetime introduces the principal connection of general linear group symmetry GL(1,$D_h$-1, R), which inspires us to generalize global inhomogeneous Lorentz-type transformation of Poincar\'e-type group symmetry PO(1,$D_h$-1) in globally flat Minkowski hyper-spacetime to {\it general coordinate transformation of general linear group symmetry }GL(1,$D_h$-1, R). The general coordinate transformation is defined as an arbitrary reparametrization of coordinate systems, i.e.:
\be \label{ARP}
x^{\fM} \to x^{'\fM} \equiv x^{'\fM}(x), 
\ee
which presents the local transformation for describing a distinct reparametrization at every point in curved Riemannian hyper-spacetime. Such an arbitrary reparametrization leads the displacement and derivative of coordinate systems to obey the following transformation laws:
\be \label{TL}
dx^{\fM} & \to &  dx^{'\fM}= T^{\fM}_{\;\; \; \fN}\, dx^{\fN}\, , \quad  T^{\fM}_{\;\; \; \fN}  \equiv \frac{\p x^{'\fM}}{\p x^{\fN}} , \nn \\
 \p_{\fM} & \to & \p'_{\fM} = T_{\fM}^{\;\, \fN}\, \p_{\fN} \, ,\quad  T_{\fM}^{\;\; \fN} \equiv \frac{\p x^{\fN}}{\p x^{'\fM}} ,
\ee
which keeps the following inner product:
\be
\langle \p'_{\fM}, dx^{'\fM'} \rangle = T_{\fM}^{\;\, \fN} T^{\fM'}_{\;\; \; \fN'}  
\langle \p_{\fN}, dx^{\fN'} \rangle = T_{\fM}^{\;\, \fN} T^{\fM'}_{\;\; \; \fN'} \eta_{\fN}^{\;\; \fN'} = \eta_{\fM}^{\;\; \fM'} = \langle \p_{\fM}, dx^{\fM'} \rangle ,
\ee
where $\p_{\fM}$ is regarded as covariant vector and $dx^{\fM}$ as contravariant vector.

With the above transformation laws, any covariant vector field $\cA_{\fM}(x)$ with a lower index and contravariant vector field $\hcA^{\fM}(x)$ with an upper index will get the following general transformation properties:
\be \label{TLTensor}
\cA'_{\fM}(x') = T_{\fM}^{\;\; \fN}\, \cA_{\fN}(x), \quad  \hcA^{'\fM}(x') = T^{\fM}_{\;\; \; \fN}\, \hcA^{\fN}(x),
\ee
which leads the scalar product of covariant vector field with contravariant vector field to be invariant, i.e.:
\be
\cA_{\fM}(x) \hcA^{\fM}(x) = \cA'_{\fM}(x') \hcA^{'\fM}(x') .
\ee

The hyper-gravigauge field $\fA_{\fM}^{\;\fA}(x)$ and its dual vector $\hfA_{\fA}^{\; \fM}(x)$ correspond to the covariant and contravariant vector fields, respectively, in hyper-spacetime. The hyper-gravimetric field $\mH_{\fM\fN}(x)$ defined  by the covariant vector field $\fA_{\fM}^{\;\fA}(x)$ is a covariant tensor field with the following transformation law:
\be
\mH'_{\fM\fN}(x')  = \frac{\p x^{\fP}}{\p x^{'\fM}} \frac{\p x^{\fQ}}{\p x^{'\fN}}  \, \mH_{\fP\fQ}(x) =  T_{\fM}^{\;\; \fP}  T_{\fN}^{\;\; \fQ}\, \mH_{\fP\fQ}(x). 
\ee
Similarly, the dual hyper-gravimetric field $\hmH^{\fM\fN}(x)$ defined  by the contravariant vector field $\hfA_{\fA}^{\;\fM}(x)$ is a contravariant tensor with the corresponding transformation law. It can be checked that the Christoffel symbols transform as follows:
\be
\fGa_{\fM\fN}^{'\fP} (x') & = &  ( \frac{\p x^{\fM'}}{\p x^{'\fM}}\frac{\p x^{\fN'}}{\p x^{'\fN}} ) \left(\, \frac{\p x^{'\fP}}{\p x^{\fP'}} \fGa_{\fM'\fN'}^{\fP'} (x) - \frac{\p^2 x^{'\fP}}{\p x^{\fM'}\p x^{\fN'}} \, \right)\nn \\
& \equiv & T_{\fM}^{\;\; \fM'}  T_{\fN}^{\;\; \fN'} T^{\fP}_{\;\; \; \fP'}  \left(\, \fGa_{\fM'\fN'}^{\fP'} (x) -  T^{\fP'}_{\;\; \, \fQ'} \p_{\fM'} T^{\fQ'}_{\;\; \, \fN'}  \right) ,
\ee
which behaves as local gauge-type transformation of general linear group symmetry GL(1,$D_h$-1, R) with appearance of inhomogeneous transformation term.

The covariant derivatives for covariant vector field $\cA_{\fM}(x)$ and contravariant vector field $\hcA^{\fM}(x)$ obey the general covariance under general coordinate transformations, i.e.:
\be  \label{GC}
& & (\nabla_{\fM}\cA_{\fN})^{'} =  ( \frac{\p x^{\fP}}{\p x^{'\fM}}\frac{\p x^{\fQ}}{\p x^{'\fN}} ) (\nabla_{\fP}\cA_{\fQ}) \equiv T_{\fM}^{\;\; \fP}  T_{\fN}^{\;\; \fQ} (\nabla_{\fP}\cA_{\fQ}) , \nn \\
& & (\nabla_{\fM}\hcA^{\fN})^{'} =   ( \frac{\p x^{\fP}}{\p x^{'\fM}})  ( \frac{\p x^{'\fQ}}{\p x^{\fN}} ) (\nabla_{\fP}\hcA^{\fQ}) \equiv T_{\fM}^{\;\; \fP}  T^{\fN}_{\;\; \fQ}  (\nabla_{\fP}\hcA^{\fQ})  ,
\ee
with the definition of covariant derivatives as follows:
\be
& & \nabla_{\fM}\cA_{\fN} = \p_{\fM} \cA_{\fN} - \fGa_{\fM\fN}^{\fP} \cA_{\fP}  , \nn \\
& & \nabla_{\fM}\hcA^{\fN} = \p_{\fM} \hcA^{\fN} + \fGa_{\fM\fP}^{\fN} \hcA^{\fP}  . 
\ee

Therefore, the scaling gauge invariant hyper-spacetime gravimetric-gauge field strength $\fR_{\fM\fN\fQ}^{\;\fP}$ defines Riemann curvature tensor which characterizes {\it curved Riemannian hyper-spacetime} with the {\it emergence of Riemann geometry}. When lowering indices with the hyper-gravimetric field, we can express the Riemann curvature tensor as follows:
\be
\fR_{\fM\fN\fP\fQ} & \equiv & \fR_{\fM\fN\fQ}^{\;\fP'} \mH_{\fP'\fP} \nn \\
& = & \frac{1}{2} (\p_{\fM}\p_{\fQ} \mH_{\fN\fP} -  \p_{\fM}\p_{\fP} \mH_{\fN\fQ} + \p_{\fN}\p_{\fP} \mH_{\fM\fQ} - \p_{\fN}\p_{\fQ} \mH_{\fM\fP}  )  \nn \\ 
& + & \hmH^{\fL\fL'}( \fGa_{\fM\fL\fQ}\fGa_{\fN\fL'\fP}-  \fGa_{\fN\fL\fQ}\fGa_{\fM\fL'\fP} ).
\ee
which has the following symmetry properties as both gauge field strength and Riemann curvature tensor:
\be \label{SPRT}
& & \fR_{\fM\fN\fP\fQ} = -\fR_{\fN\fM\fP\fQ} , \quad \fR_{\fM\fN\fP\fQ} = -\fR_{\fM\fN\fQ\fP}, \nn \\ 
& & \fR_{\fM\fN\fP\fQ} = \fR_{\fN\fM\fQ\fP}, \quad \fR_{\fM\fN\fP\fQ} = \fR_{\fP\fQ\fM\fN} ,
\ee
and satisfies two kind of Bianchi identities:
\be  \label{IRRT}
& & \nabla_{\fL}\fR_{\fM\fN\fQ}^{\;\fP} + \nabla_{\fM}\fR_{\fN\fL\fQ}^{\;\fP} + \nabla_{\fN}\fR_{\fL\fM\fQ}^{\;\fP} = 0 , \nn \\
& & \fR_{\fM\fN\fQ\fP} + \fR_{\fN\fQ\fM\fP} + \fR_{\fQ\fM\fN\fP} = 0 .
\ee

It is straightforward to verify that the action in Eq.(\ref{actionHUFTHGF}) becomes invariant under the general coordinate transformation defined as an arbitrary reparametrization in coordinate system presented in Eqs.(\ref{ARP}) and (\ref{TL}). The verification can easily be carried out from the property that the transformation laws of tensors and Christoffel symbols as well as covariant derivatives lead the scalar product between covariant tensor and contravariant tensor to be invariant under general coordinate transformation. 

It can be proved that the action in Eq.(\ref{actionHUFTHGF}) brings on the following emergent general linear group symmetry:
\be
G_S = \mbox{GL(}1,D_h-1, \mbox{R)} ,
\ee
which is thought to govern gravitational interaction in curved Riemannian hyper-spacetime with emergence of Riemann geometry. Such a group symmetry appears actually as a hidden local symmetry in the action built by following along the gauge invariance principle and scaling invariance hypothesis in biframe hyper-spacetime as shown in Eq.(\ref{actionHUFT})

In general, the hyperunified field theory formulated in hidden gauge formalism should possess the following maximal joint symmetry:
\be \label{EGLS}
G_S = \mbox{GL(}1,D_h-1, \mbox{R)} \Join \mbox{WS(}1, D_h-1),
\ee
which brings on the extension of global Poincar\'e-type group symmetry PO(1,$D_h$) in globally flat Minkowski hyper-spacetime to a general linear group symmetry GL(1,$D_h$-1,R) in curved Riemannian hyper-spacetime. The inhomogeneous hyperspin gauge symmetry WS(1,$D_h$-1) still holds for the hyperunified qubit-spinor field, while it appears as a hidden gauge symmetry in the action concerning purely bosonic interactions of hyper-spacetime gauge field and scalar field. 

Therefore, we come to the statement that {\it the laws of nature should be independent of the choice of coordinate systems}, which is indicated from emergent general linear group symmetry GL(1,$D_h$-1,R) which brings on the emergence of Riemann geometry in curved Riemannian hyper-spacetim.  So that we achieve a generalization of gauge invariance principle that the gauge invariance principle is not only essential to the fundamental symmetry of basic fields in Hilbert space but also applicable to the fundamental symmetry of coordinates in Minkowski spacetime. In this sense, we may refer to such an extension of gauge invariance principle as {\it generalized gauge invariance principle}. It is intriguing to notice that when the fundamental symmetry of basic fields in Hilbert space is gauged to be local symmetry by following along the gauge invariance principle together with scaling invariance hypothesis, the fundamental symmetry of coordinates in Minkowski spacetime automatically obeys generalized gauge invariance principle with the emergence of general linear group symmetry GL(1,$D_h$-1,R) and genesis of Riemann geometry in curved Riemannian hyper-spacetime. Such a feature implies a {\it gravity-geometry correspondence} which will be discussed below.


\subsection{Flowing unitary gauge with symmetric Goldstone-like hyper-gravigauge field and the gravity-geometry correspondence in hyper-spacetime}

In the action formulated in hidden gauge formalism shown in Eq.(\ref{actionHUFTHGF}), the hyperspin gauge symmetry SP(1,$D_h$-1) for the hyperunified qubit-spinor field $\fPsi_{\fQH}$ is preserved by the scaling gauge invariant totally antisymmetric hyper-spacetime gravigauge field $\fOm_{[\fM\fP\fQ]}$ which is characterized by the Goldstone-like hyper-gravigauge field $\fA_{\fM}^{\;\; \fA}$. The hyper-spacetime gravimetric-gauge field $ \fGa_{\fM\fN}^{\fP}$ as Christoffel symbols decouples from the hyperunified qubit-spinor field due to the self-conjugated chiral spinor representation of hyperunified qubit-spinor field and the symmetric property of Christoffel symbols $\fGa_{\fM\fN}^{\fP} =\fGa_{\fN\fM}^{\fP}$.  To see that explicitly, let us rewrite the action in Eq.(\ref{actionHUFTHGF}) into the following form:
\be  \label{actionHUFTHGF2}
& & \cS_{\mH\mU} \equiv \int [d^{D_h}x] \, \fkA(x)\, \fkL = \int [d^{D_h}x] \, \fkA(x)\, \nn \\
& & \qquad \cdot \{ \bar{\fPsi}_{\fQH} \hfSi_{-}^{\fM}i\cD_{\fM} \fPsi_{\fQH} -  \beta_Q \sinh\chi_s \bar{\fPsi}_{\fQH} \tvSi_{-}\fPsi_{\fQH}  \nn \\
& & \qquad -  \hmH^{\fM\fM'}  \hmH^{\fN\fN'}  g_H^{-2} ( \frac{1}{4} \hfR_{\fM\fN}^{\fP\fQ}  \fR_{\fM'\fN'\fP\fQ} + \frac{1}{2} \hfR_{\fM\fN}^{\fP\fQ}  \fF_{\fM'\fN'\fP\fQ} )  \nn \\
& & \qquad - \hmH^{\fM\fM'}  \hmH^{\fN\fN'} g_H^{-2}\frac{1}{4} \hfF_{\fM\fN}^{\fP\fQ}  \fF_{\fM'\fN'\fP\fQ} +  \frac{1}{2} g_H^{-2}\beta_G^2 \beta_Q^2 \sinh^2\chi_s \hmH^{\fM\fM'} \hfA_{\fM}^{\fP\fQ} \fA_{\fM'\fP\fQ}   \nn \\
& & \qquad + g_H^{-2} \fR - \hmH^{\fM\fM'}  \hmH^{\fN\fN'} \frac{1}{4} g_W^{-2} \mF_{\fM\fN} \mF_{\fM'\fN'} + \frac{1}{2} \lambda_S^2( 1 +  \sinh^2\chi_s )  \hmH^{\fM\fM'} \mW_{\fM} \mW_{\fM'}  \nn \\
& & \qquad + \frac{1}{2} \hmH^{\fM\fN} \lambda_S^2 \p_{\fM} \chi_s \p_{\fN}\chi_s  - \lambda_D^2 \cF(\chi_s)  \, \}  + 2\p_{\fM}\p_{\fN} (\fkA\, \hmH^{\fM\fN}) ,
\ee
where we have used the decomposition formulation for the hyper-spacetime gauge field strengths given in Eqs.(\ref{HSGFS}) and (\ref{GGeR}). The covariant derivative is given as follows:
\be
 i\cD_{\fM} \equiv i\p_{\fM} + (\fOm_{[\fM\fP\fQ]} +  \fA_{[\fM\fP\fQ]})\frac{1}{2}\hfSi^{\fP\fQ} .
\ee
where $\fA_{[\fM\fP\fQ]}$ is a totally antisymmetric hyper-spacetime covariant-gauge field instead of totally antisymmetric hyper-spacetime gauge field $\cA_{[\fM\fP\fQ]} $.

It is noticed that in the gravitational interaction of hyperunified qubit-spinor field, the hyper-spacetime gravimetric-gauge field $\fGa_{\fM\fP\fQ}$($\fGa_{\fM\fQ}^{\fP}$) decouples due to its symmetric property and the self-conjugated chiral feature of hyperunified qubit-spinor field. Only the totally antisymmetric hyper-spacetime covariant-gauge field $\fA_{[\fM\fP\fQ]}$ and totally antisymmetric hyper-spacetime gravigauge field $\fOm_{[\fM\fP\fQ]}$ interact with the hyperunified qubit-spinor field. Note that the hyper-spacetime gravimetric-gauge field $\fGa_{\fM\fP\fQ}$ characterized by the hyper-gravimetric field reflects the geometric property of gravitational interaction, while $\fOm_{[\fM\fP\fQ]}$ is actually characterized by the hyper-gravigauge field rather than hyper-gravimetric field. Therefore, the gravitational interaction of hyperunified qubit-spinor field is essentially governed by the Goldstone-like hyper-gravigauge field $\fA_{\fM}^{\;\;\fA}$ ($\hfA_{\fA}^{\;\; \fM}$) through the hyper-spacetime gravigauge field $\fOm_{[\fM\fP\fQ]}$ and hyper-spacetime group generators $\hfSi_{-}^{\fM}$ and $\hfSi^{\fP\fQ}$ of inhomogeneous hyperspin gauge symmetry WS(1,$D_h$-1). 

On the other hand, the bosonic hyper-spacetime gauge interaction is described by antisymmetric hyper-spacetime covariant gauge field $\fA_{\fM}^{ \fP \fQ}$ and symmetric hyper-spacetime gravimetric-gauge field $\fGa_{\fM\fN}^{\fP}$ as Christoffel symbols, which is governed by the emergent general linear group symmetry GL(1,$D_h$-1,R). Namely, the hyperspin gauge symmetry is completely transmuted into a hidden gauge symmetry in the bosonic interactions. As $\fGa_{\fM\fN}^{\fP}$ is characterized by the symmetric Goldstone-like hyper-gravimetric field $\mH_{\fM\fN}$, it concerns $N_{\mH} =D_h(D_h+1)/2$ degrees of freedom. 

As the fundamental gravitational field is essentially characterized by the Goldstone-like hyper-gravigauge field $\fA_{\fM}^{\;\;\fA}$ which has $N_{\fA} = D_h\times D_h$ degrees of freedom. In comparison to the symmetric hyper-gravimetric field $\mH_{\fM\fN}$, the hyper-gravigauge field $\fA_{\fM}^{\;\;\fA}$ contains extra degrees of freedom given by $N_E = D_h^2 - D_h(D_h+1)/2=D_h(D_h -1)/2$, which exactly reflects the equivalence classes of hyperspin gauge symmetry SP(1, $D_h$-1) and has the same numbers as those of group parameters of SP(1, $D_h$-1), i.e., $N_G=D_h(D_h -1)/2=N_E$.

In general, the presence of gauge symmetry in the action involves redundant degrees of freedom. In order to eliminate the redundant degrees of freedom arising from the gauge symmetry, it is necessary to make a gauge prescription with imposing gauge fixing condition. For the hyperspin gauge symmetry which is unlike usual internal gauge symmetry as it has a gravitational origin characterized by the $\cW_e$-spin invariant-gauge field $\fA_{\fM}^{\;\;\fA}$ as hyper-gravigauge field, we are able to realize a simple gauge prescription. That is just to take an appropriate hyperspin gauge transformation $\bar{\Lambda}_{\;\, \fA}^{\fB}(x)$ to transmute the Goldstone-like hyper-gravigauge field into symmetric one, i.e.: 
\be
& & \fA_{\fM \fA}(x) \to  \bfA_{\fM\fA}(x) = \fA_{\fM\fB}(x) \bar{\Lambda}^{\fB}_{\; \, \fA}(x)  
=  \bfA_{\fA\fM}(x) , \nn \\
& & \hfA^{\fA\fM}(x) \to  \hbfA^{\fA\fM}(x) = \hfA^{\fB\fM}(x) \bar{\Lambda}_{\fB}^{\;\; \fA}(x)  
=  \bhfA^{\fM\fA}(x) ,
\ee 
which sets an explicit gauge prescription as a gauge fixing condition. We may refer to such a gauge prescription as {\it flowing unitary gauge}, which holds locally with a chosen coordinate system at point to point in hyper-spacetime. A detailed discussion on such a flowing unitary gauge will be presented later on. 

When fixing the hyperspin gauge symmetry with flowing unitary gauge, both Goldstone-like hyper-gravigauge field $\bfA_{\fM\fA}(x)$ and hyper-gravimetric field $\mH_{\fM\fN}(x)$ are symmetric fields, they involve exactly the same degrees of freedom. Explicitly, they are correlated determinately each other with the following relation: 
\be
\mH_{\fM\fN} = \fA_{\fM\fA} \fA_{\fN\fB} \eta^{\fA\fB} = \bfA_{\fM\fA}\eta^{\fA\fB} \bfA_{\fB\fN}  \equiv  (\bfA)^2_{\;\, \fM\fN} .
\ee
By adopting a nonlinearly realized exponential representation, we can express them as matrices with the following forms:
\be
& & \bfA_{\fM\fA} = \phi (e^{G/\Mka})_{\fM\fA}\, , \quad  G_{\fM\fA} = G_{\fA\fM} , \nn \\
& & \mH = \bfA^2 = \phi^2 e^{G/\Mka} e^{G/\Mka} = \phi^2 e^{2G/\Mka} , 
\ee 
where the matrix $G_{\fM\fA}$ is regarded as nonlinearly realized {\it symmetric Goldstone-like hyper-gravigauge field} and $\Mka$ as basic mass scale.

In such a flowing unitary gauge, both hyper-spacetime gravigauge field $\fOm_{[\fM\fP\fQ]}$ and hyper-gravigauge $\Gamma$-matrices $\fGa^{\fM}$ are described by the symmetric Goldstone-like hyper-gravigauge field $\bfA_{\fM\fA}$ and its dual field $\hbfA^{\fA\fM}$,
\be
& &  \fOm_{[\fM\fP\fQ]} \to \bar{\fOm}_{\fM\fP\fQ} \equiv \frac{1}{3} ( \p_{\fM} \bfA_{\fP}^{\;\;\fC}\, \bfA_{\fQ\fC} + \p_{\fP} \bfA_{\fQ}^{\;\;\fC}\, \bfA_{\fM\fC}  + \p_{\fQ} \bfA_{\fM}^{\;\;\fC}\, \bfA_{\fP\fC}), \nn \\
& & \hfSi_{-}^{\fM} \to \bfSi_{-}^{\fM} \equiv \bfGa^{\fM} \vGa_{-}, \quad \bfGa^{\fM} \equiv \hbfA^{\fA\fM} \vGa_{\fA} = \hbfA^{\fM\fA} \vGa_{\fA} \nn \\
& & \hfSi^{\fP\fQ} \to \bfSi^{\fP\fQ} \equiv \frac{i}{4} [\bfGa^{\fP}, \bfGa^{\fQ} ] = \hbfA^{\fA\fP} \hbfA^{\fB\fQ} \varSigma_{\fA\fB} = \hbfA^{\fP\fA} \hbfA^{\fQ\fB} \varSigma_{\fA\fB} . 
\ee

In the hyperunified field theory formulated in a hidden gauge formalism, when setting the gauge fixing condition to be flowing unitary gauge, both hyper-spacetime gravigauge field $\bar{\fOm}_{\fM\fP\fQ}$ and hyper-spacetime gravimetric field $\fGa_{\fM\fQ}^{\fP}$ are basically characterized by the symmetric Goldstone-like hyper-gravigauge field $\bfA_{\fM\fA}(x)= \bfA_{\fA\fM}(x)$ (or nonlinearly realized symmetric Goldstone-like hyper-gravigauge field $G_{\fM\fA}(x)=G_{\fA\fM}(x)$). Therefore, the independent degrees of freedom concerning the hyper-spacetime gauge interaction are represented by the symmetric Goldstone-like hyper-gravigauge field $\bfA_{\fM\fA}(x)= \bfA_{\fA\fM}(x)$ (or equivalently the symmetric Goldstone-like hyper-gravimetric field $\mH_{\fM\fN}(x)= (\bfA)^2_{\; \fM\fN}(x)$) and antisymmetric hyper-spacetime covariant-gauge field $\fA_{\fM}^{\fP\fQ}(x)$. Nevertheless, the total independent degrees of freedom in the hyperunified field theory should remain unchanged as the extra degrees of freedom in a general Goldstone-like hyper-gravigauge field $\fA_{\fM}^{\;\;\fA}(x)$ are considered to be absorbed into the antisymmetric hyper-spacetime covariant-gauge field $\fA_{\fM}^{\fP\fQ}(x)$, which behaves as a {\it massive-like gauge field} with $\cQ_c$-spin scaling field regarded as mass-like term shown from the scalar interaction term $\beta_{G}$ in the action given in Eq.(\ref{actionHUFTHGF2}). 

As the symmetry groups GL(1,$D_h$-1, R) and WS(1,$D_h$-1) in Eq.(\ref{EGLS}) cannot act as a direct product group, the gauge fixing condition is actually a local one and holds only for a given reparametrization of coordinate systems in hyper-spacetime. To make such a gauge fixing condition valid at any reparametrization of coordinate systems in hyper-spacetime, the gauge fixing operation must always associate with any given reparametrization of coordinate systems and proceed with an appropriate hyperspin gauge transformation at any given reparametrization so as to preserve symmetric hyper-gravigauge field. Such a gauge fixing condition is regarded as {\it flowing unitary gauge} in the gravigauge hyper-fiber bundle structure when carrying out general transformations under symmetry groups GL($D_h$, R) and WS(1,$D_h$-1), which makes the gauge fixing procedure to run with a flowing gauge fixing condition, so that it ensures the symmetric property of hyper-gravigauge field at arbitrary point of coordinate systems in hyper-spacetime. Namely, for every general coordinate transformation defined as an arbitrary reparametrization of coordinate systems, $x' \equiv x'(x)$, one can always perform an appropriate hyperspin gauge transformation associated with a distinct local reparametrization at every point in hyper-spacetime. 

Such a gauge fixing procedure can always be realized with two-step transformations. Namely, carrying out a general coordinate transformation as follows:
\be
 & & dx^{\fM} \to dx^{'\fM} = T^{\fM}_{\;\; \, \fN} \, dx^{\fN}\, , \quad \p_{\fM} \to \p'_{\fM} =  T_{\fM}^{\;\; \fN} \p_{\fN} ,  \nn \\
 & & T_{\fM}^{\;\, \fN} , \; T^{\fM}_{\; \; \, \fN} \in GL(1, D_h-1, R)
\ee
as the first step, and then performing an appropriate hyperspin gauge transformation at point $x' \equiv x'(x)$,
 \be
& &\vGa^{\fA} \to S'(\bar{\Lambda}') \vGa^{\fA} S^{'-1}(\bar{\Lambda}') =  \bar{\Lambda}^{'\fA}_{\; \; \fB}(x') \vGa^{\fB}\, , \nn \\
& & \bar{\Lambda}^{'\fA}_{\; \; \fB}(x') \in SP(1,D_h\mbox{-}1) , 
\ee
as the second step, so that we arrive at a symmetric hyper-gravigauge field at every point in hyper-spacetime, i.e.:
\be
& & \fA_{\fM\fA}(x) = \fA_{\fA\fM}(x) \to \fA'_{\fM\fA}(x') = T_{\fM}^{\;\; \fN} \, \fA_{\fN\fA}(x) \neq \fA'_{\fA\fM}(x') ,  \nn \\
& & \fA^{''}_{\fM\fA}(x') = \bar{\Lambda}_{\fA}^{'\; \fB}(x')\, \fA'_{\fM\fB}(x') = \fA^{''}_{\fA\fM}(x') .
\ee

For the gravitational interaction, the flowing unitary gauge always enables us to take symmetric Goldstone-like hyper-gravigauge field as fundamental gravitational field at every point in hyper-spacetime. The introduction of flowing unitary gauge allows us to regard all points in hyper-spacetime to be equivalent when taking gauge fixing condition to be a unitary gauge. 

To keep Goldstone-like hyper-gravigauge field $\bfA_{\fM\fA}(x)$ be always symmetric under flowing unitary gauge for a chosen coordinate system, the action is turned out to possess only an associated symmetry in which global hyperspin symmetry SP(1,$D_h$-1) of hyperunified qubit-spinor field must correlate with global Lorentz-type group symmetry SO(1,$D_h$-1) of coordinates, so that their group transformations have to coincide with each other as indicated from their isomorphic property SP(1,$D_h$-1)$\cong$SO(1,$D_h$-1).

Therefore, when the hyperunified field theory is formulated in light of hidden gauge formalism and presented under flowing unitary gauge, the gauge interactions for both hyperunified qubit-spinor field and bosonic fields are described by the symmetric Goldstone-like hyper-gravigauge field $\bfA_{\fM\fA}(x)$ and antisymmetric hyper-spacetime covariant-gauge field $\fA_{\fM}^{\fP\fQ}(x)$. The hyper-spacetime gravimetric-gauge field $\fGa_{\fM\fN}^{\fP}(x)$ as Christoffel symbols characterized by hyper-gravimetric field is completely determined by $\bfA_{\fM\fA}(x)$ via $\mH_{\fM\fN}(x) = \bfA^2_{\;\fM\fN}(x)$. Therefore, the geometry and dynamics of hyper-spacetime are essentially governed by symmetric Goldstone-like hyper-gravigauge field $\bfA_{\fM\fA}(x)$, which leads to the {\it gravity-geometry correspondence} in hyper-spacetime.

Therefore, for a chosen gauge fixing condition under flowing unitary gauge with the symmetric hyper-gravigauge field $\bfA_{\fM\fA}(x)=\bfA_{\fA\fM}(x)$, the action of hyperunified field theory in hidden gauge formalism given in Eq.(\ref{actionHUFTHGF2}) possesses the following associated symmetry with global inhomogeneous hyperspin symmetry WS(1,$D_h$-1) in association with global Poincar\'e-type group symmetry PO(1,$D_h$-1), i.e.: 
\be
G_S = SC(1)\ltimes PO(1,D_h-1) \wtjoin WS(1,D_h-1) , \nn
\ee
which reproduces the fundamental symmetry in free-motion Minkowski hyper-spacetime, where the transformations of global Lorentz-type group symmetry SO(1,$D_h$-1) and hyperspin symmetry SP(1,$D_h$-1) have to be coincidental as indicated by the symbol `$\wtjoin$'.

As the flowing unitary gauge is valid locally at point by point, to fix the flowing property of general linear group symmetry GL(1,$D_h$-1, R), it remains to impose an additional gauge fixing condition to achieve a whole unitary gauge, which will be discussed later on.


\section{ Hyperunified field theory in hidden coordinate formalism with geometry-gauge correspondence and the dynamics of fundamental fields with hyperspin gravigauge field as auxiliary field and structure factor of non-commutative geometry in locally flat gravigauge hyper-spacetime }

The hyperunified field theory is shown to be independent of the choice of coordinate systems and hold in any coordinate systems so as to bring about the generalized gauge invariance principle, which inspires us to reformulate the hyperunified field theory in light of hidden coordinate formalism. We will apply for again the scaling gauge invariant hyper-gravigauge field as Goldstone-like boson to represent the hyperunified field theory in locally flat gravigauge hyper-spacetime, where the hyper-spacetime of coordinates becomes a hidden hyper-spacetime. It is intriguing to explore the nature of gravitational interaction in locally flat gravigauge hyper-spacetime and corroborate the geometry-gauge correspondence of gravitational interaction in hidden coordinate formalism. Relative to the gravity-geometry correspondence shown in hidden gauge formalism in which the gravitational interaction can equivalently be characterized by the dynamics of Riemann geometry in curved Riemannian hyper-spacetime, we will demonstrate that the geometry of locally flat gravigauge hyper-spacetime emerges as non-commutative geometry which is characterized by non-Abelian Lie algebra structure via non-commutative relation of hyper-gravicoordinate derivative operator. It is unlike globally flat free-motion hyper-spacetime described by Minkowski hyper-spacetime with commutative geometry, the gravitational interaction in locally flat gravigauge hyper-spacetime appears as an emergent interaction which is characterized through non-commutative geometry. In fact, we will verify that the hyperspin gravigauge field in locally flat gravigauge hyper-spacetime behaves as an auxiliary field and determines completely the structure factor of non-Abelian Lie algebra for characterizing non-commutative geometry. So that the dynamics of all fundamental fields is associated with gravitational interaction via emergent non-commutative geometry in locally flat gravigauge hyper-spacetime.


\subsection{Dimensionless hyperspin gauge field and field strength in hidden coordinate formalism with zero global and local scaling charges }

Taking the scaling gauge invariant hyper-gravigauge field as Goldstone-like boson, it is straightforward to project the hyperspin gauge field $\cA_{\fM}^{\fA\fB}$ as vector field in free-motion hyper-spacetime $\fM_h$ into the vector field in locally flat gravigauge hyper-spacetime $\fG_h$,
\be \label{HSGFGG}
 \cA_{\fC}^{\fA\fB} \equiv \hfA_{\fC}^{\;\, \fM} \cA_{\fM}^{\fA\fB} . 
\ee
The gravitational origin of hyperspin gauge symmetry shown in Eq.(\ref{HSGGF1}) enables us to express the above hyperspin gauge field into the following two parts:
\be \label{HSGFDC3}
\cA_{\fC}^{\fA\fB} =  \fOm_{\fC}^{\fA\fB} + \fA_{\fC}^{\fA\fB} ,
\ee
where $\fOm_{\fC}^{\fA\fB}$ and $\fA_{\fC}^{\fA\fB}$ define the {\it dimensionless hyperspin gravigauge field and hyperspin covariant-gauge field}, respectively, in locally flat gravigauge hyper-spacetime. $\fOm_{\fC}^{\fA\fB}$ reveals the gravitational origin of gauge symmetry, which is explicitly given by dual hyper-gravigauge fields $\fA_{\fM}^{\;\, \fA}$ and $\hfA_{\fA}^{\;\,\fM}$ as follows: 
\be \label{HSGGFHC3}
\fOm_{\fC}^{\fA\fB}  & = &  \frac{1}{2} [\, \hfA^{\fA\fM} \heth_{\fC}\fA_{\fM}^{\;\, \fB} - \hfA^{\fB\fM} \heth_{\fC}\fA_{\fM}^{\;\, \fA} - \hfA_{\fC}^{\;\,\fM} ( \heth^{\fA}\fA_{\fM}^{\;\, \fB} - \heth^{\fB}\fA_{\fM}^{\;\, \fA} ) \nn \\
& + & ( \heth^{\fA}\hfA^{\fB\fM} -  \heth^{\fB}\hfA^{\fA\fM} ) \fA_{\fM\fC}], 
\ee
where $\heth_{\fC} \equiv \hfA_{\fC}^{\;\, \fM} \p_{\fM}$ is the dimensionless hyper-gravicoordinate derivative operator defined in locally flat gravigauge hyper-spacetime. 

Correspondingly, the dimensionless field strength of hyperspin gauge field in hidden coordinate formalism can be expressed into the following two parts:
\be
\cF_{\fC\fD}^{\fA\fB} = \fR_{\fC\fD}^{\fA\fB} + \fF_{\fC\fD}^{\fA\fB}  , 
\ee
with the explicit forms given as follows: 
\be \label{HSGFS3}
& & \cF_{\fC\fD}^{\fA\fB} = \ckcD_{\fC} \cA_{\fD}^{\fA\fB} - \ckcD_{\fD} \cA_{\fC}^{\fA\fB} + ( \cA_{\fC \fB'}^{\fA} \cA_{\fD}^{\fB' \fB} -  \cA_{\fD \fB'}^{\fA} \cA_{\fC}^{\fB' \fB} )  , \nn \\
& &  \ckcD_{\fC} \cA_{\fD}^{\fA\fB}  =  \heth_{\fC} \cA_{\fD}^{\fA\fB}  - \fOm_{\fC \fD}^{\fC'} \cA_{\fC'}^{\fA \fB} , 
\ee
and
\be \label{HSGGFS3}
& & \fR_{\fC\fD}^{\fA\fB} = \ckcD_{\fC} \fOm_{\fD}^{\fA\fB} - \ckcD_{\fD} \fOm_{\fC}^{\fA\fB} + \fOm_{\fC \fB'}^{\fA} \fOm_{\fD}^{\fB' \fB} -  \fOm_{\fD \fB'}^{\fA} \fOm_{\fC}^{\fB' \fB}   \, , \nn \\
& &  \ckcD_{\fC} \fOm_{\fD}^{\fA\fB}  =  \heth_{\fC} \fOm_{\fD}^{\fA\fB}  - \fOm_{\fC \fD}^{\fC'} \fOm_{\fC'}^{\fA \fB} , \nn \\
 & & \fF_{\fC\fD}^{\fA\fB} = \ckcD_{\fC} \fA_{\fD}^{\fA\fB} - \ckcD_{\fD} \fA_{\fC}^{\fA\fB}  +  \fA_{\fC \fB'}^{\fA} \fA_{\fD}^{\fB' \fB} -  \fA_{\fD \fB'}^{\fA} \fA_{\fC}^{\fB' \fB} \nn \\
 & & \qquad \quad + \fOm_{\fC \fB'}^{\fA} \fA_{\fD}^{\fB' \fB} + \fOm_{\fC \fB'}^{\fB} \fA_{\fD}^{\fA \fB'}  -  \fOm_{\fD \fB'}^{\fA} \fA_{\fC}^{\fB' \fB} - \fOm_{\fD \fB'}^{\fB} \fA_{\fC}^{\fA \fB'}, \nn \\
& & \ckcD_{\fC}  \fA_{\fD}^{\fA\fB} = \heth_{\fC}  \fA_{\fD}^{\fA\fB}  -  \fOm_{\fC \fD}^{\fC'} \fA_{\fC'}^{\fA\fB} ,
\ee
where $\fR_{\fC\fD}^{\fA\fB}$ and $\fF_{\fC\fD}^{\fA\fB}$ are the dimensionless hyperspin gravigauge field strength and hyperspin covariant-gauge field strength, respectively. $\ckcD_{\fC}$ defines the dimensionless covariant derivative of hyper-gravicoordinate derivative $\heth_{\fC}$ under the presence of hyperspin gravigauge field in hidden coordinate formalism. For convenience, $\ckcD_{\fC}$ is mentioned as {\it dimensionless hyper-gravigauge covariant derivative}. 

Similarly, the dimensionless scaling gauge field strength is defined as follows:  
\be \label{CSGFS}
& & \mF_{\fC\fD} = \ckcD_{\fC}\mW_{\fD} - \ckcD_{\fD}\mW_{\fC}  , \quad  
\mW_{\fC} = \hfA_{\fC}^{\;\, \fM} \mW_{\fM} , \nn \\
 & & \ckcD_{\fC}\mW_{\fD} = \heth_{\fC}\mW_{\fD} - \fOm_{\fC \fD}^{\fC'} \mW_{\fC'} . 
\ee

The hyperspin gravigauge field strength $\fR_{\fC\fD}^{\fA\fB}$ defines Riemann-like tensor in locally flat gravigauge hyper-spacetime, which can be verified from the relation between the field strength of hyperspin gravigauge field defined in hidden coordinate formalism and that defined in hidden gauge formalism, i.e.:
\be
\fR_{\fC\fD \fA\fB} = \hfA^{\;\fM}_{\fC} \hfA^{\;\fN}_{\fD}  \fR_{\fM\fN \fA\fB} =  \hfA^{\;\fM}_{\fC} \hfA^{\;\fN}_{\fD} \fR_{\fM\fN \fP\fQ} \hfA^{\;\fP}_{\fA} \hfA^{\;\fQ}_{\fB} ,
\ee
which enables us to define the symmetric Ricci-like curvature tensor by contracting the Riemann-like tensor in locally flat gravigauge hyper-spacetime,
 \be \label{RicciLT}
\fR_{\fC\fB} = \fR_{\fC\fD\fB}^{\fA} \eta_{\fA}^{\, \fD} = - \fR_{\fD\fC\fB}^{\fA} \eta_{\fA}^{\, \fD} = - \fR_{\fA\fC\fB}^{\fA} = \fR_{\fB\fC} ,
\ee 
and also the Ricci-like scalar curvature given by the trace of Ricci-like curvature tensor, 
\be  \label{RicciLS}
\fR \equiv \eta^{\fC\fB} \fR_{\fC\fB} = \eta^{\fB\fC}   \fR_{\fC\fD\fB}^{\fA} \eta_{\fA}^{\, \fD}  = - \eta^{\fB\fC}   \fR_{\fC\fD\fB\fA}\eta^{\fA\fD}  .
\ee

From the symmetry properties and identities of Riemann tensors shown in Eqs.(\ref{SPRT}) and (\ref{IRRT}), we arrive at the following analogous relations:
\be \label{SP3}
& & \fR_{\fC\fD\fA\fB} = -\fR_{\fD\fC\fA\fB}  , \quad \fR_{\fC\fD\fA\fB} = -\fR_{\fC\fD\fB\fA} , \nn \\
& &  \fR_{\fC\fD\fA\fB} = \fR_{\fA\fB\fC\fD} , \quad  \fR_{\fC\fD\fA\fB} = \fR_{\fB\fA\fD\fC} ,
\ee
for the symmetry properties, and 
\be
& & \fR_{\fC\fD\fB\fA} + \fR_{\fD\fB\fC\fA} + \fR_{\fB\fC\fD\fA} = 0  , \nn \\
& & \cD_{\fE}\fR_{\fC\fD\fB}^{\fA} + \fD_{\fC}\cR_{\fD\fE\fB}^{\fA} + \fD_{\fD}\cR_{\fE\fC\fB}^{\fA} = 0 ,
\ee
for the Bianchi identities.

In general, we have the following relations between the field strength defined in locally flat gravigauge hyper-spacetime and that in biframe hyper-spacetime with gravigauge hyper-fiber bundle structure:
 \be
& & \cF_{\fC\fD}^{\fA\fB} = \hfA^{\;\fM}_{\fC} \hfA^{\;\fN}_{\fD} \cF_{\fM\fN}^{\fA\fB}  ; \quad \cF_{\fC\fD\fA\fB} =  \hfA^{\;\fM}_{\fC} \hfA^{\;\fN}_{\fD} \cF_{\fM\fN\fA\fB}  , \nn \\
& & \fR_{\fC\fD}^{\fA\fB} =  \hfA^{\;\fM}_{\fC} \hfA^{\;\fN}_{\fD} \fR_{\fM\fN}^{\fA\fB} ; \quad  \fR_{\fC\fD \fA\fB} = \hfA^{\;\fM}_{\fC} \hfA^{\;\fN}_{\fD} \fR_{\fM\fN \fA\fB}   , \nn \\
& & \fF_{\fC\fD}^{\fA\fB}  = \hfA^{\;\fM}_{\fC} \hfA^{\;\fN}_{\fD} \fF_{\fM\fN}^{\fA\fB} ; \quad \fF_{\fC\fD\fA\fB} = \hfA^{\;\fM}_{\fC} \hfA^{\;\fN}_{\fD} \fF_{\fM\fN\fA\fB} . 
\ee
By lowering and raising the indices with the constant metric matrices $\eta^{\fA\fB}$ and $\eta_{\fA\fB}$, we can define the following gauge field and field strength in the hidden coordinate formalism: 
\be \label{HSGFGGS}
& & \cA^{\fC\fA\fB} \equiv \eta^{\fC\fC'} \cA_{\fC'}^{\fA\fB} \equiv  \eta^{\fC\fC'} \eta^{\fA\fA'} \eta^{\fB\fB'}\cA_{\fC'\fA'\fB'} , \nn \\
& &  \cF^{\fC\fD\fA\fB} \equiv \eta^{\fC\fC'} \eta^{\fD\fD'}\cF_{\fC'\fD'}^{\fA\fB} \equiv \eta^{\fC\fC'} \eta^{\fD\fD'} \eta^{\fA\fA'} \eta^{\fB\fB'} \cF_{\fC'\fD'\fA'\fB'} .  
\ee

It is interesting to note that all fields defined in the hidden coordinate formalism become dimensionless with zero global and local scaling charges $\cC_s=0$ and $\check{\cC}_s=0$.


\subsection{ Hyperunified field theory in hidden coordinate formalism with geometry-gauge correspondence and the hyperspin gravigauge field as auxiliary field in locally flat gravigauge hyper-spacetime}

From the above analysis, we are able to represent the action of hyperunified field theory in hidden coordinate formalism as follows: 
\be  \label{actionHUFTHCF}
\cS_{\mH\mU} & \equiv & \int [d^{D_h}x] \, \fkA(x)\, \fkL = \int [\delta^{D_h} \vka ] \, \nn \\
& \cdot & \{\, \bar{\fPsi}_{\fQH} \vSi_{-}^{\fC} i\cD_{\fC} \fPsi_{\fQH} - \beta_Q \sinh\chi_s \, \bar{\fPsi}_{\fQH} \tvSi_{-}\fPsi_{\fQH}  \nn \\
& - & \frac{1}{4} g_H^{-2} \cF_{\fC\fD\fA\fB}  \cF^{\fC\fD\fA\fB} + g_H^{-2} \eta^{\fC\fD} \fR_{\fC\fD} \nn \\
& + & \frac{1}{2} g_H^{-2}\beta_G^2\beta_Q^2 \sinh^2\chi_s \, ( \cA_{\fC\fA\fB} - \fOm_{\fC\fA\fB} )(\cA^{\fC\fA\fB} - \fOm^{\fC\fA\fB})  \nn \\
& - & \frac{1}{4} g_W^{-2} \mF_{\fC\fD} \mF^{\fC\fD} + \frac{1}{2} \lambda_S^2( 1+ \sinh^2\chi_s ) \mW_{\fC} \mW^{\fC}  \nn \\
&  + & \frac{1}{2} \lambda_S^2 \heth_{\fC} \chi_s \heth^{\fC}\chi_s  - \lambda_D^2  \cF(\chi_s)  \, \}  
\ee
with the definitions:
\be
& & i\cD_{\fC}\equiv i\heth_{\fC} + \cA_{\fC\fA\fB}\frac{1}{2}\vSi^{\fA\fB} \equiv \hfA_{\fC}^{\;\fM} i\cD_{\fM}, \nn \\
& & [\delta^{D_h} \vka ] \equiv [d^{D_h}x] \, \fkA(x) , \quad \delta \vka^{\fA}(x) = \fA_{\fM}^{\;\;\fA} dx^{\fM}, 
\ee
where $[\delta^{D_h} \vka]$ is regarded as the integral measure in locally flat gravigauge hyper-spacetime, and $\delta \vka^{\fA}(x)$ is considered as the hyper-gravicoordinate displacement of hyper-gravicoordinate vector field $\vka^{\fA}(x)$ in locally flat gravigauge hyper-spacetime. $\vka^{\fA}(x)$ is also called as dimensionless hyper-gravivector field for short. 

The action in hidden coordinate formalism shown in Eq.(\ref{actionHUFTHCF}) gets a similar form as the action in hidden gauge formalism presented in Eq.(\ref{actionHUFTHGF}). Unlike the action in Eq.(\ref{actionHUFTHGF}), the action in Eq.(\ref{actionHUFTHCF}) does not explicitly involve the hyper-gravigauge field $\hfA_{\fA}^{\;\fM}$ and hyper-gravimetric field $\hmH^{\fM\fN}$, all Latin alphabet indices for vectors and tensors are raised and lowered by the constant metric matrices $\eta^{\fA\fB}$ or $\eta_{\fA\fB} =\diag.(1,-1,\ldots,-1)$ in locally flat gravigauge hyper-spacetime. It is interesting to note that all basic fields defined in hidden coordinate formalism become dimensionless and have zero global and local scaling charges $\cC_s=0$ and $\check{\cC}_s=0$.

The fundamental gauge interactions in hidden coordinate formalism are governed by the hyperspin gauge field $\cA_{\fC}^{\fA\fB}$ and hyperspin gravigauge field $\fOm_{\fC}^{\fA\fB}$. In comparison to the action in hidden gauge formalism given in Eq.(\ref{actionHUFTHGF}), the dimensionless hyperspin gravigauge field $\fOm_{\fC}^{\fA\fB}$ in hidden coordinate formalism replaces both hyper-gravigauge field $\fA_{\fM}^{\;\fA}$ and hyper-gravimetric field $\mH_{\fM\fN}$ to characterize the gravitational interaction in locally flat gravigauge hyper-spacetime. The dynamics of $\fOm_{\fC}^{\fA\fB}$ is characterized by the Ricci-like scalar tensor $\fR \equiv \eta^{\fC\fB} \fR_{\fC\fB} = \eta^{\fC\fB}  \fR_{\fC\fD\fA\fB}\eta^{\fA\fD}$, which is analogous to the scaling gauge invariant Einstein-Hilbert type action in hidden gauge formalism. Nevertheless, they have distinguishable features that the Ricci-like scalar tensor $\fR \equiv \eta^{\fC\fB}  \fR_{\fC\fD\fA\fB}\eta^{\fD\fA}$ in hidden coordinate formalism is governed by hyperspin gauge symmetry SP(1,$D_h$-1) via hyperspin gravigauge field $\fOm_{\fC}^{\fA\fB}$, whereas the Ricci-like scalar tensor $\fR \equiv \hmH^{\fM\fQ}  \fR_{\fM\fN\fP\fQ}\hmH^{\fN\fP}$ in hidden gauge formalism is governed by general linear group symmetry GL(1,$D_h$-1) via hyper-spacetime gravimetric-gauge field $\fGa_{\fM\fQ}^{\fP}$ as Christoffel symbols in curved Riemannian hyper-spacetime. 

In such a hidden coordinate formalism, the fundamental gravitational interaction is characterized by the gauge interaction through the hyperspin gravigauge field $\fOm_{\fC}^{\fA\fB}$. Therefore, when comparing the action in hidden coordinate formalism in Eq.(\ref{actionHUFTHCF}) with that in hidden gauge formalism in Eq.(\ref{actionHUFTHGF}), we come to the {\it geometry-gauge correspondence}. 

To show explicitly such a {\it geometry-gauge correspondence}, it is useful to corroborate the following relation:
\be
\fkA\, \fR & \equiv & \fkA\, \hmH^{\fM\fQ}  \fR_{\fM\fQ} = \fkA\, \hmH^{\fM\fQ}  \fR_{\fM\fN\fP\fQ}\hmH^{\fN\fP} \nn \\
& \equiv & \fkA\,  \eta^{\fC\fB} \fR_{\fC\fB} = \fkA\, \eta^{\fC\fB}  \fR_{\fC\fD\fA\fB}\eta^{\fD\fA} ,  \nn \\
& \equiv & -\fkA\, \fOm_{\fC\fA\fB} \fOm^{\fA\fB\fC} - \fkA\, \fOm_{\fC\fA}^{\fC} \fOm_{\fD}^{\fD\fA} + 2\p_{\fM} ( \fkA\, \hfA_{\fA}^{\;\fM} \fOm_{\fC}^{\fC\fA} ),
\ee
which enables us to rewrite the action in Eq.(\ref{actionHUFTHCF}) into the following form:
\be  \label{actionHUFTHCF2}
\cS_{\mH\mU} & \equiv &  \int [\delta^{D_h} \vka ]  \{\, \bar{\fPsi}_{\fQH} \vSi_{-}^{\fC} i\cD_{\fC} \fPsi_{\fQH} - \beta_Q \sinh\chi_s \bar{\fPsi}_{\fQH} \tvSi_{-}\fPsi_{\fQH}  \nn \\
& - & \frac{1}{4} g_H^{-2}\cF_{\fC\fD\fA\fB}  \cF^{\fC\fD\fA\fB} - g_H^{-2} ( \fOm_{\fC\fA\fB} \fOm^{\fA\fB\fC} + \fOm_{\fC\fB}^{\fC} \fOm_{\fD}^{\fD\fB} ) \nn \\
& + & \frac{1}{2}g_H^{-2}\beta_G^2\beta_Q^2 \sinh^2\chi_s \, ( \cA_{\fC\fA\fB} - \fOm_{\fC\fA\fB} )(\cA^{\fC\fA\fB} - \fOm^{\fC\fA\fB})  \nn \\
& - & \frac{1}{4} g_W^{-2} \mF_{\fC\fD} \mF^{\fC\fD} + \frac{1}{2} \lambda_S^2 ( 1 +\sinh^2\chi_s ) \mW_{\fC} \mW^{\fC}  \nn \\
& + & \frac{1}{2} \lambda_S^2 \heth_{\fC} \chi_s \heth^{\fC}\chi_s  - \lambda_D^2 \cF(\chi_s)  \, \} ,
\ee
where we have ignored the surface term $2\p_{\fM} ( \fkA\, \hfA_{\fA}^{\;\fM} \fOm_{\fC}^{\fC\fA} )$. Such an action represents the hyperunified field theory in locally flat gravigauge hyper-spacetime, which explicitly displays the geometry-gauge correspondence for gravitational interaction. It is noticed that when taking the hyperspin gauge field $\cA_{\fC}^{\fA\fB}$ as the fundamental field, there appears no kinetic term for the hyperspin gravigauge field $\fOm_{\fC}^{\fA\fB}$ in locally flat gravigauge hyper-spacetime.

Therefore, we come to the observation that the hyperunified field theory formulated in locally flat gravigauge hyper-spacetime as shown in Eq.(\ref{actionHUFTHCF2}) brings on an intriguing feature that {\it the gravitational interaction appears to decouple from the hyperunified qubit-spinor field and is solely characterized by the hyperspin gravigauge field $\fOm_{\fC}^{\fA\fB}$ which behaves as an auxiliary field }.


\subsection{ Emergence of non-commutative geometry with non-Abelian Lie algebra structure and the constraint equation of hyper-gravigauge field as auxiliary field in locally flat gravigauge hyper-spacetime }

Let us begin with the nontrivial commutation relation of hyper-gravicoordinate derivative operator which forms gravigauge hyper-basis $\{\heth_{\fC}\}$ for locally flat gravigauge hyper-spacetime $\fG_h$, i.e.:
 \be \label{NCG}
 & & [\heth_{\fC}\, , \heth_{\fD} ] = \hat{f}_{\fC\fD}^{\fA}\1 \heth_{\fA} , \nn \\
 & & \hat{f}_{\fC\fD}^{\fA} \equiv \fOm_{[\fC\fD]}^{\fA} \equiv \fOm_{\fC\1 \fD}^{\fA} - \fOm_{\fD\1 \fC}^{\fA} \equiv -\eta^{\fA\fB} \fOm_{[\fC\fD]\fB} , 
\ee
which leads to {\it non-Abelian Lie algebra} of hyper-gravicoordinate derivative operator $\heth_{\fC}$. Unlike the usual non-Abelian Lie algebra, the structure factor $\hat{f}_{\fC\fD}^{\fA}$ for such a non-Abelian Lie algebra is no longer a constant, which is determined by the hyperspin gravigauge field $\fOm_{[\fC\fD]}^{\fA}$ and brings on the {\it emergence of non-commutative geometry} in locally flat gravigauge hyper-spacetime $\fG_h$. To explore the non-commutative geometric property of locally flat gravigauge hyper-spacetime $\fG_h$, it needs to resolve the dynamics of hyperspin gravigauge field $\fOm_{[\fC\fD]}^{\fA}$ from the geometry-gauge correspondence. 

In the action of hyperunified field theory formulated in hidden coordinate formalism with geometry-gauge correspondence shown in Eq.(\ref{actionHUFTHCF2}), the gravitational interaction is completely characterized by the hyperspin gravigauge field $\fOm_{\fC\fD}^{\fA}$. Nevertheless, when taking the hyperspin gauge field $\cA_{\fC}^{\fA\fB}$ as fundamental gauge field in locally flat gravigauge hyper-spacetime, the hyperspin gravigauge field $\fOm_{\fC\fD}^{\fA}$ should not act as an independent field in order to preserve the total independent degrees of freedom in the hyperunified field theory. In fact, $\fOm_{\fC\fD}^{\fA}$ couples only to the hyperspin gauge field $\cA_{\fC}^{\fA\fB}$ and scaling gauge field via dimensionless hyper-gravigauge covariant derivative appearing in their gauge field strengths shown in Eqs.(\ref{HSGFS}) and (\ref{CSGFS}) and meanwhile possesses the $\cQ_c$-spin scalar-gauge interaction term indicated by $\beta_G^2$. To show explicitly the hyperspin gravigauge field interaction, it is useful to rewrite Eqs.(\ref{HSGFS}) and (\ref{CSGFS}) into the following forms:
\be \label{HSGFS2}
& & \cF_{\fC\fD}^{\fA\fB} \equiv  \tilde{\cF}_{\fC\fD}^{\fA\fB} + \fOm_{[\fC\fD]}^{\fE} \cA_{\fE}^{\fA \fB}  , \nn \\
& & \tilde{\cF}_{\fC\fD}^{\fA\fB} \equiv  \heth_{\fC} \cA_{\fD}^{\fA\fB}  -  \heth_{\fD} \cA_{\fC}^{\fA\fB} + ( \cA_{\fC \fB'}^{\fA} \cA_{\fD}^{\fB' \fB} -  \cA_{\fD \fB'}^{\fA} \cA_{\fC}^{\fB' \fB} ) , \nn \\
& & \mF_{\fC\fD} \equiv \tilde{\mF}_{\fC\fD} + \fOm_{[\fC\fD]}^{\fE} \mW_{\fE} , \nn \\
& & \tilde{\mF}_{\fC\fD} \equiv \heth_{\fC} \mW_{\fD}  -  \heth_{\fD} \mW_{\fC} .
\ee

To study the dynamics of non-commutative geometry through the structure factor $\hat{f}_{\fC\fD}^{\fA}\equiv \fOm_{[\fC\fD]}^{\fA}$ of non-Abelian Lie algebra for hyper-gravicoordinate derivative operator $\heth_{\fC}$, it is useful to express the hyperspin gravigauge field $\fOm_{\fC\fD\fA}$ appearing in the action (Eq.(\ref{actionHUFTHCF2}) ) as antisymmetric structure factor $\fOm_{[\fC\fD]\fA}$. For that, we should use the following relations:
\be \label{HGGFrelation}
& & \fOm_{\fC \fD\fA} \equiv \eta_{\fC\fD\fA}^{\; \fC'\fD'\fA'} \fOm_{[\fC' \fD']\fA'} , \nn \\
& & \fOm_{\fC\fA\fB} \fOm^{\fA\fB\fC} = \frac{1}{2} \fOm_{[\fC\fD]\fA} \fOm^{[\fC\fD]\fA} - \fOm _{\fC\fD\fA} \fOm^{\fC\fD\fA}, \nn \\
& &  \fOm_{\fC\fD\fA} \fOm^{\fC\fD\fA}  \equiv \fOm_{[\fC\fD]\fA} \fOm^{[\fC'\fD']\fA'} \bar{\eta}_{\fC'\fD'\fA'}^{\;\; \fC\fD\fA} ,  
\ee
with the constant tensors defined as follows:
\be \label{GGtensor}
& & \eta_{\fC\fD\fA}^{\; \fC'\fD'\fA'} = \frac{1}{2} ( \eta_{\fC}^{\;\fC'} \eta_{\fD}^{\;\fD'} \eta_{\fA}^{\;\fA'} - \eta_{\fC}^{\;\fC'} \eta_{\fD}^{\;\fA'} \eta_{\fA}^{\;\fD'} - \eta_{\fC}^{\;\fA'} \eta_{\fD}^{\;\fC'} \eta_{\fA}^{\;\fD'} ) , \nn \\
& & \bar{\eta}_{\fC'\fD'\fA'}^{\;\; \fC\fD\fA} = \frac{1}{4} ( 3 \eta_{\fC'}^{\;\fC} \eta_{\fD'}^{\;\fD} \eta_{\fA'}^{\;\fA} + 3\eta_{\fA'}^{\;\fC} \eta_{\fD'}^{\;\fD} \eta_{\fC'}^{\;\fA} +\eta_{\fA'}^{\;\fC} \eta_{\fC'}^{\;\fD} \eta_{\fD'}^{\;\fA} ) ,
\ee
which enables us to rewrite the action in Eq.(\ref{actionHUFTHCF2}) into the following form:
\be  \label{actionHUFTHCF3}
\cS_{\mH\mU} & \equiv &  \int [\delta^{D_h} \vka ] \{\, \bar{\fPsi}_{\fQH} \vSi_{-}^{\fC} i\cD_{\fC} \fPsi_{\fQH} - \beta_Q \sinh\chi_s \bar{\fPsi}_{\fQH} \tvSi_{-}\fPsi_{\fQH}  \nn \\
& - & \frac{1}{4} g_H^{-2}(\tilde{\cF}_{\fC\fD\fA\fB}\tilde{\cF}^{\fC\fD\fA\fB}  - 2 \fOm^{[\fC\fD]\fA}  \tilde{\cF}_{\fC\fD\fA'\fB'}\cA_{\fA}^{\fA'\fB'} ) \nn \\
& + & g_H^{-2} \fOm_{[\fC\fD]\fA} \fOm^{[\fC'\fD']\fA'} \tilde{\eta}_{\fC'\fD'\fA'}^{\;\; \fC\fD\fA} 
- \frac{1}{4} g_H^{-2}\fOm_{[\fC\fD]\fA} \fOm^{[\fC\fD]\fB}  \cA^{\fA\fA'\fB'} \cA_{\fB\fA'\fB'}
 \nn \\
& + & \frac{1}{2} g_H^{-2}\beta_G^2\beta_Q^2 \sinh^2\chi_s \, \left( \fOm_{[\fC\fD]\fA} \fOm^{[\fC'\fD']\fA'} + (\cA_{[\fC\fD]\fA} - 2\fOm_{[\fC\fD]\fA} ) \cA^{[\fC'\fD']\fA'} \right) \bar{\eta}_{\fC'\fD'\fA'}^{\;\; \fC\fD\fA}  \nn \\
& - & \frac{1}{4} g_W^{-2} (\tilde{\mF}_{\fC\fD} \tilde{\mF}^{\fC\fD} - 2 \fOm_{[\fC\fD]\fA} \tilde{\mF}^{\fC\fD}  \mW^{\fA}) + \frac{1}{2} \lambda_S^2( 1 + \sinh^2\chi_s ) \mW_{\fC} \mW^{\fC}  \nn \\
&  - & \frac{1}{4} g_W^{-2}\fOm_{[\fC\fD]\fA} \fOm^{[\fC\fD]\fB}  \mW^{\fA} \mW_{\fB} + \frac{1}{2} \lambda_S^2 \heth_{\fC} \chi_s \heth^{\fC}\chi_s  - \lambda_D^2 \cF(\chi_s)  \, \} ,
\ee
with the definitions,
\be \label{GGItensor}
& & \cA_{[\fC\fD]\fA}  \equiv \cA_{\fC\fD\fA}  - \cA_{\fD\fC\fA} , \nn \\
& & \tilde{\eta}_{\fC'\fD'\fA'}^{\;\; \fC\fD\fA} = \bar{\eta}_{\fC'\fD'\fA'}^{\;\; \fC\fD\fA} 
- \frac{1}{2} \eta_{\fC'}^{\;\fC} \eta_{\fD'}^{\;\fD} \eta_{\fA'}^{\;\fA} - \eta_{\fC'\fA'}\eta_{\fD'}^{\;\fD} \eta^{\fC\fA} ,\nn \\
& & \qquad \quad \; \, = \frac{1}{4} (  \eta_{\fC'}^{\;\fC} \eta_{\fD'}^{\;\fD} \eta_{\fA'}^{\;\fA} + 3\eta_{\fA'}^{\;\fC} \eta_{\fD'}^{\;\fD} \eta_{\fC'}^{\;\fA} +\eta_{\fA'}^{\;\fC} \eta_{\fC'}^{\;\fD} \eta_{\fD'}^{\;\fA} ) - \eta_{\fC'\fA'}\eta_{\fD'}^{\;\fD} \eta^{\fC\fA} , 
\ee
where the tensor $\bar{\eta}_{\fC'\fD'\fA'}^{\;\; \fC\fD\fA}$ is presented in Eq.(\ref{GGtensor}).

In the above action, the hyperspin gravigauge field $\fOm_{[\fC\fD]\fA}$ gets no kinetic term and appears as an {\it auxiliary field} in locally flat gravigauge hyper-spacetime. From the least action principle, we obtain the following relation:  
\be \label{GGCGE}
\tilde{\eta}_{\fC'\fD'\fA'}^{\;\; [\fC\fD]\fA}\, \fOm^{[\fC'\fD']\fA'} + \tilde{f}_{\fC'\fD'\fA'}^{\;\; [\fC\fD]\fA} \, \fOm^{[\fC'\fD']\fA'} = \tilde{F}^{[\fC\fD]\fA} ,
\ee
which presents a {\it constraint equation of hyperspin gravigauge field $\fOm_{[\fC\fD]\fA}$}. Where we have introduced the following definitions:
\be \label{GGCGE0}
& & \tilde{f}_{\fC'\fD'\fA'}^{\;\; [\fC\fD]\fA} = \frac{1}{2}  \bar{\eta}_{\fC'\fD'\fA'}^{\;\; [\fC\fD]\fA}  \beta_G^2 \beta_Q^2 \sinh^2\chi_s  - \frac{1}{4}\eta_{\fC'\fD'}^{\;\; [\fC\fD]} (\cA_{\fA'\fB\fB'}\cA^{\fA\fB\fB'} + \frac{g_H^2}{g_W^2}  \mW_{\fA'}\mW^{\fA} ) , \nn \\
& & \tilde{F}^{[\fC\fD]\fA} = \frac{1}{2} \bar{\eta}_{\fC'\fD'\fA'}^{\;\; [\fC\fD]\fA}  \beta_G^2 \beta_Q^2 \sinh^2\chi_s \, \cA^{[\fC'\fD']\fA'}  - \frac{1}{2} (\tilde{\cF}^{\fC\fD\fA'\fB'}\eta^{\fA\fC'} \cA_{\fC'\fA'\fB'} + \frac{g_H^2}{g_W^2}  \tilde{\mF}^{\fC\fD}\mW^{\fA} ) , \nn \\
& & \tilde{\eta}_{\fC'\fD'\fA'}^{\;\; [\fC\fD]\fA} \equiv \tilde{\eta}_{\fC'\fD'\fA'}^{\;\; \fC\fD\fA} - \tilde{\eta}_{\fC'\fD'\fA'}^{\;\; \fD\fC\fA} , \quad \eta_{\fC'\fD'}^{\;\; [\fC\fD]} \equiv \eta_{\fC'}^{\;\fC} \eta_{\fD'}^{\;\fD} - \eta_{\fC'}^{\;\fD} \eta_{\fD'}^{\;\fC} , \nn \\
& & \bar{\eta}_{\fC'\fD'\fA'}^{\;\; [\fC\fD]\fA} \equiv \bar{\eta}_{\fC'\fD'\fA'}^{\;\; \fC\fD\fA} - \bar{\eta}_{\fC'\fD'\fA'}^{\;\; \fD\fC\fA}  .
\ee

It is clear that the hyperspin gravigauge field $\fOm_{[\fC\fD]\fA}$ is no longer an independent gauge field in locally flat gravigauge hyper-spacetime when the hyperspin gauge field $\cA_{\fC}^{\fA\fB}$ is chosen as fundamental gauge field in hidden coordinate formalism. It is shown from Eqs.(\ref{GGCGE}) and (\ref{GGCGE0}) that $\fOm_{[\fC\fD]\fA}$ is in generally given by the hyperspin gauge field and $\cQ_c$-spin scaling field $\chi_s$ as well as scaling gauge field $\mW_{\fC}$. $\fOm_{[\fC\fD]\fA}$ in the constraint equation presented in Eq.(\ref{GGCGE}) appears in formal to be irrelevant to the hyperunified qubit-spinor field $\fPsi_{\fQH}$, while it is indirectly correlated to $\fPsi_{\fQH}$ through the dynamic equations of hyperspin gauge field and $\cQ_c$-spin scaling field shown as follows. 

Based on the hyperunified field theory in locally flat gravigauge hyper-spacetime with geometry-gauge correspondence as shown in Eq.(\ref{actionHUFTHCF3}), we come to the observation that the gravitational interaction behaves as an emergent interaction described by the hyperspin gravigauge field $\fOm_{[\fC\fD]\fA}$. $\fOm_{[\fC\fD]\fA}$ as auxiliary field in locally flat gravigauge hyper-spacetime is given by the fundamental bosonic fields through the constraint equation presented in Eqs.(\ref{GGCGE}) and (\ref{GGCGE0}). There are three fundamental bosonic fields in the hyperunified field theory including the hyperspin gauge field $\cA_{\fC}^{\fA\fB}$ and $\cQ_c$-spin scaling field $\chi_s$ as well as scaling gauge field $\mW_{\fC}$. The constraint equation for $\fOm_{[\fC\fD]\fA}$ enables us to determine in principle the structure factor of non-Abelian Lie algebra for hyper-gravicoordinate derivative operator $\heth_{\fC}$, which characterizes an emergent non-commutative geometry as genesis of gravitational interaction in locally flat gravigauge hyper-spacetime.


\subsection{Dynamics of fundamental fields in hyperunified field theory with emergent gravitational interaction in locally flat gravigauge hyper-spacetime} 

The hyperspin gravigauge field $\fOm_{[\fC\fD]}^{\fA}$ characterizing the geometry-gauge correspondence and emergence of non-commutative geometry is verified to be determined by the fundamental bosonic fields through the constraint equation presented in Eqs.(\ref{GGCGE}) and (\ref{GGCGE0}). The gravitational interaction in locally flat gravigauge hyper-spacetime appears to decouple in formal from the hyperunified qubit-spinor field $\fPsi_{\fQH}$ when taking the hyperspin gauge field $\cA_{\fC}^{\fA\fB}$ as fundamental gauge fields in hidden coordinate formalism of the action shown in Eq.(\ref{actionHUFTHCF3}). We will demonstrate how the dynamics of all fundamental fields gets gravitational effects through emergent gravitational interaction caused from the appearance of non-commutative geometry in locally flat gravigauge hyper-spacetime.

To derive the dynamic equations of motion for the hyperunified qubit-spinor field $\fPsi_{\fQH}$ and hyperspin gauge field $\cA_{\fC\fA\fB}$, it is useful to decompose the hyperspin gauge field $\cA_{\fC}^{\fA\fB}$ into the following two orthogonal parts:
\be \label{HSGFDC4}
& & \cA_{\fC\fA\fB} \equiv \cA_{[\fC\fA\fB]} + \cA_{(\fC\fA\fB]} , \nn \\
& & \cA_{[\fC\fA\fB]} \cA^{(\fC\fA\fB]} = 0, 
\ee
with the explicit definitions,
\be
& & \cA_{[\fC\fA\fB]} \equiv \frac{1}{3} ( \cA_{\fC\fA\fB} + \cA_{\fA\fB\fC} + \cA_{\fB\fC\fA}  ) \nn \\
& & \cA_{(\fC\fA\fB]}  \equiv  \frac{1}{3} ( 2\cA_{\fC\fA\fB} - \cA_{\fA\fB\fC} - \cA_{\fB\fC\fA} ), 
\ee
where $\cA_{[\fC\fA\fB]}$ denotes {\it totally antisymmetric hyperspin gauge field} and $\cA_{(\fC\fA\fB]}$ represents {\it symmetric-antisymmetric mixing hyperspin gauge field}. Correspondingly, the hyperspin gauge field strength $\cF_{\fC\fD\fA\fB}$ can also be decomposed into two orthogonal parts:
\be \label{HSGFSDC4}
& & \cF_{\fC\fD\fA\fB} \equiv \cF_{\fC[\fD\fA\fB]} + \cF_{\fC(\fD\fA\fB]} , \nn \\
& & \cF_{\fC[\fD\fA\fB]} \equiv \frac{1}{3} ( \cF_{\fC\fD\fA\fB}  +  \cF_{\fC\fA\fB\fD}  + \cF_{\fC\fB\fD\fA} ) \nn \\
& & \cF_{\fC(\fD\fA\fB]} \equiv \frac{1}{3} ( 2\cF_{\fC\fD\fA\fB}  -  \cF_{\fC\fA\fB\fD}  -\cF_{\fC\fB\fD\fA} ) , \nn \\
& &  \cF_{\fC[\fD\fA\fB]} \cF^{\fC(\fD\fA\fB]} = 0 . 
\ee

With the above decompositions, the action in Eq.(\ref{actionHUFTHCF2}) can be expressed into the following form,
\be  \label{actionHUFTHCF4}
\cS_{\mH\mU} & \equiv &  \int [\delta^{D_h} \vka ]  \{\, \bar{\fPsi}_{\fQH} \vSi_{-}^{\fC} i\cD_{\fC}\fPsi_{\fQH} - \beta_Q \sinh\chi_s \, \bar{\fPsi}_{\fQH} \tvSi_{-}\fPsi_{\fQH}  \nn \\
& - & \frac{1}{4} g_H^{-2} ( \cF_{\fC[\fD\fA\fB]}  \cF^{\fC[\fD\fA\fB]} +  \cF_{\fC(\fD\fA\fB]}  \cF^{\fC(\fD\fA\fB]}  ) \nn \\
& - & g_H^{-2} ( \fOm_{\fC\fA\fB} \fOm^{\fA\fB\fC} + \fOm_{\fC\fB}^{\fC} \fOm_{\fD}^{\fD\fB} ) \nn \\
& + & \frac{1}{2}g_H^{-2}\beta_G^2\beta_Q^2 \sinh^2\chi_s \, ( \cA_{[\fC\fA\fB]} - \fOm_{[\fC\fA\fB]} )(\cA^{[\fC\fA\fB]} - \fOm^{[\fC\fA\fB]})  \nn \\
& + & \frac{1}{2}g_H^{-2}\beta_G^2 \beta_Q^2 \sinh^2\chi_s \, ( \cA_{(\fC\fA\fB]} - \fOm_{(\fC\fA\fB]} )(\cA^{(\fC\fA\fB]} - \fOm^{(\fC\fA\fB]})  \nn \\
& - & \frac{1}{4} g_W^{-2} \mF_{\fC\fD} \mF^{\fC\fD} + \frac{1}{2} \lambda_S^2( 1 + \sinh^2\chi_s )\, \mW_{\fC} \mW^{\fC}  \nn \\
& + & \frac{1}{2} \lambda_S^2 \heth_{\fC} \chi_s \heth^{\fC}\chi_s  - \lambda_D^2  \cF(\chi_s)  \, \} ,
\ee
with 
\be \label{CD4}
i\cD_{\fC} \equiv i\heth_{\fC} + \cA_{[\fC\fA\fB]} \frac{1}{2}\vSi^{\fA\fB} . 
\ee

It is noticed that in hidden coordinate formalism of the action, only the totally antisymmetric hyperspin gauge field $\cA_{[\fC\fA\fB]}$ couples to the hyperunified qubit-spinor field $\fPsi_{\fQH}$ due to its self-conjugated chiral property. 

Based on the least action principle, we are able to derive the dynamic equations of motion for hyperspin gauge field in locally flat gravigauge hyper-spacetime, which can be written into the following two types:
\be \label{EMHSGF2}
& & \cD_{\fC} \cF^{\fC[\fD\fA\fB]} = - g_H^2\cJ^{[\fD\fA\fB]} , \nn \\
& & \cD_{\fC} \cF^{\fC(\fD\fA\fB]} = - g_H^2 \cJ^{(\fD\fA\fB]} , \nn \\
& & \cD_{\fC}\cF^{\fC[\fD\fA\fB]} \equiv \frac{1}{3} ( \cD_{\fC}\cF^{\fC\fD\fA\fB}  +  \cD_{\fC}\cF^{\fC\fA\fB\fD}  + \cD_{\fC}\cF^{\fC\fB\fD\fA} ) \nn \\
& & \cD_{\fC}\cF^{\fC(\fD\fA\fB]} \equiv \frac{1}{3} ( 2\cD_{\fC}\cF^{\fC\fD\fA\fB}  - \cD_{\fC} \cF^{\fC\fA\fB\fD}  - \cD_{\fC}\cF^{\fC\fB\fD\fA} ) ,
\ee 
where $\cJ^{[\fD\fA\fB]}$ represents the totally antisymmetric covariant tensor current and $\cJ^{(\fD\fA\fB]}$ the symmetric-antisymmetric mixing covariant tensor current. Their explicit forms are given as follows:
\be \label{CTC2}
& & \cJ^{[\fD\fA\fB]} =  \frac{1}{4} \bar{\fPsi}_{\fQH} \{ \vSi_{-}^{\fD}, \varSigma^{\fA\fB} \} \fPsi_{\fQH} -  g_H^{-2}\beta_G^2\beta_Q^2 \sinh^2\chi_s \, (\cA^{[\fD\fA\fB]} - \fOm^{[\fD\fA\fB]}) , \nn \\
& & \cJ^{(\fD\fA\fB]} =   -  g_H^{-2}\beta_G^2 \beta_Q^2 \sinh^2\chi_s \, (\cA^{(\fD\fA\fB]} - \fOm^{(\fD\fA\fB]}) . 
\ee
The covariant derivative in Eq.(\ref{EMHSGF2}) is defined as follows:
\be
& & \cD_{\fC} \cF^{\fC\fD\fA\fB} \equiv  \ckcD_{\fC} \cF^{\fC\fD\fA\fB} +  \cA_{\fC\fA'}^{\fA}\cF^{\fC\fD\fA'\fB} + \cA_{\fC\fB'}^{\fB}\cF^{\fC\fD\fA\fB'}, \nn \\
 & & \ckcD_{\fC} \cF^{\fC\fD\fA\fB} \equiv \heth_{\fC} \cF^{\fC\fD\fA\fB} + \fOm_{\fC\fC'}^{\fC}\cF^{\fC'\fD\fA\fB} + \fOm_{\fC\fD'}^{\fD}\cF^{\fC\fD'\fA\fB} , \nn \\
& & \cF^{\fC\fD\fA\fB} \equiv \tilde{\cF}^{\fC\fD\fA\fB} - \fOm^{[\fC\fD]\fE} \cA_{\fE}^{\fA \fB} , 
\ee
where the hyper-gravigauge covariant derivative $\ckcD_{\fC}$ and additional field strength term $\fOm^{[\fC\fD]\fE} \cA_{\fE}^{\fA \fB}$ reflect the geometry-gauge correspondence and the emergence of non-commutative geometry characterized by the hyperspin gravigauge field $\fOm_{[\fC\fD]}^{\fA}$ in locally flat gravigauge hyper-spacetime. 

In general, the above equations of motion in locally flat gravigauge hyper-spacetime can also be formulated from Eq.(\ref{EMHSGF}) obtained in biframe hyper-spacetime.

For the hyperunified qubit-spinor field $\fPsi_{\fQH}$, the equation of motion is simply given by,
\be \label{EMfQH2}
\vSi_{-}^{\fC}\left(i\cD _{\fC} - i\fV_{\fC} \right) \fPsi_{\fQH} 
 -\beta_Q \sinh\chi_s \, \tvSi_{-} \fPsi_{\fQH} =   0 , 
\ee
with the covariant derivative $\cD _{\fC}$ given in Eq.(\ref{CD4}), which appears to be irrelevant to the gravitational interaction. While its quadratic form is found to have the following form:
\be \label{EMfQH4}
& & (\ckcD_{\fC} -\fV_{\fC} ) ( \cD ^{\fC} -\fV^{\fC} ) \Psi_{\fQH} \nn \\
& & \; \; = - \left( \beta_Q^2 \sinh^2\chi_s \, +  \vGa^{\fC} \tvGa  \cosh\chi_s \, (i\heth_{\fC}\chi_s) \,  \right) \Psi_{\fQH} \nn \\
& & \; \; +  \varSigma^{\fC\fD} \left( \cF_{\fC\fD\fA\fB}\frac{1}{2}\vSi^{\fA\fB} - (\cA_{[\fC\fD]}^{\fA}- \fOm_{[\fC\fD]}^{\fA}) i ( \cD _{\fA} - \fV_{\fA} ) - i \fV_{\fC\fD}\, \right) \Psi_{\fQH} ,
 \ee
 which indicates that the quadratic form of equation of motion for the hyperunified qubit-spinor field does generate gravitational effects characterized by the hyperspin gravigauge field $\fOm_{[\fC\fD]}^{\fA}$ due to the emergence of non-commutative geometry in locally flat gravigauge hyper-spacetime. Where $\fV_{\fC}$ and $\fV_{\fC\fD}$ are graviscaling induced gauge invariant field and field strength defined as follows: 
\be
& & \fV_{\fC} \equiv \frac{1}{2} (\cA_{\fD\fC}^{\fD} - \fOm_{\fD\fC}^{\fD} ), \nn \\
& &  \fV_{\fC\fD}  = \ckcD_{\fC}\fV_{\fD}-  \ckcD_{\fD}\fV_{\fC}=  \heth_{\fC}\fV_{\fD}-  \heth_{\fD}\fV_{\fC} - \fOm_{[\fC\fD]}^{\fA} \fV_{\fA}.  
\ee

The equation of motion for the scaling gauge field $\mW_{\fC}$ can simply be expressed as follows:
\be   \label{EMSGF1}
 & & \ckcD_{\fC}  \mF^{\fC\fD}  =  g_W^{2}  \chi_s^2 \mW^{\fD} , \nn \\
 & & \ckcD_{\fC} \mF^{\fC\fD} \equiv \heth_{\fC} \mF^{\fC\fD} + \fOm_{\fC\fC'}^{\fC}\mF^{\fC'\fD} + \fOm_{\fC\fD'}^{\fD}\mF^{\fC\fD'} .
\ee

It is straightforward to write down the equation of motion for the $\cQ_c$-spin scaling field,  
\be   \label{EMCSF1}
& & \lambda_S^2 \ckcD_{\fC}\heth^{\fC}\chi_s =  \mJ_s , \nn \\
& & \ckcD_{\fC}\heth^{\fC} \equiv \heth_{\fC}\heth^{\fC} + \fOm_{\fC\fC'}^{\fC} \heth^{\fC'} ,
\ee
where the scalar current is given by,
\be
\mJ_s & = & - \beta_Q\cosh\chi_s \, \bar{\fPsi}_{\fQH} \tvSi_{-}\fPsi_{\fQH}   +  \beta_Q^2 \sinh\chi_s\cosh\chi_s \, \mW_{\fC} \mW^{\fC}  -  \lambda_D^2 \cF'(\chi_s) \nn \\
& + & g_H^{-2}\beta_G^2 \beta_Q^2 \sinh\chi_s\cosh\chi_s \, ( \cA_{\fC\fA\fB} - \fOm_{\fC\fA\fB} )(\cA^{\fC\fA\fB} - \fOm^{\fC\fA\fB})   .
\ee

To associate with the relation given in Eq.(\ref{GGCGE}) for the hyperspin gravigauge field $\fOm_{[\fC\fD]\fA}$, we can adopt the following identities:
\be
& & \fOm_{[\fC\fD\fA]} \equiv \eta_{[\fC\fD\fA]}^{\; \fC'\fD'\fA'} \fOm_{[\fC'\fD']\fA}, \quad \fOm_{(\fC\fD\fA]} \equiv \eta_{(\fC\fD\fA]}^{\; \fC'\fD'\fA'} \fOm_{[\fC'\fD']\fA} , \nn \\
& & \eta_{[\fC\fD\fA]}^{\; \fC'\fD'\fA'} \equiv \frac{1}{3} ( \eta_{\fC}^{\;\fC'} \eta_{\fD}^{\;\fD'} \eta_{\fA}^{\;\fA'} + \eta_{\fD}^{\;\fC'} \eta_{\fA}^{\;\fD'} \eta_{\fC}^{\;\fA'} + \eta_{\fA}^{\;\fC'} \eta_{\fC}^{\;\fD'} \eta_{\fD}^{\;\fA'} ) , \nn \\
& & \eta_{(\fC\fD\fA]}^{\; \fC'\fD'\fA'}\equiv \frac{1}{3} ( 2\eta_{\fC}^{\;\fC'} \eta_{\fD}^{\;\fD'} \eta_{\fA}^{\;\fA'} - \eta_{\fD}^{\;\fC'} \eta_{\fA}^{\;\fD'} \eta_{\fC}^{\;\fA'} - \eta_{\fA}^{\;\fC'} \eta_{\fC}^{\;\fD'} \eta_{\fD}^{\;\fA'} ) .
\ee

Therefore, the hyperunified field theory formulated in light of hidden coordinate formalism exhibits the gravitational geometry-gauge correspondence. When taking the hyperspin gauge field $\cA_{\fC}^{\fA\fB}$ as fundamental gauge field, the gravitational interaction in locally flat gravigauge hyper-spacetime appears in formal to decouple from the hyperunified qubit-spinor field. Meanwhile,  the hyperspin gravigauge field $\fOm_{\fC}^{\fA\fB}$ or $\fOm_{[\fC\fD]}^{\fA}$ behaves as an auxiliary field, which is shown to be presented by fundamental bosonic fields through the constraint equation given in Eq.(\ref{GGCGE}). Nevertheless, the dynamics of all fundamental fields turns out to get gravitational effects due to the emergence of non-commutative geometry characterized by the hyperspin gravigauge field via the non-Abelian Lie algebra of hyper-gravicoordinate derivative operator in locally flat gravigauge hyper-spacetime.


\section{ Hyperunified field theory in framework of GQFT with fundamental symmetry under entirety unitary gauge and the dynamics of electromagnetic-like gravitational field with gauge-type and Einstein-type gravitational equations and first order gravitational differential equation based the gauge-geometry duality}

In this section, we are going to further investigate the foundation of the hyperunified field theory within the framework of gravitational quantum field theory (GQFT)\cite{GQFT} by making an appropriate gauge prescription under entirety unitary gauge. With the demonstration of gauge-geometry duality in biframe hyper-spacetime, we will present a detailed discussion on the dynamics of hyper-gravigauge field and hyper-gravimetric gauge field with electric-like and magnetic-like gravitational interactions in both gauge-type and geometric Einstein-type gravitational equations. 
  

\subsection{ Hyperunified field theory within framework of GQFT and gauge-geometry duality with respect to hyper-gravigauge field and hyper-gravimetric gauge field }

It has been shown that the $\cW_e$-spin gauge group symmetry W$^{1,D_h-1}$ that causes a sign flip in chirality for the transformed hyperunified qubit-spinor field brings on the $\cW_e$-spin invariant-gauge field $\fA_{\fM}^{\;\;\fA}$. Such an entanglement-correlated translation-like $\cW_e$-spin gauge group W$^{1,D_h-1}$ represents the Abelian subgroup of inhomogeneous hyperspin gauge symmetry WS(1,$D_h$-1), so that the $\cW_e$-spin invariant-gauge field $\fA_{\fM}^{\;\;\fA}$ as hyper-gravigauge field reveals the genesis of gravitational interaction and elaborates the gauge-gravity correspondence. In the hyperunified field theory with the action formulated in hidden gauge formalism, the gravitational interaction is shown to be described by the symmetric hyper-gravimetric field $\mH_{\fM\fN}$ which is composed of hyper-gravigauge field, i.e., $\mH_{\fM\fN} = \fA_{\fM}^{\;\;\fA} \fA_{\fN}^{\;\;\fB}\eta_{\fA\fB}$. When making the gauge prescription to flowing unitary gauge, the symmetric hyper-gravimetric field is determined directly by the symmetric hyper-gravigauge field $\bfA_{\fM\fA}=\bfA_{\fA\fM}$, i.e., $\mH_{\fM\fN} = (\bfA)^2_{\fM\fN}$, with $\bfA_{\fM\fA}$ representing physical degrees of freedom under flowing unitary gauge, which exposits the gravity-geometry correspondence in curved Riemannian hyper-spacetime. Furthermore, in the hyperunified field theory with the action formulated in hidden coordinate formalism, the gravitational interaction appears to be an emergent one characterized via the emergent non-commutative geometry in locally flat gravigauge hyper-spacetime. Such a non-commutative geometry is described by the hyperspin gravigauge field $\fOm_{\fC\fD}^{\fA}$ which behaves as an auxiliary field determined through the constraint equation concerning only basic bosonic fields, which elucidates the geometry-gauge correspondence.

Let us now turn to consider the hyperunified field theory within the framework of GQFT. By noticing the following relations,
\be 
& & \thmH_{\fA\fA'}^{\fM\fN\fM'\fN'}  \mF_{\fM\fN}^{\fA}\mF_{\fM'\fN'}^{\fA'} = 
\hmH^{\fM\fM'}  \hmH^{\fN\fN'} \mF_{\fM\fN\fA} \mF_{\fM'\fN'}^{\fA}  \nn \\
& & \;\; + \hmH^{\fM\fM'}( 2 \hfA_{\fA'}^{\;\;\fN} \hfA_{\fA}^{\;\;\fN'}  \mF_{\fM\fN}^{\fA} \mF_{\fM'\fN'}^{\fA'} - 4 \hfA_{\fA}^{\;\;\fN} \hfA_{\fA'}^{\;\;\fN'}  \mF_{\fM\fN}^{\fA} \mF_{\fM'\fN'}^{\fA'} )  \nn \\
& & \hmH^{\fM\fM'}\hfA_{\fA}^{\;\;\fN} \hfA_{\fA'}^{\;\;\fN'}  \mF_{\fM\fN}^{\fA} \mF_{\fM'\fN'}^{\fA'}  \equiv  \hat{\fkA}^2 \p_{\fM}(\fkA\, \hfA_{\fA}^{\;\;\fM} ) \p_{\fN}(\fkA\, \hfA^{\fA\fN} ) ,
\ee
we are able to rewrite the basic action for the hyperunified field theory as follows: 
\be  \label{actionHUFTGQFT1}
\cS_{\mH\mU} & \equiv & \int [d^{D_h}x] \, \fkA(x) \{\, \hmH^{\fM\fN} \bar{\fPsi}_{\fQH}  \fA_{\fM\fC}  \vSi_{-}^{\fC} i\cD_{\fN}  \fPsi_{\fQH} -  \beta_Q \sinh\chi_s \, \bar{\fPsi}_{\fQH} \tvSi_{-}\fPsi_{\fQH}  \nn \\
& - & g_H^{-2} \frac{1}{4} \hmH^{\fM\fM'} \hmH^{\fN\fN'}   \left( \cF_{\fM\fN\fA\fB}  \cF_{\fM'\fN'}^{\fA\fB}  -  \mF_{\fM\fN\fA}\mF_{\fM'\fN'}^{\fA} \right) \nn \\
& + & g_H^{-2} \frac{1}{2} \hmH^{\fM\fM'} \hfA_{\fA'}^{\;\;\fN} \hfA_{\fA}^{\;\;\fN'}  \mF_{\fM\fN}^{\fA} \mF_{\fM'\fN'}^{\fA'} - g_H^{-2}\hat{\fkA}^2 \p_{\fM}(\fkA\, \hfA_{\fA}^{\;\;\fM} ) \p_{\fN}(\fkA\, \hfA^{\fA\fN} ) \nn \\
& + & \frac{1}{2} g_H^{-2}\beta_G^2 \beta_Q^2 \sinh^2\chi_s \,  \hmH^{\fM\fM'} ( \cA_{\fM\fA\fB} -  \fOm_{\fM\fA\fB} )(\cA_{\fM'}^{\fA\fB} -  \fOm_{\fM'}^{\fA\fB} ) \nn \\
& - & \hmH^{\fM\fM'} \hmH^{\fN\fN'} \frac{1}{4} g_W^{-2} \mF_{\fM\fN} \mF_{\fM'\fN'} + \frac{1}{2} \lambda_S^2 ( 1 + \sinh^2\chi_s) \hmH^{\fM\fM'}  \mW_{\fM} \mW_{\fM'}  \nn \\
&  + &  \frac{1}{2} \hmH^{\fM\fN} \lambda_S^2 \p_{\fM} \chi_s \p_{\fN}\chi_s  - \lambda_D^2  \cF(\chi_s)  \, \} ,
\ee
with the definitions,
\be
& &  i\cD_{\fM} = i\p_{\fM} + \cA_{\fM}^{\fA\fB} \frac{1}{2}\vSi_{\fA\fB},  \nn \\
& &  \fOm_{\fM\fA\fB} \equiv \hmH_{\fM\fA\fB\fC}^{\; \; \fP\fQ} \mF_{\fP\fQ}^{\fC} ,\quad \mF_{\fM\fN}^{\fA} = \p_{\fM}\fA_{\fN}^{\; \fA} - \p_{\fN}\fA_{\fM}^{\; \fA}, \nn \\
& & \hmH_{\fM\fA\fB\fC}^{\; \; \fP\fQ} = \frac{1}{2} \fA_{\fM}^{\; \; \fE } \left( \hfA_{\fE}^{\;\; \fP} \hfA_{\fA}^{\;\; \fQ}\eta_{\fB\fC} - \hfA_{\fE}^{\;\; \fP} \hfA_{\fB}^{\;\; \fQ}\eta_{\fA\fC} - \hfA_{\fA}^{\;\; \fP} \hfA_{\fB}^{\;\; \fQ}\eta_{\fE\fC}  \right) ,
\ee
where the hyperspin gauge field $\cA_{\fM}^{\fA\fB}$ and hyper-gravigauge field $\fA_{\fM}^{\; \; \fA }$ are taken to be the fundamental gauge fields, which is governed by the inhomogeneous hyperspin gauge symmetry and also exposits the gauge-gravity correspondence.  
 
It is interesting to notice that the gravitational interaction of all fundamental bosonic fields involves solely the dual Goldstone-like hyper-gravimetric field $\hmH^{\fM\fN}$. Inspired from the gauge-gravity and gravity-geometry correspondences, it is useful to take the hyper-gravimetric field $\mH_{\fM\fN}$ and hyperspin gauge field $\cA_{\fM}^{\fA\fB}$ as physical degrees of freedom in the gravigauge hyper-fiber bundle structure of biframe hyper-spacetime. In general, we can prove the following relations:
\be \label{GGD}
& & \frac{1}{4} \fkA\, \thmH^{\fM\fN\fM'\fN'}_{\fA\fA'} \mF_{\fM\fN}^{\fA}\mF_{\fM'\fN'}^{\fA'}   \equiv  \fkA\, \hmH^{\fM\fN} \fR_{\fM\fN} -  2 \p_{\fM} (\fkA \hmH^{\fM\fP} \hfA_{\fA}^{\;\fQ} \mF_{\fP\fQ}^{\fA} ) , 
\ee
and
\be \label{GGD1}
& &  \frac{1}{4} \fkA\, \thmH^{\fM\fN\fM'\fN'}_{\fA\fA'} \mF_{\fM\fN}^{\fA}\mF_{\fM'\fN'}^{\fA'}   \equiv \fkA \frac{1}{8}  \hmH^{\fM\fM'} \hmH^{\fN\fN'} \hmH_{\fP\fP'} \mF_{\fM\fN}^{\fP} \mF_{\fM'\fN'}^{\fP'} 
 \nn \\
& & -\fkA \frac{1}{4} \hmH^{\fM\fN} \p_{\fM}(\ln \fkA) \p_{\fN}(\ln \fkA) + \fkA \frac{1}{4} \hmH_{\fP\fQ} \fkA^2 \hp^{\fM}(\hat{\fkA} \mH_{\fM}^{\; \fP} ) \hp^{\fN}( \hat{\fkA} \mH_{\fN}^{\; \fQ})  \nn \\
& & + \frac{1}{2} \fkA^2 \hmH_{\fQ}^{\; \fM} \p_{\fM}(\ln\fkA) \hp^{\fN} (\hat{\fkA} \mH_{\fN}^{\; \fQ} ) - \frac{1}{4} \fkA^2 \hmH^{\fN\fP} \hp^{\fM} \p_{\fN} (\hat{\fkA} \mH_{\fM\fP} ) \nn \\
& & -\frac{1}{4} \p_{\fM} \left( \hat{\fkA} \mH_{\fP\fQ} \p_{\fN} (\fkA \hmH^{\fM \fP} \fkA \hmH^{\fN\fQ}) - \p_{\fN} (\hat{\fkA} \mH_{\fP\fQ} ) \fkA \hmH^{\fM \fP} \fkA \hmH^{\fN\fQ} \right) \nn \\
& & + \p_{\fM} \left( \fkA \hmH^{\fM\fN} \p_{\fN} \ln \fkA \right)  - 2 \p_{\fM} \left(\fkA \hfA_{\fA}^{\;\; \fM}\p_{\fN}\hfA^{\fA\fN} \right)  + \frac{3}{2} \p_{\fM}\p_{\fN} \left( \fkA\hmH^{\fM\fN} \right) ,
\ee
with $\hat{\fkA}=1/\fkA$ and $\hp^{\fM}\equiv \hmH^{\fM\fM'}\p_{\fM'}$. Note that the last four terms appear as surface terms in the action. Where we have introduced the following Abelian-type field strength: 
\be
\mF_{\fM\fN}^{\fP} = \p_{\fM}\mH_{\fN}^{\; \fP} - \p_{\fN}\mH_{\fM}^{\; \fP} ,
\ee
which indicates that the hyper-gravimetric field $\mH_{\fM}^{\; \fP}$ behaves as an Abelian-type gauge field. To be physically more meaningful, the hyper-gravimetric field $\mH_{\fM}^{\; \fP}$ and field strength $\mF_{\fM\fN}^{\fP}$ may be referred to as {\it hyper-gravimetric gauge field} and {\it hyper-gravimetric gauge field strength}, which brings on useful duality relations with respective to the hyper-gravigauge field $\fA_{\fM}^{\; \fA}$ and field strength $\mF_{\fM\fN}^{\fA}$. 

The relations in Eqs.(\ref{GGD}) and (\ref{GGD1}) not only explicate the gauge-gravity and gravity-geometry as well as geometry-gauge correspondences,  but also corroborate that the geometric hyper-gravimetric field $\mH_{\fM}^{\; \fP}$ in itself can be regarded as an Abelian-type gauge field, which reveals in general {\it gauge-geometry duality}. 

With such a gravitational gauge-geometry duality, we can take the hyperspin gauge field $\cA_{\fM}^{\fA\fB}$ and hyper-gravimetric gauge field $\mH_{\fM}^{\; \fP}$ as fundamental gauge fields to express the action of the hyperunified field theory in light of gravigauge hyper-fiber bundle structure of biframe hyper-spacetime. Explicitly, such an action can be represented within the framework of GQFT as follows:
\be  \label{actionHUFTGQFT2}
\cS_{\mH\mU} & \equiv & \int [d^{D_h}x] \, \fkA(x) \{\, \hmH^{\fM\fN} \bar{\fPsi}_{\fQH} \fA_{\fM\fC}  \vSi_{-}^{\fC} i\cD_{\fN}  \fPsi_{\fQH} \nn \\
& - &  \beta_Q \sinh\chi_s \,  \bar{\fPsi}_{\fQH} \tvSi_{-}\fPsi_{\fQH}  -  \hmH^{\fM\fM'} \hmH^{\fN\fN'}  g_H^{-2} \frac{1}{4} \cF_{\fM\fN\fA\fB}  \cF_{\fM'\fN'}^{\fA\fB}  \nn \\
& + & \frac{1}{2} g_H^{-2}\beta_G^2 \beta_Q^2 \sinh^2\chi_s \,  \hmH^{\fM\fM'} ( \cA_{\fM\fA\fB} -  \fOm_{\fM\fA\fB} )(\cA_{\fM'}^{\fA\fB} -  \fOm_{\fM'}^{\fA\fB} )  \nn \\
& + & \frac{1}{8}  \hmH^{\fM\fM'} \hmH^{\fN\fN'} \hmH_{\fP\fP'} g_H^{-2} \mF_{\fM\fN}^{\fP} \mF_{\fM'\fN'}^{\fP'} \nn \\
& - & \frac{1}{4} g_H^{-2} \fkA \hmH^{\fN\fP} \hp^{\fM} \p_{\fN} (\hat{\fkA} \mH_{\fM\fP} )  - \frac{1}{4} g_H^{-2} \hp^{\fN}(\ln \fkA) \p_{\fN}(\ln \fkA) \nn \\
& + & \frac{1}{4} g_H^{-2} \hmH_{\fP\fQ} \fkA^2 \hp^{\fM}(\hat{\fkA} \mH_{\fM}^{\; \fP} ) \hp^{\fN}( \hat{\fkA} \mH_{\fN}^{\; \fQ}) + \frac{1}{2} g_H^{-2} \fkA \hp_{\fQ}(\ln\fkA) \hp^{\fN} (\hat{\fkA} \mH_{\fN}^{\; \fQ} ) \nn \\
& - & \hmH^{\fM\fM'} \hmH^{\fN\fN'} \frac{1}{4} g_W^{-2} \mF_{\fM\fN} \mF_{\fM'\fN'} + \frac{1}{2} \lambda_S^2 ( 1 + \sinh^2\chi_s ) \hmH^{\fM\fN}  \mW_{\fM} \mW_{\fN}  \nn \\
&  + &  \frac{1}{2} \hmH^{\fM\fN} \lambda_S^2 \p_{\fM} \chi_s \p_{\fN}\chi_s  - \lambda_D^2 \cF(\chi_s)  \, \} ,
\ee
where we have ignored the surface terms. The hyperspin gravigauge field can be expressed as follows:
\be
 \fOm_{\fM\fA\fB} = \fGa_{\fM[\fP\fQ]} \hfA_{\fA}^{\;\;\fP}\hfA_{\fB}^{\;\;\fQ} - \frac{1}{2} \mH_{\fP\fQ} [\, \p_{\fM}(\hfA_{\fA}^{\;\;\fP}) \hfA_{\fB}^{\;\;\fQ} -  \p_{\fM}(\hfA_{\fB}^{\;\;\fP}) \hfA_{\fA}^{\;\;\fQ} \, ] ,
\ee
which indicates that $\fOm_{\fM\fA\fB}$ concerns both hyper-gravimetric gauge field $\mH_{\fP}^{\; \fM}$ and hyper-gravigauge field $\fA_{\fM}^{\; \fA}$. It is also manifest that the gravitational interaction of hyperqubit-spinor field always involves the hyper-gravigauge field $\fA_{\fM}^{\; \fA}$. It is natural to take flowing unitary gauge so as to relate directly the symmetric hyper-gravigauge field to the symmetric hyper-gravimetric field, $\mH_{\fM\fN} = (\bfA)^2_{\fM\fN}$ with $\bfA_{\fM\fA}=\bfA_{\fA\fM}$.

Therefore, within the framework of GQFT based on the basic structure of biframe hyper-spacetime, the dynamics of gravitational interaction can be described either by the symmetric hyper-gravigauge field or equivalently by the symmetric hyper-gravimetric gauge field under flowing unitary gauge with maintaining physical degrees of freedom. Such an equivalent description exposits the gauge-geometry duality for the gravitational interaction.

 
\subsection{ Hyperunified field theory with maximal joint symmetry and the entirety unitary gauge in gravitational quantum field theory with fundamental symmetry of nature  }

The hyperunified field theory in the framework of GQFT is presented to show the gravitational gauge-geometry duality. Before discussing such an action, let us briefly outline the main properties of hyperunified field theory formulated in various formalisms.

As demonstrated in the hidden gauge formalism of the action given in Eq.(\ref{actionHUFTHGF}), such an action of hyperunified field theory holds in any coordinate system due to the emergence of general linear group symmetry GL(1,$D_h$-1,R), which does indicate that the laws of nature should be independent of the choice of coordinate systems. Geometrically, the gravitational interactions concerning fundamental bosonic fields are described by the dynamics of emergent Riemann geometry in curved Riemannian hyper-spacetime, which is characterized by Christoffel symbols $\fGa_{\fM\fN}^{\fP}$ defined from the hyper-gravimetric field $\mH_{\fM\fN}$. Such a gravitational interaction is governed by the emergent general linear group symmetry GL(1,$D_h$-1,R) and meanwhile the hyperspin gauge symmetry becomes a hidden gauge symmetry, which shows the gravity-geometry correspondence. As a consequence, the basic action of hyperunified field theory possesses emergent general linear group symmetry GL(1,$D_h$-1,R) under general coordinate transformation, 
\be
G_S=  \mbox{GL(}1,D_h-1,\mbox{R)} , \nn
\ee
which represents the real Lie group of dimension $N_D= D_h^2 $ and consists of matrices that have non-zero determinant. It is known that such a general linear group symmetry lays the foundation of Einstein's general theory of relativity in four dimensional spacetime.  

It is proved in the hidden coordinate formalism of the action that, locally, at every point of hyper-spacetime, the basic action of hyperunified field theory possesses inhomogeneous hyperspin gauge symmetry WS(1,$D_h$-1). Whereas the general linear group symmetry GL($D_h$,R) goes to be a hidden symmetry. The gravitational interaction is characterized by the hyperspin gravigauge field $\fOm_{[\fC\fD]}^{\fA}$ which behaves as an auxiliary field determined through the constraint equation concerning only basic bosonic fields. Therefore, the gravitational interaction acts as an emergent interaction in correspondence to the emergence of non-commutative geometry in locally flat gravigauge hyper-spacetime. The hyperspin gauge field $\cA_{\fC\fA\fB}$ as fundamental gauge field is taken to be dynamic field, so that the basic action exhibits the inhomogeneous hyperspin gauge symmetry as a maximal gauge symmetry, 
\be
G_S =  \mbox{WS(1},D_h\mbox{-1)} , \nn
\ee
which has been applied to construct the basic action by following along the gauge invariance principle. 

It is straightforward to justify that when the action is formulated in light of gravigauge hyper-fiber bundle structure of biframe hyper-spacetime as shown in Eqs.(\ref{actionHUFTGQFT1}) and (\ref{actionHUFTGQFT2}), it gets a bimaximal gauge symmetry,
\be
G_S= \mbox{GL(}1,D_h-1, \mbox{R)} \Join \mbox{WS(}1, D_h-1),
\ee
which is regarded as a joint Lie group gauge symmetry since they cannot be operated as a direct product group. In general, such a joint gauge symmetry can be adopted as generalized gauge invariance principle to construct the basic action of hyperunified field theory. One is general linear group symmetry GL(1,$D_h$-1, R) operating on the coordinates in curved Riemannian hyper-spacetime, and the other is inhomogeneous hyperspin gauge symmetry WS(1,$D_h$-1) operating on the entangled hyperqubit-spinor field and gauge field in locally flat gravigauge hyper-spacetime.

In practice, the basic action of hyperunified field theory is built within the framework of quantum field theory based on the gauge invariance principle and scaling invariance hypothesis in light of gravigauge hyper-fiber bundle structure of biframe hyper-spacetime. Where the free-motion hyper-spacetime described by globally flat Minkowski hyper-spacetime is taken as a base spacetime in gravigauge hyper-fiber bundle structure, so that the basic action of hyperunified field theory is built to possess global Poincar\'e-type group symmetry PO(1,$D_h$-1) and inhomogeneous hyperspin gauge symmetry WS(1,$D_h$-1) as well as global and local scaling symmetries, 
\be
G_S = \mbox{SC(1)}\ltimes  \mbox{PO(1},D_h\mbox{-1)}\Join \mbox{WS(1},D_h\mbox{-1)}\rtimes \mbox{SG(1)} ,
\ee
which has a total group dimension $N_D = 2\times D_h(D_h-1)/2 + 2D_h + 2 = D_h(D_h+1) +2 $. Where the Lorentz-type group SO(1,$D_h$-1) is a subgroup of GL($D_h$, R). In such a consideration, the global and local scaling symmetries play an essential role in the construction of the basic action for the hyperunified field theory.

From the gauge invariance principle, the gauge field is introduced to reveal the dynamics of fundamental building block of nature. Locally, the gauge symmetry means that the laws of nature should be independent of the choice of field configurations. Namely, the gauge symmetry leads each physically distinct configuration of the system to be as an equivalence class of detailed local field configurations, which indicates that for any two detailed configurations that are in the same equivalence class, they are related by a gauge transformation. Therefore, the gauge symmetry concerns redundant unphysical degrees of freedom.

Physically, all observables should be gauge independent.  In order to remove unphysical degrees of freedom arising from local gauge symmetry, it is necessary to fix gauge symmetry by making appropriate gauge prescription. Therefore, a gauge fixing is required to provide a mathematical procedure for dealing with redundant degrees of freedom in field variables caused by gauge symmetry. 

In the hidden gauge formalism of the action, it has been shown that it is appropriate to choose a gauge prescription under flowing unitary gauge. This is because there exists always a hyperspin gauge transformation $\bar{\Lambda}_{\;\, \fA}^{\fB}(x)$ to transform the hyper-gravigauge field into a symmetric bicovariant vector field at any chosen point of hyper-spacetime,  $\bfA_{\fM\fA}(x) = \fA_{\fM\fB}(x) \bar{\Lambda}^{\fB}_{\; \, \fA}(x)=\bfA_{\fA\fM}(x)$, which is referred to as flowing unitary gauge.  In such a flowing unitary gauge, both symmetric hyper-gravigauge field $\bfA_{\fM\fA}(x)$ and hyper-gravimetric field $\mH_{\fM\fN}(x)$ involve exactly the same degrees of freedom with a simple relation $\mH_{\fM\fN} \equiv (\bfA)^2_{\;\, \fM\fN}$. 

Let us turn to the basic action of the hyperunified field theory with the gauge-geometry duality shown in Eq.(\ref{actionHUFTGQFT1}) or (\ref{actionHUFTGQFT2}). Such an action built in biframe hyper-spacetime with a gravigauge hyper-fiber bundle structure gets a maximal joint symmetry of two groups GL(1,$D_h$-1, R)$\Join$WS(1,$D_h$-1) which cannot form a direct product group. It is noticed that only when setting up flowing unitary gauge to keep the symmetric property of hyper-gravigauge field at arbitrary point of coordinate systems in hyper-spacetime, we are able to obtain the same physical degrees of freedom to bring on both gauge-gravity and gravity-geometry correspondences. Namely, once the inhomogeneous hyperspin gauge symmetry WS(1,$D_h$-1) which governs the dynamics of hyperunified quibit-spinor field and gauge field in locally flat gravigauge hyper-spacetime is fixed appropriately to unitary gauge, the general linear symmetry group GL(1,$D_h$-1, R) which characterizes the dynamics of Riemann geometry in curved Riemannian hyper-spacetime should also be fixed simultaneously to ensure the symmetric property of hyper-gravigauge field. Namely, under the flowing unitary gauge, the basic action (\ref{actionHUFTGQFT1}) or (\ref{actionHUFTGQFT2}) formulated in light of gravigauge hyper-fiber bundle structure of biframe hyper-spacetime possesses only the following associated symmetry: 
\be
G_S = \mbox{SC(1)}\ltimes  \mbox{PO(1},D_h\mbox{-1)}\wtjoin \mbox{WS(1},D_h\mbox{-1)}, \nn
\ee
which brings about the fundamental symmetry achieved from the so-called $\cM_c$-spin charge of entangled hyperqubit-spinor field in free-motion hyper-spacetime. 

To freeze over the flowing feature of unitary gauge for the hyperspin gauge symmetry SP(1,$D_h$-1) and provide a whole gauge prescription, we should further make a gauge prescription to fix the general linear group symmetry GL(1,$D_h$-1, R). From the basic action presented in Eq.(\ref{actionHUFTGQFT2}) with gauge-geometry duality, it is reasonable to take the following gauge prescription:
\be \label{DG}
(D_h-2)^{-1}\hp^{\fM} ( \hat{\fkA} \mH_{\fM \fP}  ) - D_h^{-1} \p_{\fP} \hat{\fkA} =0, 
\;\; \mbox{i.e.}, \;\;  \hp^{\fM}\mH_{\fM\fP} - \frac{2}{D_h} \p_{\fP}\ln \fkA = 0 ,
\ee
which leads the flowing unitary gauge to be frozen into a usual unitary gauge.

With the above gauge prescription, the action presented in Eq.(\ref{actionHUFTGQFT2}) within the framework of GQFT is modified to be as follows:
\be  \label{actionHUFTGQFT3}
\cS_{\mH\mU} & \equiv & \int [d^{D_h}x] \, \fkA(x) \{\, \hmH^{\fM\fN} \bar{\fPsi}_{\fQH}  \fA_{\fM\fC}  \vSi_{-}^{\fC} i\cD_{\fN}  \fPsi_{\fQH} \nn \\
& - &  \beta_Q \sinh\chi_s \,  \bar{\fPsi}_{\fQH} \tvSi_{-}\fPsi_{\fQH}  -  \hmH^{\fM\fM'} \hmH^{\fN\fN'}  g_H^{-2} \frac{1}{4} \cF_{\fM\fN\fA\fB}  \cF_{\fM'\fN'}^{\fA\fB}  \nn \\
& + &  \frac{1}{2} g_H^{-2}\beta_G^2\beta_Q^2 \sinh^2\chi_s \,  \hmH^{\fM\fM'} ( \cA_{\fM\fA\fB} -  \fOm_{\fM\fA\fB} )(\cA_{\fM'}^{\fA\fB} -  \fOm_{\fM'}^{\fA\fB} )  \nn \\
& + &  \frac{1}{8}  \hmH^{\fM\fM'} \hmH^{\fN\fN'} \hmH_{\fP\fP'}  g_H^{-2} \mF_{\fM\fN}^{\fP} \mF_{\fM'\fN'}^{\fP'} \nn \\
& - & \frac{1}{4} g_H^{-2} \fkA \hmH^{\fN\fP} \hp^{\fM} \p_{\fN} (\hat{\fkA} \mH_{\fM\fP} )  - \frac{1}{4} g_H^{-2} \hp^{\fN}(\ln \fkA) \p_{\fN}(\ln \fkA) \nn \\
& + & \frac{1}{4} g_H^{-2} \fkA^2\hmH_{\fP\fQ} \hp^{\fM}(\hat{\fkA} \mH_{\fM}^{\; \fP} ) \hp^{\fN}( \hat{\fkA} \mH_{\fN}^{\; \fQ}) + \frac{1}{2} g_H^{-2} \fkA \hp_{\fQ}(\ln\fkA) \hp^{\fN} (\hat{\fkA} \mH_{\fN}^{\; \fQ} ) \nn \\
& + & \frac{\lambda_G}{8}   g_H^{-2}  \hmH^{\fP\fQ} [ \fkA\hp^{\fM}(\hat{\fkA}\mH_{\fM \fP}) + \frac{D_h-2}{D_h} \p_{\fP} \ln \fkA  ]  [\fkA\hp^{\fN}(\hat{\fkA}\mH_{\fN\fQ}) + \frac{D_h-2}{D_h} \p_{\fQ} \ln \fkA ]  \nn \\
& - & \hmH^{\fM\fM'} \hmH^{\fN\fN'} \frac{1}{4} g_W^{-2} \mF_{\fM\fN} \mF_{\fM'\fN'} + \frac{1}{2} \lambda_S^2( 1 + \sinh^2\chi_s ) \hmH^{\fM\fM'}  \mW_{\fM} \mW_{\fM'}  \nn \\
&  + &  \frac{1}{2} \hmH^{\fM\fN} \lambda_S^2 \p_{\fM} \chi_s \p_{\fN}\chi_s  - \lambda_D^2 \cF(\chi_s)  \, \}  ,
\ee
where we have introduced an additional term with arbitrary constant $\lambda_G$ to provide the gauge fixing condition.

The above action of hyperunified field theory with imposing a gauge prescription on the general linear group symmetry GL(1,$D_h$-1, R) shown in Eq.(\ref{DG}) exhibits a manifest gauge-geometry duality within the framework of GQFT. For short, we may call such a gauge prescription as {\it duality gauge} prescription in hyper-spacetime. 

It is clear that the combination of duality gauge for the group symmetry GL(1,$D_h$-1, R) in curved Riemannian hyper-spacetime with flowing unitary gauge for the gauge symmetry SP(1,$D_h$-1) in Hilbert space leads to a full gauge prescription, which may be referred to as {\it entirety unitary gauge} prescription for convenience. By choosing the gauge fixing conditions under entirety unitary gauge, we should be able to remove in principle all unphysical degrees of freedom. 

Therefore, to remove all unphysical degrees of freedom, we should take the {\it entirety unitary gauge} which combines duality gauge and flowing unitary gauge together. Consequently, the fundamental symmetry of hyperunified field theory turns out to be an associated symmetry in which the global inhomogeneous hyperspin symmetry WS(1,$D_h$-1) is always in association with global Poincar\'e-type group symmetry PO(1,$D_h$-1) together with global scaling symmetry SC(1), i.e.:
\be
G_S = \mbox{SC(1)}\ltimes  \mbox{PO(1},D_h\mbox{-1)}\wtjoin \mbox{WS(1},D_h\mbox{-1)}, \nn
\ee
where the isomorphic group SP(1,$D_h$-1)$\cong$SO(1,$D_h$-1) must keep coincidental symmetry transformations.


\subsection{ Gauge-type and geometric Einstein-type gravitational equations and beyond for the dynamics of gravitational field } 

Let us first revisit the equation of motion for the hyper-gravigauge field $\fA_{\fM}^{\;\; \fA}$. From the hyperunified field theory presented in Eq.(\ref{actionHUFTGQFT1}) within the framework of GQFT, we obtain the following equation of motion:
\be  \label{GGEb}
\p_{\fN} \hmF_{\fA}^{\fM\fN}  =  \hmJ_{\fA}^{\;\;\fM} , 
\ee
with the definition,
\be \label{HGGFST}
& & \hmF_{\fA}^{\fM\fN} \equiv \thmH^{[\fM\fN]\fM'\fN'}_{\fA\fA'} \mF_{\fM'\fN' }^{\fA'}  = -  \hmF_{\fA}^{\fN\fM}, \nn \\
& & \mF_{\fM'\fN' }^{\fA'} = \p_{\fM'}\fA_{\fN'}^{\;\; \fA'} - \p_{\fM'}\fA_{\fN'}^{\;\; \fA'} , 
\ee
where the tensor $\thmH^{[\fM\fN]\fM'\fN'}_{\fA\fA'}$ is defined in Eq.(\ref{Tensors}).

We may refer to $\hmF_{\fA}^{\fM\fN}$ as {\it hyper-gravigauge field strength tensor}. $\hmJ_{\fA}^{\;\;\fM}$ is called {\it hyper-gravigauge field current} with the following form: 
\be \label{mJ1}
 \hmJ_{\fA}^{\;\; \fM} \equiv \fkA  \mJ_{\fA}^{\;\; \fM} - \hmF_{\fA}^{\fM\fN} \p_{\fN}\ln \fkA  \equiv \mG_{\fA}^{\;\; \fM} + g_H^{2} \mT_{\fA}^{\;\; \fM}, 
 \ee
with $\mJ_{\fA}^{\;\;\fM}$ the bicovariant vector current defined in Eq.(\ref{mJ}). $\mG_{\fA}^{\;\; \fM}$ and $\mT_{\fA}^{\;\; \fM}$ are explicitly given as follows:
 \be
 \mG_{\fA}^{\;\; \fM} & \equiv &  ( \hfA_{\fA}^{\; \; \fP} \thmH^{[\fM\fN]\fM'\fN'}_{\fA''\fA'}  - \frac{1}{4} \hfA_{\fA}^{\;\;\fM} \thmH^{\fP\fN\fM'\fN'}_{\fA''\fA'})  \mF_{\fP\fN}^{\fA''}  \mF_{\fM'\fN'}^{\fA'} - \thmH^{[\fM\fN]\fM'\fN'}_{\fA\fA'} \mF_{\fM'\fN' }^{\fA'} \p_{\fN}\ln \fkA , \nn \\
\mT_{\fA}^{\;\; \fM} & \equiv &  - \fkA\, \hfA_{\fA}^{\;\;\fM} \tilde{\fkL} +  \fkA\, \hfA_{\fA}^{\; \; \fP}  \{\,  \hfA_{\fC}^{\;\; \fM}  \bar{\fPsi}_{\fQH} \vSi_{-}^{\fC} i \cD_{\fP} \fPsi_{\fQH}     \nn \\
& - &\hmH^{[\fM\fN]\fM'\fN'}_{\fA''\fB\; \fA'\fB'} g_H^{-2} \cF_{\fP\fN}^{\fA''\fB} \cF_{\fM'\fN'}^{\fA'\fB'}  +  \bhmH^{[\fM\fN]\fM'\fN'}_{\fA''\fA'}g_H^{-2}\beta_G^2\beta_Q^2 \sinh^2\chi_s \,  \fF_{\fP\fN}^{\fA''} \fF_{\fM'\fN'}^{\fA'}   \nn \\
& - & g_W^{-2}\hmH^{\fM\fM'} \hmH^{\fN\fN'}\mF_{\fP\fN} \mF_{\fM'\fN'} + \lambda_S^2( 1 + \sinh^2\chi_s ) \hmH^{\fM\fN} \mW_{\fP} \mW_{\fN}  \nn \\
& + & \lambda_S^2 \hmH^{\fM\fN} \p_{\fP}\chi_s  \p_{\fN} \chi_s  - \cD_{\fN} ( \fkA\, \bhmH^{[\fM\fN]\fM'\fN'}_{\fA\fA'} g_H^{-2}\beta_G^2\beta_Q^2 \sinh^2\chi_s \, \fF_{\fM'\fN' }^{\fA'} )   \, \}, 
\ee
where $\mG_{\fA}^{\;\; \fM}$ is referred to as {\it pure hyper-gravigauge field current} caused solely by the hyper-gravigauge field itself. $\mT_{\fA}^{\;\; \fM}$ is referred to as {\it hyper-gravigauge field bicovariant vector current} resulting from all other basic field in the hyperunified field theory.

We may refer to Eq.(\ref{GGEb}) as {\it gauge-type gravitational equation} for the hyper-gravigauge field. Such a gauge-type gravitational equation leads to the following conserved current:
\be
\p_{\fM}\hmJ_{\fA}^{\;\; \fM} = 0 ,
\ee
due to the antisymmetry property of the hyper-gravigauge field strength tensor. Therefore, in the hyperunified field theory with the action presented in the framework of GQFT, the hyper-gravigauge field does behave as an Abelain gauge field.

To obtain Einstein-type equation in hyper-spacetime, we should utilize Eqs.(\ref{GGD}) and (\ref{GGD1}) and take into account the gauge-gravity and gravity-geometry correspondences. It can be verified that the gauge-type gravitational equation presented in Eq.(\ref{GGEb}) can be reformulated into the following formalism in light of Ricci curvature tensor $\fR_{\fM\fN}$: 
\be
& & -2 g_H^{-2}\fkA \fR_{\fM'\fN'}\hfA_{\fA}^{\; \fM'}\hmH^{\fN'\fM}  = \mJ_{\fA}^{\; \fM}   - \fkA g_H^{-2} \hfA_{\fA}^{\;\fP} \tilde{\hmH}^{[\fM\fQ]\fM'\fN'}_{\fA''\fA'}  \mF_{\fP\fQ}^{\fA''} \mF_{\fM'\fN'}^{\fA'} .
\ee
When projecting the above bicovariant vector tensors into hyper-spacetime tensors by adopting the hyper-gravigauge field as Goldstone-like boson, we are able to obtain two types of dynamic equations of motion. 

The first dynamic equation of motion is resulted from the symmetric tensor, which has the following explicit form: 
\be \label{GGEE}
& & \fR_{\fM\fN} - \frac{1}{2} \mH_{\fM\fN} \fR = -\frac{1}{2} g_H^2 \mT_{\fM\fN} ,
\ee
and the second dynamic equation of motion is caused from the antisymmetric tensor, which gets the following explicit form: 
\be \label{GGEB}
g_H^{-2}\beta_G^2 \beta_Q^2 \sinh^2\chi_s \hat{\nabla}_{\fP} \hfA^{\fP}_{\;\; \; \; \fM\fN} =  \mT_{[\fM\fN]} .
\ee

It can be checked that the symmetric tensor $\mT_{\fM\fN}$ presents a generalized Einstein-type symmetric hyper-stress energy-momengtum tensor in hyper-spacetime, which is determined by the hyper-gravigauge field bicovariant vector current $\mT_{\fA}^{\;\; \fM}$ as follows:
\be \label{EEMT}
\mT_{\fM\fN} & \equiv & \hat{\fkA}  \fA_{\fM}^{\;\; \fA} \mT_{\fA}^{\;\; \fP} \mH_{\fP\fN} \nn \\
& = & \frac{1}{2} \left( \bar{\fPsi}_{\fQH} \vSi_{-}^{\fC} \fA_{\fM\fC } i\cD_{\fN}  \fPsi_{\fQH}  
+ \bar{\fPsi}_{\fQH} \vSi_{-}^{\fC} \fA_{\fN\fC } i\cD_{\fM}  \fPsi_{\fQH}\right) \nn \\
& - &  g_H^{-2} \left( \cF_{\fM\fP}^{\fA\fB} \cF_{\fN\fQ\fA\fB}  - \mH_{\fM\fN} \frac{1}{4} \hmH^{\fM'\fN'}\cF_{\fM'\fP}^{\fA\fB} \cF_{\fN'\fQ}^{ \fA\fB} \right)  \hmH^{\fP\fQ} \nn \\
& + & g_H^{-2}\beta_G^2\beta_Q^2 \sinh^2\chi_s \,  \left(\hfA_{\fM}^{\fP\fQ} \fA_{\fN\fP\fQ}  - \mH_{\fM\fN}  \frac{1}{2} \hmH^{\fM'\fN'}    \hfA_{\fM'}^{\fP\fQ} \fA_{\fN'\fP\fQ}\right) \nn \\
& + & \frac{1}{2} g_H^{-2}\beta_G^2\beta_Q^2 \sinh^2\chi_s\, \left( \hfA_{\;\;\;\; \;\; \;\;\fM}^{(\fP\fQ)} \fA_{(\fP\fQ)\fN}  - \hfA_{\;\;\;\; \;\; \;\;\fM}^{[\fP\fQ]} \fA_{[\fP\fQ]\fN}\right) \nn \\
& + & \frac{1}{2} g_H^{-2}\beta_G^2\beta_Q^2 \sinh^2\chi_s \,  \left( \hfA_{\fM}^{\fP\fQ} \fA_{[\fP\fQ]\fN}  + \hfA_{\fN}^{\fP\fQ} \fA_{[\fP\fQ]\fM} + 2 \hat{\nabla}_{\fP} \fA_{(\fM\fN)}^{\fP} \right) \nn \\
& - &  g_W^{-2}\left( \mF_{\fM\fP} \mF_{\fN\fQ}  -  \mH_{\fM\fN} \frac{1}{4} \hmH^{\fM'\fN'}\mF_{\fM'\fP} \mF_{\fN'\fQ} \right) \hmH^{\fP\fQ} \nn \\
& + & \left(W_{\fM}W_{\fN} - \mH_{\fM\fN}  \frac{1}{2} \hmH^{\fP\fQ}W_{\fP}W_{\fQ} \right) \nn \\
& + & \lambda_S^2 \left(\p_{\fM}\chi_s\p_{\fN}\chi_s - \mH_{\fM\fN}  \frac{1}{2} \hmH^{\fP\fQ} \p_{\fP}\chi_s\p_{\fQ}\chi_s  \right) \nn \\
& + & \left(\beta_Q \sinh\chi_s \,  \bar{\fPsi}_{\fQH} \tvSi_{-}\fPsi_{\fQH} + \lambda_D^2 \cF(\chi_s) \right) \mH_{\fM\fN}, 
\ee
where $\fA_{\fP\fM\fN}$ is the hyper-spacetime covariant-gauge field defined in Eq.(\ref{HSTGF}). We have used the following notations and definitions, 
\be
& & \fA_{\fP\fM\fN} \equiv  \mH_{\fM\fQ} \fA_{\fP\fN}^{\fQ} = - \fA_{\fP\fN\fM}, \quad \fA_{\fP}^{\fM\fN} \equiv  \hmH^{\fN\fQ} \fA_{\fP\fQ}^{\fM} = - \fA_{\fP}^{\fN\fM}, \nn \\
& & \fA_{\fP\fM\fN}  \equiv \fA_{\fM\fA} ( \p_{\fP}\fA_{\fN}^{\; \fA}  + \cA_{\fP\fB}^{\fA} \fA_{\fN}^{\;\fB} - \fGa_{\fP\fN}^{\fQ}\fA_{\fQ}^{\; \fA} ) \equiv \cA_{\fP\fM\fN} - \fGa_{\fP\fM\fN}  \nn \\
& & \qquad \quad  = \frac{1}{2} (\fA_{\fM}^{\; \fA} \p_{\fP}\fA_{\fN\fA} - \fA_{\fN}^{\; \fA} \p_{\fP}\fA_{\fM\fA}) - \fGa_{\fP[\fM\fN]} + \cA_{\fP}^{\fA\fB} \fA_{\fM\fA} \fA_{\fN\fB}, \nn \\
& & \fA_{[\fP\fQ]\fM} \equiv \fA_{\fP\fQ\fM} - \fA_{\fQ\fP\fM} , \quad \fA_{(\fP\fQ)\fM} \equiv \fA_{\fP\fQ\fM} + \fA_{\fQ\fP\fM} , \nn \\
& & \hfA_{\;\;\;\; \;\; \fM}^{[\fP\fQ]} \equiv  \hfA_{\;\;\;\; \;\; \fM}^{\fP\fQ} -  \hfA_{\;\;\;\; \;\; \fM}^{\fQ\fP}, \quad \hfA_{\;\;\;\; \;\; \fM}^{(\fP\fQ)} \equiv  \hfA_{\;\;\;\; \;\; \fM}^{\fP\fQ} +  \hfA_{\;\;\;\; \;\; \fM}^{\fQ\fP}, \nn \\
& & \hfA_{\;\;\;\; \;\; \fM}^{\fP\fQ} \equiv \hmH^{\fQ'\fQ}  \hfA^{\fP}_{\;\; \;  \fQ'\fM}  \equiv \hmH^{\fP\fP'} \hmH^{\fQ\fQ'} \fA_{\fP'\fQ'\fM} , \nn \\
& &  \hfA^{\fP}_{\;\; \;  \fM\fN} \equiv \hmH^{\fP\fQ} \fA_{\fQ\fM\fN} \equiv \hmH^{\fP\fQ}\mH_{\fM\fQ'} \fA_{\fQ\fN}^{\fQ'} ,  \nn \\
& & \hat{\nabla}_{\fP} \fA_{\fM\fN}^{\fP} \equiv \p_{\fP}  \fA_{\fM\fN}^{\fP} + \fGa_{\fP\fQ}^{\fP} \fA_{\fM\fN}^{\fQ} -   \cA_{\fP\fM}^{\fQ} \fA_{\fQ\fN}^{\fP} -  \cA_{\fP\fN}^{\fQ} \fA_{\fM\fQ}^{\fP} . 
\ee

The antisymmetric tensor $\mT_{[\fM\fN]}$ defines antisymmetric hyper-stress tensor in hyper-spacetime with the following explicit form:
\be
 \mT_{[\fM\fN]} \equiv  \bar{\fPsi}_{\fQH} \vSi_{-}^{\fC} \fA_{\fM\fC } i\cD_{\fN}  \fPsi_{\fQH}  - \bar{\fPsi}_{\fQH} \vSi_{-}^{\fC} \fA_{\fN\fC } i\cD_{\fM}  \fPsi_{\fQH} , 
\ee

Therefore, two equations of motion presented in Eqs.(\ref{GGEE}) and (\ref{GGEB}) are referred to as {\it geometric gravitational equations}. Eq.(\ref{GGEE}) with symmetric tensor is a generalized Einstein-type equation, which is regarded as the extension of general theory of relativity in four-dimensional spacetime to hyperunified field theory in hyper-spacetime. Eq.(\ref{GGEB}) with antisymmetric tensor brings on an additional dynamic equation beyond Einstein-type equation, which arises from hyper-gravigauge field, instead of hyper-gravimetric field, as fundamental gravitational field in the hyperunified field theory with inhomogeneous hyperspin gauge symmetry that governs the fundamental interaction of nature.


\subsection{ The first oder gravitational differential equation of hyper-gravigauge field from the constraint equation of hyperspin gravigauge field as auxiliary field in locally flat gravigauge hyper-spacetime }

Let us now turn to the geometry-gauge correspondence in the hidden coordinate formalism of the action for the hyperunified field theory, which leads to the following relations:
\be
& & \hmH^{\fM\fM'}  ( \hmH^{\fN\fN'} \mF_{\fM\fN\fA} \mF_{\fM'\fN'}^{\fA} + 2 \hfA_{\fA'}^{\;\;\fN} \hfA_{\fA}^{\;\;\fN'}  \mF_{\fM\fN}^{\fA} \mF_{\fM'\fN'}^{\fA'} ) = 4 \fOm_{\fC\fA\fB} \fOm^{\fA\fB\fC} , \nn \\
& & \hmH^{\fM\fM'}  \hfA_{\fA}^{\;\;\fN} \hfA_{\fA'}^{\;\;\fN'}  \mF_{\fM\fN}^{\fA} \mF_{\fM'\fN'}^{\fA'} = -  \fOm_{\fC\fA}^{\fC} \fOm_{\fD}^{\fD\fA},  \nn \\
& & \frac{1}{4} \fkA\, \thmH^{\fM\fN\fM'\fN'}_{\fA\fA'} \mF_{\fM\fN}^{\fA}\mF_{\fM'\fN'}^{\fA'}  \equiv 
\fkA\, \eta^{\fC\fD} \fR_{\fC\fD} - 2\p_{\fM} ( \fkA\, \hfA_{\fA}^{\;\fM} \fOm_{\fC}^{\fC\fA} ), \nn \\
& & \qquad \qquad \qquad \qquad \qquad \;\; \equiv \fkA\, (\fOm_{\fC\fA\fB} \fOm^{\fA\fB\fC} + \fOm_{\fC\fA}^{\fC} \fOm_{\fD}^{\fD\fA} ) .
\ee
In such a formalism, the gravitational interaction is solely characterized by the hyperspin gravigauge field $\fOm_{[\fC\fD]}^{\fA}$ which behaves as an auxiliary field in locally flat gravigauge hyper-spacetime, so that $\fOm_{[\fC\fD]}^{\fA}$ is determined through the constraint equation which concerns only basic bosonic fields as shown in Eqs.(\ref{GGCGE}) and (\ref{GGCGE0}). Meanwhile, the hyperspin gravigauge field $\fOm_{[\fC\fD]}^{\fA}$ is shown in Eq.(\ref{NCG}) to present as the structure factor of non-Abelian Lie group for the hyper-gravicoordinator derivative operator, which brings on the emergence of non-commutative geometry in locally flat gravigauge hyper-spacetime. Therefore, the gravitational interaction acts as an emergent interaction in locally flat gravigauge hyper-spacetime characterized by non-commutative geometry.

The hyperspin gravigauge field $\fOm_{[\fC\fD]}^{\fA}$ is actually presented by Abelian-type gauge field strength $\mF_{\fM\fN}^{\fA}$ as follows:
\be
\fOm_{[\fC\fD]}^{\fA} \equiv -\hfA_{\fC}^{\;\; \fM} \hfA_{\fD}^{\;\; \fN} \mF_{\fM\fN}^{\fA} , \quad \mF_{\fM\fN }^{\fA} = \p_{\fM}\fA_{\fN}^{\;\; \fA} - \p_{\fM}\fA_{\fN}^{\;\; \fA} .
\ee
When projecting Eq.(\ref{GGCGE}) into hyper-spacetime of coordinates by using the hyper-gravigauge field as Goldstone-like bicovariant vector field in biframe hyper-spacetime, we obtain the following constraint equation:
\be \label{GGCDE}
\mF_{\fM\fN \fA} - \chi_{[\fM\fN]\fA}^{\fM'\fN'\fA'} \mF_{\fM'\fN'\fA'} = - \mT_{[\fM\fN] \fA} ,
\ee
where we have introduced the following definitions:
\be \label{GGCDE0}
\mT_{[\fM\fN] \fA} & \equiv &   \beta_G^2 \beta_Q^2 \sinh^2\chi_s [  \frac{3}{2} (\mF_{\fM\fN \fA} - \frac{1}{3} \tchi_{[\fM\fN]\fA}^{\fM'\fN'\fA'} \mF_{\fM'\fN'\fA'} )  \nn \\
& + &  \cA_{\fM\fA\fB} \fA_{\fN}^{\; \fB} - \cA_{\fN\fA\fB} \fA_{\fM}^{\; \fB} + \fA_{\fM}^{\; \fA'} \fA_{\fN}^{\; \fB'}   \hfA_{\fA}^{\; \fM'} \cA_{\fM'\fA'\fB'} ] \nn \\
& - & \mF_{\fM\fN \fB} \hfA_{\fA}^{\fM'} \hfA^{\fB\fN'} (\cA_{\fM'}^{\fA'\fB'} \cA_{\fN'\fA'\fB'} + g_H^2g_W^{-2} \mW_{\fM'}\mW_{\fN'} )  \nn \\
& + &  ( \tcF_{\fM\fN}^{\fA'\fB'} \cA_{\fM'\fA'\fB'}   + g_H^2g_W^{-2} \tcF_{\fM\fN} \mW_{\fM'}  ) \hfA_{\fA}^{\; \fM'} ,
\ee
and 
\be
& & \chi_{[\fM\fN]\fA}^{\fM'\fN'\fA'} \equiv \eta_{\fM}^{\; \fN'} ( \fA_{\fN}^{\; \fA'} \hfA_{\fA}^{\fM'} - 2 \fA_{\fN\fA} \hfA^{\fA'\fM'} ) - \eta_{\fN}^{\; \fN'} ( \fA_{\fM}^{\; \fA'} \hfA_{\fA}^{\fM'} - 2 \fA_{\fM\fA} \hfA^{\fA'\fM'} ) , \nn \\
& & \tchi_{[\fM\fN]\fA}^{\fM'\fN'\fA'} \equiv \eta_{\fM}^{\; \fN'} \fA_{\fN}^{\; \fA'} \hfA_{\fA}^{\fM'} - \eta_{\fN}^{\; \fN'} \fA_{\fM}^{\; \fA'} \hfA_{\fA}^{\fM'} .
\ee
In four-dimensional spacetime, such a constraint equation in Eq.(\ref{GGCDE}) brings about twenty four first order partial differential equations.    

Therefore, we come to the observation that the constraint equation given in Eq.(\ref{GGCGE}) for hyperspin gravigauge field as auxiliary field in locally flat gravigauge hyper-spacetime brings on the first order partial differential equation of hyper-gravigauge field in biframe hyper-spacetime, which is regarded as the {\it first order gravitational differential equation}.

In the special case with keeping only hyperspin gravigauge field by turning away all other fields, the first order gravitational differential equation presented in Eq.(\ref{GGCDE}) is simplified to be as follows:
\be \label{GGCDE0}
\mF_{\fM\fN \fA} - \chi_{[\fM\fN]\fA}^{\fM'\fN'\fA'} \mF_{\fM'\fN'\fA'} = 0 ,
\ee
which provides the first order gravitational differential equation purely for the hyper-gravigauge field. 


\subsection{ Dynamic equation of hyper-gravimetric gauge field as geometric gauge-type gravitational equation  and the electric-like and magnetic-like gravitational interactions based on gauge-geometry duality }

From the gauge-geometry duality, the gravitational interaction is described by Abelian-type hyper-gravigauge field or equivalently by hyper-gravimetric field which can also be regarded as hyper-gravimetric gauge field. We are inspired to investigate the electric-like and magnetic-like gravitational interactions in the hyperunified field theory within the framework of GQFT. 

Once regarding $\mH_{\fN}^{\; \fP}$ as hyper-gravimetric gauge field, we are able to derive from the action of hyperunified field theory presented in Eq.(\ref{actionHUFTGQFT3}) the following {\it geometric gauge-type gravitational equation}: 
\be \label{GGEh}
\p_{\fN} \hmF_{\fP}^{\fM\fN} = \hat{\mJ}_{\fP}^{\; \fM} , 
\ee
where $\hmF_{\fP}^{\fM\fN} $ is defined as follows:
\be
 & & \hmF_{\fP}^{\fM\fN} \equiv  \hmH^{\fM\fM'}\hmH^{\fN\fN'}\hmH_{\fP\fP'}\mF_{\fM'\fN'}^{\fP'} .
\ee
which is referred to as {\it hyper-gravimetric gauge field strength tensor}. $\hat{\mJ}_{\fP}^{\; \fM}$ defines the corresponding tensor current,
\be
\hat{\mJ}_{\fP}^{\; \fM} \equiv 2 \hat{\mG}_{\fP}^{\; \fM} + 2 g_H^2 \hmT_{\fP}^{\; \fM} , 
\ee
which is referred to as {\it hyper-gravimetric gauge field current}. Where $\hmT_{\fP}^{\; \fM}$ is given by the generalized Einstein-type hyper-stress energy-momentum tensor $\mT_{\fM\fN}$ (Eq.(\ref{EEMT})) as follows:
\be \label{ETEMT}
\hmT_{\fP}^{\; \fM} \equiv  \hmH_{\fP}^{\; \fQ} \mT_{\fQ\fN} \hmH^{\fN\fM} = \hfA_{\;\;\fP}^{\fA} \mT_{\fA}^{\;\;\fM} , 
\ee
and $\hat{\mG}_{\fP}^{\; \fM}$ has the following explicit form:
\be \label{PGFC}
\hat{\mG}_{\fP}^{\; \fM} & \equiv & - \hmH^{\fM\fM'}\hmH^{\fN\fN'}\hmH_{\fP\fP'}\mF_{\fM'\fN'}^{\fP'} \p_{\fN}\ln \fkA + \frac{1}{4} \hmH^{\fM\fN} \hmH_{\fP}^{\fN'} \hmH^{\fQ\fM'} \hmH_{\fP'\fQ'}  \mF_{\fQ\fN}^{\fQ'} \mF_{\fM'\fN'}^{\fP'} \nn \\
& + & \frac{1}{8} \hmH_{\fQ'}^{\fM} \hmH_{\fP\fP'}  \hmH^{\fQ\fM'} \hmH^{\fN\fN'} \mF_{\fQ\fN}^{\fQ'} \mF_{\fM'\fN'}^{\fP'} +  \frac{1}{4}  \hmH^{\fM\fN}  \hmH_{\fP}^{\fQ}   \p_{\fN}(\ln \fkA) \p_{\fQ}(\ln \fkA)  \nn \\
& + & \frac{1}{4} \fkA^2 \hmH^{\fM\fQ}  \hmH_{\fP}^{\fQ'}  \hp^{\fN}(\hat{\fkA} \mH_{\fN\fQ} )  \hp^{\fN'}( \hat{\fkA} \mH_{\fN'\fQ'}) +  \frac{1}{2}  \hmH^{\fM\fN}  \hmH_{\fP}^{\fQ}  \fkA \p_{\fN} \p_{\fQ}(\ln \fkA) \nn \\
& + & \frac{1}{2} \fkA (  \hmH^{\fM\fQ}  \hp_{\fP}\ln \fkA  + \hmH_{\fP}^{\fQ}  \hp^{\fM}\ln \fkA  ) \hp^{\fN}(\hat{\fkA} \mH_{\fN\fQ} ) \nn \\
& +& \frac{1}{4} \fkA (  \hmH^{\fM\fQ}  \hp_{\fP} + \hmH_{\fP}^{\fQ}  \hp^{\fM} ) \hp^{\fN}(\hat{\fkA} \mH_{\fN\fQ} ) \nn \\
& - & \frac{1}{8} \fkA (  \hmH^{\fM\fQ}  \hmH_{\fP}^{\fQ'}  +  \hmH^{\fM\fQ'} \hmH_{\fP}^{\fQ} ) \hp^{\fN}\p_{\fQ'} (\hat{\fkA} \mH_{\fN\fQ} ) \nn \\
& - & \frac{1}{8} \fkA  \hmH^{\fQ\fQ'} (  \hmH^{\fM\fN} \hp_{\fP}  +  \hmH_{\fP}^{\fN} \hp^{\fM}) \p_{\fQ'} (\hat{\fkA} \mH_{\fN\fQ} ) \nn \\
& - & \frac{1}{8} \hmH_{\fP}^{\fM} [ \, \hmH^{\fQ\fM'} \hmH^{\fN\fN'} \hmH_{\fQ'\fP'}  \mF_{\fQ\fN}^{\fQ'} \mF_{\fM'\fN'}^{\fP'} -2  \hmH^{\fQ\fN}  \p_{\fQ}(\ln \fkA) \p_{\fN}(\ln \fkA) \nn \\
& +  & 2 \fkA^2  \hmH^{\fP'\fQ'} \hp^{\fQ}(\hat{\fkA} \mH_{\fQ\fP'} ) \hp^{\fN}( \hat{\fkA} \mH_{\fN\fQ'})  + 4 \fkA \hp^{\fQ}(\ln\fkA) \hp^{\fN}(\hat{\fkA} \mH_{\fN\fQ}) )  \nn \\
& - & 4 \hmH^{\fQ\fN}  \p_{\fQ} \p_{\fN}(\ln \fkA) -  \fkA \hmH^{\fQ\fQ'} \hp^{\fN} \p_{\fQ'} (\hat{\fkA} \mH_{\fN\fQ} )  \nn \\
& + & \hat{\fkA}^2 \mH_{\fN\fQ} \p_{\fP'} \p_{\fQ'} ( \fkA \hmH^{\fQ\fQ'} \fkA \hmH_{\fN\fP'} ) \, ] ,
\ee
which defines the {\it pure hyper-gravimetric gauge field current} resulting just from the hyper-gravimetric gauge field itself. 

Such an introduced hyper-gravimetric gauge field current $\hat{\mJ}_{\fP}^{\; \fM}$ is a conserved current,
\be
\p_{\fM} \hat{\mJ}_{\fP}^{\; \fM} = 0,
\ee
which is attributed to the antisymmetric property of hyper-gravimetric gauge field strength tensor $\hmF_{\fP}^{\fM\fN} = - \hmF_{\fP}^{\fN\fM}$.

As either geometric gauge-type gravitational equation (\ref{GGEh}) or gauge-type gravitational equation (\ref{GGEb}) is Abelian-type gauge equation in correspondence to hyper-gravimetric gauge field $\mH_{\fM}^{\;\; \fP}$ or hyper-gravigauge field $\fA_{\fM}^{\;\; \fA}$, we are allowed to introduce electric-like and magnetic-like gravitational field strengths in analogous to Maxwell electromagnetic field strength. For the hyper-gravimetric gauge field strength tensor, let us make the following definitions:
\be
& & \hmF_{\fP}^{0\fI} =  \hmH^{0\fM'}\hmH^{\fI\fN'}\hmH_{\fP\fP'}\mF_{\fM'\fN'}^{\fP'} = - \hmF_{\fP}^{\fI 0} \equiv  -\mE_{\fP}^{\fI}, \nn \\
& & \hmF_{\fP}^{\fI\fJ} =  \hmH^{\fI\fM'}\hmH^{\fJ\fN'}\hmH_{\fP\fP'}\mF_{\fM'\fN'}^{\fP'} \equiv f^{\fI\fJ}_{\;\; \; \fK\fL} \mB_{\fP}^{\fK\fL}, 
\ee
where $f^{\fI\fJ}_{\;\;\fK\fL}$ is a constant tensor with the following property:
\be
 & & f^{\fI\fJ}_{\;\;\fK\fL} = -  f^{\fJ\fI}_{\;\;\fK\fL} = -  f^{\fI\fJ}_{\;\;\fL\fK} , \nn \\
 & &   f^{\fI\fJ}_{\;\;\fK\fL}  f^{\fK\fL}_{\;\;\fI'\fJ'} = \frac{1}{2} (\eta^{\fI}_{\; \fI'} \eta^{\fJ}_{\; \fJ'} - \eta^{\fJ}_{\; \fI'} \eta^{\fI}_{\; \fJ'} ) .
\ee
Similar definitions hold for the hyper-gravigauge field strength tensor,
\be
& & \hmF_{\fA}^{0\fI} = \thmH^{[0\fI]\fM'\fN'}_{\fA\fA'} \mF_{\fM'\fN' }^{\fA'}  = - \hmF_{\fA}^{\fI 0}  \equiv - \mE_{\fA}^{\fI}, \nn \\
& & \hmF_{\fA}^{\fI\fJ} = \thmH^{[\fI\fJ]\fM'\fN'}_{\fA\fA'} \mF_{\fM'\fN' }^{\fA'}  \equiv  f^{\fI\fJ}_{\;\;\fK\fL} \mB_{\fA}^{\fK\fL} .
\ee
The gravitational field strengths $\mE_{\fA}^{\fI}$ and $\mE_{\fP}^{\fI}$ are regarded as electric-like hyper-gravigauge field strength and hyper-gravimetric gauge field strength, the antisymmetric gravitational field strengths $\mB_{\fA}^{\fK\fL}= -\mB_{\fA}^{\fL\fK} $ and $\mB_{\fP}^{\fK\fL}= -\mB_{\fP}^{\fL\fK}$ are viewed as generalized magnetic-like hyper-gravigauge field strength and hyper-gravimetric gauge field strength, respectively. 

In terms of the electric-like and magnetic-like gravitational field strengths, the geometric gauge-type gravitational equations can be expressed as follows:
\be
& & \p_{\fI}\mE_{\fP}^{\fI} = - \hmJ_{\fP}^{0}, \nn \\
& & f^{\fI[\fJ\fK\fL]} \p_{\fJ}\mB_{\fP \fK\fL} + f^{\fI(\fJ\fK\fL]} \p_{\fJ}\mB_{\fP \fK\fL} + \p_{0}\fE_{\fP}^{\fI} = \hat{\mJ}_{\fP}^{\fI} , 
\ee  
and similarly, the gauge-type gravitational equation can be written as follows:
\be
& & \p_{\fI}\mE_{\fA}^{\fI} = - \hmJ_{\fA}^{0}, \nn \\
& & f^{\fI[\fJ\fK\fL]} \p_{\fJ}\mB_{\fA \fK\fL} + f^{\fI(\fJ\fK\fL]} \p_{\fJ}\mB_{\fA \fK\fL} + \p_{0}\fE_{\fA}^{\fI} = \hat{\mJ}_{\fA}^{\fI}. 
\ee  
We have introduced the following definitions:
\be
& & f^{\fI[\fJ\fK\fL]} \equiv \frac{1}{3} ( f^{\fI\fJ\fK\fL} + f^{\fI\fK\fL\fJ} + f^{\fI\fL\fJ\fK} ), \nn \\ 
& & f^{\fI(\fJ\fK\fL]} \equiv \frac{1}{3} ( 2 f^{\fI\fJ\fK\fL} - f^{\fI\fK\fL\fJ} - f^{\fI\fL\fJ\fK} ),
\ee
where $f^{\fI[\fJ\fK\fL]}$ and $f^{\fI(\fJ\fK\fL]}$ denote the total antisymmetric tensor and symmetric-antisymmetric mixing tensor, respectively, for the superscripts `$\fJ\fK\fL$' in the brackets.

As a simple consideration, let us suppose that the hyperunified field theory in hyper-spacetime evolves dynamically into an effective theory in four dimensional spacetime, where the gauge-geometry duality leads the dynamics of gravigauge field to be equivalent to Einstein's general theory of relativity characterized by gravimetric gauge field. So that, the geometric gauge-type gravitational equation in four dimensional spacetime is simplified to be:
\be
& & \p_{i}\mE_{\rho}^{i} = -\hmJ_{\rho}^{0} , \nn \\
& & \varepsilon^{ij}_{\;\;\; k} \p_{j}\mB_{\rho}^{k} +\p_{0}\mE_{\rho}^{i} = \hat{\mJ}_{\rho}^{i} ,  
\ee  
with $i, j, k=1,2,3$ and $\varepsilon^{ijk}$ the totally antisymmetric constant tensor $\varepsilon^{123}=1$. The electric-like gravimetric field $\mE_{\rho}^{i}$ and magnetic-like gravimetric field $\mB_{\rho}^{i}$ are given as follows:
\be 
& & \hmF_{\rho}^{0i} =  \hmH^{0\mu'}\hmH^{i\nu'}\hmH_{\rho\rho'}\mF_{\mu'\nu'}^{\rho'} \equiv - \mE_{\rho}^{i}, \nn \\
& & \hmF_{\rho}^{ij} =  \hmH^{i\mu'}\hmH^{j\nu'}\hmH_{\rho\rho'}\mF_{\mu'\nu'}^{\rho'} \equiv \varepsilon^{ij}_{\;\; \; k} \mB_{\rho}^{k} .
\ee

By fixing the scaling gauge symmetry to Einstein basis at low energy, the dual gravimetric fields can be expressed as follows:
\be \label{DGMF}
& & \mH_{\mu\nu} \equiv \Mka^2 ( \eta_{\mu\nu} + \frac{1}{M_{p}}h_{\mu \nu} ), \quad M_p \equiv \sqrt{2} \Mka/g_h  , \nn \\
& & \hmH^{\mu\nu} =( \mH_{\mu\nu} )^{-1} = \Mka^{-2} ( \eta^{\mu\nu} - \frac{1}{M_{p}} \eta^{\mu\mu'} \eta^{\nu\nu'} h_{\mu' \nu'} + O(\frac{1}{M_{p}^2}) ), 
\ee 
where $\Mka$ is regarded as {\it fundamental mass scale}, $M_{p}$ and $g_h$ are considered as the basic mass scale and coupling constant in four-dimensional spacetime. 

In an approximation that the inverse of gravimetric gauge field is taken at the leading-order, we obtain the following relations:
\be
& & \hmF_{\rho}^{0i} \simeq \Mka^{-6} \eta^{0\mu'}\eta^{i\nu'}\eta_{\rho\rho'}\mF_{\mu'\nu'}^{\rho'} = \Mka^{-6} \mF^{0i}_{\rho} \equiv - \Mka^{-6} E_{\rho}^{i}, \nn \\
& & \hmF_{\rho}^{ij} \simeq \Mka^{-6} \eta^{i\mu'}\eta^{j\nu'}\eta_{\rho\rho'} \mF_{\mu'\nu'}^{\rho'} = \Mka^{-6} \mF^{ij}_{\rho} \equiv \Mka^{-6} \varepsilon^{ij}_{\;\; \; k} B_{\rho}^{k} ,
\ee
for the electric-like gravimetric field $E_{\rho}^{i}$ and magnetic-like gravimetric field $B_{\rho}^{i}$, and 
\be
\hat{\mJ}_{\rho}^{\; \mu} & \equiv & 2 \hat{\mG}_{\rho}^{\; \mu} + g_h^2 2 \hmT_{\rho}^{\; \mu} \simeq \Mka^{-6} \hJ_{\rho}^{\; \mu}, \quad \hJ_{\rho}^{\; \mu} = \hG_{\rho}^{\; \mu} + g_h^2 2 \hT_{\rho}^{\; \mu} , 
\ee
for the hyper-gravimetric gauge field current $\hJ_{\rho}^{\; \mu}$. Where $\hT_{\rho}^{\; \mu}$ represents Einstein-type energy-momentum tensor of matter field in four-dimensional spacetime, which can be deduced from $\hmT_{\fP}^{\; \fM}$ defined in Eq.(\ref{ETEMT}) via the hyper-stress energy-momentum tensor $\mT_{\fM\fN}$ given in Eq.(\ref{EEMT}). While $\hG_{\rho}^{\; \mu}$ denotes pure gravimetric field tensor, which can be resulted from the tensor $\hat{\mG}_{\fP}^{\; \fM}$ given in Eq.(\ref{PGFC}) by fixing the scaling gauge symmetry to Einstein basis shown in Eq.(\ref{DGMF}).

In light of the vector notations for both electromagnetic-like gravitational field and current, i.e., $\fE_{\rho} \equiv (E_{\rho}^1, E_{\rho}^2, E_{\rho}^3)$, $\fB_{\rho} \equiv  (B_{\rho}^1, B_{\rho}^2, B_{\rho}^3)$ and $\hfJ_{\rho} \equiv (\hJ_{\rho}^1, \hJ_{\rho}^2, \hJ_{\rho}^3)$, the geometric gauge-type gravitational equations can be rewritten as follows:
\be \label{EMTGGE1}
& & \nabla \cdot \fE_{\rho} \simeq -\hJ_{\rho}^{0} , \nn \\
& & \nabla \times \fB_{\rho} - \p_{0}\fE_{\rho} \simeq -\hfJ_{\rho} . 
\ee  
From the Bianchi identity,
\be
\p_{\mu} \mF_{\rho}^{\nu\sigma}  + \p_{\nu} \mF_{\rho}^{\sigma\mu} + \p_{\sigma} \mF_{\rho}^{\mu\nu} = 0, \nn
\ee
we obtain the following gauge-type gravitational equations:
\be \label{EMTGGE2}
& & \nabla \cdot \fB_{\rho} = 0 , \nn \\
& & \nabla \times \fE_{\rho} + \p_{0}\fB_{\rho} = 0 . 
\ee  

Therefore, the Abelian-like geometric gauge-type gravitational equations presented in Eqs.(\ref{EMTGGE1}) and (\ref{EMTGGE2}) in four dimensional spacetime are truly analogous to Maxwell equations of electromagnetic field, which provide electric-like and magnetic-like gravimetric field equations in an appropriate approximation. Unlike the electromagnetic field with source current arising solely from electric charged field, the gravimetric gauge field current $\hat{\mJ}_{\rho}^{\; \mu}$ as source current contains not only the current tensor $\hat{\mT}_{\rho}^{\; \mu}$ arising from all basic matter fields but also from the pure gravimetric gauge field current tensor $\hat{\mG}_{\rho}^{\; \mu}$.


\section{ Conformally flat gravigauge hyper-spacetime with evolving graviscalar field as nonsingularity background for evolution of early universe and scaling gauge field and $\cQ_c$-spin scaling field as basic constituents of dark universe and quantum cosmic matter in hyperunified field theory }

In the previous sections, we have demonstrated that the entirety unitary gauge as an appropriate gauge prescription allows us to remove all unphysical degrees of freedom which are caused from inhomogeneous hyperspin gauge symmetry WS(1,$D_h$-1) and hidden general linear group symmetry GL(1,$D_h$-1, R) in hyperunified field theory built within the framework of gravitational quantum field theory (GQFT)\cite{GQFT}. In this section, we shall discuss various gauge prescriptions for the scaling gauge symmetry in the hyperunified field theory under entirety unitary gauge. In Einstein-type basis, we will show that the conformally flat gravigauge hyper-spacetime with evolving graviscalar field provides nonsingularity background spacetime to describe the evolution of the early universe\cite{IE1,IE2,IE3,IE4}. Furthermore, we will demonstrate how the early universe undergoes an inflationary expansion in light of the co-moving cosmic time. In particular, the scaling gauge field is found to be as a dark matter candidate when inflationary universe comes to an end. It is interesting to corroborate that once the potential of $\cQ_c$-spin scaling field falls into the minimum, it not only stops inflationary expansion of universe but also provides cosmic energy density as dark energy candidate.

\subsection{ Hyperunified field theory under entirety unitary gauge with scaling gauge fixing and conformally flat gravigauge hyper-spacetime under background basis with fundamental mass scale }

It has been shown that the entanglement-correlated and translation-like $\cW_e$-spin group symmetry W$^{1,D_h-1}$ is an Abelian-type subgroup of inhomogeneous hyperspin symmetry WS(1,$D_h$-1) of entangled qubit-spinor field. The $\cW_e$-spin invariant-gauge field $\fA_{\fM}^{\;\;\fA}$ as genesis of hyper-gravigauge field reveals the gauge-gravity correspondence. Meanwhile, the gravity-geometry correspondence in curved Riemannian hyper-spacetime indicates that the gravitational interaction can equivalently be described by the hyper-gravimetric gauge field when fixing the hyperspin gauge symmetry under flowing unitary gauge, so that the hyper-gravimetric gauge field is determined from the symmetric hyper-gravigauge field via the simple relation $\mH_{\fM\fN} \equiv (\bfA)^2_{\fM\fN}$ ($\bfA_{\fM\fA}=\bfA_{\fA\fM}$) in which the symmetric hyper-gravigauge field has the same physical degrees of freedom as hyper-gravimetric gauge field. To remove all unphysical degrees of freedom, the duality gauge prescription for general linear group symmetry GL(1,$D_h$-1, R) is proposed to freeze over the flowing unitary gauge for hyperspin gauge symmetry SP(1,$D_h$-1), which leads to the entirety unitary gauge prescription and corroborates the gauge-geometry duality. When taking the gauge fixing conditions under entirety unitary gauge as basic gauge prescription, we are allowed to remove in principle all unphysical degrees of freedom. Eventually, the fundamental symmetry of hyperunified field theory turns out to be an associated symmetry characterized by the global inhomogeneous hyperspin symmetry WS(1,$D_h$-1) in association with global Poincar\'e-type group symmetry PO(1,$D_h$-1) together with scaling symmetry SC(1),  where the symmetry transformations of groups SP(1,$D_h$-1) and SO(1,$D_h$-1) must be coincidental so as to keep the hyperunified field theory under entirety unitary gauge prescription.

The scaling gauge symmetry appears to be a hidden gauge symmetry in the hyperunified field theory expressed in terms of hidden scaling gauge formalism, which indicates that we can in principle choose any convenient scaling gauge fixing condition via appropriate scaling gauge transformation. Namely, under the entirety unitary gauge, we are allowed to further make a typical scaling gauge transformation $\xi_u(x)$ for both graviscalar field $\phi(x)$ and dimensionless symmetric hyper-gravigauge field $\bchi_{\fM\fA}(x)$, i.e.: 
\be
& & \phi(x) \to \phi_u(x) \equiv \xi_u(x) \phi(x) , \quad \bchi_{\fM\fA}(x) \to \bchi_{\fM\fA}^u(x) \equiv \xi_u^{-1}(x)\bchi_{\fM\fA}(x) , \nn \\
& & \bfA_{\fM\fA}=\bfA_{\fA\fM} = \phi(x)\bchi_{\fM\fA}(x) \equiv \phi_u(x)\bchi_{\fM\fA}^u(x) , \nn \\
& & \mH_{\fM\fN}= (\bfA)^2_{\fM\fN} = \phi^2(x)\chi_{\fM\fN}(x) \equiv \phi_u^2(x)\chi_{\fM\fN}^u(x) , 
\ee
so that the determinant of $\bchi_{\fM\fA}^u(x)$ or $\chi_{\fM\fN}^u(x)$ is fixed to be the unit,
\be
\chi^u \equiv \det ( \bchi_{\fM\fB}^{u}(x) \eta^{\fB\fA} ) = \sqrt{ \det \left( (-1)^{D_h-1} \chi_{\fM\fN}^u(x)\right) }  = 1  , 
\ee 
which is referred to as {\it unitary basis} for the scaling gauge symmetry.

When fixing the scaling gauge symmetry to be in unitary basis under the entirety unitary gauge, we are able to express the action of the hyperunified field theory given in Eq.(\ref{actionHUFTHSG}) or Eq.(\ref{actionHUFTGQFT1}) and (\ref{actionHUFTGQFT2}) into the following form:
\be  \label{actionHUFTGQFTU}
\cS_{\mH\mU} & \equiv & \int [d^{D_h}x] \, \{\, \chih^{u \fM\fN} \bar{\fPsi}_{\fQH} \bchi_{\fM\fC}^u  \vSi_{-}^{\fC} i\cD_{\fN}  \fPsi_{\fQH} - \beta_Q \phi_u \sinh(\chi_s) \,  \bar{\fPsi}_{\fQH} \tvSi_{-}\fPsi_{\fQH}  
\nn \\
& + & \phi_u^{D_h-4} [ -  \chih^{u \fM\fM'} \chih^{u \fN\fN'} \frac{1}{4g_H^2} \cF_{\fM\fN\fA\fB}  \cF_{\fM'\fN'}^{\fA\fB} \nn \\
& + & \chih_{\fC}^{u \fM}\chih_{\fD}^{u \fN} \chih_{\fC'}^{u \fM'}\chih_{\fD'}^{u \fN'} \tilde{\eta}^{\fC\fD\fC'\fD'}_{\fA\fA'}  \frac{1}{4g_H^{2}}  \mF_{\fM\fN}^{\fA}\mF_{\fM'\fN'}^{\fA'} \nn \\
& + & \frac{1}{2g_H^{2}} \beta_G^2 \beta_Q^2 \phi_u^2 \sinh^2\chi_s \,  \chih^{u \fM\fM'} ( \cA_{\fM\fA\fB} - \fOm_{\fM\fA\fB} )(\cA_{\fM'}^{\fA\fB} - \fOm_{\fM'}^{\fA\fB} )  \nn \\
& - & \chih^{u \fM\fM'} \chih^{u \fN\fN'} \frac{1}{4g_W^2} \mF_{\fM\fN} \mF_{\fM'\fN'} + \frac{1}{2}  \lambda_S^2 \phi_u^2 ( 1 + \sinh^2\chi_s ) \chih^{u \fM\fN}  \mW_{\fM} \mW_{\fN}  \nn \\
&  + &  \frac{1}{2} \chih^{u \fM\fN} \phi_u^2\lambda_S^2 \p_{\fM} \chi_s \p_{\fN}\chi_s - \lambda_D^2 \phi_u^4 \cF(\chi_s)  ] \, \} \nn \\
& \equiv & \int [d^{D_h}x] \, \{\, \chih^{u \fM\fN} \bar{\fPsi}_{\fQH} \bchi_{\fM\fC}^u  \vSi_{-}^{\fC} i\cD_{\fN}  \fPsi_{\fQH} -  \beta_Q \phi_u \sinh(\chi_s) \,  \bar{\fPsi}_{\fQH} \tvSi_{-}\fPsi_{\fQH}  \nn \\
& + & \phi_u^{D_h-4} [ -  \chih^{u \fM\fM'} \chih^{u \fN\fN'} \frac{1}{4g_H^2} \cF_{\fM\fN\fA\fB}  \cF_{\fM'\fN'}^{\fA\fB}  
\nn \\
& + & g_H^{-2} \chih^{u \fM\fN} \left( \phi_u^2  \mR_{\fM\fN}^u - (D_h-1)(D_h-2) \p_{\fM}\phi_u \p_{\fN}\phi_u \right) \nn \\
& + & \frac{1}{2g_H^2} \beta_G^2  \beta_Q^2 \phi_u^2 \sinh^2\chi_s \, \chih^{u \fM\fM'} ( \cA_{\fM\fA\fB} - \fOm_{\fM\fA\fB} )(\cA_{\fM'}^{\fA\fB} -  \fOm_{\fM'}^{\fA\fB} )  \nn \\
& - & \chih^{u \fM\fM'} \chih^{u \fN\fN'} \frac{1}{4g_W^2} \mF_{\fM\fN} \mF_{\fM'\fN'} + \frac{1}{2} \lambda_S^2 \phi_u^2 ( 1 + \sinh^2\chi_s) \chih^{u \fM\fN}  \mW_{\fM} \mW_{\fN}  \nn \\
&  + &  \frac{1}{2} \chih^{u \fM\fN} \phi_u^2\lambda_S^2 \p_{\fM} \chi_s \p_{\fN}\chi_s - \lambda_D^2 \phi_u^4 \cF(\chi_s)  ] \, \} ,
\ee
where we have used the following redefinitions:
\be
& &  \fPsi_{\fQH} \to \phi_u^{-\frac{D_h-1}{2}} \fPsi_{\fQH} , \nn \\
& & \fOm_{\fM\fA\fB} \equiv \chi_{\fM\fN}^{[\fA\fB]}\, \p^{\fN} \ln \phi^u + \mOm_{\fM\fA\fB} , \nn \\
& & \chi_{\fM\fN}^{[\fA\fB]} \equiv (\bchi_{\fM}^{u\; \fA} \bchih^{u \fP \fB} -  \bchi_{\fM}^{u\; \fB} \bchih^{u \fP\fA} )\eta_{\fP\fN} , 
\ee 
and 
 \be 
\chi_s(x) & \equiv &  \sinh^{-1}\frac{\lambda_Q\phi_1(x)}{\lambda_S \phi_u(x)},  \nn \\
\frac{\phi_1(x)}{\phi_u(x)}  & \equiv & \beta_Q \sinh\chi_s(x) , \quad \beta_Q \equiv \frac{\lambda_S}{\lambda_Q} .
\ee
Where $\mOm_{\fM\fA\fB}$ is the hyperspin gravigauge field governed by the symmetric hyper-gravigauge field $\bchi_{\fM\fA}(x)$ in unitary basis. The rank-two tensor $\mR^u_{\fM\fN}$ is Ricci-type curvature tensor defined from the hyper-gravimetric field $\chi_{\fM\fN}^u(x)$ in unitary basis. In unitary basis, the duality gauge fixing condition given in Eq.(\ref{DG}) is simplified to be,
\be
\hp^{\fM} \chi_{\fM\fN}^u = 0 .
\ee 
Such a unitary basis is considered to be a physically meaningful basis in calculating quantum effects within the framework of GQFT. 

Alternatively, the scaling gauge invariance also allows us to make a special scaling gauge transformation $\xi_s(x)$ so as to relate the graviscalar field $\phi(x)$ with {\it fundamental mass scale} $\Mka$ as follows:
\be
\label{GFCE}
\phi(x) \to  \Mka \equiv \xi_s(x) \phi(x) . 
\ee
When making the same scaling gauge transformation for $\bchi_{\fM\fA}(x)$, 
\be
& & \bchi_{\fM\fA}(x) \to \bchi_{\fM\fA}^e(x) \equiv \xi_s^{-1}(x)\bchi_{\fM\fA}(x), 
\ee
we can rewrite the scaling gauge invariant symmetric hyper-gravigauge field and hyper-gravimetric gauge field as follows:
\be
\bfA_{\fM\fA}(x) & = & \bfA_{\fA\fM}(x) \equiv  \Mka \bchi_{\fM\fA}^{e}(x) , \nn \\
\mH_{\fM\fN}(x) & = & (\bfA)^2_{\fM\fN}(x) \equiv \Mka^2 \chi_{\fM\fN}^e(x) ,
\ee
which is referred to as Einstein-type basis for the scaling gauge symmetry. Where $\bchi_{\fM\fA}^{e}(x)$ and $\chi_{\fM\fN}^e(x)$ denote the symmetric hyper-gravigauge field and hyper-gravimetric field in {\it Einstein-type basis}. 

When the scaling gauge fixing condition is chosen to be in Einstein-type basis, the action of the hyperunified field theory presented in Eq.(\ref{actionHUFTGQFTU}) can be rewritten into the following form:
\be  \label{actionHUFTGQFTE}
\cS_{\mH\mU} & \equiv & \int [d^{D_h}x] \chi^e  \Mka^{D_h-4}\{ \chih^{e \fM\fN} \bar{\fPsi}_{\fQH} \bchi^e_{\fM\fC}  \vSi_{-}^{\fC} i\cD_{\fN}  \fPsi_{\fQH} -  M_S \sinh(\chi_s) \,  \bar{\fPsi}_{\fQH} \tvSi_{-}\fPsi_{\fQH}  \nn \\
& - &  \chih^{e \fM\fM'} \chih^{e \fN\fN'} \frac{1}{4g_H^2} \cF_{\fM\fN\fA\fB}  \cF_{\fM'\fN'}^{\fA\fB}  
+ \frac{1}{2} \frac{\Mka^2}{g_H^2} \chih^{e \fM\fN} \mR^e_{\fM\fN} \nn \\
& + & \frac{1}{2g_H^2} \beta_G^2  M_S^2 \sinh^2\chi_s \,  \chih^{e \fM\fM'} ( \cA_{\fM\fA\fB} - \fOm_{\fM\fA\fB} )(\cA_{\fM'}^{\fA\fB} - \fOm_{\fM'}^{\fA\fB} )  \nn \\
& - & \chih^{e \fM\fM'} \chih^{e \fN\fN'} \frac{1}{4g_W^2} \mF_{\fM\fN} \mF_{\fM'\fN'} + \frac{1}{2} \lambda_Q^2 M_S^2 ( 1+ \sinh^2\chi_s ) \chih^{e \fM\fN}  \mW_{\fM} \mW_{\fN}  \nn \\
&  + &  \frac{1}{2} \chih^{e \fM\fN}\lambda_Q^2 M_S^2 \p_{\fM} \chi_s \p_{\fN}\chi_s - \lambda_D^2 \Mka^4 \cF(\chi_s)  \, \} ,
\ee
where $\mR^e_{\fM\fN}$ is Ricci-type curvature tensor defined from the hyper-gravimetric field $\chi_{\fM\fN}^e(x)$ in Einstein-type basis. We have also used the following redefinitions:
\be 
 & & \chi_s(x)  \equiv  \sinh^{-1}\frac{\phi_1(x)}{M_S},  \nn \\
 & & \frac{\phi_1(x)}{M_S}   \equiv  \sinh\chi_s(x) ,  \nn \\
 & & M_S \equiv  \frac{\lambda_S\Mka}{\lambda_Q} = \beta_Q \Mka , \quad \beta_Q \equiv \frac{\lambda_S}{\lambda_Q} ,
\ee
with $M_S$ regarded as a {\it fundamental graviscaling mass scale}. Where $\chi_s(x)$ is characterized by the $\cQ_c$-spin scalar field $\phi_1(x)$ in Einstein-type basis.


\subsection{ Geometry of gravigauge hyper-spacetime and the dynamics of graviscalar and $\cQ_c$-spin scalar fields with conformally flat gravigauge hyper-spacetime as background spacetime}

To understand the evolution of universe, let us turn to study the geometry of locally flat gravigauge hyper-spacetime $\fG_h$. In such a locally flat gravigauge hyper-spacetime, the gravitational interaction is considered as an emergent interaction via non-commutative geometry characterized by the hyperspin gravigauge field $\fOm_{[\fC\fD]}^{\fA}$ shown in Eq.(\ref{NCG}), i.e.:
 \be 
 & & [\heth_{\fC}\, , \heth_{\fD} ] = \fOm_{[\fC\fD]}^{\fA}\1 \heth_{\fA} , \nn 
\ee
with $\heth_{\fC}$ the hyper-gravicoordinate derivative operator in locally flat gravigauge hyper-spacetime $\fG_h$. The hyperspin gravigauge field $\fOm_{[\fC\fD]}^{\fA}$ as structure factor of non-Abelian Lie algebra is shown to behave as an auxiliary field in the hyperunified field theory formulated in terms of hidden coordinate formalism, which is determined through the constraint equation of hyperspin gravigauge field $\fOm_{[\fC\fD]}^{\fA}$ as shown in Eq.(\ref{GGCGE}).

Let us now define the gauge and scaling invariant line element in locally flat gravigauge hyper-spacetime with the gravigauge hyper-basis $\{\delta\vka^{\fA}\}$ or $\{\delta\zeta^{\fA}\}$ as follows:
\be
\delta^2 \cS \equiv \Mka^{-2} \delta\vka^{\fA} \eta_{\fA\fB} \delta\vka^{\fB} = \frac{\phi^2}{\Mka^2} \delta\zeta^{\fA} \eta_{\fA\fB} \delta\zeta^{\fB},
\ee
where the fundamental mass scale $\Mka$ as multiplying factor is introduced to make the gauge and scaling invariant line element to have the length dimension. Based on the biframe hyper-spacetime structure and biframe displacement correspondence from which the hyper-gravicoordinate displacement for hyper-gravivector field $\vka^{\fA}(x)$ in locally flat gravigauge hyper-spacetime is associated to the ordinary coordinate displacement in globally flat Minkowski hyper-spacetime through $\cW_e$-spin invariant-gauge field as shown in Eq.(\ref{HGCDP}), we can express the above invariant line element in terms of the coordinate hyper-basis as follows:
\be
\delta^2\cS  & = & \Mka^{-2}dx^{\fM} \fA_{\fM}^{\; \fA} \eta_{\fA\fB} \fA_{\fN}^{\; \fB}  dx^{\fN}= \Mka^{-2} dx^{\fM} \mH_{\fM\fN} dx^{\fN} \nn \\
& = & \frac{\phi^2}{\Mka^2} dx^{\fM}  \chi_{\fM}^{\; \fA} \eta_{\fA\fB} \chi_{\fN}^{\; \fB}  dx^{\fN}= \frac{\phi^2}{\Mka^2} dx^{\fM} \chi_{\fM\fN}  dx^{\fN} \equiv d^2l_S ,
\ee
with $\mH_{\fM\fN}$ (or $\chi_{\fM\fN}$) the hyper-gravimetric tensor field,
\be
\mH_{\fM\fN} =  \fA_{\fM}^{\; \fA} \eta_{\fA\fB} \fA_{\fN}^{\; \fB} = \phi^2 \chi_{\fM}^{\; \fA}\eta_{\fA\fB}  \chi_{\fN}^{\; \fB} = \phi^2 \chi_{\fM\fN} . \nn
\ee

By taking the special scaling gauge transformation $\xi_s(x)$ as scaling gauge fixing condition, we can relate the graviscalar field $\phi(x)$ to the fundamental mass scale $\Mka$ as follows:
\be
\phi(x) \to \xi_s(x)\phi(x) = \Mka , \nn 
\ee
and meanwhile, the dimensionless symmetric hyper-gravigauge field under the entirety unitary gauge is fixed to be in Einstein-type basis,
\be
\bchi_{\fM}^{\; \fA}(x) \to \xi^{-1}_s(x) \bchi_{\fM}^{\; \fA}(x) = \bchi_{\fM}^{e\; \fA}(x). \nn
\ee
So that the line element can be rewritten into the following form:
\be
& & \delta^2\cS  \equiv d^2l_S =  \chi_{\fM\fN}^e dx^{\fM} dx^{\fN}, \nn \\
& &  \chi_{\fM\fN}^e (x) =\bchi_{\fM}^{e\, \fA}(x) \eta_{\fA\fB} \bchi_{\fN}^{e\, \fB}(x)  ,
\ee
which defines the line element in Einstein-type basis.

To study the evolution of early universe with the hypothesis of conformally flat hyper-spacetime as background gravigauge hyper-spacetime, we shall analyze the basic properties of background fields and find out their solutions by solving their field equations of motion. It is familiar and useful to work in Einstein-type basis with the following background structure: 
\be 
& & \langle \fPsi_{\fQH}(x) \rangle = 0 , \quad  \langle \mW_{\mu} \rangle = 0 , \quad \langle \chi_s(x) \rangle = \nu_S(x) , \nn \\
& &  \langle \bfA_{\fM\fA}(x) \rangle \equiv \Mka \langle \bchi^e_{\fM\fA}(x) \rangle = \Mka a_s(x) \eta_{\fM\fA} ,  \nn \\
& & \langle \mH_{\fM\fN}(x) \rangle \equiv \Mka^2 \langle \chi_{\fM\fN}^e(x) \rangle = \Mka^2 a_s^2(x) \eta_{\fM\fN} , \nn \\
& &  \langle \cA_{\fM}^{\fA\fB}(x) \rangle  = \eta_{\fM\fP}^{[\fA\fB]}  \mA^{\fP} , \quad  \eta_{\fM\fP}^{[\fA\fB]} \equiv  \eta_{\fM}^{\; \fA} \eta_{\fP}^{\; \fB} - \eta_{\fM}^{\; \fB} \eta_{\fP}^{\; \fB} , 
\ee
where $a_s(x)$ is regarded to be a {\it background graviscaling field} and $\nu_S(x)$ characterizes the scaling invariant {\it background $\cQ_c$-spin scaling field}. From the gravitational origin of hyperspin gauge symmetry through the hyperspin gravigauge field $\fOm_{\fM}^{\fA\fB}(x)$, we can obtain the following relations:
\be
& &  \langle \fOm_{\fM}^{\fA\fB}(x) \rangle  =  \eta_{\fM\fP}^{[\fA\fB]} \mOm^{\fP}, \quad \mOm^{\fP} \equiv \p^{\fP}\ln a_s(x), \nn \\
& & \mA^{\fP} \equiv \mOm^{\fP} + \tmA^{\fP} = \p^{\fP}\ln a_s(x) + \tmA^{\fP}, 
\ee
where $\mA_{\fP}$ reflects the background feature of hyperspin gauge field and $\tmA^{\fP}$ denotes a possible difference from the background feature of hyperspin gravigauge field $\mOm^{\fP}$. 

Such a background structure forms {\it background gravigauge hyper-spacetime} with the following background line element: 
\be
\langle d^2l_S \rangle =  \eta_{\fA\fB}  \langle \bchi_{\fM}^{e\, \fA}(x) \bchi_{\fN}^{e\, \fB}(x) \rangle  dx^{\fM} dx^{\fN}  = a_s^2(x)\, \eta_{\fM\fN} \, dx^{\fM} dx^{\fN}  , 
\ee
which brings on {\it conformally flat Minkowski hyper-spacetime} governed by the background graviscaling field $a_s(x)$. Such a background gravigauge hyper-spacetime distinguishes from globally flat Minkowski hyper-spacetime introduced as inertial reference frame of coordinates.  Both Ricci-type tensor and Ricci-type scalar tensor are found to have the following forms:
\be
\langle \mR_{\fM\fN}^e \rangle & = & (D_h-2) d_{\fM}\mOm_{\fN} + \left( d_{\fP}\mOm^{\fP} + (D_h-1) \mOm_{\fP}\mOm^{\fP} \right) \eta_{\fM\fN} \nn \\
 & = & (D_h-2) \left( \p_{\fM}\p_{\fN}\ln a_s - \p_{\fM}\ln a_s \p_{\fN}\ln a_s \right) \nn \\ 
& + & \left(\p^2\ln a_s +  (D_h-2) \p_{\fP}\ln a_s \p^{\fP}\ln a_s \right) \eta_{\fM\fN} , \nn \\
\langle \mR^e \rangle & = & 2(D_h-1) d_{\fP}\mOm^{\fP} + D_h(D_h-1) \mOm_{\fP}\mOm^{\fP} \nn \\
& = & 2 (D_h-2) \p^2\ln a_s +  (D_h-1)(D_h-2) \p_{\fP}\ln a_s \p^{\fP}\ln a_s , \nn \\
d_{\fM}\mOm_{\fN} & \equiv & \p_{\fM}\mOm_{\fN} - \mOm_{\fM}\mOm_{\fN} = \p_{\fM}\p_{\fN} \ln a_s - 
 \p_{\fM}\ln a_s \p_{\fN}\ln a_s ,
\ee
and the background field strength of hyperspin gauge field is given by,
\be
& & \langle \cF_{\fM\fN}^{\fA\fB} \rangle =  d_{\fM}\mA^{\fP} \eta_{\fN\fP}^{[\fA\fB]} - d_{\fN}\mA^{\fP} \eta_{\fM\fP}^{[\fA\fB]} - \mA_{\fP}\mA^{\fP} \eta_{\fM\fN}^{[\fA\fB]} ,\nn \\
& & \langle \cF_{\fM\fN}^{\fA\fB} \cF^{\fM\fN}_{\fA\fB} \rangle  = 4(D_h-2) d_{\fM}\mA_{\fN} d^{\fM}\mA^{\fN} + 
4 (d_{\fM}\mA^{\fM})^2 \nn \\
& & \qquad \qquad \quad\; \;  + 8(D_h-1) d_{\fM}\mA^{\fM} \mA_{\fN}\mA^{\fN} + 2D_h(D_h-1)  (\mA_{\fN}\mA^{\fN})^2, \nn \\
& & d_{\fM}\mA_{\fN} \equiv \p_{\fM}\mA_{\fN} - \mA_{\fM}\mA_{\fN} .
\ee

From the equation of motion presented in Eq.(\ref{EMHSGF}) for the hyperspin gauge field $\cA_{\fM}^{\fA\fB}(x)$, we arrive at the following equation of motion for the background field $\langle \cA_{\fM}^{\fA\fB}(x) \rangle = \eta_{\fM\fP}^{[\fA\fB]} \mA^{\fP}$: 
\be \label{BHSGF}
& & [d_{\fQ}\mA^{\fQ} \mA^{\fP} - d_{\fQ}d^{\fQ}\mA^{\fP} - 2\mA_{\fQ}d^{\fP}\mA^{\fQ} - (D_h-4) \mOm_{\fQ}d^{\fQ}\mA^{\fP} + (D_h-4) \mA_{\fQ}\mA^{\fQ} \tmA^{\fP}  ]  \eta_{\fM\fP}^{[\fA\fB]} \nn \\
& & +  [ d^{\fN}d_{\fM}\mA^{\fP} - \mA_{\fM}d^{\fN}\mA^{\fP} - (D_h-4) d_{\fM}\mA^{\fP} \tmA^{\fN}  ] \eta_{\fN\fP}^{[\fA\fB]} = \beta_G^2M_S^2 a_s^2  \sinh^{2}\nu_S \, \tmA^{\fP} \eta_{\fM\fP}^{[\fA\fB]}. 
\ee
Following along the equation of motion presented in Eq.(\ref{EMfA}) or Eq.(\ref{GGEb}) for the hyper-gravigauge field $\fA_{\fM}^{\; \fA}(x)$ and also the alternative equation of motion shown in Eq.(\ref{GGEE}) or Eq.(\ref{GGEh}) for the hyper-gravimetric field $\mH_{\fM\fN}(x)$, we obtain the following equation of motion for the background field:
\be \label{BHGGF}
& & \Mka^2 a_s^2 \{ (D_h-2) d_{\fM}\mOm_{\fN} - [ (D_h-2) d_{\fP}\mOm^{\fP} + \frac{1}{2}(D_h-1)(D_h-2) \mOm_{\fP}\mOm^{\fP} ] \eta_{\fM\fN}  \} \nn \\ 
& & \quad =  (D_h-3) d_{\fM}\mA_{\fP} d_{\fN}\mA^{\fP} - d_{\fP}\mA_{\fM} d^{\fP}\mA_{\fN}  +  ( d_{\fM}\mA_{\fN} + d_{\fN}\mA_{\fM}) \left( d_{\fP}\mA^{\fP} + (D_h-2) \mA_{\fP}\mA^{\fP} \right) \nn \\ 
& &  \quad - [\, \frac{1}{2}(D_h-4) d_{\fP}\mA_{\fQ} d^{\fP}\mA^{\fQ} + (D_h-2) d_{\fP}\mA^{\fP} \mA_{\fQ}\mA^{\fQ} ) + \frac{1}{4}(D_h +1)(D_h - 4)  (\mA_{\fP}\mA^{\fP})^2  \nn \\
 & & \quad + \frac{1}{2} (d_{\fP}\mA^{\fP} )^2  ] \eta_{\fM\fN}  + \frac{1}{2} \beta_G^2 M_S^2 \sinh^{2}\nu_S \, [ \left(  2d_{\fP}\tmA^{\fP} + (D_h-1) \tmA_{\fP}\tmA^{\fP} + (D_h-2) \mOm_{\fP}\tmA^{\fP} \right) \eta_{\fM\fN} 
 \nn \\
& & \quad - d_{\fM} \tmA_{\fN}  - d_{\fN} \tmA_{\fM} - (D_h-1) \left( \mOm_{\fM} \tmA_{\fN} + \mOm_{\fN} \tmA_{\fM} - 2\tmA_{\fM} \tmA_{\fN}  \right) ] \nn \\
& &\quad  - \frac{1}{2}g_H^2 \Mka^2 a_s^2 [ \lambda_S^2 \p_{\fM}\nu_S \p_{\fN} \nu_S   - (\frac{1}{2} \lambda_S^2 \p_{\fP}\nu_S \p^{\fP} \nu_S  - \lambda_D^2 \Mka^2 a_s^2 \cF(\nu_S) ) \eta_{\fM\fN} ] , 
\ee
where $\cF(\nu_S)$ represents the total dimensionless potential function. 

The equation of motion for the dimensionless scaling field $\chi_s(x)$ presented in Eq.(\ref {EMCSF}) leads to the following equation of motion for the background $\cQ_c$-spin scaling field $\nu_S(x)$:
\be
\p^2 \nu_S + (D_h-2) \mOm^{\fP}\p_{\fP}\nu_S & = & -\lambda_D^2 \lambda_S^{-2}\Mka^2 a_s^2 \frac{\p}{\p \nu_S} \cF(\nu_S) \nn \\
& + & 2 (D_h-1)g_H^{-2} \beta_G^2\lambda_Q^{-2} \sinh\nu_S\cosh\nu_S \, \tmA^2 . 
\ee

It is still hard to solve analytically the above equations of motion in the general case. In particular, the equations of motion for the background hyperspin gauge field and hyper-gravigauge field remain to be quite complicated. We should examine the most interesting case by firstly concentrating on the equation of motion for the background $\cQ_c$-spin scaling field $\nu_S(x)$. For that, let us begin with considering rational hypotheses for the total dimensionless potential function and background hyperspin gauge field $\mA^{\fP}$. 

For the potential function $\cF(\nu_S)$ of background $\cQ_c$-spin scaling field $\nu_S(x)$, it is proposed to have the following feature:
\be \label{PF}
\frac{\p}{\p \nu_S} \cF(\nu_S) \to 0, \quad \cF(\nu_S) \to 1, \quad  \nu_S \gg 1 , 
\ee
which indicates that $\cF(\nu_S)$ is expected to vary slowly as the function of background $\cQ_c$-spin scaling field $\nu_S(x)$ for $\nu_S \gg 1$, so that the potential function can approximately be regarded as a slowly varying cosmological constant. 

For the background hyperspin gauge field $\mA^{\fP}$, it is simply supposed to be the same as the background hyperspin gravigauge field $\mOm^{\fP}$ based on the gravitational origin of hyperspin gauge field $\cA_{\fM}^{\fA\fB} \equiv \fOm_{\fM}^{\fA\fB} + \fA_{\fM}^{\fA\fB}$, which indicates the following relation:
\be
  \langle \cA_{\fM}^{\fA\fB} -\fOm_{\fM}^{\fA\fB} \rangle = \langle \fA_{\fM}^{\fA\fB} \rangle \to 0, \quad \mbox{i.e.} \quad \mA^{\fP} - \mOm^{\fP} \equiv \tmA^{\fP} \to 0 .
\ee 

From the above hypotheses and analyses, we arrive at the following simplified equation for the background $\cQ_c$-spin scaling field $\nu_S(x)$:
\be
\p^2 \nu_S + (D_h-2) \mOm^{\fP}\p_{\fP}\nu_S \simeq 0 ,
\ee
which brings on the following solution in a good approximation,
\be
\p_{\fP}\nu_S \simeq 0 . 
\ee
Meanwhile, the equation of motion for the background hyperspin gauge field $\langle \cA_{\fM}^{\fA\fB}(x) \rangle$ presented in Eq.(\ref{BHSGF}) is simplified for $\tmA^{\fP} \to 0$ to be as follows:
\be
 & & d_{\fQ}\mOm^{\fQ} \mOm^{\fP} - d_{\fQ}d^{\fQ}\mOm^{\fP} - 2\mOm_{\fQ}d^{\fP}\mOm^{\fQ} - (D_h-4) \mOm_{\fQ}d^{\fQ}\mOm^{\fP} \to 0 , 
\ee
which can be rewritten into the following form:
\be \label{BHSGFS}
3\p^2 a_s \p_{\fP} a_s - a_s\p^2 \p_{\fP} a_s - (D_h-3) ( a_s\p_{\fP}\p_{\fQ} a_s - 2\p_{\fP}a_s \p_{\fQ}a_s ) \p^{\fQ}\ln a_s = 0 .
\ee

It can be checked that the equation of motion for the background hyper-gravigauge field presented in Eq.(\ref{BHGGF}) can be simplified in a good approximation into the following two equations:
\be \label{BHGGFS}
& & \Mka^2 a_s^2 (D_h-2) d_{\fM}\mOm_{\fN}  \simeq  (D_h-3) d_{\fM}\mOm_{\fP} d_{\fN}\mOm^{\fP} - d_{\fP}\mOm_{\fM} d^{\fP}\mOm_{\fN}  \nn \\
& & \quad +  ( d_{\fM}\mOm_{\fN} + d_{\fN}\mOm_{\fM}) \left( d_{\fP}\mOm^{\fP} + (D_h-2) \mOm_{\fP}\mOm^{\fP} \right) \nn \\
& &  \Mka^2 a_s^2 [ (D_h-2) d_{\fP}\mOm^{\fP} + \frac{1}{2}(D_h-1)(D_h-2) \mOm_{\fP}\mOm^{\fP} ]  \nn \\ 
& & \quad  \simeq \frac{1}{2}(D_h-4) d_{\fP}\mOm_{\fQ} d^{\fP}\mOm^{\fQ} + (D_h-2) d_{\fP}\mOm^{\fP} \mOm_{\fQ}\mOm^{\fQ}  \nn \\ 
& & \quad + \frac{1}{4}(D_h +1)(D_h - 4)  (\mOm_{\fP}\mOm^{\fP})^2  + \frac{1}{2} (d_{\fP}\mOm^{\fP} )^2  +  \frac{1}{2}g_H^2 \lambda_D^2 \Mka^4 a_s^4  . 
\ee

To obtain an analytic solution, let us first consider the simplified equation (\ref{BHSGFS}). Suppose that Eq.(\ref{BHSGFS}) holds in any dimension $D_h$ of hyper-spacetime, which leads the background scaling factor $a_s(x)$ to satisfy the following equations:
\be
& &  a_s\p_{\fP}\p_{\fQ} a_s - 2\p_{\fP}a_s \p_{\fQ}a_s = 0 , \nn \\
& & 3\p^2 a_s \p_{\fP} a_s - a_s\p^2 \p_{\fP} a_s = 0 , 
\ee
which are found to be equivalent to the following two identities:
\be
& & d_{\fP}\mOm_{\fQ} = 0, \quad  2\mOm^{\fP} \p_{\fQ}\mOm^{\fQ}  - \p^2\mOm^{\fP} =0 .
\ee
With these two identities, two equations in Eq.(\ref{BHGGFS}) can be simplified further into the following single equation:  
\be \label{BHFS}
\Mka^2 a_s^2 (D_h-1)(D_h-2) \mOm_{\fP}\mOm^{\fP} \simeq g_H^2 \lambda_D^2 \Mka^4 a_s^4 + \frac{1}{2}(D_h +1)(D_h - 4)  (\mOm_{\fP}\mOm^{\fP})^2 .
\ee

A general solution for the background scaling factor $a_s(x)$ is found to be as follows:
\be \label{GSGSF}
a_s(x) = \pm \frac{ \apk m_{\kappa}/\Mka }{ 1 \mp \kappa_{\fM}( x^{\fM} - x_o^{\fM} )  } ,
\ee
where $\kappa_{\fM}$ is regarded as {\it basic cosmic vector}, which defines {\it fundamental cosmic mass scale} $m_{\kappa}$ via the Lorentz-type invariant scalar product as follows: 
\be
& & \kappa^2 = \kappa_{\fM}\kappa^{\fM} \equiv m_{\kappa}^2 \cka^2,
\ee 
where $\cka$ may be referred to as {\it invariant proper speed} for convenience of mention (we will take it as a basic unit with the convention $\cka=1$). $x_o^{\fM}$ is viewed as a reference coordinate position chosen appropriately by an observer in conformally flat background hyper-spacetime, which leads the solution to be translational invariant, 
\be
& & x^{\fM} \to x^{' \fM} =  x^{\fM} + \varpi^{\fM},  \quad x_o^{\fM} \to x_o^{' \fM} =  x_o^{\fM} + \varpi^{\fM}, \nn \\
& & a_s(x) \to a_s (x + \varpi ) = a_s(x) .
\ee
It can be verified that the general solution presented in Eq.(\ref{GSGSF}) becomes consistent when the constant $\apk$ meets to the constraint obtained from the equation given in Eq.(\ref{BHFS}), i.e.: 
\be \label{ECS}
& & 4g_H^2\lambda_D^2 \apk^4  - 4 (D_h-1)(D_h-2) \apk^2 +  (D_h+1)(D_h-4) = 0 , 
\ee  
which implies that the constant $\apk$ depends not only on the coupling constants but also on the dimension of hyper-spacetime. For convenience of mention, $\apk$ is referred to as {\it conformal scaling constant}. 

It is easy to check that the equations of motion have a discrete symmetry under $Z_2$ transformation: $a_s(x) \to - a_s(x)$, and also reflection symmetry for coordinates: $x^{\fM}\to - x^{\fM}$, $a_s(x)\to a_s(-x)$. So that there are in general four solutions for the background graviscalar field,
\be  \label{BGSF}
\phi_o(x) \equiv \Mka a_s(x) \to \pm \phi_o^{\pm}(x) = \pm \frac{ \apk m_{\kappa} }{ 1 \mp \kappa_{\fM}( x^{\fM} - x_o^{\fM} )  } ,
\ee
where the conformal scaling constant $\apk$ is easily obtained from Eq.(\ref{ECS}) as follows:
\be \label{TSL}
\apk^2 = \frac{(D_h-1)(D_h-2)}{2g_H^2\lambda_D^2} \{ 1 \pm \left( 1- \frac{ (D_h+1)(D_h-4) g_H^2 \lambda_D^2}{(D_h-1)^2(D_h-2)^2 } \right)^{1/2} \} . 
\ee

It is obvious that when hyper-spacetime is reduced to be four-dimensional one, the conformal scaling constant $\apk$ gets the usual solution in four dimensional spacetime\cite{GQFT},
\be
 \apk  = \frac{\sqrt{6}}{(g_H\lambda_D)} .
\ee
It is interesting to notice that for hyper-spacetime beyond four dimensions ($D-h>4$), the general solution Eq.(\ref{GSGSF}) for the background scaling factor is still valid even if the potential of background $\cQ_c$-spin scaling field is absent ($\lambda_Q =0$). In this case, the conformal scaling constant $\apk$ is simply given by,
\be
 \apk^2 =  \frac{(D_h+1)(D_h-4)}{4  (D_h-1)(D_h-2)} .
\ee
In fact, for $g_H\lambda_D \alt O(1)$, the two solutions in Eq.(\ref{TSL}) can be expressed in a good approximation as follows:
\be \label{APK}
\alpha_{\kappa +}^{2} & \simeq & \frac{(D_h-1)(D_h-2)}{g_H^2\lambda_D^2} [ \, 1 - \frac{ (D_h+1)(D_h-4) g_H^2 \lambda_D^2}{2(D_h-1)^2(D_h-2)^2 } \, ] , \nn \\
\alpha_{\kappa -}^{2} & \simeq &  \frac{(D_h+1)(D_h-4)}{4  (D_h-1)(D_h-2)} [ \, 1 + \frac{ (D_h+1)(D_h-4) g_H^2 \lambda_D^2}{2 (D_h-1)^2(D_h-2)^2  } \, ]  .
\ee
For the hyper-spacetime dimension $D_h=19$ in the hyperunified field theory, we can write down the conformal scaling constant $\apk$ as follows:
 \be
\alpha_{\kappa +} & \simeq & \frac{3\sqrt{34}}{g_H\lambda_D} \left(\, 1 - \frac{ 150 g_H^2 \lambda_D^2}{18^2\cdot 17^2 } \, \right)^{1/2} , \nn \\
\alpha_{\kappa -} & \simeq &  \frac{5\sqrt{3}}{3\sqrt{34}} \left(\, 1 + \frac{ 150 g_H^2 \lambda_D^2}{18^2\cdot 17^2 }  \, \right)^{1/2}  .
\ee


\subsection{ Evolution of early universe with evolving graviscalar and $\cQ_c$-spin scalar field in conformally flat gravigauge hyper-spacetime as nonsingularity background spacetime }

To study the evolution of early universe, let us demonstrate firstly the basic property of background gravigauge hyper-spacetime with the following background line element: 
\be
& & \langle d^2l_S \rangle =  a_s^2(x)\, dx^{\fM} \eta_{\fM\fN}  dx^{\fN}  , \nn \\
& & a_s^2 (x) \to (a_s^{\pm}(x))^2 = \frac{ \apk^2 m_{\kappa}^2/\Mka^2 }{ \left( 1 \mp \kappa_{\fM}( x^{\fM} - x_o^{\fM} ) \right)^2 } , 
\ee
which is invariant under the following Poincar\'e-type transformation: 
\be
& & x^{\fM} \to x^{'\fM} = L^{\fM}_{\;\; \fN}\, x^{\fN}, \quad x_o^{\fM} \to x_o^{'\fM} = L^{\fM}_{\;\; \fN}\, x_o^{\fN}, \quad \kappa_{\fM} \to \kappa'_{\fM}= L_{\fM}^{\;\; \fN}\, \kappa_{\fN}, \nn \\
& & x^{\fM} \to x^{'\fM} = x^{\fM} + \varpi^{\fM} , \quad x_o^{\fM} \to x_o^{'\fM} = x_o^{\fM} + \varpi^{\fM} .
\ee
In general, the existence of basic cosmic vector $\kappa_{\fM}$ indicates that the background gravigauge hyper-spacetime is not isometric and homogeneous. 

The Poincar\'e-type invariance of the scalar product  $\kappa_{\fM} ( x^{\fM} - x_o^{\fM} )$ enables us to define a {\it conformal proper time} $\htt$ as follows:
\be
\kappa_{\fM} ( x^{\fM} - x_o^{\fM} ) \equiv \hmk \cka^2 (\htt - \htt_o) ,
\ee 
where $\hmk$ is regarded as {\it conformal proper cosmic mass scale}, $\htt_o$ is considered as {\it reference conformal proper time}. In general, the definitions of $\htt$ ($\htt_o$) and $\hmk$ allow to have the following intrinsic conformal scaling invariant transformation: 
\be
\htt \to \htt' = \lambda^{-\alpha}\, \htt ,  \quad \htt_o \to \htt'_o = \lambda^{-\alpha}\, \htt_o,  \quad \hmk  \to \hmk'  = \lambda^{\alpha} \hmk ,
\ee
where the power $\alpha$ is constant parameter. For $\kappa^2 = \kappa_{\fM}\kappa^{\fM} = m_{\kappa}^2 > 0$, one can always choose a typical reference frame by making a special Lorentz-type transformation, so that 
\be \label{CCLT}
& & \kappa_{\fI} =0, \quad \fI = 1,2,\cdots, D_h , \nn \\
& & \hmk = \kappa_0 = \mk , \quad \htt = \eta, \; \htt_o = \eta_o, \quad \alpha =0 , 
\ee
which is referred to as {\it cosmic co-moving reference frame}. Where $\eta$ is referred to as {\it cosmic co-moving Lorentz time} and $\eta_o$ as {\it cosmic co-moving reference Lorentz time}. Only in such a cosmic co-moving reference frame, the intrinsic conformal scaling transformation is fixed with $\alpha=0$. 

In light of conformal proper time $\htt$, the temporal coordinate $x^0$ can be expressed as follows: 
\be 
& & x^0 = \frac{\hmk\cka}{\kappa_0}\, \cka \htt - \frac{\kappa_I}{\kappa_0} x^I \equiv  \frac{1}{u_0} ( \cka \htt - u_I x^I )  , \nn \\
& & u_0 \equiv \frac{\kappa_0}{\hmk\cka }\, , \quad  u_I \equiv  \frac{\kappa_I }{\hmk\cka} ,
\ee  
so that the coordinates $x^{\fM}$ and displacement $dx^0$ are replaced by,
\be 
& &  x^{\fM} \to \hx^{\fM} = (\cka \htt, x^I), \nn \\
& &  dx^0 = \frac{1}{u_0} ( \cka d\htt - u_I dx^I ) .
\ee 
The line element can be rewritten as follows:
\be
& & \langle d^2l_{S} \rangle = a_s^2(\htt)  d\hx^{\fM} \chi_{\fM\fN}^{c} d\hx^{\fN}   , \nn \\
& & a_s^2(x) \to \left(a_s^{\pm}(\htt)\right)^2 = \frac{ \apk^2 m_{\kappa}^2/\Mka^2 }{ \left( 1 \mp \hmk (\htt - \htt_o) \cka^2 \right)^2 } ,
\ee
where $a_s(\htt)$ evolves as the graviscaling field of conformal proper time $\htt$ with $\hx^{\fM}$ as the coordinates in terms of conformal proper time $\htt$.  $\chi_{\fM\fN}^c$ is the {\it conformal proper time background metric} with the following form: 
\be
\chi_{\fM\fN}^c = \frac{1}{u_0^2}
\left(
\begin{array}{cc}
 1 &  -u_J     \\
 -u_I  &    u_0^2  \chi_{IJ}^c
\end{array}
\right)   , \quad 	 \chi_{IJ}^c = \eta_{IJ} + \frac{u_I u_J}{u_0^2 } , 
\ee
which is determined by the basic cosmic vector $\kappa_{\fM}$. 

In general, the conformal proper time background metric is non-diagonal. Only in the cosmic co-moving reference frame, the background gravigauge hyper-spacetime becomes isotropic and homogenous with the following features:  
\be
& & \kappa_{\fM} \to \kappa_0 =\mk\cka, \quad \htt \to \eta , \quad  \hx^{\fM} \to x_{\eta}^{\fM} = (\cka\eta, x^{I}) , \quad  d\hx^{\fM} \to dx_{\eta}^{\fM} , \nn \\
& &  \chi_{\fM\fN}^c \to  \eta_{\fM\fN} , \quad \langle d^2l_S \rangle = a_s^2(\htt)  d\hx^{\fM}  \chi_{\fM\fN}^c d\hx^{\fN} = a_s^2(\eta) dx_{\eta}^{\fM} \eta_{\fM\fN} dx_{\eta}^{\fN}   , 
\ee
where the Lorentz-type symmetry is fixed to specify both Lorentz time $t$ and conformal proper time $\htt$ to be the same as cosmic co-moving Lorentz time, i.e., $\htt = t = \eta$. 

Let us now investigate the feature of background graviscaling field in the case $\nu_S(x) \gg 1$. Formally, it appears to have a singularity when the conformal proper time or cosmic co-moving Lorentz time relative to its reference time approaches to the epoch given by the inverse of conformal proper cosmic mass scale or fundamental cosmic mass scale. Namely, when the conformal proper time or cosmic co-moving Lorentz time goes to the following critical point: 
\be
& & \htt \to \htt_c^{\pm} \equiv \htt_o \pm \frac{1}{\hmk\cka^2} \equiv \htt_o \pm \htk , \quad \htk \equiv \frac{1}{\hmk \cka^2}, \quad \mbox{or} \nn \\
& & \eta \to \eta_c^{\pm} \equiv \eta_o \pm \frac{1}{\mk\cka^2} \equiv \eta_o \pm \eta_{\kappa} , \quad \eta_{\kappa} \equiv  \frac{1}{\mk \cka^2 }, 
\ee
the conformal size of background gravigauge hyper-spacetime becomes infinitely large,
\be
& & a_s(\eta)\to a_s^{\pm}(\eta_c^{\pm} ) , \;\; a_s^{\pm}(\htt_c^{\pm} ) \rightarrow  \infty .
\ee
In this situation, the distance relative to a reference point traveled by a field with invariant proper speed $\cka$ is given as follows:  
\be
\Lk = \eta_{\kappa} \cka = \frac{1}{\mk\cka}, \quad \mbox{or} \quad \hLk = \htk \cka  = \frac{1}{\hmk\cka}  , 
\ee
where $\Lk$ is referred to as {\it co-moving cosmic horizon} or $\hLk$ as {\it conformal proper cosmic horizon} in background gravigauge hyper-spacetime. 

Nevertheless, we will show that in view of {\it conformal proper cosmic time} or {\it co-moving cosmic time}, it takes an infinitely long time to reach the conformal proper cosmic horizon $\hLk$ or co-moving cosmic horizon $\Lk$. Let us now introduce {\it conformal proper cosmic time} $\htau$ and {\it co-moving cosmic time} $\tau$ as follows:
\be
& & d\htau \equiv a_s(\htt) d\htt ,  \quad \mbox{or} \quad d\tau = a_s(\eta) d\eta , \nn \\
& & a_s(\htt ) \to a_s^{\pm}(\htt ) = \frac{\apk m_{\kappa} /\Mka}{ 1 \mp \hmk \cka^2 (\htt - \htt_o) }  = \frac{\apk m_{\kappa} /\Mka}{ 1 \mp \cka(\htt - \htt_o)/\hLk } , \nn \\
& & a_s(\eta) \to a_s^{\pm}(\eta) = \frac{\apk m_{\kappa} /\Mka}{ 1 \mp \mk \cka^2 (\eta - \eta_o) }  = \frac{\apk m_{\kappa} /\Mka}{ 1 \mp \cka (\eta- \eta_o)/\Lk } ,
\ee
which enables us to obtain the following relation between the conformal proper time $\htt$ and conformal proper cosmic time $\htau$ or between the cosmic co-moving Lorentz time $\eta$ and co-moving cosmic time $\tau$:
\be
 & & \frac{1}{1 - \hmk \cka^2 (\htt - \htt_o)} = e^{\hHk \cka^2 (\htau - \htau_o)} , \;\; -\infty \leq \cka (\htt-\htt_o) \leq \hLk  , \;\;  -\infty \leq \htau-\htau_o \leq \infty  , \nn \\
 & & \frac{1}{1 - \mk \cka^2 (\eta - \eta_o)} = e^{\Hka \cka^2 (\tau - \tau_o)} , \;\; -\infty \leq \cka (\eta-\eta_o) \leq \Lk  , \;\;  -\infty \leq \tau-\tau_o \leq \infty  , \nn \\
 & & \frac{1}{1 + \hmk \cka^2 (\htt - \htt_o) } = e^{-\hHk \cka^2 (\htau - \htau_o)} , \;\; -\hLk \leq \cka (\htt-\htt_o) \leq \infty  , \;\;  -\infty \leq \htau-\htau_o \leq \infty   , \nn \\
 & & \frac{1}{1 + \mk \cka^2 (\eta - \eta_o) } = e^{-\Hka \cka^2 (\tau - \tau_o)} , \;\; -\Lk \leq \cka (\eta-\eta_o) \leq \infty  , \;\;  -\infty \leq \tau-\tau_o \leq \infty   ,
 \ee
where the normalization is fixed by the integration condition at initial points, i.e, $\htt = \htt_o$ and $\htau = \htau_o$ or $\eta = \eta_o$ and $\tau = \tau_o$ when carrying out the integrations over $\htt$ and $\htau$ or $\eta$ and $\tau$. $\tau_o$ is referred to as {\it co-moving reference cosmic time} and $\htau_o$ as {\it conformal proper reference cosmic time}. 

The conformal scaling factor in light of conformal proper cosmic time or co-moving cosmic time is found to have the following form:
\be
& &  a_s(\htau) \to a_s^{\pm}(\htau) = a_o e^{\pm \hHk c(\htau-\htau_o)} = a_o e^{\pm \Hka c (\tau - \tau_o)} = a_s^{\pm}(\tau),  \nn \\
& & a_o = \apk \frac{\mk}{\Mka} \equiv \frac{\mk}{\Hka} = \frac{\lka}{\Lk} \equiv \frac{\hmk}{\hHk} = \frac{\hlka}{\hLk}  ,
\ee
where $a_o$ is regarded as {\it reference conformal scaling factor} at {\it co-moving reference cosmic time}. We have introduced the following definitions:
 \be \label{HMS}
 & & \Hka \equiv \frac{\Mka}{\apk}, \quad \lka \equiv \frac{1}{\Hka} , \nn \\
 & & \hHk \equiv \frac{\Mka}{\apk} \frac{\hmk}{\mk}  ,  \quad \hlka \equiv \frac{1}{\hHk}  ,
 \ee
where $\Hka$ is referred to as {\it primordial Hubble-type mass scale} and $\hHk$ as {\it conformal proper primordial Hubble-type mass scale}, and $\lka$ is referred to as {\it primordial Hubble-type length scale} and $\hlka$ as {\it conformal proper primordial Hubble-type length scale}. 

It is noted that the conformal scaling factor $a_s(x)$ is regarded as {\it intrinsic scaling invariant quantity} with $a_s(\htau) = a_s(\tau)$, while the conformal proper primordial Hubble-type mass scale $\hHk$ is not an intrinsic scaling invariant quantity as its value depends on the choice of conformal proper cosmic time $\htau$. Only in the cosmic co-moving reference frame, the conformal proper primordial Hubble-type mass scale $\hHk$ turns to be the primordial Hubble-type mass scale $\Hka = \Mka/\apk$ which is regarded as {\it primordial Hubble-type constant}.

It is clear that when the distance traveled via the invariant proper speed $\cka$ in background spacetime is closing to the co-moving cosmic horizon $\cka |\eta -\eta_o| \to \Lk$, it needs to take an infinitely large co-moving cosmic time $|\tau -\tau_o| \to \infty$. 

The line element in light of the conformal proper cosmic time can be expresses as follows:
\be
& & \langle d^2l_S \rangle =  \chi_{\fM\fM}^c(\htau) d\hx_{\tau}^{\fM} d\hx_{\tau}^{\fN}  , \quad  \hx_{\tau}^{\fM} = (c\htau, x^I) ,
\ee
where the metric tensor $\chi_{\fM\fN}^c (\htau)$ is given by, 
\be
& & \chi_{\fM\fN}^c (\htau) = \frac{1}{u_0^2}
\left(
\begin{array}{cc}
 1 &  - a_s(\htau)\, u_J     \\
 - a_s(\htau)\, u_I  &     a_s^2(\htau)\, u_0^2\, \chi_{IJ}^c
\end{array}
\right)  , \quad 	 \chi_{IJ}^c = \eta_{IJ} + \frac{u_I u_J}{u_0^2 } , 
\ee
which indicates that the background gravigauge hyper-spacetime is in general not isotropic in terms of the conformal proper cosmic time $\htau$. 

Only in the cosmic co-moving reference frame with $u_I =0$ and $u_0=1$, the co-moving cosmic time $\tau$ defined from the cosmic co-moving Lorentz time $\eta$ describes an isotropic and homogeneous background gravigauge hyper-spacetime with the following line element:
\be
& & \langle d^2l_S \rangle =  \chi_{\fM\fN}^c(\tau) dx_{\tau}^{\fM}dx_{\tau}^{\fN} =  \cka^2 d\tau^2 -  a_s^2(\tau)  (dx^{I})^2 , 
\ee
with the definitions,
\be
& & x_{\tau}^{\fM} = (\cka\tau, x^I), \quad 
\chi_{\fM\fN}^c (\tau) = 
\left(
\begin{array}{cc}
 1 &     0   \\
 0  &     a_s^2(\tau) \eta_{IJ}
\end{array}
\right)  . 
\ee
For the positive graviscaling factor,
\be
a_s^{+}(\tau) = a_o\, e^{\Hka \cka^2 (\tau-\tau_o)} ,
\ee
which brings on the well-known Friedmann-Lema\^{i}tre-Robertson-Walker (FLRW) metric for describing inflationary expansion of early universe as the time arrow from the past or reference point to the future. Obviously, the negative solution for the graviscaling factor $a_s^{-}(\tau) = a_o\, e^{-\Hka \cka^2 (\tau-\tau_o)}$ characterizes deflationary expansion of early universe when the time arrow from the past or reference point to the future. Whereas once regarding such a negative solution as a time reversal one, it reflects inflationary expansion of universe for the time arrow from the future or reference point to the past.    

In general, such an inflationary expansion of early universe is going to break down and end up eventually only when the potential $\cF(\nu_S)$ of background $\cQ_c$-spin scaling field $\nu_S(x)$ falls into a minimum value, where the background field $\nu_S(x)$ decreases and approaches to its minimum point $\nu_S(x) \to \nu_0 $, i.e.:
\be \label{MC}
& & \cF'(\nu_S) \equiv \frac{\p}{\p \nu_S} \cF(\nu_S) = 0 , \quad \mbox{at} \quad \nu_S = \nu_0 , \nn \\
& & \cF(\nu_S = \nu_0 ) \ll 1 .
\ee
So that the background graviscaling field as conformal scaling factor will not get into an infinitely large value since an inflationary expansion of early universe will stop before the conformal proper time (or cosmic co-moving Lorentz time) relative to its reference time reaches to the singularity point characterized by the conformal proper cosmic horizon (or co-moving cosmic horizon). 

Let us suppose that the universe evolves from conformally flat gravigauge hyper-spacetime as nonsingularity background spacetime in which the potential of background $\cQ_c$-spin scaling field varies slowly for a large value $\langle \chi_s(x)\rangle = \nu_S(x) \gg 1$. Begin with such a nonsingularity background hyper-spacetime as inflationary early universe, the hyperunified qubit-spinor field and hyperspin gauge field as well as $\cQ_c$-spin scalar field and scaling gauge field behave as very heavy massive particles, so that all fundamental particles except hyper-gravigauge field are considered to decouple from interactions at the very beginning of universe. Nevertheless, when approaching to the end of inflationary expansion of early universe once the potential reaches to the minimum point $\nu_S = \nu_0$ with $\nu_0 \ll 1$, the hyperunified qubit-spinor field and hyperspin gauge field are expected to receive negligible masses in proportional to an extra small factor $ \sinh\nu_S \simeq \nu_S=\nu_0 \ll 1$.  

It is noted that the above analytic solution for the background graviscaling field $a_s(x)$ is obtained in a good approximation for the slowly varying potential of background $\cQ_c$-spin scaling field, which describes the evolution of early universe and will be discussed further below. Similar inflationary scenarios with two or more general scalar fields based on the scaling gauge invariance have been studied extensively in refs.\cite{TW1,TW2,TW3}, which demonstrated that such kind of inflationary scenarios can provide a consistent explanation for the observed data. We will examine below an explicit potential of $\cQ_c$-spin scaling field to verify all required slow-roll conditions.


\subsection{  Evolution of early universe with inflationary expansion governed by the potential energy of $\cQ_c$-spin scalar field}

At the very beginning of inflationary expansion of universe, all fundamental fields except the hyper-gravigauge field appears as very heavy particles with masses characterized by the fundamental graviscaling mass scale $M_S$, such an early universe is governed by the slowly varying potential energy of background scaling field $\nu_S$. The kinetic terms of all heavy massive particles are expected to be negligible smaller in comparison with their mass-like terms, so that all such basic fields are thought to decouple from interactions at the beginning of early universe. The slow-roll conditions are generally characterized by the following slow-roll parameters:
\be
\epsilon & = &  \frac{1}{2}\frac{\Mka^2}{g_H^2\lambda_Q^2 M_S^2} \left( \frac{\cF'(\nu_S)}{\cF(\nu_S)} \right)^2 \ll 1  , \nn \\
\eta & = & \frac{\Mka^2}{g_H^2\lambda_Q^2 M_S^2} \frac{\cF''(\nu_S)}{\cF(\nu_S)} \ll 1 .
\ee
When the background field $\nu_S$ goes through down to the magnitude which makes $\epsilon \sim 1$ and $|\eta| \sim 1$, the kinetic energy of $\cQ_c$-spin scaling field becomes compatible with the potential energy, which invalidates the slow-roll conditions. In this situation, the inflationary expansion of early universe starts to break down and the potential of $\cQ_c$-spin scaling field is quickly falling into the minimum point at $\nu_S=\nu_0$ with $\nu_0 \ll 1$. Meanwhile, the varying masses of hyperunified qubit-spinor field and hyperspin gauge field as well as $\cQ_c$-spin scalar field become negligible small as their masses are proportional to the small factor $\sinh\nu_0 \simeq \nu_0 \ll 1$, only the scaling gauge field obtains a finite mass characterized by the fundamental mass scale $\Mka$. 

To see explicitly, let us rewrite the $\cQ_c$-spin scaling field around the minimum point $\nu_0$ as follows:
\be
& & \chi_s(x) \equiv \nu_0 + \frac{\Phi(x)}{M_S} , 
\ee
where $\Phi(x)$ is regarded as quantized $\cQ_c$-spin scalar field. In this case, the action in Einstein-type basis is easily read from the action given in Eq.(\ref{actionHUFTGQFTE}) to be as follows:
\be  \label{actionHUFTGQFTS}
\cS_{\mH\mU} & \equiv &  \int [d^{D_h}x] \, \chi^e \Mka^{D_h-4} \cL_{D_h} \equiv  \int [d^{D_h}x] \, \chi^e \Mka^{D_h-4} \{\, \chih^{e \fM\fN} \bar{\fPsi}_{\fQH} \bchi^e_{\fM\fC}  \vSi_{-}^{\fC} i\cD_{\fN}  \fPsi_{\fQH}  \nn \\
& - &  m_{\fQH} \bar{\fPsi}_{\fQH} \tvSi_{-}\fPsi_{\fQH} 
- 2M_S \sinh( \frac{\Phi}{2M_S} ) \cosh(\nu_0 + \frac{\Phi}{2M_S} ) \,  \bar{\fPsi}_{\fQH} \tvSi_{-}\fPsi_{\fQH}  \nn \\
& - &  \chih^{e \fM\fM'} \chih^{e \fN\fN'} \frac{1}{4g_H^2} \cF_{\fM\fN\fA\fB}  \cF_{\fM'\fN'}^{\fA\fB}  
+ \frac{1}{2g_H^2} \Mka^2 \chih^{e \fM\fN} \mR^e_{\fM\fN} \nn \\
& + & \frac{1}{2g_H^2} \mu_{\fA}^2  ( \cA_{\fM\fA\fB} -  \fOm_{\fM\fA\fB} )(\cA_{\fM'}^{\fA\fB} -\fOm_{\fM'}^{\fA\fB} ) \chih^{e \fM\fM'}  \nn \\
& + & \frac{1}{2g_H^2} \beta_G^2  M_S^2 [\sinh^2(\frac{\Phi}{M_S})  +  2 \sinh(\nu_0)\sinh(\frac{\Phi}{M_S}) \cosh (\nu_0 + \frac{\Phi}{M_S} ) \, ] \nn \\   
& \cdot & ( \cA_{\fM\fA\fB} -  \fOm_{\fM\fA\fB} )(\cA_{\fM'}^{\fA\fB} -  \fOm_{\fM'}^{\fA\fB} ) \chih^{e \fM\fM'}  \nn \\
& - & \chih^{e \fM\fM'} \chih^{e \fN\fN'} \frac{1}{4g_W^2} \mF_{\fM\fN} \mF_{\fM'\fN'} + \frac{1}{2} \mu_{\mW}^2  \mW_{\fM} \mW_{\fN}   \chih^{e \fM\fN}  \nn \\
& + &  \frac{1}{2}\lambda_Q^2 M_S^2 [\sinh^2(\frac{\Phi}{M_S})  +  2 \sinh(\nu_0)\sinh(\frac{\Phi}{M_S}) \cosh (\nu_0 + \frac{\Phi}{M_S} ) \, ]  \, \mW_{\fM} \mW_{\fN} \chih^{e \fM\fN}, \nn \\
&  + &  \frac{1}{2} \chih^{e \fM\fN}\lambda_Q^2 \p_{\fM} \Phi \p_{\fN}\Phi -  \Lambda_Q^4 
- \frac{1}{2} \mu_{Q}^2 \Phi^2  \nn \\
& - &  \frac{\lambda_D^2}{\beta_Q^4} M_S^4 [\, \cF(\nu_0 + \frac{\Phi}{M_S})  - \cF(\nu_0)  - \frac{1}{2} \cF''(\nu_0) \frac{\Phi^2}{M_S^2} \, ] \} ,
\ee
with the following definitions:
\be
& & m_{\fQH} = M_S \sinh(\nu_0) , \quad   \mu_{\fA} = \beta_G M_S \sinh(\nu_0) , \nn \\
& & \mu_{\mW} = \lambda_Q M_S  (1 + \sinh^2(\nu_0) )^{1/2}, \quad \Lambda_Q^4 = \lambda_D^2 \Mka^4 \cF(\nu_0) , \nn \\
& &  \mu_{Q}^2 = \frac{\lambda_D^2}{\beta_Q^4} M_S^2 \cF''(\nu_0) , \quad \cF''(\nu_0) \equiv \frac{ \p^2 \cF(\chi_s)}{\p^2 {\chi_s}} |_{\chi_s=\nu_0} , 
\ee
where $m_{\fQH}$ is regarded as the mass of hyperunified qubit-spinor field in Einstein-type basis. $\mu_{\fA}$, $\mu_{\mW}$ and $\mu_{Q}$ characterize the masses of hyperspin gauge field, scaling gauge field and $\cQ_c$-spin scalar field in hyper-spacetime, respectively. $\cF(\nu_0 + \frac{\Phi}{M_S})$ is the potential of $\cQ_c$-spin scalar field. The $\Lambda_Q^4$ term which arises from the minimum potential reflects {\it cosmic energy density} in Einstein-type basis after inflation. 

For a more concrete and realistic consideration, let us examine the following simple potential $\cF(\chi_s)$ of $\cQ_c$-spin scaling field $\chi_s(x)$:
\be
& & \cF(\chi_s) = \frac{\sinh^4(\chi_s) + \lak^2}{( \lachi^{-2} + \sinh^{2}(\chi_s) )^2 } , 
\ee
with $\lak$ and $\lachi$ two constant parameters. The minimal condition of the potential is found to be as follows: 
\be
& & \frac{\p}{\p \chi_s} \cF(\chi_s) =   \frac{ 2 \sinh(2\chi_s) ( \lachi^{-2} \sinh^{2}(\chi_s) - \lak^2 ) }{ ( \lachi^{-2} + \sinh^{2}(\chi_s) )^3} = 0,  
\ee
which leads to the following solution on the minimum point $\chi_s \equiv \nu_0$:
\be
\sinh^2 \chi_s = \lachi^2 \lak^2 , \quad \chi_s \equiv \nu_0 = \pm \sinh^{-1}(\lachi\lak) . 
\ee

For the background $\cQ_c$-spin scaling field $\langle \chi_s(x)\rangle = \nu_S(x)$, it can be checked that the above simple potential can satisfy all required properties shown in Eqs.(\ref{PF}) and (\ref{MC}) for inflationary expansion of early universe as long as the constant parameters $\lak$ and $\lachi$ meet to the following conditions:
\be
\lak \ll 1 , \quad \lachi > 1 , 
\ee
which brings on the needed feature as follows:
\be 
& & \frac{\p}{\p \nu_S} \cF(\nu_S) \to 0, \;\;  \cF(\nu_S) \to 1,  \quad \nu_S \gg 1, \nn \\
& & \cF(\nu_0) = \frac{\lachi^4\lak^2}{1 + \lachi^4\lak^2 } \ll 1 . 
\ee

In general, let us verify the slow-roll conditions by considering the so-called slow-roll parameters $\epsilon$ and $\eta$ (it should not be confused with the cosmic co-moving Lorentz time $\eta$ defined in Eq.(\ref{CCLT}) for $\nu_S \gg 1$, 
\be \label{SRC1}
\epsilon & = &  \frac{1}{2}\frac{\Mka^2}{g_H^2\lambda_Q^2 M_S^2} \left( \frac{\cF'(\nu_S)}{\cF(\nu_S)} \right)^2 = \frac{2}{g_H^2\lambda_S^2} \frac{\sinh^{2}(2\nu_S) \left( \sinh^{2}\nu_S -  \lachi^{2}\lak^2\right)^2 }{\left(1 + \lachi^2\sinh^{2}\nu_S \right)^2 \left( \sinh^{4}(\nu_S) + \lak^2 \right)^2},  \nn \\
& \simeq & \frac{8\cosh^{2}(\nu_S) }{g_H^2\lambda_S^2 \left(1 + \lachi^2\sinh^{2}\nu_S \right)^2 \sinh^{2}\nu_S } \simeq  \frac{8 \coth^2 \nu_S} {g_H^2\lambda_S^2\lachi^4 (1 + \lachi^2 \sinh^2\nu_S)^2 }
\sim 128 g_H^{-2}\lambda_S^{-2} \lachi^{-4} e^{- 4\nu_S} ,  \\
\eta & = & \frac{\Mka^2}{g_H^2\lambda_Q^2 M_S^2} \frac{\cF''(\nu_S)}{\cF(\nu_S)} \nn \\
& = & \frac{ 4 (1 + \lachi^2 \sinh^2(\nu_S) )\cosh(2\nu_S) ( \sinh^2(\nu_S) -\lachi^2\lak^2 ) + 2\sinh^{2}(2\nu_S) ( 1 - 2\lachi^2\sinh^2(\nu_S) + 3\lachi^4\lak^2 ) } { g_H^2\lambda_S^2 ( 1+ \lachi^2 \sinh^2(\nu_S) )^2 ( \sinh^{4}(\nu_S) + \lak^2 )}, \nn \\
& \simeq & \frac{ 4 \cosh(2\nu_S) (1 + \lachi^2 \sinh^2(\nu_S) ) + 8 \cosh^2(\nu_S) ( 1 - 2 \lachi^2 \sinh^2(\nu_S) ) }{ g_H^2\lambda_S^2 (1 + \lachi^2 \sinh^2(\nu_S) )^2 \sinh^2(\nu_S) }  \nn \\
& \simeq &  \frac{4\lachi^2 \cosh(2\nu_S) - 16 \lachi^2\cosh^2\nu_S } {g_H^2\lambda_S^2 (1 + \lachi^2 \sinh^{2}\nu_S)^2} \sim -32 g_H^{-2}\lambda_S^{-2} \lachi^{-2} e^{- 2\nu_S} , 
\ee
The slow-roll process holds for the case $\epsilon \ll 1$ and $|\eta| \ll 1$, which corresponds to the following conditions:
\be \label{SRC2}
& & e^{2\nu_S} \lachi^2  g_H\lambda_S \gg  8\sqrt{2},  \quad  \epsilon \ll 1 ,  \nn \\
& &  e^{\nu_S} \lachi  g_H\lambda_S  \gg  4\sqrt{2} , \quad |\eta| \ll 1 , 
\ee
which show that when the slow-roll parameter $\eta$ satisfies the slow-roll condition $|\eta| \ll 1$, the slow-roll parameter $\epsilon$ automatically fits to the condition $\epsilon \ll 1$ as $e^{\nu_S}\lachi > 1$. 
 
When the slow-roll parameters increase to close to the order of one, i.e., $\epsilon \sim 1$ and $|\eta| \sim 1$, the slow-roll conditions become invalid and the inflationary expansion of early universe starts to break down. As a consequence, the potential of $\cQ_c$-spin scaling field quickly runs into the minimum value around the minimum point $\nu_S = \nu_0 \ll 1$. In this case, the relevant cosmic energy density $\Lambda_Q^4$ after inflation is given by,
\be \label{CED}
& & \Lambda_Q^4 \equiv \lambda_D^2 \Mka^4 \cF(\nu_0) = \frac{ \lachi^4\lak^2}{1 + \lachi^4\lak^2 } \lambda_D^2 \Mka^4  \ll \Mka^4 , \quad \mbox{for} \quad \lachi^2\lak \ll 1 .
\ee
The relevant masses of hyperunified qubit-spinor field and hyperspin gauge field as well as $\cQ_c$-spin scalar field and scaling gauge field in hyper-spacetime are characterized as follows:
\be
& & m_{\fQH} = \lachi \lak M_S \ll M_S, \nn \\
& & \mu_{\fA} =  \lachi \lak \beta_G M_S \ll  M_S, \nn \\
& &  \mu_{Q} = 2\sqrt{2}\frac{ \lachi^2 \lak ( 1 + \lachi^2 \lak^2)^{1/2} }{ (1+\lachi^4\lak^2 )}  \frac{\lambda_D}{\beta_Q^2} M_S  \simeq 2\sqrt{2} \lachi^2 \lak \frac{\lambda_D}{\beta_Q^2} M_S \ll M_S , \nn \\
& & \mu_{\mW} = \lambda_Q M_S  (1 + (\lachi \lak)^2 )^{1/2} \simeq \lambda_Q M_S \sim M_S, 
\ee
where we have used the following relation in obtaining the relevant mass of $\cQ_c$-spin scalar boson:
\be
\cF''(\nu_0) \equiv \frac{\p^2 \cF(\nu_S) }{\p \nu_S} |_{\nu_S =  \nu_0 } =
 \frac{ 8\lachi^4 \lak^2( 1 + \lachi^2 \lak^2) }{ (1+\lachi^4\lak^2 )^{2} } .
 \ee
It is seen that the hyperunified qubit-spinor field and hyperspin gauge field as well as $\cQ_c$-spin scalar field get relevant masses much smaller than the fundamental graviscaling mass scale $M_S$, while the scaling gauge field can receive a very heavy mass characterized by the mass scale $M_S$. 

Therefore, when the inflation ends up and the potential falls into the minimum, the initial high potential energy characterized by the fundamental mass scale $\Mka$ is transmuted into kinetic energies of fundamental fields which have much smaller masses in comparison with the fundamental mass scale $\Mka$. So that the gauge interactions of inhomogeneous hyperspin gauge field and hyperunified qubit-spinor field together with the interactions of $\cQ_c$-spin scalar field become active as fundamental interactions of nature, which are thought to play a significant role in radiative and matter dominated evolutions of the universe. In general, the scaling gauge field may still be out of the thermal equilibrium due to its very heavy mass.

\subsection{ Evolving universe with scaling gauge field as dark matter candidate and $\cQ_c$-spin scalar field as quantum cosmic matter and dark energy candidate }

Let us suppose that all fundamental fields as functionals of coordinates in hyper-spacetime have {\it localizing distribution amplitudes} in extra spatial dimensions beyond four-dimensional spacetime. In general, the fundamental fields as functionals of coordinates are distributed with different {\it localizing length scales} $l_{A}$ ($A=5,\cdots, D_h$) in different extra spatial dimensions of hyper-spacetime. These localizing length scales in extra spatial dimensions are thought to form a manifold with a finite integrable volume $\cV_{D_h-4}\propto l_1\times l_2\times \cdots \times l_{D_h-4}$. As the present consideration is focusing on the evolution of universe, we shall not investigate extensively on the localizing distribution amplitudes of fundamental fields. 

To simplify the discussion, let us consider the case that all fundamental fields will run from the high energy scale characterized by the fundamental mass scale $\Mka$ down to the mass scale $\mu_M$ determined by the {\it maximal localizing length scale} $\l_{M}=\mbox{Max}.\{ \l_{A} \} =1/\mu_M$. When the running energy scale undergoes to be below the {\it minimal localizing energy scale} $\mu_M$, i.e., $\mu < \mu_M$, all fundamental fields as effective fields appear to have motions and interactions in four-dimensional spacetime, which leads in principle the hyperunified field theory in hyper-spacetime to be effective hyperunified field theory at low energies in four-dimensional spacetime. In formal, we may display such an effective hyperunified field theory in Einstein-type basis as follows:
\be  \label{actionHUFTGQFTS0}
\cS_{\mH\mU} & \equiv & \int [d^{D_h}x] \, \chi^e(x) \Mka^{D_h-4} \cL_{D_h} = \int [d^{4}x] \, \chi^e(x)  \cV_{D_h-4}  \Mka^{D_h-4} \cL_{4D} \nn \\
& \equiv & \int [d^{4}x] \, \chi^e(x) \cL_{4D}^{eff} ,
\ee
where $\cL_{4D}^{eff}$ represents the effective Lagrangian at low energies in four-dimensional spacetime. 

It is useful to introduce a {\it characteristic localizing length scale} $\la$ or {\it characteristic localizing mass scale} $\Ma = 1/\la$ to characterize the whole volume $\cV_{D_h-4}$ as follows:
\be
& & \cV_{D_h-4} \equiv \la^{D_h-4} = \frac{1}{\Ma^{D_h-4}} \propto l_1\times l_2\times \cdots \times l_{D_h-4} , \nn \\
& &  \cV_{D_h-4}  \Mka^{D_h-4} \equiv \left(\frac{\Mka}{\Ma}\right)^{D_h-4} \equiv \xi_{\alpha}^{-(D_h-4)} , \quad \xia \equiv \frac{\Ma}{\Mka} ,
\ee
where $\xia$ is regarded as {\it dimension-reduction scaling factor}, which enables us to define, from the basic coupling constants $g_H$, $g_W$, $\lambda_S$ and $\lambda_Q$ of hyperunified field theory in hyper-spacetime, the effective coupling constants $g_h$, $g_w$, $\lambda_s$ and $\lambda_q$ in four-dimensional spacetime as follows:
\be
& & g_h^2 \equiv g_H^2 \left(\frac{\Ma}{\Mka}\right)^{D_h-4} \equiv g_H^2 \xia^{D_h-4}  , \nn \\
& & g_w^2 \equiv g_W^2 \left(\frac{\Ma}{\Mka}\right)^{D_h-4} \equiv g_W^2 \xia^{D_h-4} , \nn \\
& & \lambda_s^2 \equiv \lambda_S^2 \left(\frac{\Mka}{\Ma}\right)^{D_h-4} \equiv \lambda_S^2 \xia^{-(D_h-4)}, \nn \\
& & \lambda_q^2 \equiv \lambda_Q^2 \left(\frac{\Mka}{\Ma}\right)^{D_h-4} \equiv \lambda_S^2 \xia^{-(D_h-4)}, \nn \\
& & \lambda_d^2 \equiv \lambda_D^2 \left(\frac{\Mka}{\Ma}\right)^{D_h-4} \equiv \lambda_D^2 \xia^{-(D_h-4)} . 
\ee
It is manifest that the following combinations of coupling constants are independent of the dimension-reduction scaling factor:
\be
& & g_H\lambda_S = g_h\lambda_s, \quad g_H\lambda_Q = g_h\lambda_q, \quad g_H\lambda_D = g_h\lambda_d, \nn \\
& & g_W\lambda_S = g_w\lambda_s , \quad g_W\lambda_Q = g_w\lambda_q , \nn \\
& &  \beta_Q\equiv \frac{\lambda_S}{\lambda_Q} = \frac{\lambda_s}{\lambda_q}, \quad \frac{\lambda_D}{\lambda_Q} = \frac{\lambda_d}{\lambda_q} ,
\ee
which are referred to as {\it combined dimension-independent coupling constant}.

It is noticed that the slow-roll conditions presented in Eqs.(\ref{SRC1})-(\ref{SRC2}) are spacetime dimension independent as they are given by the combined dimension-independent coupling constant $g_H\lambda_S = g_h\lambda_s$.

To formulate the effective Lagrangian $\cL_{4D}^{eff}$ in four-dimensional spacetime, the kinetic terms of hyperspin gauge field and scaling gauge field as well as $\cQ_c$-spin scalar field are normalized via the following replacements: 
\be
& & \cA(x) \to g_h \cA(x) , \quad \mW(x) \to g_w \mW(x) , \nn \\
& & \chi_s(x) \to \lambda_s^{-1} \chi_s(x) \quad \text{or} \;\; \Phi(x) \to \lambda_q^{-1}\Phi(x) , 
\ee
and the gravitational coupling $G_N$ in four-dimensional spacetime is defined as follows:
\be
\frac{1}{16\pi G_N} \equiv \frac{\Mka^2}{g_h^2} = \frac{\Mka^2}{g_H^2} \xia^{4-D_h} \equiv \frac{M_P^2}{16\pi}  = \frac{1}{2}M_p^2 ,  
\ee
which indicates that the gravitational coupling $G_N$ or Planck mass $M_P$ ( reduced Planck mass $M_p$) presented in four-dimensional spacetime depends in general on the dimension-reduction scaling factor $\xia$ and hyper-spacetime dimension $D_h$ when relating Planck mass $M_P$ to the fundamental mass scale $\Mka$ and basic coupling constant $g_H$ in the hyperunified field theory in hyper-spacetime.

With the above considerations and analyses, we now turn to study the scaling gauge field and $\cQ_c$-spin scalar field as basic constituents of universe and discuss their roles for the evolution of universe in an effective hyperunified field theory at low energies in four-dimensional spacetime. 

Let us discuss firstly the property of scaling gauge field $\mW_{\fM}$. It is noticed that the scaling gauge field has no direct interactions with hyperunified qubit-spinor field and hyperspin gauge field. In the hidden scaling gauge formalism of hyperunified field theory as shown in Eq.(\ref{actionHUFTHSG}) with the redefinition $\mW_{\fM} \to \mW_{\fM} - \frac{1}{2}\p_{\fM}\ln (\lambda_S^2 \phi^2 + \lambda_Q^2\phi_1^2)$, the scaling gauge field interacts only with the scaling field $\chi_s(x)$ and hyper-gravigauge field. As the action of Eq.(\ref{actionHUFTHSG}) or Eq.(\ref{actionHUFTGQFTS}) possesses a discrete symmetry under $Z_2$ transformation, i.e., $\mW_{\fM} \to - \mW_{\fM}$, so that $\mW_{\fM}$ becomes stable and survives to present once it  is produced in the early universe. Therefore, the scaling gauge field is expected to be as {\it dark matter candidate} in the hyperunified field theory.  

It can be checked that the primordial Hubble-type mass scale $\Hka$ and scaling gauge boson mass $M_W$ are given by, 
\be
 \Hka & = & \frac{\Mka}{\apk} \simeq (g_H \lambda_D) \left((D_h-1)(D_h-2)\right)^{-1/2} \Mka \nn \\ 
 & = & (g_h \lambda_d) \left((D_h-1)(D_h-2)\right)^{-1/2} \Mka , \nn \\
M_W & = & (g_w \lambda_q)( 1+ \lachi^2\lak^2 )^{1/2} M_S  \simeq (g_w \lambda_q) M_S \nn \\
& = & (g_W \lambda_Q) M_S =  (g_W \lambda_S) \Mka =  (g_w \lambda_s)\Mka, 
\ee
which indicates that both $\Hka$ and $M_W$ are independent of the dimension-reduction scaling factor $\xia$ as they are given by the combined dimension-independent coupling constants $g_h\lambda_q = g_H\lambda_Q$ and $g_w \lambda_q = g_W \lambda_Q$ or  $g_w \lambda_s = g_W \lambda_S$, respectively. It is clear that the magnitudes for both $\Hka$ and $M_W$ are characterized by the fundamental mass scale $\Mka$.

Let us examine a possible mechanism to produce the heavy scaling gauge boson $\mW(x)$ via the $\cQ_c$-spin scalar annihilation, i.e., $\Phi + \Phi\rightarrow \mW +\mW $, after inflation in the early universe. As analyzed in ref.\cite{TW3}, the cross section of the annihilation process dominated by the longitudinal mode of scaling gauge boson is given by, 
\be
\sigma \sim \frac{g_w^4}{E^2}\frac{E^4}{M_W^4} = \frac{E^2}{\Mka^4\lambda_s^4}, \nn
\ee
with $E$ as the typical energy of final scaling gauge boson $\mW(x)$. 

To obtain the correct relic density, a high reheating temperature $T_R$ is expected to be reached when the potential energy is transmuted into the kinetic energies of fundamental fields which get very small masses after inflation. It was demonstrated in ref.\cite{TW3} that in the instantaneous reheating limit, the reheating temperature $T_R$ is related to the primordial Hubble-type mass scale $\Hka$ through the relation, $T_R\sim \sqrt{\Hka M_p}$. The reheating temperature $T_R$ can maximally reach to be $T_R\sim 4\times 10^{15}\GeV$, which can produce abundantly the heavy scaling gauge boson $\mW(x)$ with the mass up to $M_W \sim 5\times 10^{16}\GeV$. 

It is attributed to the discrete symmetry $Z_2$ with $\mW(x) \to -\mW(x)$ that the scaling gauge boson $\mW(x)$ becomes stable, which brings naturally such a heavy scaling gauge boson to be as {\it dark matter candidate} in the hyperunified field theory.

We now come to investigate basic properties arising from $\cQ_c$-spin scalar field. Let us consider a simple case that the smallness of constant $\lak$ is characterized just by two fundamental mass scales $M_S$ and $\mk$ in the present theory with the following relation:
\be
\lak = \frac{\mk}{M_S} =   \frac{\lambda_Q\mk}{\lambda_S \Mka} =  \frac{\lambda_q\mk}{\lambda_s \Mka}, 
\ee
where the fundamental graviscaling mass scale $M_S$ reflects {\it ultraviolet energy scale} in early universe and the fundamental cosmic mass scale $\mk$ describes {\it infrared energy scale} of present universe. Such a constant $\lak$ is regarded as {\it basic cosmic constant}. 

The cosmic energy density $\Lambda_{\cQ_c}^4$ in four-dimensional spacetime is given as follows:
\be \label{CED}
\Lambda_{\cQ_c}^4 & \equiv &\xia^{D_h-4} \Lambda_Q^4 = \lambda_d^2 \Mka^4 \cF(\nu_0) =  \frac{\lambda_d^2 \Mka^4\lachi^4\lak^2}{ 1+\lachi^4\lak^2 } \nn \\
& \simeq & \frac{\lambda_q^2}{\lambda_s^2} \lambda_d^2 \lachi^4\Mka^2 \mk^2 =  \frac{1}{2}\frac{\lambda_q^2}{\lambda_s^2} (\lambda_d g_h)^2 \lachi^4 M_p^2 \mk^2 \equiv \frac{\lambda_q^2}{\lambda_s^2} \lambda_d^2 \lachi^4\Mka^2 \Lk^{-2} \nn \\ 
& = & \frac{1}{2}\frac{\lambda_q^2}{\lambda_s^2} (\lambda_d g_h)^2\lachi^4 M_p^2 \Lk^{-2} =  \frac{1}{2}\frac{\lambda_Q^2}{\lambda_S^2} (\lambda_D g_H)^2\lachi^4 M_p^2 \Lk^{-2} \equiv 3 \dla^2 M_p^2 \Lk^{-2}, 
\ee
which indicates that the cosmic energy density $\Lambda_{\cQ_c}^4$ in four dimensional spacetime depends on the dimension-reduction scaling factor $\xia$. Such a dependence is the same as that of Planck mass square in four-dimensional spacetime. Note that $\lambda_d g_h = \lambda_Dg_H$ and $\lambda_q/\lambda_s = \lambda_Q/\lambda_S$ are the combined dimension-independent coupling constants.  

As the fundamental cosmic mass scale $\mk$ reflects the co-moving cosmic horizon of the universe, i.e., $\mk=\frac{1}{\Lk}= \frac{1}{\cka \eka}$, it is natural to expect that the cosmic energy density $\Lambda_{\cQ_c}^4$ brings on the basic constituent of universe as {\it dark energy candidate}, which can be achieved when the cosmic energy scale $\Lambda_{\cQ_c}$ fits to the current experimental observation with the magnitude $\Lambda_{\cQ_c} \sim 10^{-2.5}$eV.  

In fact, it was shown extensively in ref.\cite{eHDE} that the dark energy characterized by the total co-moving cosmic horizon is viewed as {\it holographic dark energy}, which can provide a consistent explanation on the accelerated expansion of present universe and meanwhile overcome the so-called fine-tuning problem and coincidence problem\cite{SW}. When applying a similar analysis to the present consideration, the cosmic energy density $\Lambda_{\cQ_c}^4$ can present a consistent dark energy candidate as long as the combination of coupling constants satisfies the following condition:
\be \label{DLA}
 \dla \equiv \frac{1}{\sqrt{6}} \frac{\lambda_q}{\lambda_s} (\lambda_d g_h) \lachi^2 \agt 10^1\sim10^2, 
\ee
which may be referred to as {\it $\cQ_c$-spin cosmic dark energy} for a short of mention. 

As the scaling gauge field can serve as dark matter candidate in the hyperunified field theory, so that $\cQ_c$-spin cosmic dark energy not only becomes compatible with the so-called $\Lambda$CDM model but also enables us to understand both the fine-tuning problem and coincidence problem directly in connection with the inflationary expansion of early universe.

When the $\cQ_c$-spin cosmic dark energy $\Lambda_{\cQ_c}^4$ becomes dominant as it is just the case in the present universe, the cosmic energy scale $\Lambda_{\cQ_c}$ can be regarded as an ultraviolet (UV) cutoff of effective hyperunified field theory at low energies of present universe, which is proposed to meet a relationship between UV cutoff and infrared (IR) cutoff ($1/L$) with $L$ the size of volume $L^3$ such that an effective field theory should be a good description of Nature\cite{CKN}. Such a relationship indicates that all physical states except those that have already collapsed to a black hole should be described by conventional quantum field theory in four-dimensional spacetime. From such a relationship, we arrive at the following condition in our present considerations: 
\be
\Lambda_{\cQ_c}^4 = 3 \dla^2 M_p^2 \Lk^{-2} \equiv M_p^2 L_{\cQ_c}^{-2} < M_p^2 L^{-2} , 
\ee
which implies that the IR cutoff $1/L$ is constrained to be,
\be
 & & L < L_{\cQ_c} \equiv \frac{M_p}{\Lambda_{\cQ_c}^2} \sim 10^{33} \, \mbox{eV}^{-1} , \nn \\
 & & L_{\cQ_c} \equiv \frac{\Lk}{\sqrt{3}\dla}  \alt ( 10^{-1} -10^{-2}) \Lk ,
\ee
where $L_{\cQ_c}$ is regarded as the size of visible universe within the gravitational quantum field theory, which is shown to be less than the co-moving cosmic horizon $\Lk$ by one to two orders of magnitude. In general,  $L_{\cQ_c}$ may be referred to as {\it visible co-moving cosmic horizon} of present universe. It is clear that the effective hyperunified field theory as effective local gravitational quantum field theory can provide a good approximate description of present universe within the visible co-moving cosmic horizon $L_{\cQ_c}$. Meanwhile, such a cosmic energy density $\Lambda_{\cQ_c}^4$ as $\cQ_c$-spin cosmic dark energy matches to the spirit of holographic principle\cite{HP1,HP2} in four-dimensional spacetime. 

As the cosmic energy scale $\Lambda_{\cQ_c} \sim 10^{-2.5}$eV presents an UV cutoff of present universe, we are able to determine the fundamental cosmic mass scale $\mk$ characterized by the co-moving cosmic horizon $\Lk^{-1}= (\cka \eka)^{-1}=\mk $ from both $\cQ_c$-spin cosmic dark energy and coupling constant parameter presented in Eqs.(\ref{CED}) and (\ref{DLA}). The magnitude value of $\mk$ is given as follows:
\be
& & \mk = \frac{1}{\Lk} = \frac{1}{c\eka} = \frac{\Lambda_{\cQ_c}^2}{\sqrt{3}\dla M_p } \alt (10^{-34} - 10^{-35}) \mbox{eV}  .
\ee 

The mass of $\cQ_c$-spin scalar boson $\Phi(x)$ is found to be, 
\be
m_{\cQ_c} & = &  2\sqrt{2}\frac{\lambda_d}{\lambda_s} \Mka^2 \sqrt{\cF''(\nu_0)} =  
2\sqrt{2} \frac{\lambda_d}{\lambda_s}  \frac{ \lachi^2\lak (1 + \lachi^2 \lak^2)^{1/2}}{(1+\lachi^4\lak^2)} \Mka , \nn \\
& \simeq &  2\sqrt{2} \frac{\lambda_d\lambda_q}{\lambda_s^2} \lachi^2 \mk =  2\sqrt{2} \frac{\lambda_D\lambda_Q}{\lambda_S^2} \lachi^2 \mk ,
\ee
which is independent of the dimension-reduction scaling factor $\xia$ as it is given by the combined dimension-independent coupling constants $\lambda_d/\lambda_s = \lambda_D/\lambda_S$ and $\lambda_q/\lambda_s = \lambda_Q/\lambda_S$. The magnitude of $\cQ_c$-spin scalar boson mass is determined by the fundamental cosmic mass scale $\mk$ and combined dimension-independent coupling constants as follows:
\be
& & m_{\cQ_c} \simeq 2\sqrt{2} \frac{\lambda_d \lambda_q }{\lambda_s^2} \lachi^2 \mk = \frac{4\Lambda_{\cQ_c}^2}{\lambda_s g_hM_p} \sim \frac{4}{\lambda_s g_h} \times 10^{-33} \mbox{eV} .
\ee
which indicates that the $\cQ_c$-spin scalar boson $\Phi(x)$ receives a tiny {\it cosmic mass}. 

We may refer to $\cQ_c$-spin scalar boson $\Phi(x)$ as {\it quantum cosmic matter}, which is expected to play an essential role on the evolution of universe. This is because the $\cQ_c$-spin scalar boson $\Phi(x)$ as quantum cosmic matter not only has universal gravitational interaction with the hyper-gravigauge field but also interacts with the scaling gauge boson as dark matter candidate, and meanwhile it couples directly to the hyperspin gauge field and hyperunified qubit-spinor field. 

It is interesting to notice that the hyperunified qubit-spinor field and hyperspin gauge field also receive tiny masses:
\be
& & m_{\fQH} = M_S \sinh\nu_0 =   M_S \lachi \lak = \lachi \mk \leq \lachi (10^{-34} - 10^{-35}) \mbox{eV} , \nn \\
& &  m_{\fA} = \beta_G  M_S \sinh\nu_0  = \beta_G  M_S \lachi\lak =  \beta_G \lachi \mk \leq \beta_G \lachi (10^{-34} - 10^{-35}) \mbox{eV}  , 
\ee
which are independent of dimension-reduction scaling factor. 

We would like to address that all vector-like leptons and quarks contained in the hyperunified qubit-spinor field obtain a tiny cosmic mass $m_{\fQH}$ determined by the fundamental cosmic mass scale $\mk$ and coupling constant $\lachi$. Similarly, the hyperspin gauge field also acquires a tiny cosmic mass $m_{\fA}$ governed by the fundamental cosmic mass scale $\mk$ and combined coupling constant $\beta_G \lachi$, which means that all gauge fields including electromagnetic field and strong gauge fields receive tiny cosmic masses. 

It is intriguing that the gluons and photon in such a hyperunified field theory become massive gauge bosons although their tiny cosmic masses appear unobservable small. The current experiment only provides the upper limits on the mass of photon to be around\cite{PDG}:
\be
 m_{\gamma} < 10^{-18} \mbox{eV} \sim 10^{-26} \mbox{eV} .
\ee

Therefore, the evolving $\cQ_c$-spin scaling field in the hyperunified field theory plays a significant role for the evolution of universe. The initial high energy potential of $\cQ_c$-spin scaling field is characterized by the fundamental mass scale $\Mka$ and brings on an inflationary expansion of early universe when the background $\cQ_c$-spin scaling field undertakes slowly varying extra large values. Meanwhile, after the $\cQ_c$-spin scaling field falls into the minimum point of the potential, the resulting cosmic energy density as $\cQ_c$-spin cosmic dark energy leads to an accelerated expansion of present universe. In addition, the $\cQ_c$-spin scalar boson $\Phi(x)$ as quantum cosmic matter with tiny cosmic mass is expected to be still essential for the future evolution of universe. Furthermore, the heavy scaling gauge field produced by the $\cQ_c$-spin scalar boson $\Phi(x)$ is regarded as dark matter candidate. In particular, all vector-like leptons and quarks contained in the hyperunified qubit-spinor field and all gauge fields including strong gauge field (gluons) and electromagnetic field (photon) all acquire tiny cosmic masses and interact with the $\cQ_c$-spin scalar boson $\Phi(x)$.


\section{ Symmetry structure and symmetry breaking mechanism in hyperunified field theory and comprehension on the presence of Higgs boson and three families of leptons and quarks }

To realize the SM in four dimensional spacetime from the hyperunified field theory in hyper-spacetime, it is inevitable to understand the presence of Higgs boson and three families of leptons and quarks. Following along the gauge invariance principle and scaling invariance hypothesis, we have demonstrated in previous sections how the inhomogeneous hyperspin gauge symmetry and scaling gauge symmetry govern the fundamental interactions of nature. We have also verified that when making gauge fixing conditions to remove all unphysical degrees of freedom for gauge fields, the fundamental symmetry of hyperunified field theory in 19-dimensional hyper-spacetime keeps only global inhomogeneous hyperspin symmetry WS(1,18)=SP(1,18)$\rtimes$ W$^{1,18}$ in association with inhomogeneous Lorentz-type/Poincar\'e-type group symmetry PO(1,18)=P$^{1,18}\ltimes$SO(1,18) together with global scaling symmetry SC(1). As such a hyperunified qubit-spinor field contains four families of vector-like lepton-quark states with both chiral type lepton-quark states and corresponding mirror lepton-quark states, the subgroup symmetry SP(5)$\cong$SO(5) of WS(1,18) reflects the maximal family-spin symmetry of four families. Nevertheless, the current experiments only observed three families of chiral type leptons and quarks with a single light Higgs boson in SM and provided the lower mass limits up to 1370 GeV for the vector-like fourth family quarks\cite{PDG}, which motivates us to examine the hyperunified symmetry structure with the presence of Higgs-like scalar fields and illustrate the appearance of three families of leptons and quarks in the chiral spinor representation with negative U-parity and the existence of the fourth family of vector-like lepton-quark state.

\subsection{ Hyperunified field theory with inhomogeneous hyperspin gauge symmetry WS(1,18) as hyperunified symmetry in gravigauge hyper-spacetime }

To demonstrate the hyperunified symmetry structure with Higgs-like fields and possible symmetry breaking mechanisms for realizing three families of leptons and quarks in light of the qubit-spinor structure of hyperunified qubit-spinor field $\fPsi_{\fQH}(x)$ presented in Eqs.(\ref{ffQH})-(\ref{WEQH}), let us express the action in gravigauge hyper-spacetime as follows:
\be  \label{actionHUFTGHS}
\cS_{\mH\mU} & \equiv &  \int [\delta^{D_h} \vka ]  \{\, \bar{\fPsi}_{\fQH} \vSi_{-}^{\fC} i\cD_{\fC}\fPsi_{\fQH} - \beta_Q \sinh\chi_s \, \bar{\fPsi}_{\fQH} \tvSi_{-}\fPsi_{\fQH}  \nn \\
& - & \frac{1}{4} g_H^{-2} ( \cF_{\fC[\fD\fA\fB]}  \cF^{\fC[\fD\fA\fB]} +  \cF_{\fC(\fD\fA\fB]}  \cF^{\fC(\fD\fA\fB]}  ) - g_H^{-2} \eta^{\fC\fD} \fR_{\fC\fD} \nn \\
& + & \frac{1}{2}g_H^{-2}\beta_G^2\beta_Q^2 \sinh^2\chi_s \, ( \cA_{[\fC\fA\fB]} - \fOm_{[\fC\fA\fB]} )(\cA^{[\fC\fA\fB]} - \fOm^{[\fC\fA\fB]})  \nn \\
& + & \frac{1}{2}g_H^{-2}\beta_G^2 \beta_Q^2 \sinh^2\chi_s \, ( \cA_{(\fC\fA\fB]} - \fOm_{(\fC\fA\fB]} )(\cA^{(\fC\fA\fB]} - \fOm^{(\fC\fA\fB]})  \nn \\
& - & \frac{1}{4} g_W^{-2} \mF_{\fC\fD} \mF^{\fC\fD} + \frac{1}{2} \lambda_S^2( 1 + \sinh^2\chi_s )\, \mW_{\fC} \mW^{\fC}  \nn \\
& + & \frac{1}{2} \lambda_S^2 \heth_{\fC} \chi_s \heth^{\fC}\chi_s  - \lambda_D^2  \cF(\chi_s)  \, \} ,
\ee
with $\fA, \fM= 0,1,2,3,5, \cdots, 19$. Where the definitions for gauge fields and field strengths in gravigauge hyper-spacetime are presented in Eqs.(\ref{HSGFGG})-(\ref{HSGFGGS}) and Eqs.(\ref{HSGFDC4})-(\ref{HSGFSDC4}). The covariant derivative is given by,
\be \label{CDGGHST}
& & i\cD_{\fC}\equiv i\heth_{\fC} + \cA_{[\fC\fA\fB]}\frac{1}{2}\vSi^{\fA\fB} \equiv \hbfA_{\fC}^{\;\fM} ( i\heth_{\fM} + \cA_{[\fM\fA\fB]}\frac{1}{2}\vSi^{\fA\fB} ), \nn \\
& & \cA_{[\fC\fA\fB]} \equiv \frac{1}{3} ( \cA_{\fC\fA\fB} + \cA_{\fA\fB\fC} + \cA_{\fB\fC\fA} ), \quad  \cA_{\fC\fA\fB} \equiv \hbfA_{\fC}^{\;\fM} \cA_{\fM\fA\fB} ,
\ee
which implies that the hyperspin gauge interaction with hyperunified qubit-spinor field only concerns the totally antisymmetric hyperspin gauge field $\cA_{[\fC\fA\fB]}$ in gravigauge hyper-spacetime.

It can be verified from Eq.(\ref{GMffQH2}) that all $\vGa$-matrices $\vGa^{\fA}$ with $\fA= 0, 1, 2, 3, 5, \cdots, 20$ are anticommuting, which can form in general a symmetry group SP(1,19)$\cong$SO(1,19) with group generators given directly by the commutators of $\vGa$-matrices,
\be
\varSigma^{\fA\fB} = \frac{i}{4} [ \vGa^{\fA}, \vGa^{\fB} ], \quad \fA, \fB= 0,1,2,3,5,\cdots, 20 . 
\ee
which can be decomposed into the following subgroup symmetries: 
\be
 SP(1,19) & \supset &  SP(1,3)\times SP(16)  \supset SP(1,3) \times SP(10) \times SP(6) \nn \\
&  \cong & SO(1,3)\times SO(16) \supset SO(1,3) \times SO(10)\times SO(6) \nn \\
& \supset & SO(1,3) \times SU_C(4)\times SU_L(2)\times SU_R(2) \times SU_F(4), 
\ee
with subgroup algebras corresponding to the following group generators: 
\be
& & \varSigma^{\fA\fB}  ( \fA, \fB= 0,1,2,3 ) \in sp(1,3)\cong so(1,3)  \nn \\
& &  \varSigma^{\fA\fB}  (\fA, \fB= 5,\cdots, 20 )  \in sp(16) \cong so(16), \nn \\
& &  \varSigma^{\fA\fB}  (\fA, \fB= 5,\cdots, 14 )  \in sp(10) \cong so(10), \nn \\
& &  \varSigma^{\fA\fB}  (\fA, \fB= 15,\cdots, 20 )  \in sp(6) \cong so(6)\cong su_F(4), \nn \\
& & \varSigma^{\fA\fB}  (\fA, \fB= 5,\cdots,10 )  \in sp(6) \cong so(6)\cong su_C(4) , \nn \\
& &  \varSigma^{\fA\fB}  ( \fA, \fB= 11,\cdots14) \in sp(4)\cong so(4)\cong su_L(2)\times su_R(2) . 
\ee
Where the subgroup symmetry SU$_F$(4) appears in formal as family-spin symmetry of four families of lepton-quark and mirror lepton-quark states. Nevertheless, the $\vGa$-matrix $\vGa^{20}\equiv \tvGa$ is actually a motion-irrelevant $\cQ_c$-matrix rather than motion-correlation $\cM_c$-matrix, which leads the action in Eq.(\ref{actionHUFTGHS}) to have only a reduced hyperspin symmetry SP(1,18) in association with the Lorentz-type group symmetry SO(1,18), which indicates that the hyperunified qubit-spinor field $\fPsi_{\fQH}(x)$ cannot hold a maximal family-spin symmetry SO(6)$\cong$SU$_F$(4) for four-families of lepton-quark and mirror lepton-quark states. It remains not sufficient to understand directly why there are only three families of leptons and quarks in SM. This is because the hyperunified qubit-spinor field $\fPsi_{\fQH}(x)$ still possesses the family-spin symmetry SO$_F$(5) which contains the maximal subgroup family-spin symmetry SO$_F$(4),
\be
SO_F(5) \supset SO_F(4) \cong SU_{F}(2)\times SU_{\hat{F}}(2).
\ee
In the above qubit-spinor structure, SU$_{F}$(2) characterizes the family-spin symmetry between the first two families and SU$_{\hat{F}}$(2) reflects the family-spin symmetry between the third and fourth families. 

In general, the action presented in Eq.(\ref{actionHUFTGHS}) with gauge fixing conditions possesses the global inhomogeneous hyperspin symmetry WS(1,18) in association with inhomogeneous Lorentz-type/Poincar\'e-type group symmetry PO(1,18) in hidden coordinate Minkowski hyper-spacetime,
\be
G_S & = & PO(1,18) \wtjoin WS(1,18)  \nn \\
& = & P^{1,18}\ltimes SO(1,18) \wtjoin  SP(1,18) \rtimes W^{1,18},
\ee
where the group transformation of hyperspin symmetry SP(1,18) in Hilbert space should be coincidental to that of Lorentz-type group symmetry SO(1,18) in Minkowski hyper-spacetime as indicated by the symbol ``$\wtjoin$". Note that WS(1,18)=SP(1,18)$\rtimes$ W$^{1,18}$ and PO(1,18) = P$^{1,18}\ltimes$ SO(1,18) are semidirect product group symmetries, where W$^{1,18}$ is translation-like $\cW_e$-spin Abelian group symmetry in Hilbert space and P$^{1,18}$ is ordinary translational Abelian group symmetry in Minkowski hyper-spacetime. The group generators of inhomogeneous hyperspin symmetry WS(1,18) are directly given by the commutators of $\vGa$-matrices, 
\be
& & \varSigma^{\mA\mB} = \frac{i}{4} [ \vGa^{\mA}, \vGa^{\mB} ] \in sp(1,18), \nn \\
& &  \vSi_{-}^{\mA} = \frac{1}{2} \vGa^{\mA}\vGa_{-} \in w^{1,18} , 
\ee
with $\mA, \mB= 0,1,2,3,5,\cdots, 19$. 

The hyperspin symmetry SP(1,18) of the action in Eq.(\ref{actionHUFTGHS}) as the subgroup symmetry of SP(1,19) has a natural decomposition into the following subgroup symmetry: 
\be
SP(1,18) & \supset & SP(1,3)\times SP(15)  \supset SP(1,3) \times SP(10) \times SP(5) \nn \\
&  \cong & SO(1,3)\times SO(15) \supset SO(1,3) \times SO(10)\times SO_F(5) \nn \\
& \supset & SO(1,3)\times SO(6) \times SO(4)\times SO_F(4) \nn \\
& \cong & SO(1,3) \times SU_C(4)\times SU_L(2)\times SU_R(2) \times SU_F(2)\times SU_{\hat{F}}(2) \nn \\
& \supset & SO(1,3)\times SU_C(3) \times SU_L(1)\times U_Y(1), 
\ee
where the four color-spin symmetry SU$_C$(4), left-right isospin symmetry SU$_L$(2)$\times$SU$_R$(2) and heavy-light family-spin symmetry SU$_F$(2)$\times$SU$_{\hat{F}}$(2) should be broken down eventually so as to reproduce the SM with Higgs-like boson and three families of leptons and quarks.


\subsection{ Hyperspin gauge fields beyond four-dimensional gravigauge spacetime as the Higgs-like scalar fields and presence of Higgs-like bosons in hyperunified field theory }

It can be seen from the action shown in Eq.(\ref{actionHUFTGHS}) and covariant derivative given in Eq.(\ref {CDGGHST}) that the hyperspin gauge field defined in gravigauge hyper-spacetime couples to the self-conjugated hyperunified qubit-spinor field only via its totally antisymmetric tensor form $\cA_{[\fC\fA\fB]}$. When taking scaling gauge fixing condition to be in Einstein-type basis, we can express such a totally antisymmetric hyperspin gauge field as follows:
\be
\cA_{[\fC\fA\fB]} & \equiv & \frac{1}{3} ( \cA_{\fC\fA\fB} + \cA_{\fA\fB\fC} + \cA_{\fB\fC\fA}  )\nn \\
& \to & \frac{1}{3} ( \chih_{\fC}^{e \fM} \cA_{\fM\fA\fB} + \chih_{\fA}^{e \fM} \cA_{\fM\fB\fC} + \chih_{\fB}^{e \fM} \cA_{\fM\fC\fA}) ,
\ee 
where $\chih_{\fA}^{e \fM}$ is hyper-garvigauge field in Einstein-type basis and also regarded as Goldstone-like bicovariant vector field, and $\cA_{\fM\fA\fB}$ is hyperspin gauge field in biframe hyper-spacetime.

It can be verified that the action presented in Eq.(\ref{actionHUFTGHS}) with scaling gauge prescription in Einstein-type basis possesses inhomogeneous hyperspin gauge symmetry WS(1,18) in Hilbert space jointly together with inhomogeneous Lorentz-type/Poincar\'e-type group symmetry PO(1,18)  in hidden coordinate Minkowski hyper-spacetime,
\be
G_S & = & PO(1,18) \Join WS(1,18) = P^{1,18}\ltimes SO(1,18) \Join  SP(1,18) \rtimes W^{1,18},
\ee
where WS(1,18)=SP(1,18)$\rtimes$ W$^{1,18}$ and PO(1,18) = P$^{1,18}\ltimes$ SO(1,18) are semidirect product group symmetries with W$^{1,18}$ the translation-like $\cW_e$-spin Abelian group symmetry in Hilbert space and P$^{1,18}$ the ordinary translational Abelian group symmetry in Minkowski hyper-spacetime. 

The hyperspin gauge symmetry SP(1,18) contains the following maximal subgroup symmetry:
\be
SP(1,18) \supset SP(1,3)\times SP(10)\times SP(5) \cong SO(1,3)\times SO(10)\times SO(5), 
\ee
where SP(1,3) is relevant to the usual spin and boost-spin symmetry in four-dimensional gravigauge spacetime, while SP(10)$\times$SP(5) are regarded as internal symmetries besides four-dimensional gravigauge spacetime. In light of the above subgroup symmetry, it is useful to decompose the totally antisymmetric hyperspin gauge field $\cA_{[\fC\fA\fB]}$ into the following forms:
\be
\cA_{[\fC\fA\fB]} & = & (\cA_{[cab]} , \cA_{[c\bA\bB]},  \cA_{[c\tA\tB]}, \cA_{[\bC ab]},  \cA_{[\bC \bA\bB]}, 
 \nn \\ 
 & & \cA_{[\bC \tA\tB]}, \cA_{[\tC ab]}, \cA_{[\tC \bA\bB]},  \cA_{[\tC \tA\tB]},   \cA_{[c \bA \tA]}),
\ee
with the notations:
\be
& & \fA, \fB, \fC \equiv ( a,\bA, \tA ), (b, \bB, \tB), (c, \bC, \tC ) , \nn \\
& &  a,b,c = 0,1,2,3, \quad \bA, \bB, \bC = 5, 6, \cdots, 14, \nn \\
& &  \tA, \tB, \tC = 15,\cdots,19 .
\ee

It is noticed that as the decomposed hyperspin gauge fields $\cA_{[\bC \bA\bB]}$,  $\cA_{[\bC \tA\tB]}$, $\cA_{[\tC \bA\bB]}$ and $\cA_{[\tC \tA\tB]}$ concern only the internal hyperspin gauge symmetries SP(10)$\times$SP(5) beyond four-dimensional gravigauge spacetime, they are regarded as Higgs-like scalar fields in view of four dimensional gravigauge spacetime. To be more explicit, let us denote the totally antisymmetric hyperspin gauge field in gravigauge hyper-spacetime as follows:
\be
& & \cA_{[c \bA\bB]} \equiv \cA_{[c\; I+4\; J+4 ]} \equiv \cA_{c I J} , \nn \\
& & \cA_{[c \tA\tB]} \equiv \cA_{[c\; \ta+14\; \tb+14 ]} \equiv \cA_{c \ta \tb} , \nn \\
& & \cA_{[c \bA \tA]} \equiv \cA_{[c\; I+4\; \ta +14]} \equiv \cV_{c I \ta} , \nn \\
& & \cA_{[\bC a b]} \equiv \cA_{[K+4\; a b]} \equiv \cT_{K a b} , \nn \\
& & \cA_{[\tC a b]} \equiv \cA_{[\tc+14\; a b]} \equiv \cT_{\tc\, a b} , \nn \\
& & \cA_{[\bC \bA\bB]} \equiv \cA_{[K+4\; I+4\; J+4 ]} \equiv \cH_{K I J} , \nn \\
& &  \cA_{[\bC \tA\tB]} \equiv \cA_{[K+4\; \ta+14\; \tb+14 ]} \equiv \cH_{K \ta\tb} , \nn \\
& &  \cA_{[\tC \bA\bB]} \equiv \cA_{[\tc+14\; I+4\; J+4 ]} \equiv \varPhi_{\tc I J} , \nn \\
& &  \cA_{[\tC \tA\tB]} \equiv \cA_{[\tc+14\; \ta+14\; \tb+14 ]} \equiv \varPhi_{\tc\ta\tb} , 
\ee
with the definitions:
\be
& & \bA , \bB, \bC \equiv I +4 , \, J + 4, \, K+4, \quad I, J, K = 1, 2, \cdots, 10 , \nn \\
& & \tA , \tB, \tC \equiv \ta + 14 , \, \tb + 14, \, \tc+14, \quad \ta, \tb, \tc = 1, 2, \cdots, 5 .
\ee
In view of four-dimensional gravigauge spacetime, $\cA_{c I J}$, $\cA_{c \ta \tb}$ and $\cV_{c I \ta}$ are regarded as vector-like hyperspin gauge fields, $\cT_{K a b}$ and $\cT_{\tc\, a b}$ represent tensor-like hyperspin gauge fields, while $\cH_{K I J}$, $\cH_{K \ta\tb}$, $\varPhi_{\tc I J}$ and $\varPhi_{\tc\ta\tb}$ are considered as Higgs-like scalar fields. The representation dimensions of all totally antisymmetric hyperspin gauge fields are listed in the table 1.

\begin{table}[htp]
\caption{Representation dimension of totally antisymmetric hyperspin gauge fields according to subgroup decomposition SP(1,3)$\times$SP(10)$\times$SP(5) of hyperspin symmetry group SP(1,18) }
\begin{center}
\begin{tabular}{|c|c|c|c|c|}
\hline
Hyperspin symmetry SP(1,18) & Subgroups & SP(1,3)$\cong$SO(1,3)  & SP(10)$\cong$SO(10) & SP(5)$\cong$SO(5)  \\
\hline
Hyperspin gauge fields & Classification &  Rep. Dim. & Rep. Dim. & Rep. Dim. \\
\hline
$\cA_{[cab]}$ & $\epsilon^{abcd}\cA_d $  & 4 & 0 & 0  \\
$\cA_{[c \bA\bB]}$ & $\cA_{c}^{I J}$  & 4 & 45 & 0 \\
$\cA_{[c \tA\tB]}$ & $\cA_{c}^{\ta \tb}$ & 4 & 0 & 10 \\
$\cA_{[c \bA\tA]}$ & $\cV_{c}^{I \ta}$  & 4 & 10 & 5 \\
$\cA_{[\bC a b]}$ & $\cT_{K}^{ab}$  & 6 & 10 & 0 \\
$\cA_{[\tC a b]}$ & $\cT_{\tc}^{ab}$  & 6 & 0 & 5 \\
$\cA_{[\bC \bA \bB]}$ & $\cH_{KIJ}$  & 0 & 120 & 0 \\
$\cA_{[\bC \tA \tB]}$ & $\cH_{K}^{\ta\tb}$ & 0 & 10 & 10 \\
$\cA_{[\tC \bA \bB]}$ & $\varPhi_{\tc}^{IJ}$  & 0 & 45 & 5 \\
$\cA_{[\tC \tA \tB]}$ & $\varPhi_{\tc\ta\tb}$  & 0 & 0 & 10 \\
\hline
\end{tabular}
\end{center}
\label{Table 1}
\end{table}%

From the above decomposition of hyperspin gauge field, the action for the hyperspin gauge interaction of hyperunified qubit-spinor field can be expressed as follows:
\be
\cS_G & \equiv & \int d^{19}x \chi^e(x) \,\{  \bar{\fPsi}_{\fQH} [\, \vSi_{-}^{cab} \cA_{[cab]} + \vSi_{-}^{c\bA\bB}\cA_{[c \bA\bB]} +  \vSi_{-}^{c\tA\tB}\cA_{[c \tA\tB]}  \, ] \fPsi_{\fQH} \nn \\
& + & \bar{\fPsi}_{\fQH} [\, +  \vSi_{-}^{c\bA\tA}\cA_{[c \bA\tA]} + \vSi_{-}^{\bC ab} \cA_{[\bC ab]} + \vSi_{-}^{\bC\bA\bB}\cA_{[\bC \bA\bB]} +  \vSi_{-}^{\bC\tA\tB}\cA_{[\bC \tA\tB]}  \, ] \fPsi_{\fQH} \nn \\
& + & \bar{\fPsi}_{\fQH} [\, \vSi_{-}^{\tC ab} \cA_{[\tC ab]} + \vSi_{-}^{\tC\bA\bB}\cA_{[\tC \bA\bB]} +  \vSi_{-}^{\tC\tA\tB}\cA_{[\tC \tA\tB]}  \, ] \fPsi_{\fQH}  \},
\ee
with the definitions:
\be
& & \vSi_{-}^{cab} \equiv \vSi_{-}^c \vSi^{ab} , \quad \vSi_{-}^{c\bA\bB} \equiv \vSi_{-}^c \vSi^{\bA\bB} ,   \quad \vSi_{-}^{c\tA\tB} \equiv \vSi_{-}^c \vSi^{\tA\tB} ,  \quad \vSi_{-}^{c\bA\tA} \equiv \vSi_{-}^c \vSi^{\bA\tA}, \nn \\
& & \vSi_{-}^{\bC ab} \equiv \vSi_{-}^{\bC} \vSi^{ab} , \quad \vSi_{-}^{\bC\bA\bB} \equiv \vSi_{-}^{\bC}\vSi^{\bA\bB} ,   \quad \vSi_{-}^{\bC\tA\tB} \equiv \vSi_{-}^{\bC} \vSi^{\tA\tB} , \nn \\
& & \vSi_{-}^{\tC ab} \equiv \vSi_{-}^{\tC} \vSi^{ab} , \quad \vSi_{-}^{\tC\bA\bB} \equiv \vSi_{-}^{\tC}\vSi^{\bA\bB} ,   \quad \vSi_{-}^{\tC\tA\tB} \equiv \vSi_{-}^{\tC} \vSi^{\tA\tB} .
\ee

In light of grand unified qubit-spinor fields for chiral type lepton-quark states and mirror lepton-quark states, the above action can be rewritten as follows:
\be
\cS_G & \equiv & \int d^{19}x \chi^e(x) \frac{1}{2} \{\, \sum_{f=1}^{4} [ \bar{\Psi}_{\fQGfn}\vGa^{c} ( \cA_{[cab]}\frac{1}{2}\vSi^{ab} + \cA_{c}^{I J} \frac{1}{2}\vSi_{IJ} )  \Psi_{\fQGfn}   \nn \\
& + & \bar{\Psi}_{\fQGfp}\vGa^{c} ( \cA_{[cab]}\frac{1}{2}\vSi^{ab} + \cA_{c}^{I J} \frac{1}{2}\vSi_{IJ} )  \Psi_{\fQGfp}  ] \nn \\ 
 & + & \sum_{f,f'=1}^{4}  [ \bar{\Psi}_{\fQGfn} \vGa^{c} \cA_{c}^{p} \frac{1}{4}(\lambda_{p})_{ff'} \Psi_{\fQGfbn}+ \bar{\Psi}_{\fQGfp} \vGa^{c} \cA_{c}^{p} \frac{1}{4} (\lambda_{p})_{ff'} \Psi_{\fQGfbp} \nn \\
 & + & \bar{\Psi}_{\fQGfn} \vGa^{K} \cH_{K}^{p} \frac{1}{4}(\lambda_2\lambda_{p})_{ff'} \Psi_{\fQGfbn}+ \bar{\Psi}_{\fQGfbp} \vGa^{K} \cH_{K}^{p} \frac{1}{4} (\lambda_2\lambda_{p})_{ff'}\Psi_{\fQGfbp} \nn \\
 & + & (\lambda_2)_{ff'} \bar{\Psi}_{\fQGfn} \vGa^{K} ( \cT_{K}^{ab}\frac{1}{2}\vSi_{ab} + \cH_{KIJ} \frac{1}{2}\vSi_{IJ} )  \Psi_{\fQGfbn} \nn \\
 & + & (\lambda_2)_{ff'} \bar{\Psi}_{\fQGfp}\vGa^{K} ( \cT_{K}^{ab}\frac{1}{2}\vSi_{ab} + \cH_{KIJ} \frac{1}{2}\vSi_{IJ} )   \Psi_{\fQGfbp}   \nn \\
 & + & \bar{\Psi}_{\fQGfn} \vGa^{c} \vGa_{I}\cV_{c}^{I\ta} \frac{1}{4}(\lambda_2\eta_{\ta})_{ff'} \Psi_{\fQGfbp} + \bar{\Psi}_{\fQGfp} \vGa^{c} \vGa_{I}\cV_{c}^{I\ta} \frac{1}{4} (\lambda_2\eta_{\ta})_{ff'} \Psi_{\fQGfbn}  \nn \\
 & + & \bar{\Psi}_{\fQGfn}\eta^{\tc}_{ff'} ( \cT_{\tc}^{ab}\frac{1}{2}\vSi_{ab} + \varPhi_{\tc}^{IJ} \frac{1}{2}\vSi_{IJ} )  \Psi_{\fQGfbp}  + \bar{\Psi}_{\fQGfp} \eta^{\tc}_{ff'} ( \cT_{\tc}^{ab}\frac{1}{2}\vSi_{ab} + \varPhi_{\tc}^{IJ} \frac{1}{2}\vSi_{IJ} )  \Psi_{\fQGfbn}  \nn \\
 & - & \bar{\Psi}_{\fQGfn} \varPhi_{p} \frac{1}{4}(\tlam_{p})_{ff'}  \Psi_{\fQGfbp} - 
\bar{\Psi}_{\fQGfp}\varPhi_{p} \frac{1}{4}(\tlam_{p})_{ff'}  \Psi_{\fQGfbn} ] \, \} ,
\ee
with $\vGa$-matrices $\vGa^a$ and $\vGa^I$ given as follows: 
\be \label{GMffQGa}
& & \vGa^0 =\;\; \;  \sigma_0 \otimes \sigma_0 \otimes \sigma_0 \otimes \sigma_0 \otimes\sigma_0 \otimes \sigma_1 \otimes \sigma_0, \nn \\
& &  \vGa^1 =\;\;  i  \sigma_0 \otimes \sigma_0 \otimes \sigma_0 \otimes \sigma_0 \otimes \sigma_0\otimes \sigma_2\otimes \sigma_1, \nn \\
& & \vGa^2 = \;\;  i  \sigma_0 \otimes \sigma_0 \otimes \sigma_0 \otimes  \sigma_0 \otimes\sigma_0\otimes  \sigma_2\otimes \sigma_2, \nn \\
& & \vGa^3 = \;\; i  \sigma_0 \otimes \sigma_0 \otimes \sigma_0 \otimes \sigma_0 \otimes\sigma_0\otimes  \sigma_2\otimes \sigma_3, 
\ee
for $\vGa^a$ ($a=0,1,2,3$), and 
\be \label{GMffQGI}
& &  \vGa^1 = i  \sigma_1 \otimes \sigma_0 \otimes \sigma_1 \otimes \sigma_0 \otimes \sigma_2\otimes  \sigma_3\otimes \sigma_0, \nn \\
& & \vGa^2 = i  \sigma_1 \otimes \sigma_0 \otimes \sigma_2 \otimes  \sigma_3 \otimes\sigma_2\otimes  \sigma_3\otimes \sigma_0, \nn \\
& & \vGa^3 = i  \sigma_1 \otimes \sigma_0 \otimes \sigma_1 \otimes \sigma_2 \otimes\sigma_3\otimes  \sigma_3\otimes \sigma_0 , \nn \\
& &  \vGa^4 = i  \sigma_1 \otimes \sigma_0 \otimes \sigma_2 \otimes \sigma_2 \otimes  \sigma_0\otimes \sigma_3\otimes \sigma_0, \nn \\
& &  \vGa^5 = i  \sigma_1 \otimes \sigma_0 \otimes \sigma_1 \otimes  \sigma_2 \otimes \sigma_1\otimes \sigma_3\otimes \sigma_0 , \nn \\
& &  \vGa^{6} =  i  \sigma_1 \otimes \sigma_0 \otimes \sigma_2 \otimes \sigma_1 \otimes \sigma_2\otimes \sigma_3\otimes \sigma_0 , \nn \\
& & \vGa^{7} =  i \sigma_2 \otimes \sigma_0 \otimes \sigma_0 \otimes  \sigma_0 \otimes \sigma_0\otimes \sigma_3\otimes \sigma_0 , \nn \\
& & \vGa^{8} =  i \sigma_1 \otimes \sigma_1 \otimes \sigma_3 \otimes  \sigma_0 \otimes \sigma_0\otimes \sigma_3\otimes \sigma_0 , \nn \\
& & \vGa^{9} =  i \sigma_1 \otimes \sigma_2 \otimes \sigma_3 \otimes  \sigma_0 \otimes \sigma_0\otimes \sigma_3\otimes \sigma_0 , \nn \\
& & \vGa^{10} =  i  \sigma_1 \otimes \sigma_3 \otimes  \sigma_3 \otimes  \sigma_0 \otimes \sigma_0\otimes \sigma_3\otimes \sigma_0 , 
\ee
for $\vGa^I$ ($I=1,2,\cdots, 10$). $\varSigma^{ab}$ and $\varSigma^{IJ}$ are defined as follows:
\be
\varSigma^{ab} = \frac{i}{4} [ \vGa^{a}, \vGa^{b} ] \in sp(1,3), \quad \varSigma^{IJ} = \frac{i}{4} [ \vGa^{I}, \vGa^{J} ] \in sp(10) .
\ee
We have also introduced the following definitions:
\be
& &  \vGa_{\bA}= \vGa_{I + 4} \to \vGa_{I}, \quad  \vGa_{\tA} = \vGa_{\ta + 14} \to \vGa_{\ta} \equiv \eta_{\ta} , \nn \\
& & \cA_{c}^{\ta \tb}\frac{1}{2}\vSi_{\ta\tb} \to \cA_{c}^{p}\frac{1}{4}\lambda^p, \quad  \cA_{K}^{\ta \tb}\frac{1}{2}\vSi_{\ta\tb} \to \cH_{K}^{p} \frac{1}{4} \lambda^{p}, \quad \lambda^{p} \in sp(5) , \nn \\
& &  \varPhi^{\tc\ta\tb} \vSi_{[\tc\ta\tb]} \to -\varPhi_{p} \frac{1}{4}\tilde{\lambda}_{p} , 
\ee
where $\lambda_{p}$ and $\tilde{\lambda}_{p}$ are defined as follows:
\be
& & \vSi_{\ta\tb} = ( \vSi_{23}, \vSi_{31}, \vSi_{12}, \vSi_{41}, \vSi_{42}, \vSi_{43}, \vSi_{15}, \vSi_{25}, \vSi_{35}, \vSi_{45} ) \nn \\
& & \quad \quad \equiv \frac{1}{2} ( \lambda_1 ,  \lambda_2,  \lambda_3, \lambda_4 ,  \lambda_5,  \lambda_6, \lambda_7 ,  \lambda_8,  \lambda_9, \lambda_{10} ) \equiv \frac{1}{2} \lambda_p ,  \nn \\
& & \vSi_{[\tc\ta\tb]} \equiv \frac{1}{2} \{ \vGa_{\tc}, \vSi_{\ta\tb} \} = ( \eta_1 \vSi_{23}, \eta_1\vSi_{24}, \eta_1 \vSi_{25}, \eta_1\vSi_{34},  \eta_1 \vSi_{35}, \eta_1\vSi_{45}, \eta_2\vSi_{34}, \eta_2\vSi_{35}, \nn \\
& & \qquad \quad \eta_2\vSi_{45}, \eta_3\vSi_{45} )  \equiv \frac{1}{2} ( \tlam_1 , \tlam_2, \tlam_3, \tlam_4 , \tlam_5 , \tlam_6 , \tlam_7 , \tlam_8, \tlam_9, \tlam_{10} ) \equiv \frac{1}{2} \tlam_p , 
\ee
and $\eta_{\ta}$ are expressed as follows: 
\be \label{ETAM}
& & \eta_1 = \sigma_2 \otimes \sigma_1 \tilde{\gamma}_5 , \quad \eta_2 = \sigma_2 \otimes \sigma_2 \gamma_{15}, \quad \eta_3 = \sigma_2 \otimes \sigma_3 \tilde{\gamma}_5 , \nn \\
& & \eta_4 = \sigma_1 \otimes \sigma_0 \gamma_{15}, \quad \eta_5 = \sigma_3 \otimes \sigma_0 \gamma_{15} , \nn \\
& &  \tilde{\gamma}_5 =  \sigma_0 \otimes \sigma_0 \otimes  \sigma_0 \otimes  \sigma_0 \otimes \sigma_0\otimes \gamma_5 , \nn \\
& & \gamma_{15} =  \sigma_3 \otimes \sigma_0 \otimes  \sigma_0 \otimes  \sigma_0 \otimes \sigma_0\otimes \sigma_3\otimes \sigma_0 ,
\ee
The matrices $\lambda_p$ and $\tlam_p$ ($p=1,2,\cdots,10$) have the following explicit forms: 
\be
& & \lambda_1 = \sigma_0 \otimes \sigma_1 \tilde{\gamma}_{11} , \quad  \lambda_2 = \sigma_0 \otimes \sigma_2 , \quad \lambda_3 = \sigma_0 \otimes \sigma_3 \tilde{\gamma}_{11} , \nn \\
& & \lambda_4 = \sigma_3 \otimes \sigma_1 \tilde{\gamma}_{11} , \quad  \lambda_5 = \sigma_3 \otimes \sigma_2 , \quad \lambda_6 = \sigma_3 \otimes \sigma_3 \tilde{\gamma}_{11} , \nn \\
& & \lambda_7 = \sigma_1 \otimes \sigma_1 \tilde{\gamma}_{11} , \quad  \lambda_8 = \sigma_1 \otimes \sigma_2 , \quad \lambda_9 = \sigma_1 \otimes \sigma_3 \tilde{\gamma}_{11} , \nn \\
& & \lambda_{10} = - \sigma_2 \otimes \sigma_0 , \quad \tilde{\gamma}_{11} =  \sigma_3 \otimes \sigma_0 \otimes  \sigma_0 \otimes  \sigma_0 \otimes \sigma_0\otimes \sigma_0\otimes \sigma_0 ,
\ee
and
\be
& & \tlam_1 = \sigma_2 \otimes \sigma_0 \gamma_{15} , \quad  \tlam_2 = \sigma_1 \otimes \sigma_3 \tilde{\gamma}_{5}, \quad  \tlam_3 = \sigma_3 \otimes \sigma_3 \tilde{\gamma}_{5} , \nn \\
& & \tlam_4 = -\sigma_1 \otimes \sigma_2 \gamma_{15}  , \quad  \tlam_5 = -\sigma_3 \otimes \sigma_2\gamma_{15} , \quad \tlam_6 = -\sigma_0 \otimes \sigma_1 \tilde{\gamma}_{5} , \nn \\
& & \tlam_7 = \sigma_1 \otimes \sigma_1 \tilde{\gamma}_{5} , \quad  \tlam_8 = \sigma_3 \otimes \sigma_1 \tilde{\gamma}_{5}, \quad \tlam_9 = -\sigma_0 \otimes \sigma_2 \gamma_{15} , \nn \\
& & \tlam_{10} = - \sigma_0 \otimes \sigma_3\tilde{\gamma}_{5} . 
\ee
Where $\eta_{\tc}$, $\lambda_p$ and $\tlam_p$ are presented as $4\times4$ matrices in four family-spin representation associated with the grand unified qubit-spinor representation in 128-dimensional Hilbert space. The ten matrices $\lambda_p$ form SP(5) family-spin symmetry of four families. Note that the ten matrices $\lambda_2\lambda_{p}$ become all symmetric and the five matrices $\lambda_2\eta_{\ta}$ are all antisymmetric, i.e.:
\be
(\lambda_2\lambda_{p})^T = \lambda_2\lambda_{p}, \quad (\lambda_2\eta_{\ta})^T = - \lambda_2\eta_{\ta} .
\ee

It is clear that the decomposed hyperspin gauge fields $\cA_{[cab]}$, $\cA_{c}^{I J}$ and $\cA_{c}^{p}$ are in correspondence to the subgroup gauge symmetries SP(1,3), SP(10) and SP(5) of hyperspin gauge symmetry SP(1,18) in gravigauge hyper-spacetime. Whereas the decomposed hyperspin gauge fields $\cA_{[\bC \tA\tB]}\equiv \cH_{K}^{\ta\tb}\equiv \cH_K^p $ and $\cA_{[\bC \bA\bB]}\equiv \cH_{K I J}$ are Higgs-like scalar fields which provide scalar-type interactions either for lepton-quark states as grand unified qubit-spinor fields with negative U-parity or for mirror lepton-quark states as grand unified qubit-spinor fields with positive U-parity. The decomposed hyperspin gauge fields $\cA_{[\tC \bA\bB]}\equiv\varPhi_{\tc}^{I J}$ and $\cA_{[\tC \tA\tB]}\equiv \varPhi_{[\tc\ta\tb]}\equiv \varPhi_p $ are regarded as scalar-like fields, they bring on scalar-type interactions between lepton-quark and mirror lepton-quark states which are regarded as grand unified qubit-spinor fields with negative and positive U-parities, respectively.

As the subgroup hyperspin symmetry SP(10) leads to the usual grand unified symmetry SO(10)$\cong$SP(10) of leptons and quarks in SM, it contains the well-known subgroup symmetries as follows:
\be
SP(10) \supset SP(6)\times SP(4)\cong SO(6)\times SO(4) \cong SU_C(4)\times SU_L(2)\times SU_R(2), 
\ee 
with SU$_C$(4) the four color-spin symmetry of lepton-quark states and SU$_L$(2)$\times$ SU$_R$(2) the left-right isospin symmetry of up-type and down-type lepton-quark states. 

For our current purpose, let us pay attention to the properties of hyperspin gauge fields as Higgs-like scalar fields. From the above decomposition of SP(10) symmetry, the Higgs-like scalar field $\cA_{[\bC \bA\bB]}\equiv \cH_{K I J}$ belongs to the 120-dimensional representation of SO(10) and concerns the left-handed and right-handed triplet Higgs-like scalar fields of subgroups SU$_L$(2) and SU$_R$(2), respectively, which can bring about Majorana masses of neutrinos once the symmetry gets breaking down. The Higgs-like scalar field $\cA_{[\bC \tA\tB]}\equiv \cH_{K}^{\ta\tb}\equiv \cH_K^p$ is in both basic representation of SP(10) symmetry with $K=1, \cdots, 10$ and adjoint representation of SP(5) family-spin symmetry with $p=1, \cdots, 10$. Since each basic representation of grand unified symmetry SP(10)$\cong$SO(10) contains one Higgs doublet, there are in general ten Higgs doublets corresponding to ten elements of SP(5) family-spin symmetry in 19-dimensional hyper-spacetime. 

To be more explicit, let us write down the Yukawa-type coupling for ten Higgs-like scalar fields $\cH_K^p$ as follows:
\be
\cL_Y =  \frac{1}{4}\bar{\Psi}_{\fQGfn} \vGa^{I} (\fH_{I})_{ff'}\Psi_{\fQGfbn} , 
\ee
where the Higgs-like scalar matrix $(\fH_{I})_{ff'}$ ($I=1,2,\cdots, 10$) is given by
\be
 \fH_{I} \equiv 
 \begin{pmatrix}
 \tcH_{I}^2 - i\tcH_{I}^1 \gamma_{11} & i\tcH_{I}^3 \gamma_{11} & \cH_{I}^8-i\cH_{I}^7 \gamma_{11} & \cH_{I}^{10} + i\cH_{I}^9 \gamma_{11} \\
  i\tcH_{I}^3 \gamma_{11} & \tcH_{I}^2 + i\tcH_{I}^1 \gamma_{11} & - \cH_{I}^{10} + i\cH_{I}^9 \gamma_{11} & \cH_{I}^8 + i\cH_{I}^7 \gamma_{11} \\
  \cH_{I}^8-i\cH_{I}^7 \gamma_{11} & - \cH_{I}^{10} + i\cH_{I}^9 \gamma_{11} & \tcH_{I}^5 - i\tcH_{I}^4 \gamma_{11} & i\tcH_{I}^6 \gamma_{11} \\
  \cH_{I}^{10} + i\cH_{I}^9 \gamma_{11} & \cH_{I}^8 + i\cH_{I}^7 \gamma_{11} & i\tcH_{I}^6 \gamma_{11} & \tcH_{I}^5 - i\tcH_{I}^4 \gamma_{11}
 \end{pmatrix} ,
\ee
with the definitions
\be
& & \tcH_{I}^1 = \cH^1 + \cH^4 , \quad \tcH_{I}^4 = \cH^1 - \cH^4 , \nn \\
& & \tcH_{I}^2 = \cH^2 + \cH^5 , \quad \tcH_{I}^5 = \cH^2 - \cH^5 , \nn \\
& & \tcH_{I}^3 = \cH^3 + \cH^6 , \quad \tcH_{I}^6 = \cH^3 - \cH^6 .
\ee 

To be more explicit, let us express the grand unified qubit-spinor fields for the chiral type lepton-quark states appearing in SM as the westward entangled hyperqubit-spinor fields $\Psi_{W_f}(x)$ presented in Eq.(\ref{WEQH}),
\be \label{WQH}
& & \Psi_{\fQ_{\mG_f}^{-}}^{T}(x) \equiv  \Psi_{W_f}^{T}(x)  \equiv (\Psi_{f L }, \Psi_{f R}  )^T \\
& & \Psi_{f L}^T \equiv (U_f^{r}, U_f^{b}, U_f^{g}, U_f^{w}, D^{r}_{fc}, D^{b}_{fc}, D^{g}_{fc}, D^{w}_{fc}, D_f^{r}, D_f^{b}, D_f^{g}, D_f^{w}, -U^{r}_{fc}, -U^{b}_{fc}, -U^{g}_{fc}, -U^{w}_{fc})_L^T ,  \nn \\
& & \Psi_{f R}^T \equiv (U_f^{r}, U_f^{b}, U_f^{g}, U_f^{w}, D^{r}_{fc}, D^{b}_{fc}, D^{g}_{fc}, D^{w}_{fc}, D_f^{r},  D_f^{b}, D_f^{g}, D_f^{w}, -U^{r}_{fc}, -U^{b}_{fc}, -U^{g}_{fc}, -U^{w}_{fc})_R^T , \nn
\ee
so that the above Yukawa-type coupling for Higgs-like scalar fields can be rewritten into the following form: 
\be
\cL_Y =  \frac{1}{4} \bar{\Psi}_{f L} \Gamma_{I} (\mH_{I})_{ff'}\Psi_{f' R}  + H.c.  ,
\ee
with the Higgs-like scalar matrix in four-family Hilbert space,
\be
 \mH_{I} \equiv 
 \begin{pmatrix}
 \tcH_{I}^1 - i\tcH_{I}^2  & -\tcH_{I}^3 & \cH_{I}^7-i\cH_{I}^8 & -\cH_{I}^{9} - i\cH_{I}^{10} \\
  -\tcH_{I}^3 & -\tcH_{I}^1 - i\tcH_{I}^2  & - \cH_{I}^{9} + i\cH_{I}^{10} & -\cH_{I}^7 - i\cH_{I}^8  \\
  \cH_{I}^7-i\cH_{I}^8  & - \cH_{I}^{9} + i\cH_{I}^{10} & \tcH_{I}^4 - i\tcH_{I}^5  & -\tcH_{I}^6  \\
  -\cH_{I}^{9} - i\cH_{I}^{10}  & -\cH_{I}^7 - i\cH_{I}^8 & -\tcH_{I}^6  & -\tcH_{I}^4 - i\tcH_{I}^5 
 \end{pmatrix} .
\ee
Where $\Gamma$-matrices $\Gamma_I$ ($I=1,2,\cdots, 10$) in 64-dimensional Hilbert space are defined as follows:
\be \label{GMffQGI}
& &  \Gamma^1 =  \sigma_0 \otimes \sigma_1 \otimes \sigma_0 \otimes \sigma_2\otimes  \sigma_0\otimes \sigma_0, \nn \\
& & \Gamma^2 = \sigma_0 \otimes \sigma_2 \otimes  \sigma_3 \otimes\sigma_2\otimes  \sigma_0\otimes \sigma_0, \nn \\
& & \Gamma^3 = \sigma_0 \otimes \sigma_1 \otimes \sigma_2 \otimes\sigma_3\otimes  \sigma_0\otimes \sigma_0 , \nn \\
& &  \Gamma^4 =  \sigma_0 \otimes \sigma_2 \otimes \sigma_2 \otimes  \sigma_0\otimes \sigma_0\otimes \sigma_0, \nn \\
& &  \Gamma^5 = \sigma_0 \otimes \sigma_1 \otimes  \sigma_2 \otimes \sigma_1\otimes \sigma_0\otimes \sigma_0 , \nn \\
& &  \Gamma^{6} =  \sigma_0 \otimes \sigma_2 \otimes \sigma_1 \otimes \sigma_2\otimes \sigma_0\otimes \sigma_0 , \nn \\
& & \Gamma^{7} =  \sigma_1 \otimes \sigma_3 \otimes  \sigma_0 \otimes \sigma_0\otimes \sigma_0\otimes \sigma_0 , \nn \\
& & \Gamma^{8} =  \sigma_2 \otimes \sigma_3 \otimes  \sigma_0 \otimes \sigma_0\otimes \sigma_0\otimes \sigma_0 , \nn \\
& & \Gamma^{9} =   \sigma_3 \otimes  \sigma_3 \otimes  \sigma_0 \otimes \sigma_0\otimes \sigma_0\otimes \sigma_0 , \nn \\
& & \Gamma^{10} =  -i  \sigma_0 \otimes \sigma_0 \otimes  \sigma_0 \otimes \sigma_0\otimes \sigma_0\otimes \sigma_0 .
\ee
It can be checked that the Higgs-like scalar fields $\cH_{I}^{p}(x)$ with $I =7,8,9,10$ bring on ten Higgs-like isospin doublets ($p=1,2,\cdots, 10$). Where $\cH_{I}^{p}(x)$ with $I =7,8$ form charged Higgs-like bosons and $\cH_{I}^{p}$ with $I =9,10$ represent neutral Higgs-like bosons. 

Note that as the Higgs-like scalar fields $\cH_{I}^{p}(x)$ arise from hyperspin gauge field with vector representations beyond four-dimensional spacetime in 19-dimensional locally flat gravigauge hyper-spacetime, it is useful to revisit the action of hyperspin gauge field in locally flat gravigauge hyper-spacetime, which can easily be read from Eq.(\ref{actionHUFTHCF3}) as follows:
\be  \label{actionHUFTHCFA}
\cS_{\mH\mU}^{\cA} & \equiv &  \int [\delta^{D_h} \vka ] \{\, - \frac{1}{4} g_H^{-2} \tilde{\cF}_{\fC\fD\fA\fB}\tilde{\cF}^{\fC\fD\fA\fB} \nn \\
& - & \frac{1}{4} g_H^{-2}\fOm_{[\fC\fD]\fA} \fOm^{[\fC\fD]\fB}  \cA^{\fA\fA'\fB'} \cA_{\fB\fA'\fB'}  + \frac{1}{2} g_H^{-2} \fOm^{[\fC\fD]\fA}  \tilde{\cF}_{\fC\fD\fA'\fB'}\cA_{\fA}^{\fA'\fB'}    \, \}
\ee
which shows that such an action differs from the action of ordinary gauge field in globally flat Minkowski hyper-spacetime with absence of gravitational interaction. It is clear that once the locally flat gravigauge hyper-spacetime approaches to a globally flat hyper-spacetime in the limit case, i.e., $\fA_{\fM}^{\; \fA}\to \eta_{\fM}^{\;\fA}$ and $\fOm_{[\fC\fD]\fA} \to 0$, the above action gets to have only the first term which turns to be an action of ordinary gauge field in globally flat Minkowski hyper-spacetime.

It is verified in Eq.(\ref{actionHUFTHCF3}) that the hyperspin gravigauge field $\fOm_{[\fC\fD]\fA}$ appears as an auxiliary field due to the absence of its kinetic term in the action formulated in locally flat gravigauge hyper-spacetime. So that $\fOm_{[\fC\fD]\fA}$ satisfies a constraint equation presented in Eq.(\ref{GGCGE}). Let us suppose that the hyperspin gravigauge field $\fOm_{[\fC\fD]\fA}$ is resolved from the constraint equation to be as a function of hyperspin gauge field $\cA_{\fC}^{\fA\fB}$ together with other basic bosonic fields, so that we may formally express its solution as follows:
\be 
\fOm_{[\fC\fD]\fA}  \equiv \Phi_{[\fC\fD]\fA} (\cA, \mW, \chi_s) .
\ee
The action of hyperunified field theory in Eq.(\ref{actionHUFTHCF3}) can be rewritten into the following form:
\be  \label{actionHUFTHCF4}
\cS_{\mH\mU} & \equiv &  \int [\delta^{D_h} \vka ] \{\, \bar{\fPsi}_{\fQH} \vSi_{-}^{\fC} i\cD_{\fC} \fPsi_{\fQH} - \beta_Q \sinh\chi_s \bar{\fPsi}_{\fQH} \tvSi_{-}\fPsi_{\fQH}  \nn \\
& - & \frac{1}{4} g_H^{-2}\tilde{\cF}_{\fC\fD\fA\fB}\tilde{\cF}^{\fC\fD\fA\fB}  + \frac{1}{4} g_H^{-2}\Phi_{[\fC\fD]\fA}(\cA, \mW, \chi_s) \Phi^{[\fC\fD]\fB}(\cA, \mW, \chi_s)  \cA^{\fA\fA'\fB'} \cA_{\fB\fA'\fB'} \nn \\
& - & g_H^{-2}(1 + \frac{1}{2} \beta_G^2\beta_Q^2 \sinh^2\chi_s) ( \tilde{\eta}_{\fC'\fD'\fA'}^{\;\; \fC\fD\fA}  + \bar{\eta}_{\fC'\fD'\fA'}^{\;\; \fC\fD\fA})\Phi_{[\fC\fD]\fA}(\cA, \mW, \chi_s) \Phi^{[\fC'\fD']\fA'}(\cA, \mW, \chi_s)  \nn \\
& + & \frac{1}{2} g_H^{-2}\beta_G^2\beta_Q^2 \sinh^2\chi_s \,  \cA_{[\fC\fD]\fA}\cA^{[\fC'\fD']\fA'} \bar{\eta}_{\fC'\fD'\fA'}^{\;\; \fC\fD\fA} \nn \\
&  - & \frac{1}{4} g_W^{-2}\Phi_{[\fC\fD]\fA}(\cA, \mW, \chi_s) \Phi^{[\fC\fD]\fB}(\cA, \mW, \chi_s)  \mW^{\fA} \mW_{\fB}  + \frac{1}{2} \lambda_S^2( 1 + \sinh^2\chi_s ) \mW_{\fC} \mW^{\fC} \nn \\
& - & \frac{1}{4} g_W^{-2} \tilde{\mF}_{\fC\fD} \tilde{\mF}^{\fC\fD} + \frac{1}{2} \lambda_S^2 \heth_{\fC} \chi_s \heth^{\fC}\chi_s  - \lambda_D^2 \cF(\chi_s)  \, \} ,
\ee
which indicates that the hyperunified field theory formulated in locally flat gravigauge hyper-spacetime does generate an effective potential of hyperspin gauge field $\cA_{\fC}^{\fA\fB}$ when integrating out hyperspin gravigauge field $\fOm_{[\fC\fD]\fA}$. Such an action appears to involve no dynamical gravitational field, all dynamical fields get emergent gravitational interaction via non-commutative geometry described by the hyperspin gravigauge field $\fOm_{[\fC\fD]\fA}$ which behaves as auxiliary field and emerges as structure factor of non-Abelian Lie algebra for hyper-gravicoordinate derivative operator $\heth_{\fC}$ as shown in Eq.(\ref{NCG}). 

Such an emergent gravitational interaction due to non-commutative geometry in 19-dimensional locally flat gravigauge hyper-spacetime is expected to bring on a distinguishing dynamical symmetry breaking mechanism for the Higgs-like bosons arising from hyperspin gauge field associated to vector representations of extra spatial dimensions. To study further the possible quantum contribution to effective potential of Higgs-like bosons, it needs to develop a new mathematical method for calculating quantum effect in the hyperunified field theory formulated in locally flat gravigauge hyper-spacetime with emergent non-commutative geometry. 
 
On the other hand, let us turn to the hyperunified field theory formulated equivalently within the framework of gravitational quantum field theory based on the concept of biframe hyper-spacetime as shown in Eqs.(\ref{actionHUFTGQFT1})-(\ref{actionHUFTGQFT3}), where the hyperspin gauge field $\cA_{\fM}^{\fA\fB}$ and hyper-gravigauge field $\fA_{\fM}^{\; \fA}$ (or hyper-gravimetric field $\mH_{\fM\fN} = \fA_{\fM}^{\; \fA} \fA_{\fN \fA}$) become independent fundamental gauge fields. In such a framework of hyperunified field theory, the Higgs-like scalar fields are composed of two gauge fields $\cA_{\fM}^{\fA\fB}$ and $\fA_{\fM}^{\; \fA}$. To be explicit, let us examine the Higgs-like bosons $\cH_{K I J}$ and $\cH_{K \ta\tb}\equiv \cH_{K}^p$ which are presented as totally antisymmetric hyperspin gauge field in locally flat gravigauge hyper-spacetime. In gravitational quantum field theory based on the biframe hyper-spacetime, the Higgs-like bosons $\cH_{K I J}$ and $\cH_{K \ta\tb}$  can be expressed as follows:
\be \label{HLB1}
& & \cH_{K I J} \equiv \cA_{[K+4\; I+4\; J+4 ]} \equiv  \cA_{[\bC \bA\bB]}  \nn \\
& & = \frac{1}{3} ( \hfA_{\bC}^{\;\fM} \cA_{\fM\bA\bB} + \hfA_{\bA}^{\; \fM} \cA_{\fM\bB\bC} + \hfA_{\bB}^{\; \fM} \cA_{\fM\bC\bA}) \nn \\
& & \equiv \frac{1}{3} [\, \hfA_{K}^{\;\mu} \cA_{\mu IJ} + \hfA_{I}^{\; \mu} \cA_{\mu J K} + \hfA_{J}^{\; \mu} \cA_{\mu KI} \nn \\
& & \quad +  \hfA_{K}^{\;\bM} \cA_{\bM I J} + \hfA_{I}^{\; \bM} \cA_{\bM J K} + \hfA_{J}^{\; \bM} \cA_{\bM K I} \nn \\
& & \quad +  \hfA_{K}^{\;\tM} \cA_{\tM IJ} + \hfA_{I}^{\; \tM} \cA_{\tM J K} + \hfA_{J}^{\; \tM} \cA_{\tM K I} \, ] ,
\ee
and
\be  \label{HLB2}
& & \cH_{K \ta\tb} \equiv \cA_{[K+4\; \ta+14\; \tb+14 ]} \equiv \cA_{[\bC \tA\tB]}  \nn \\
& & = \frac{1}{3} ( \hfA_{\bC}^{\;\fM} \cA_{\fM\tA\tB} + \hfA_{\tA}^{\; \fM} \cA_{\fM\tB\bC} + \hfA_{\tB}^{\; \fM} \cA_{\fM\bC\tA}) \nn \\
& & \equiv \frac{1}{3} [\, \hfA_{K}^{\;\mu} \cA_{\mu\ta\tb} + \hfA_{\ta}^{\; \mu} \cA_{\mu\tb K} + \hfA_{\tb}^{\; \mu} \cA_{\mu K\ta} \nn \\
& & \quad + \hfA_{K}^{\;\bM} \cA_{\bM\ta\tb} + \hfA_{\ta}^{\; \bM} \cA_{\bM\tb K} + \hfA_{\tb}^{\; \bM} \cA_{\bM K\ta} \nn \\
& & \quad +  \hfA_{K}^{\;\tM} \cA_{\tM\ta\tb} + \hfA_{\ta}^{\; \tM} \cA_{\tM\tb K} + \hfA_{\tb}^{\; \tM} \cA_{\tM K \ta} \, ],
\ee 
with
\be
\mu = 0,1,2,3; \quad \bM = 5,\cdots, 14;\quad \tM = 15,\cdots, 19 .
\ee
which shows that the Higgs-like bosons may be regarded as composed fields of two fundamental fields $\cA_{\fM}^{\fA\fB}$ and $\hfA_{\fA}^{\; \fM}$ (dual to $\fA_{\fM}^{\; \fA}$) within the framework of gravitational quantum field theory. In such a theoretical framework, it allows us to analyze how hyperspin gauge field $\cA_{\fM}^{\fA\fB}$ and hyper-gravigauge field $\fA_{\fM}^{\; \fA}$ play the roles in connection to different extra dimensions when they are associated with Higgs-like bosons. As the hyper-gravigauge field $\fA_{\fM}^{\; \fA}$ not only characterizes geometric properties of hyper-spacetime via the hyper-gravimetric field $\mH_{\fM\fN} = \fA_{\fM}^{\; \fA} \fA_{\fN \fA}$ but also governs hyperspin gauge transformations based on the gravitational origin of hyperspin gauge symmetry as shown in Eqs.(\ref{HSGFDC})-(\ref{HSGFGT1}), the gauge symmetries associated with Higgs-like bosons can be made to be broken down through geometric symmetry breaking mechanisms\cite{GGFT6D}. Therefore, from the gravity-gauge and gauge-geometry correspondences shown in the hyperunified field theory, some geometric symmetry breaking scenarios proposed in extra dimension model\cite{RS} may also be applicable to the gauge symmetry breaking associated with Higgs-like bosons, which is usually carried out by first compacting the extra dimensions and then taking proper boundary conditions for both hyper-gravigauge and hyperspin gauge fields associated with Higgs-like bosons. 
 
To produce the mass spectrum and mixings of leptons and quarks determined in SM, it is essential to investigate the gauge symmetry breaking in connection with grand unified gauge symmetry SP(10) and family-spin gauge symmetry SP(5), which are the subgroup symmetries of hyperspin gauge symmetry SP(1,18), i.e., SP(1,3)$\times$SP(10)$\times$SP(5)$\in$SP(1,18). It is noted that the hyperspin gauge symmetry is a unified gauge symmetry which involves only one gauge coupling constant, thus a basic question arises naturally how the Higgs-like bosons associated with gauge fields obtain heavy masses and meanwhile the leptons and quarks in SM achieve correct hierarchical mass spectrum through Yukawa-type couplings of Higgs-like bosons, in particular, for the heavy top quark mass. To make a general issue for such a question, it is interesting to observe the fact that the extra dimensions of hyper-spacetime in the hyperunified field theory are all physically meaningful since they are correlated to the hyperspin charges which represent the intrinsic quantum numbers of hyperunified qubit-spinor field, such a feature is different from the usual extra dimension models and superstring theory. So that the hierarchical mass spectrum and mixings of leptons and quarks in SM should be related to the localizing distribution amplitudes of fundamental fields and characterized from gauge symmetry breaking of family-spin gauge symmetry SP(5) as well as reflected by geometric properties of extra dimensions described by relevant hyper-gravimetric field $\hmH^{\tM \tN} = \hfA_{\fA}^{\; \tM} \hfA_{\fA}^{\; \tN}\eta^{\fA\fB}$ ($\tM,\tN = 15,\cdots, 19$) and also by relevant dual hyper-gravigauge fields $\hfA_{\tA}^{\; \fM}$ and $\hfA_{\bA}^{\; \fM}$. As shown explicitly in Eqs.(\ref{HLB1}) and (\ref{HLB2}), the Higgs-like bosons $\cH_{K \ta\tb}\equiv \cH_{K}^p$ and $\cH_{K I J}$ rely on the dual hyper-gravigauge fields $\hfA_{\tA}^{\; \fM}\equiv \hfA_{\ta}^{\; \fM}$ and $\hfA_{\bA}^{\; \fM} \equiv \hfA_{I}^{\; \fM}$. 

In general, the hyperspin gauge symmetry including family-spin gauge symmetry and isospin gauge symmetry is expected to be broken down dynamically and geometrically via the effective potential of gauge fields associated with Higgs-like bosons and localizing distribution amplitudes of fundamental fields as well as the compactification of bulk space in hyper-spacetime, so as to generate the standard model and achieve the correct masses and mixings of leptons and quarks as well as sources of CP violation\cite{SCPV1a,SCPV1b,SCPV2} including CP-violating KM phase\cite{KM} in SM. Theoretically, to reproduce the observed Higgs boson of SM in four-dimensional spacetime, it requires to solve the dynamical evolutions of hyperspin gauge field and hyper-gravigauge field as well as hyperunified qubit-spinor field in hyper-spacetime. As such dynamical evolutions concern highly nonlinear equations and intrinsic properties of 15-dimensional compact bulk space in 19-dimensional hyper-spacetime, it is beyond the scope of present paper. Nevertheless, it should be interesting to figure out some reasonable phenomenological models at low energies, which is going to be investigated in detail elsewhere based on the above analyses and discussions.


\subsection{ Appearance of three families of chiral type leptons and quarks in SM and property of the fourth family of lepton-quark state with chirality-correlation discrete symmetry of bulk space }

The standard model with three families of chiral type leptons and quarks has been tested up to 1.4 TeV energy scale at LHC, while the hyperunified qubit-spinor field in 19-dimensional hyper-spacetime contains four families of vector-like lepton-quark states. Namely, there exist in general four families of chiral type lepton-quark states with negative U-parity and the corresponding four families of mirror lepton-quark states with positive U-parity in 19-dimensional hyper-spacetime. To comprehend current experimental observations on three families of chiral type leptons and quarks in SM, the inhomogeneous hyperspin symmetry and Poincar\'e-type symmetry must be broken down to the symmetry of SM in four-dimensional spacetime at low energies. In general, one may consider several alternative scenarios in treating the fourth family of vector-like lepton-quark states. A naive scenario is that there might exist the fourth family of chiral type leptons and quarks with very heavy masses and small mixings to other leptons and quarks in three families, so that they are still beyond the current experimental detections. Another simple scenario is that the fourth family is composed of vector-like lepton-quark states with masses above the current experimental lower limit around 1370 GeV\cite{PDG}. We will illustrate a possible scenario with the appearance of three families of chiral type leptons and quarks in SM and the presence of the fourth family of vector-like lepton-quark states. 

To realize three families of chiral type leptons and quarks in SM, it needs to appropriately explore symmetry breaking mechanisms and scenarios of hyperunified qubit-spinor field. For such a purpose, we are going to focus on the action of hyperunified qubit-spinor field after inflation and investigate the property of hyperunified qubit-spinor field in bulk space. From the actions presented in Eq.(\ref{actionHUFTGHS}) and Eq.(\ref{actionHUFTGQFTS}), it is straightforward to write down the action of hyperunified qubit-spinor field after inflation in Einstein type basis as follows:  
\be  \label{actionHUFTHQSF}
\cS_{\mH\mU} & \equiv &  \int [d^{19}x] \, \chi^e  \{ \bar{\fPsi}_{\fQH} \vSi_{-}^{\fC} i\cD_{\fC}  \fPsi_{\fQH}  -  m_{\fQH} \bar{\fPsi}_{\fQH} \tvSi_{-}\fPsi_{\fQH} \nn \\
& - & [ M_S\sqrt{1+\lachi^2\lak^2}\sinh \frac{\Phi}{M_S} + 2\lachi\mk \sinh^2\frac{\Phi}{2M_S} ] \,  \bar{\fPsi}_{\fQH} \tvSi_{-}\fPsi_{\fQH}  \} ,
\ee
with the covariant derivative:
\be
& & i\cD_{\fC} \equiv i\heth_{\fC} + \cA_{[\fC\fA\fB]}\frac{1}{2}\vSi^{\fA\fB} \equiv \bchi_{\fC}^{e \fM} i\cD_{\fM}  =  \bchi_{\fC}^{e \fM} ( i \p_{\fM} +  \cA_{[\fM\fA\fB]}\frac{1}{2}\vSi^{\fA\fB} ) .\nn
\ee

When the potential energy of $\cQ_c$-spin scaling field is transmuted into kinetic energies of fundamental fields after inflation, the universe is going to undergo the epoch of radiative and matter dominated expansions. The hyperunified qubit-spinor field and hyperspin gauge field as well as hyper-gravigauge field and $\cQ_c$-spin scalar field will evolve from high energy scale characterized by the fundamental mass scale $\Mka$ to low energies. The localizing distribution amplitudes of fundamental fields in hyper-spacetime are considered to appear as effective fields at low energies. Meanwhile, the inhomogeneous hyperspin symmetry is expected to be broken down to its subgroup symmetries, so that the standard model can consistently be realized in four-dimensional spacetime.

For illustration, we will examine a simple symmetry breaking scenario for hyperspin symmetry. It has been shown in ref.\cite{FHUFT-I} that the dimension of hyper-spacetime is correlated to the hyperspin charge of entangled hyperqubit-spinor field, so that the local distribution amplitudes of entangled hyperqubit-spinor field not only exhibit the hyperspin symmetry in Hilbert space but also reflect the motion-correlation symmetry in Minkowski hyper-spacetime. To arrive at three families of chiral type leptons and quarks in SM, it is necessary to bring local distribution amplitudes corresponding to three families of chiral type leptons and quarks of hyperunified qubit-spinor field in hyper-spacetime to be distinguishable from three families of mirror lepton-quark states and also from the fourth family of lepton-quark states. For simplicity, let us take the nineteenth spatial dimension of hyper-spacetime as a bulk dimension with its coordinate regarded as bulk-coordinate, so that the coordinates in 19-dimensional hyper-spacetime are decomposed into two parts:
\be
& & x^{\fM} = (x^{\mM}, x^{19}) \equiv (x, z ), \quad \mM =0, 1, 2, 3, 5,\cdots, 18 ,
\ee
with $x^{\mM}$ denoting the coordinates in 18-dimensional hyper-spacetime and $x^{19}\equiv z$ representing the bulk-coordinate. 

Let us consider chirality-correlation discrete symmetry of bulk dimension under $Z_2$ reflection operation. Namely, the local distribution amplitudes of grand unified qubit-spinor fields for four families of vector-like lepton-quark states $\Psi_{\fQGf}(x, z)$ ($f=1,2,3,4$) are proposed to obey the following chirality-correlation discrete $Z_2$-parity under reflection operation on bulk-coordinates $z\to -z$: 
\be
& & \Psi_{\fQGi}(x, - z) = -\gamma_{15} \Psi_{\fQGi}(x, z), \quad i=1,2,3, \nn \\
& & \Psi_{\fQGft}(x, - z) = - \Psi_{\fQGft}(x, z),  \quad \mbox{or} \quad \Psi_{\fQGft}(x, -z) = \Psi_{\fQGft}(x, z) ,\nn \\
& & \gamma_{15} =  \sigma_3 \otimes \sigma_0 \otimes  \sigma_0 \otimes  \sigma_0 \otimes \sigma_0\otimes \sigma_3\otimes \sigma_0 . 
\ee
Explicitly, we can express the chirality-correlation discrete $Z_2$-parity for four families of chiral type lepton-quark states $\Psi_{\fQGfn}(x,  z)$ and mirror lepton-quark states $\Psi_{\fQGfp}(x,  z)$ as follows: 
\be
& & \Psi_{\fQGni}(x, -  z) = + \Psi_{\fQGni}(x,  z), \;\;  \Psi_{\fQGpi}(x, -  z) = -\Psi_{\fQGpi}(x,  z), \;\;  i =1,2,3 , \nn \\
& & \Psi_{\fQGpft}(x, -  z) = \pm \Psi_{\fQGpft}(x,  z), \quad \Psi_{\fQGnft}(x, - z) = \pm \Psi_{\fQGnft}(x,  z) , \nn
\ee
which indicates that the distribution amplitudes for three families of chiral type lepton-quark states with negative U-parity are even functions under $Z_2$ reflection operation on the bulk coordinate $z$ and meanwhile the distribution amplitudes for three families of mirror lepton-quark states with positive U-parity are odd functions. Whereas the distribution amplitudes for both chiral type lepton-quark states and mirror lepton-quark states in the fourth family are considered to have the same $Z_2$-parity under $Z_2$ reflection operation. 

The symmetric hyper-gravigauge field $\bchih_{\fA}^{e \fM}(x,  z)$ under entirety unitary gauge and $\cQ_c$-spin scalar field $\Phi(x,  z)$ are considered to have the following properties under $Z_2$ reflection operation on the bulk coordinate $z$, i.e.:
\be
& & \bchih_{\mA}^{e\, \mM}(x, - z) = + \bchih_{\mA}^{e\, \mM}(x,  z), \quad \bchih_{19}^{e\, 19}(x, - z) = + \bchih_{19}^{e\, 19}(x,  z) , \nn \\
& & \bchih_{19}^{e\, \mM}(x, - z) = -\bchih_{19}^{e\, \mM}(x,  z), \quad \bchih_{\mA}^{e\, 19}(x, - z) = - \bchih_{\mA}^{e\, 19}(x,  z) , \nn \\
& &  \bchi^e(x,-z) = + \bchi^e(x,-z),  \quad \Phi(x, - z) = + \Phi(x,  z) , \nn \\
& & \mA, \mM = 0, 1, 2, 3, 5,\cdots, 18 .
\ee

It is useful to decompose the hyperunified qubit-spinor field $\fPsi_{\fQH}(x,  z)$, according to the properties of $Z_2$-parity and family-spin charge, into the sum of three parts, i.e.:
\be 
& & \fPsi_{\fQH}(x,  z) \equiv \fPsi_{\fQH}^{(-e)}(x,  z) +  \fPsi_{\fQH}^{(+o)}(x,  z) + \fPsi_{\fQH}^{(4o)}(x,  z), \quad \mbox{or} \nn \\
& & \fPsi_{\fQH}(x,  z) \equiv \fPsi_{\fQH}^{(-e)}(x,  z) +  \fPsi_{\fQH}^{(+o)}(x,  z) + \fPsi_{\fQH}^{(4e)}(x,  z),
\ee
which have the following explicit forms:
\be \label{HUQSFZ2}
& & \fPsi_{\fQH}^{(-e)}(x,  z) \equiv  \begin{pmatrix}
\Psi_{\fQGnf}^e(x,  z) \\ \Psi_{\fQGns}^e(x,  z) \\ \Psi_{\fQGnt}^e(x,  z) \\ 0  \\ 
0 \\ 0 \\ 0 \\  0
\end{pmatrix} , \quad 
 \fPsi_{\fQH}^{(+o)}(x,  z) \equiv  \begin{pmatrix}
0 \\ 0 \\ 0\\ 0 \\ 
\Psi_{\fQGpf}^o(x,  z) \\ \Psi_{\fQGps}^o(x,  z) \\ \Psi_{\fQGpt}^o(x,  z) \\  0
\end{pmatrix} , \nn \\
& & \fPsi_{\fQH}^{(4o)}(x,  z)  \equiv  \begin{pmatrix}
0 \\ 0 \\ 0 \\ \Psi_{\fQGnft}^{o}(x,  z)  \\ 
0 \\ 0 \\ 0 \\  \Psi_{\fQGpft}^{o}(x,  z) 
\end{pmatrix}, \quad 
\fPsi_{\fQH}^{(4e)}(x,  z)  \equiv  \begin{pmatrix}
0 \\ 0 \\ 0 \\ \Psi_{\fQGnft}^{e}(x,  z)  \\ 
0 \\ 0 \\ 0 \\  \Psi_{\fQGpft}^{e}(x,  z) 
\end{pmatrix}, 
\ee
where $\fPsi_{\fQH}^{(-e)}(x,  z)$ represents the first three families of lepton-quark states with negative U-parity and even $Z_2$-parity, and $\fPsi_{\fQH}^{(+o)}(x,  z)$ denotes the first three families of mirror lepton-quark states with positive U-parity and odd $Z_2$-parity. $\fPsi_{\fQH}^{(4e)}(x,  z)$ and $\fPsi_{\fQH}^{(4o)}(x,  z)$ denote the fourth family of vector-like lepton-quark states with even $Z_2$-parity and odd $Z_2$-parity, i.e.:
\be
& & \fPsi_{\fQH}^{(4e)}(x, - z) = \fPsi_{\fQH}^{(4e)}(x,  z) ,  \nn \\
& & \fPsi_{\fQH}^{(4o)}(x, - z) = - \fPsi_{\fQH}^{(4o)}(x,  z) .
\ee

The invariance of $Z_2$ reflection discrete symmetry leads the action for the kinetic term and $\cQ_c$-spin scalar coupling term of hyperunified qubit-spinor field in Eq.(\ref{actionHUFTHQSF}) to have the following form:
\be  \label{actionHUFTHQSF1o}
\cS_{\mH\mU}^{\ast} & \equiv & \int d^{19}x \bchi^e(x,  z)\,  \{ \bar{\fPsi}_{\fQH}^{(-e)}(x,  z) \vSi_{-}^{A} ( \bchih_{A}^{e\, \mM} i\p_{\mM}  + \bchih_{A}^{e\, 19} i\p_{z} ) \fPsi_{\fQH}^{(-e)}(x,  z)   \nn \\
& + & \bar{\fPsi}_{\fQH}^{(+o)}(x,  z)  \vSi_{-}^{A} (\bchih_{A}^{e\, \mM} i\p_{\mM}+ \bchih_{A}^{e\, 19} i\p_{z} ) \fPsi_{\fQH}^{(+o)}(x,  z) \nn \\
& + &  \bar{\fPsi}_{\fQH}^{(4o)}(x,  z)  \vSi_{-}^{a} (\bchih_{a}^{e\, \mM} i\p_{\mM} + \bchih_{a}^{e\, 19} i\p_{z} ) \fPsi_{\fQH}^{(4o)}(x,  z) \nn \\
& + & \bar{\fPsi}_{\fQH}^{(-e)}(x,  z) \vSi_{-}^{19} (\bchih_{19}^{e\, \mM} i\p_{\mM}+ \bchih_{19}^{e\, 19}  i\p_{z}) \fPsi_{\fQH}^{(+o)}(x,  z)  \nn \\
& + & \bar{\fPsi}_{\fQH}^{(+o)}(x,  z)  \vSi_{-}^{19} (\bchih_{19}^{e\, \mM} i\p_{\mM}+ \bchih_{19}^{e\, 19}  i\p_{z}) \fPsi_{\fQH}^{(-e)}(x,  z) 
\nn \\
& + & \bar{\fPsi}_{\fQH}^{(4o)}(x,  z) \vSi_{-}^{\wtA} ( \bchih_{\wtA}^{e\, \mM}  i\p_{\mM} +  \bchih_{\wtA}^{e\, 19}  i\p_{z} )  \fPsi_{\fQH}^{(+o)}(x,  z)  \nn \\
& + &  \bar{\fPsi}_{\fQH}^{(+o)}(x,  z)  \vSi_{-}^{\wtA}  ( \bchih_{\wtA}^{e\, \mM}  i\p_{\mM} +  \bchih_{\wtA}^{e\, 19}  i\p_{z} ) \fPsi_{\fQH}^{(4o)}(x,  z)  \nn \\
& - & [ M_S\sqrt{1+\lachi^2\lak^2}\sinh \frac{\Phi(x,  z)}{M_S} + 2\lachi\mk \sinh^2\frac{\Phi(x,  z)}{2M_S} ] \, \bar{\fPsi}_{\fQH}^{(4o)}(x,  z) \tvSi_{-}  \fPsi_{\fQH}^{(4o)}(x,  z) \nn \\
& - & m_{\fQH} \bar{\fPsi}_{\fQH}^{(4o)}(x,  z) \tvSi_{-}  \fPsi_{\fQH}^{(4o)}(x,  z) ,
\ee
for the case of $\fPsi_{\fQH}^{(4o)}(x,  z)$ with odd $Z_2$-parity, and 
\be  \label{actionHUFTHQSF1e}
\cS_{\mH\mU}^{\ast} & \equiv & \int d^{19}x \bchi^e(x,  z)\,  \{ \bar{\fPsi}_{\fQH}^{(-e)}(x,  z)  \vSi_{-}^{A} ( \bchih_{A}^{e\, \mM} i\p_{\mM}  + \bchih_{A}^{e\, 19} i\p_{z} ) \fPsi_{\fQH}^{(-e)}(x,  z)   \nn \\
& + & \bar{\fPsi}_{\fQH}^{(+o)}(x,  z)   \vSi_{-}^{A} ( \bchih_{A}^{e\, \mM} i\p_{\mM}  + \bchih_{A}^{e\, 19} i\p_{z} ) \fPsi_{\fQH}^{(+o)}(x,  z) \nn \\
& + & \bar{\fPsi}_{\fQH}^{(4e)}(x,  z) \vSi_{-}^{a} ( \bchih_{a}^{e\, \mM} i\p_{\mM}  + \bchih_{a}^{e\, 19} i\p_{z} ) \fPsi_{\fQH}^{(4e)}(x,  z)    \nn \\
& + & \bar{\fPsi}_{\fQH}^{(-e)}(x,  z) \vSi_{-}^{19} (\bchih_{19}^{e\, \mM} i\p_{\mM}+ \bchih_{19}^{e\, 19}  i\p_{z})  \fPsi_{\fQH}^{(+o)}(x,  z)  \nn \\
& + &  \bar{\fPsi}_{\fQH}^{(+o)}(x,  z)  \vSi_{-}^{19} (\bchih_{19}^{e\, \mM} i\p_{\mM}+ \bchih_{19}^{e\, 19}  i\p_{z})  \fPsi_{\fQH}^{(-e)}(x,  z) \nn \\
& + & \bar{\fPsi}_{\fQH}^{(4e)}(x,  z) \vSi_{-}^{\wtA} ( \bchih_{\wtA}^{e\, \mM}  i\p_{\mM} +  \bchih_{\wtA}^{e\, 19}  i\p_{z} ) \fPsi_{\fQH}^{(-e)}(x,  z)  \nn \\
& + & \bar{\fPsi}_{\fQH}^{(-e)}(x,  z)  \vSi_{-}^{\wtA} ( \bchih_{\wtA}^{e\, \mM}  i\p_{\mM} +  \bchih_{\wtA}^{e\, 19}  i\p_{z} ) \fPsi_{\fQH}^{(4e)}(x,  z)  \nn \\
& - & [ M_S\sqrt{1+\lachi^2\lak^2}\sinh \frac{\Phi(x,  z)}{M_S} + 2\lachi\mk \sinh^2\frac{\Phi(x,  z)}{2M_S} ] \bar{\fPsi}_{\fQH}^{(4e)}(x,  z)\tvSi_{-} \fPsi_{\fQH}^{(4e)}(x,  z)  \nn \\
& - & m_{\fQH} \bar{\fPsi}_{\fQH}^{(4e)}(x,  z)\tvSi_{-} \fPsi_{\fQH}^{(4e)}(x,  z) , 
\ee
for the case of $\fPsi_{\fQH}^{(4e)}(x,  z)$ with even $Z_2$-parity. We have introduced the following notations:
\be
& & \fA \equiv (A, \tA,19) \equiv (a, \bA, \tA, 19)\equiv (a, \wtA, 19), \quad A= 0,1,2,3,5,\cdots, 14, \nn \\
& & a = 0,1,2,3, \quad \bA= 5,\cdots, 14, \quad \tA = 15, \cdots, 18, \quad  \wtA = 5, \cdots, 18 .
\ee

The distribution amplitudes of grand unified qubit-spinor fields in bulk dimension are thought to be localized in a finite size, so that the grand unified qubit-spinor fields with even $Z_2$-parity are expected to have zero modes and can be rewritten into the following two parts:
\be
& & \Psi_{\fQGni}(x,  z) \equiv \vPsi_{\fQGni}(x)f_{\fQGni}( z) + \Psi_{\fQGni}^e(x,  z), \quad i=1,2,3, 
\ee
where the first term $\vPsi_{\fQGni}(x)$ represents zero modes which are regarded as three families of chiral type leptons and quarks in SM. The second term $\Psi_{\fQGni}^e(x,  z)$ denotes all possible Kaluza-Klein excited states (KK-modes) corresponding to bulk dimension\cite{TK,OK}.

It is straightforward to express the above actions in terms of the grand unified qubit-spinor fields for the chiral type lepton-quark states $\Psi_{\fQGfn}(x,  z)$ and mirror lepton-quark states $\Psi_{\fQGfp}(x,  z)$ ($f=1,2,3,4$). For illustration, let us write down explicitly the above action for the case with odd $Z_2$-parity fourth family of vector-like lepton-quark state as follows:
\be  \label{actionHUFTHQSF2o}
\cS_{\mH\mU}^{\ast} & \equiv & \int d^{19}x \chi^e(x, z)\, \frac{1}{2} \{\, 
\sum_{i=1}^{3} [ \, f_{\fQGni}^2(z) \bar{\vPsi}_{\fQGni}(x) \vGa^{a} \bchih_{a}^{e \mM} i\p_{\mM} \vPsi_{\fQGni}(x) 
 \nn \\
&+ & \bar{\Psi}_{\fQGni}^e(x, z) \vGa^{a} (\bchih_{a}^{e \mM}i\p_{\mM} + \bchih_{a}^{e 19} i\p_{z} ) \Psi_{\fQGni}^e(x, z)  \nn \\
& + &  f_{\fQGni}(z) \bchih_{a}^{e \mM}  \left( \bar{\vPsi}_{\fQGni}(x) \vGa^{a}i\p_{\mM} \Psi_{\fQGni}^e(x, z)
+ \bar{\Psi}_{\fQGni}^e(x, z) \vGa^{a} i\p_{\mM} \vPsi_{\fQGni}(x) \right) \nn \\
& + & \bchih_{a}^{e 19}  \left( f_{\fQGni}(z)\bar{\vPsi}_{\fQGni}(x) \vGa^{a}i\p_{z} \Psi_{\fQGni}^e(x, z) + \bar{\Psi}_{\fQGni}^e(x, z) \vGa^{a} \vPsi_{\fQGni}(x) i\p_{z} f_{\fQGni}( z) \right)  \nn \\
&+ & \bar{\Psi}_{\fQGpi}^o(x, z) \vGa^{a} (\bchih_{a}^{e \mM}i\p_{\mM} + \bchih_{a}^{e 19} i\p_{z} ) \Psi_{\fQGpi}^o(x, z)  ] \nn \\
& - & (\bar{\vPsi}_{\fQGnf}(x)f_{\fQGnf}( z) +  \bar{\Psi}_{\fQGnf}^e(x, z) ) \vGa^{\bA} (\bchih_{\bA}^{e \mM}i\p_{\mM} + \bchih_{\bA}^{e 19} i\p_{z} ) (\vPsi_{\fQGns}(x)f_{\fQGns}( z) +  \Psi_{\fQGns}^e(x, z) ) \nn \\
& + & (\bar{\vPsi}_{\fQGns}(x)f_{\fQGns}( z)+ \bar{\Psi}_{\fQGns}^e(x, z) ) \vGa^{\bA} (\bchih_{\bA}^{e \mM}i\p_{\mM} + \bchih_{\bA}^{e 19} i\p_{z} ) (\vPsi_{\fQGnf}(x)f_{\fQGnf}( z) +  \Psi_{\fQGnf}^e(x, z) ) \nn \\
&- & \bar{\Psi}_{\fQGpf}^o(x, z) \vGa^{\bA} (\bchih_{\bA}^{e \mM}i\p_{\mM} + \bchih_{\bA}^{e 19} i\p_{z} ) \Psi_{\fQGps}^o(x, z) \nn \\
& + & \bar{\Psi}_{\fQGps}^o(x, z) \vGa^{\bA} (\bchih_{\bA}^{e \mM}i\p_{\mM} + \bchih_{\bA}^{e 19} i\p_{z} ) \Psi_{\fQGpf}^o(x, z)\nn \\
& + & \sum_{i}^{3}  \bar{\delta}_{ii} [  \bar{\Psi}_{\fQGpi}^o(x,  z)  (\bchih_{19}^{e\, \mM} \p_{\mM} + \bchih_{19}^{e\, 19} \p_{z})  ( \vPsi_{\fQGni}(x)f_{\fQGni}( z) + \Psi_{\fQGni}^e(x,  z) ) \nn \\
& - & ( \bar{\vPsi}_{\fQGni}(x)f_{\fQGni}( z)+  \bar{\Psi}_{\fQGni}^{e}(x,  z) )(\bchih_{19}^{e\, \mM} \p_{\mM} + \bchih_{19}^{e\, 19} \p_{z}) \Psi_{\fQGpi}^{o}(x,  z)   ] \nn \\
& + & \bar{\Psi}_{\fQGft}^{o}(x,  z)  \vGa^{a} (\bchih_{a}^{e \mM}i\p_{\mM} + \bchih_{a}^{e 19} i\p_{z} )   \Psi_{\fQGft}^{o}(x,  z) \nn \\
& + & \sum_{i}^{2}  [\bar{\Psi}_{\fQGpi}^{o}(x,  z) \eta^{\tA}_{i4}   (\bchih_{\tA}^{e \mM}i\p_{\mM} + \bchih_{\tA}^{e 19} i\p_{z} ) \Psi_{\fQGft}^{o}(x,  z)  \nn \\
& + & \bar{\Psi}_{\fQGft}^o(x,  z)  \eta^{\tA}_{4i}   (\bchih_{\tA}^{e \mM}i\p_{\mM} + \bchih_{\tA}^{e 19} i\p_{z} ) \Psi_{\fQGpi}^o(x,  z)  ] \nn \\
& +  & \bar{\Psi}_{\fQGft}^o(x, z)  \vGa^{\bA}  (\bchih_{\bA}^{e \mM}i\p_{\mM} + \bchih_{\bA}^{e 19} i\p_{z} )\Psi_{\fQGpt}^o(x, z)\nn \\
& - & \bar{\Psi}_{\fQGpt}^o(x, z) \vGa^{\bA} (\bchih_{\bA}^{e \mM}i\p_{\mM} + \bchih_{\bA}^{e 19} i\p_{z} )  \Psi_{\fQGft}^o(x, z) \nn \\
& - & [ M_S\sqrt{1+\lachi^2\lak^2}\sinh \frac{\Phi(x, z)}{M_S} + 2\lachi\mk \sinh^2\frac{\Phi(x, z)}{2M_S} ] \bar{\Psi}_{\fQGft}^o(x, z) \Psi_{\fQGft}^o(x, z) \nn \\
& - & m_{\fQH} \bar{\Psi}_{\fQGft}^o(x, z) \Psi_{\fQGft}^o(x, z)  \} ,
\ee
with $\bar{\delta}_{ii} = (1,1,-1)$ ($i=1,2,3$). Similarly, one can rewrite the action in Eq.(\ref{actionHUFTHQSF1e}) for the case with even $Z_2$-parity fourth family of vector-like lepton-quark state. In such a formalism of the action, the relevant matrices $\eta^{\tA} \equiv \eta^{\ta + 14} \to \eta_{\ta}$ with $\tA=15,16,17,18$ (i.e., $\ta=1,2,3,4$) are presented in Eq.(\ref{ETAM}) and the $\vGa$-matrices $\vGa^{A} \equiv (\vGa^{a}, \vGa^{\bA} )$ are defined as follows:
\be \label{GMffQG}
& & \vGa^0 =\;\; \;  \sigma_0 \otimes \sigma_0 \otimes \sigma_0 \otimes \sigma_0 \otimes\sigma_0 \otimes \sigma_1 \otimes \sigma_0, \nn \\
& &  \vGa^1 =\;\;  i  \sigma_0 \otimes \sigma_0 \otimes \sigma_0 \otimes \sigma_0 \otimes \sigma_0\otimes \sigma_2\otimes \sigma_1, \nn \\
& & \vGa^2 = \;\;  i  \sigma_0 \otimes \sigma_0 \otimes \sigma_0 \otimes  \sigma_0 \otimes\sigma_0\otimes  \sigma_2\otimes \sigma_2, \nn \\
& & \vGa^3 = \;\; i  \sigma_0 \otimes \sigma_0 \otimes \sigma_0 \otimes \sigma_0 \otimes\sigma_0\otimes  \sigma_2\otimes \sigma_3, \nn \\
& &  \vGa^5 = i  \sigma_1 \otimes \sigma_0 \otimes \sigma_1 \otimes \sigma_0 \otimes \sigma_2\otimes  \sigma_3\otimes \sigma_0, \nn \\
& & \vGa^6 = i  \sigma_1 \otimes \sigma_0 \otimes \sigma_2 \otimes  \sigma_3 \otimes\sigma_2\otimes  \sigma_3\otimes \sigma_0, \nn \\
& & \vGa^7 = i  \sigma_1 \otimes \sigma_0 \otimes \sigma_1 \otimes \sigma_2 \otimes\sigma_3\otimes  \sigma_3\otimes \sigma_0 , \nn \\
& &  \vGa^8 = i  \sigma_1 \otimes \sigma_0 \otimes \sigma_2 \otimes \sigma_2 \otimes  \sigma_0\otimes \sigma_3\otimes \sigma_0, \nn \\
& &  \vGa^9 = i  \sigma_1 \otimes \sigma_0 \otimes \sigma_1 \otimes  \sigma_2 \otimes \sigma_1\otimes \sigma_3\otimes \sigma_0 , \nn \\
& &  \vGa^{10} =  i  \sigma_1 \otimes \sigma_0 \otimes \sigma_2 \otimes \sigma_1 \otimes \sigma_2\otimes \sigma_3\otimes \sigma_0 , \nn \\
& & \vGa^{11} =  i \sigma_2 \otimes \sigma_0 \otimes \sigma_0 \otimes  \sigma_0 \otimes \sigma_0\otimes \sigma_3\otimes \sigma_0 , \nn \\
& & \vGa^{12} =  i \sigma_1 \otimes \sigma_1 \otimes \sigma_3 \otimes  \sigma_0 \otimes \sigma_0\otimes \sigma_3\otimes \sigma_0 , \nn \\
& & \vGa^{13} =  i \sigma_1 \otimes \sigma_2 \otimes \sigma_3 \otimes  \sigma_0 \otimes \sigma_0\otimes \sigma_3\otimes \sigma_0 , \nn \\
& & \vGa^{14} =  i  \sigma_1 \otimes \sigma_3 \otimes  \sigma_3 \otimes  \sigma_0 \otimes \sigma_0\otimes \sigma_3\otimes \sigma_0 .
\ee 

To be specific, let us propose that the hyperunified qubit-spinor field $\fPsi_{\fQH}(x, z)$ and $\cQ_c$-spin scalar field as well as dual hyper-gravigauge field $\bchih_{\fA}^{e\, \fM}(x, z)$ have a periodic feature in the $z$-bulk dimension,
\be
& & \fPsi_{\fQH}(x, z-l_c) = \fPsi_{\fQH}(x, z+l_c), \nn \\
& & \Phi(x, z-l_c) = \Phi(x, z + l_c), \quad \bchih_{\fA}^{e\, \fM}(x, z-l_c) = \bchih_{\fA}^{e\, \fM}(x, z+l_c) ,
\ee
so that the distribution amplitudes with boundary conditions get the following explicit forms:
\be \label{HUQSFZ2L1}
\Psi_{\fQGfn}(x, z) & = & \sum_{n=0}\Psi_{\fQGfn}^{n}(x) \frac{1}{\sqrt{l_c}} \cos (\frac{z}{l_c}n\pi ) \nn \\
& \equiv & \frac{1}{\sqrt{2l_c}}\vPsi_{\fQGfn}(x) +\sum_{n=1}\Psi_{\fQGfn}^{n}(x)\frac{1}{\sqrt{l_c}}\cos (\frac{z}{l_c}n\pi ), \nn \\
\Psi_{\fQGfp}(x, z) & = & \sum_{n=1}\Psi_{\fQGfp}^{n}(x)\frac{1}{\sqrt{l_c}}\sin (\frac{z}{l_c}n\pi ),  \nn \\
 \Psi_{\fQGnft}(x, z) & = & \sum_{n=1}\Psi_{\fQGnft}^{n}(x)\frac{1}{\sqrt{l_c}}\sin (\frac{z}{l_c}n\pi ), \nn \\ 
 \Psi_{\fQGpft}(x, z) & = & \sum_{n=1}\Psi_{\fQGpft}^{n}(x)\frac{1}{\sqrt{l_c}}\sin (\frac{z}{l_c}n\pi ) ,  
\ee
and 
\be \label{HUQSFZ2L2}
 \Phi(x, z) & = & \sum_{n=0} \Phi^{n}(x) \cos (\frac{z}{l_c}n\pi )\equiv \varPhi(x) +\sum_{n=1}\Phi^{n}(x)\cos (\frac{z}{l_c}n\pi ), \nn \\
\bchih_{\mA}^{e\, \mM}(x, z) & = & \sum_{n=0} \bchih_{\mA}^{\; \mM\, n}(x)\cos (\frac{z}{l_c}n\pi )\equiv \bchih_{\mA}^{\;\mM}(x) +\sum_{n=1} \bchih_{\mA}^{\; \mM\, n}(x) \cos (\frac{z}{l_c}n\pi ), \nn \\
\bchih_{19}^{e\, 19}(x, z) & = & \sum_{n=0} \bchih_{19}^{\; 19\, n}(x)\cos (\frac{z}{l_c}n\pi )\equiv \bchih_{19}^{\; 19}(x) +\sum_{n=1} \bchih_{19}^{\; 19\, n}(x) \cos (\frac{z}{l_c}n\pi ), \nn \\
\bchih_{19}^{e\, \mM}(x, z) & = & \sum_{n=1} \bchih_{19}^{\; \mM\, n}(x)\sin (\frac{z}{l_c}n\pi ), \nn \\
\bchih_{\mA}^{e\, 19}(x, z) & = & \sum_{n=1} \bchih_{\mA}^{\; 19\, n}(x)\sin (\frac{z}{l_c}n\pi ) , 
\ee
with $f=1,2,3$ and $\mA, \mM = 0, 1, 2, 3, 5,\cdots, 18$. Where $\vPsi_{\fQGfn}(x)$ ($f=1,2,3$) are the {\it zero-mode grand unified qubit-spinor fields}, which correspond to the three families of leptons and quarks in SM. $\Psi_{\fQGfpn}^n(x)$ ($f=1,2,3,4$; $n=1,\cdots$) are referred to as the {\it KK-mode grand unified qubit-spinor fields}, $\varPhi(x)$ and $\Phi^{n}(x)$ are called the zero-mode and KK-mode $\cQ_c$-spin scalar fields. Similarly, $\bchih_{\mA}^{\;\mM}(x)$ and $\bchih_{19}^{\; 19}(x)$ are the zero mode hyper-gravigauge fields and others are the KK-mode hyper-gravigauge fields.

When applying the explicit expressions in Eqs.(\ref{HUQSFZ2L1}) and (\ref{HUQSFZ2L2}) for the decomposed fundamental fields to the action in Eq.(\ref{actionHUFTHQSF2o}), we arrive at the following action in 18-dimensional hyper-spacetime after integrating over the bulk coordinate $z$,
\be  \label{actionHUFTHQSF2o2}
\cS_{\mH\mU}^{\ast} & \equiv & \int d^{18}x\, \bchi(x)\bchi_{19}^{\, 19}(x) \, \frac{1}{2} \{\, 
\sum_{i=1}^{3} \bar{\vPsi}_{\fQGni}(x) \vGa^{a} \bchih_{a}^{\, \mM}(x) i\p_{\mM} \vPsi_{\fQGni}(x) 
 \nn \\
& - & [ \bar{\vPsi}_{\fQGnf}(x) \vGa^{\bA} \bchih_{\bA}^{e \mM}(x)i\p_{\mM} \vPsi_{\fQGns}(x)  -  \bar{\vPsi}_{\fQGns}(x) \vGa^{\bA} \bchih_{\bA}^{e \mM}(x) i\p_{\mM} \vPsi_{\fQGnf}(x) ] \nn \\
&+ & \sum_{n=1} \sum_{i=1}^{3} [ \bar{\Psi}_{\fQGni}^n(x) \vGa^{a} \bchih_{a}^{e \mM}(x)i\p_{\mM} \Psi_{\fQGni}^n(x)  +\bar{\Psi}_{\fQGpi}^n(x) \vGa^{a} \bchih_{a}^{e \mM}(x)i\p_{\mM}\Psi_{\fQGpi}^n(x)  ] \nn \\
& - & \sum_{n=1} [ \bar{\Psi}_{\fQGnf}^n(x) \vGa^{\bA} \bchih_{\bA}^{e \mM}(x)i\p_{\mM} \Psi_{\fQGns}^n(x)  -  \bar{\Psi}_{\fQGns}^n(x) \vGa^{\bA} \bchih_{\bA}^{e \mM}(x) i\p_{\mM} \Psi_{\fQGnf}^n(x) ] \nn \\
& - & \sum_{n=1} [ \bar{\Psi}_{\fQGpf}^n(x) \vGa^{\bA} \bchih_{\bA}^{e \mM}(x)i\p_{\mM} \Psi_{\fQGps}^n(x)  -  \bar{\Psi}_{\fQGps}^n(x) \vGa^{\bA} \bchih_{\bA}^{e \mM}(x) i\p_{\mM} \Psi_{\fQGpf}^n(x) ] \nn \\
& - & \sum_{n=1}^{\infty}  M_{n}^q \sum_{i=1}^{3} \chi_{19}^{\; 19}(x)[ \,  \bar{\Psi}_{\fQGni}^{n}(x) \Psi_{\fQGpi}^{n}(x) + \bar{\Psi}_{\fQGpi}^{n}(x)\Psi_{\fQGni}^{n}(x)\,  ] \nn \\
& + & \sum_{n=1}^{\infty} [ \,  \bar{\Psi}_{\fQGnft}^{n}(x) \bchih_{a}^{\; \mM}(x)  \vGa^{a} i\p_{\mM} \Psi_{\fQGnft}^{n}(x) + \bar{\Psi}_{\fQGpft}^{n}(x) \delta_{a}^{\; \mu}  \vGa^{a} i\p_{\mu} \Psi_{\fQGpft}^{n}(x) \, ] \nn \\
& +  & \sum_{n=1}^{\infty} [ \,  \bar{\Psi}_{\fQGpft}^{n}(x) \bchih_{\bA}^{\; \mM}(x)  \vGa^{\bA} i\p_{\mM} \Psi_{\fQGpt}^{n}(x) - \bar{\Psi}_{\fQGpt}^{n}(x) \bchih_{\bA}^{\; \mM}(x)  \vGa^{\bA} i\p_{\mM} \Psi_{\fQGpft}^{n}(x) \, ]  \nn \\
& + & \sum_{i}^{2}  [\bar{\Psi}_{\fQGpi}^{n}(x) \eta^{\tA}_{i4} \bchih_{\tA}^{\;\mM}(x)i\p_{\mM} \Psi_{\fQGft}^{n}(x) +\bar{\Psi}_{\fQGft}^n(x)  \eta^{\tA}_{4i} \bchih_{\tA}^{\;\mM}(x)i\p_{\mM} \Psi_{\fQGpi}^n(x)  ] \nn \\
& - & [ M_S\sqrt{1+\lachi^2\lak^2}\sinh \frac{\Phi(x)}{M_S} + 2\lachi\mk \sinh^2\frac{\Phi(x)}{2M_S} ] \bar{\Psi}_{\fQGft}^n(x) \Psi_{\fQGft}^n(x) \nn \\
& - & m_{\fQH} \bar{\Psi}_{\fQGft}^n(x) \Psi_{\fQGft}^n(x)  \} + \cL(\bchih_{\mA}^{\; \mM\, n}(x), \Phi^n(x) ) ,
\ee
with $\vGa^a$ ($a,\mu=0,1,2,3$) and $\vGa^{\bA}$ ($\bA, \bM = 5, \cdots, 14$). We have used the definition,
\be
& & M_{n}^q = \frac{n\pi}{l_c} , \quad n=1,\cdots, 
\ee
which provide the masses for the three family KK-mode grand unified qubit-spinor fields $\Psi_{\fQGfpn}^{n}(x)$ ($f=1,2,3$, $n=1,\cdots $). The $\cQ_c$-spin scalar field $\Phi(x)$ brings on the mass-like scalar interaction term for the fourth family grand unified qubit-spinor field with all KK-modes $\Psi_{\fQGfpn}^{n}(x)$ ($f=4$,  $n=1,\cdots $). In the above action, we only explicitly write down the leading action that concerns the zero modes of hyper-gravigauge field $\bchih_{\mA}^{\; \mM}(x)$ and $\cQ_c$-spin scalar field $\Phi(x)$. Where $\cL(\bchih_{\mA}^{\; \mM\, n}(x), \Phi^n(x) )$ represents the remaining part of the action that involves the KK-modes of hyper-gravigauge field $\bchih_{\mA}^{\; \mM\, n}(x)$ and $\cQ_c$-spin scalar field $\Phi^n(x)$.

It can be verified that the actions in Eqs.(\ref{actionHUFTHQSF1o}) and (\ref{actionHUFTHQSF1e}) or Eqs.(\ref{actionHUFTHQSF2o}) and (\ref{actionHUFTHQSF2o2}) possess the following hyperspin symmetries: 
\be
G_S & = & SP(1,13) \times SP(3) \cong  SO(1,13) \times SO(3) \nn \\
& \supset & SP(1,3)\times SP(10) \times SP(3) \cong SO(1,3)\times SO(10) \times SU_F(2) ,
\ee
for the first two families of grand unified qubit-spinor fields, and 
\be
G_S = SP(1,3)\times SP(10) \cong SO(1,3)\times SO(10) ,
\ee
for the third and fourth families of grand unified qubit-spinor fields.  It is seen that the first two families of grand unified qubit-spinor fields still possess the ultra-grand unified hyperspin symmetry SP(1,13)$\cong$SO(1,13)\cite{UM}, while the third and fourth families have only the grand unified symmetry SP(10)$\cong$SO(10)\cite{GUT1,GUT2}.

So far we just illustrate a simple symmetry breaking scenario of family-spin symmetry SP(5) as subgroup symmetry of hyperspin symmetry SP(1,18) by considering chirality-correlation $Z_2$ discrete symmetry of bulk dimension, which provides instructively a possible mechanism to comprehend the basic issue why there appear only three families of chiral type leptons and quarks in SM with maximum parity violation\cite{PV1,PV2,PV3}. Nevertheless, to obtain explicitly the localizing distribution amplitudes of grand unified qubit-spinor fields with even $Z_2$-parity and odd $Z_2$-parity in bulk dimension and realize eventually the standard model in four-dimensional spacetime, it needs to solve the equations of motion for grand unified qubit-spinor fields and hyper-gravigauge field as well as hyperspin gauge field. As those equations involve highly nonlinear dynamics of fundamental fields, it is not an easy task to solve the nonlinear dynamics without making further assumption, which is beyond the scope of present paper.



\section{ Conclusions and discussions}

As in part II of the foundation of the hyperunified field theory, we have made an extended analysis and investigation by following on the previous paper as the part I of the foundation of the hyperunified field theory\cite{FHUFT-I}. In the part I, we have mainly focused on the studies about fundamental building block and fundamental symmetry based on the maximum coherence motion principle and maximum entangled-qubits motion principle as guiding principles. A detailed investigation showed in part I that the entangled hyperqubit-spinor field brings about the fundamental building block with inhomogeneous hyperspin symmetry as the fundamental symmetry. Nevertheless, we have only investigated the free motion and symmetry property of entangled hyperqubit-spinor field and meanwhile paid to the categorization of entangled hyperqubit-spinor field and spacetime dimension without discussing any dynamical interaction.

To investigate the dynamics of entangled qubit-spinor field via interactions, we have proposed in this paper the gauge invariance principle as guiding principle, which states that the laws of nature should be independent of the choice of local field configurations. So that all global symmetries in Hilbert space are supposed to be local symmetries, which leads us to take inhomogeneous $\cM_c$-spin/hyperspin symmetry and $\cQ_c$-spin symmetry as well as global scaling symmetry of entangled qubit-spinor field in Hilbert space to be gauge symmetries and meanwhile introduce the corresponding gauge fields including bicovariant vector field $\chih_{\fA}^{\;\; \fM}(x)$ to build the gauge invariant action of entangled qubit-spinor field in category-$q_c$ based on the gauge invariance principle. The least action principle enables us to derive the gauge covariant equation of motion for category-$q_c$ entangled hyperqubit-spinor field with scaling gauge invariance and its quadratic form has been verified to be governed by the field strengths of gauge fields, which describes the gravitational relativistic quantum theory of entangled hyperqubit-spinor field in the presence of $\cQ_c$-spin scalar field and bicovariant vector field as hyper-gravigauge field. Furthermore, the dual pair of bicovariant vector fields $\chih_{\fA}^{\;\; \fM}(x)$ and $\chi_{\fM}^{\; \fA}(x)$ have been demonstrated to form locally flat gravigauge hyper-spacetime and bring on the basic concept of biframe hyper-spacetime, which provides the hyper-fiber bundle structure in mathematics with locally flat gravigauge hyper-spacetime characterized by non-commutative geometry. In particular, the hyper-gravigauge field $\chi_{\fM}^{\; \fA}(x)$ has been identified to the $\cW_e$-spin invariant-gauge field associated with graviscalar field, which arises initially from the entanglement-correlated $\cW_e$-spin gauge symmetry as subgroup gauge symmetry of inhomogeneous hyperspin gauge symmetry. Therefore, the gravitational interaction can truly be regarded as gauge interaction which originates from the translation-like $\cW_e$-spin Abelian-type gauge symmetry in locally flat gravigauge hyper-spacetime. Meanwhile, it is the $\cW_e$-spin invariant-gauge field as hyper-gravigauge field that brings about the gravitational origin of gauge symmetry in hyper-spacetime. In addition, the biframe displacement correspondence in biframe hyper-spacetime is proposed to relate the hyper-gravivector field as hyper-gravicoordinate vector in locally flat gravigauge hyper-spacetime to the coordinate vector in globally flat Minkowski hyper-spacetime, which enables us to define the hyper-gravicoordinate displacement and derivative with respect to hyper-gravivector field in locally flat gravigauge hyper-spacetime.

We have briefly provided an overall and intuitive description on the entangled decaqubit-spinor field as hyperunified qubit-spinor field, which was shown in the part I\cite{FHUFT-I} to unify all known leptons and quarks into a single fundamental building block. Its action in 19-dimensional hyper-spacetime possesses an associated symmetry which is given by the inhomogeneous hyperspin symmetry WS(1,18)=SP(1,18)$\rtimes$W$^{1,18}$ in association with inhomogeneous Lorentz-type/Poincar\'e-type group symmetry PO(1,18)=P$^{1,18}\ltimes$SO(1,18) together with global scaling symmetry SC(1). Following along the gauge invariance principle and scaling invariance hypothesis in biframe hyper-spacetime, we have demonstrated that a whole hyperunified field theory can generally be constructed by taking the entangle decaqubit-spinor field as hyperunified qubit-spinor field $\fPsi_{\fQH}(x)$ for characterizing the fundamental building block and the inhomogeneous hyperspin gauge interaction governed by inhomogeneous hyperspin gauge symmetry for describing the fundamental interaction. The gauge and scaling invariant action for whole hyperunified field theory is explicitly built in 19-dimensional hyper-spacetime with including the dynamics of scaling gauge field and $\cQ_c$-spin scalar field. The scaling invariance hypothesis under both global and local scaling symmetry transformations is shown to play an essential role for the construction of the action with hyper-gravigauge field as gravitational interaction. Such an action has been demonstrated to possess a joint symmetry in biframe hyper-spacetime, i.e., SC(1)$\ltimes$PO(1,18)$\Join$WS(1,18)$\rtimes$SG(1), with SC(1) and SG(1) corresponding to global scaling symmetry and local scaling gauge symmetry. WS(1,18) represents the inhomogeneous hyperspin gauge symmetry and PO(1,18) remains to be global inhomogeneous Lorentz-type/Poincar\'e-type group symmetry. When making locally flat gravigauge hyper-spacetime becomes globally flat one in the absence of gauge interactions, the motion-correlation inhomogeneous hyperspin symmetry WS(1,18) must be required from symmetry invariance of the action to associate directly with the inhomogeneous Lorentz-type/Poincar\'e-type group symmetry PO(1,18), so that the group transformations of SP(1,18) and SO(1,18) have to be coincidental each other. Such a hyperunified field theory can be expressed in terms of the hidden scaling gauge formalism to demonstrate the gauge-gravity correspondence, which enables us to derive the equations of motion for all fundamental fields to characterize their dynamics and meanwhile to obtain various conserved currents with respect to gauge symmetries. As such a hyperunified field theory is built in biframe hyper-spacetime with globally flat Minkowski hyper-spacetime regarded as free-motion hyper-spacetime, the action possesses global Poincar\'e-type group symmetry PO(1,18) and scaling symmetry SC(1), which allows us to derive the conservation laws and dynamic evolution equations by applying for Noether's theorem in the presence of gravitational interaction.

The $\cW_e$-spin invariant-gauge field as scaling gauge invariant hyper-gravigauge field is regarded as Goldstone-like boson, which has been shown to play an essential role as projection operator to define the hyper-spacetime gauge field from hyperspin gauge field. So that we are able to obtain an equivalent action for a whole hyperunified field theory in hidden gauge formalism. In such a hidden gauge formalism, the dynamics of hyper-gravigauge field can equivalently be characterized by the dynamics of Riemann geometry in terms of Einstein-Hilbert type action which is governed by the emergent general linear group symmetry GL(1,18, R). We have investigated the profound correlation between gravitational interaction and Riemann geometry of hyper-spacetime and demonstrated the gravity-geometry correspondence in curved Riemannian hyper-spacetime, which brings on the generalized gauge invariance principle that is applicable to global Lorentz-type group symmetry. By adopting the scaling gauge invariant hyper-gravigauge field as projection operator and Goldstone-like boson, we are able to rewrite the whole hyperunified field theory into an equivalent action in hidden coordinate formalism. As such a whole hyperunified field theory is completely expressed in locally flat gravigauge hyper-spacetime and holds in any coordinate systems, which indicates that the laws of nature are independent of the choice of coordinate systems. In locally flat gravigauge hyper-spacetime, a non-commutative geometry emerges via the non-Abelian Lie algebra structure of hyper-gravicoordinate derivative operator, such a Lie algebra structure factor is given by the hyperspin gravigauge field that characterizes the field strength of hyper-gravigauge field. As a consequence, the gravitational interaction in locally flat gravigauge hyper-spacetime appears as an emergent interaction characterized by the hyperspin gravigauge field that behaves as an auxiliary field. Therefore, the dynamics of all fundamental fields in locally flat gravigauge hyper-spacetime are verified to associate with the gravitational interaction through the emergence of non-commutative geometry, which enables us to demonstrate the geometry-gauge correspondence. 

We have explicitly presented a detailed analysis on the whole hyperunified field theory within the framework of gravitational quantum field theory (GQFT) based on the biframe hyper-spacetime. It has been shown that the dynamics of gravitational interaction can be described either by symmetric hyper-gravigauge field or equivalently by symmetric hyper-gravimetric gauge field under flowing unitary gauge with keeping equal degrees of freedom for gravitational interaction. In particular, the hyper-gravimetric field is shown to behave as Abelian-type gauge field $\mH_{\fM}^{\; \fP}$ with field strength $\mF_{\fM\fN}^{\fP}$, which are referred to as hyper-gravimetric gauge field and field strength with the duality relations in correspondence to hyper-gravigauge field $\fA_{\fM}^{\; \fA}$ and field strength $\mF_{\fM\fN}^{\fA}$. Such an equivalent description reveals the gauge-geometry duality for the gravitational interaction. To corroborate the gauge-geometry duality within the framework of GQFT, we have made a duality gauge as gauge prescription for the hidden general linear group symmetry GL(1,$D_h$-1, R) in hyper-spacetime. The combination of the duality gauge for general linear group symmetry GL(1,$D_h$-1, R) together with the flowing unitary gauge for hyperspin gauge symmetry SP(1,$D_h$-1) brings about a full gauge prescription, which is referred to as entirety unitary gauge. The gauge fixing conditions in such an entirety unitary gauge allow us to remove in principle all unphysical degrees of freedom, so that the fundamental symmetry of hyperunified field theory turns to be an associated symmetry for which the global inhomogeneous hyperspin symmetry WS(1,$D_h$-1) and scaling symmetry SG(1) in Hilbert space are in association with global Poincar\'e-type group symmetry PO(1,$D_h$-1) and scaling symmetry SC(1) in Minkowski hyper-spacetime, where the symmetry groups SC(1)$\ltimes$SP(1,$D_h$-1) and SO(1,$D_h$-1)$\rtimes$SC(1) should have coincidental symmetry transformations attributed to the isomorphic property, SC(1)$\ltimes$SP(1,$D_h$-1)$\cong$SO(1,$D_h$-1)$\rtimes$SC(1). 

Based on the gauge-geometry duality for gravitational interaction, we have derived the gauge-type gravitational equation with conserved current tensor and also the equivalent geometric gravitational equation in which the equation with symmetric tensor corresponds to geometric Einstein-type equation and the equation concerning antisymmetric tensor is beyond Einstein equation. The geometric Einstein-type equation is regarded as the extension of general relativity in four-dimensional spacetime to hyperunified field theory in hyper-spacetime. The dynamic evolution equation with antisymmetric tensor arises from the hyper-gravigauge field as fundamental gravitational gauge field instead of hyper-gravimetric field. The gauge-geometry duality indicates that the hyper-gravimetric field defined as the square of hyper-gravigauge field appears as Abelian-type gauge field, which is referred to as hyper-gravimetric gauge field. As the gravitational interaction is described by Abelian-type gauge field strength, which allows us to investigate electric-like and magnetic-like gravitational interactions in the hyperunified field theory within the framework of GQFT. It has been shown that the Abelian gauge-type gravitational equations in four dimensional spacetime are analogous to Maxwell equations of electromagnetic field and lead to the electromagnetic-like gravimetric gauge field equations. Unlike the electromagnetic field with source current solely determined by electric charged fields, the gravimetric gauge field current contains not only the current tensor arising from all basic matter fields but also the pure gravimetric gauge field current tensor due to highly nonlinear interactions of gravitational field. In addition, it distinguishes from Maxwell equations of electromagnetic field due to the absence of Bianchi identity for Abelian gauge-type gravitational equation.

Various scaling gauge fixing prescriptions have been analyzed in the whole hyperunified field theory under entirety unitary gauge since the scaling property becomes essential for the evolution of universe. We have demonstrated in Einstein-type basis for the scaling gauge fixing condition that the conformally flat gravigauge hyper-spacetime as nonsingularity background spacetime enables us to describe the evolution of early universe with evolving graviscalar field. Where the evolving $\cQ_c$-spin scaling field in the hyperunified field theory is shown to play a significant role for the evolution of universe. For the slowly varying $\cQ_c$-spin scaling field with extra large values in the genesis of universe, the conformally flat gravigauge hyper-spacetime as early universe is governed by an expanding graviscaling field characterized by the basic cosmic vector $\kappa_{\fM}$. The initial high energy potential of $\cQ_c$-spin scaling field determined by the fundamental mass scale $\Mka$ brings on an expansion of early universe when the background $\cQ_c$-spin scaling field undertakes slowly varying extra large values, which appears to undergo an inflationary expansion in terms of the co-moving cosmic time. When the inflationary early universe comes to an end, the heavy scaling gauge field is shown to be as a dark matter candidate. In particular, we have verified that the presence of minimum potential for $\cQ_c$-spin scaling field not only ends up the inflationary expansion of early universe but also provides a tiny cosmic energy density as the dark energy which is characterized by the fundamental cosmic mass scale $\mk$. Meanwhile, the $\cQ_c$-spin scalar boson $\Phi(x)$ as quantum cosmic matter with tiny cosmic mass is expected to remain active for the evolution of universe.

The whole hyperunified field theory contains hyperunified qubit-spinor field as fundamental building block and inhomogeneous hyperspin gauge field mediating fundamental interaction as well as scaling gauge field as dark matter candidate and $\cQ_c$-spin scaling field as energy source for both inflationary early universe and accelerated expanding universe. Such a whole hyperunified field theory appears no manifest Higgs boson and meanwhile the hyperunified qubit-spinor field concerns four families of vector-like lepton-quark states in hyper-spacetime. To be compatible with SM, it has been shown that the hyperspin gauge field with vector representations beyond four-dimensional spacetime in 19-dimensional locally flat gravigauge hyper-spacetime can be regarded as scalar-like fields in view of four-dimensional spacetime. In particular, the scalar-like fields $\cH_{I}^{p}(x)$ with subscript $I=1,\cdots, 10$ corresponding to basic isospin and color-spin charges and superscript $p=1,\cdots, 10$ reflecting adjoint representation of family-spin symmetry appear as ten Higgs-like bosons with $I =7,8,9,10$ characterizing isospin doublet. We have examined a simple geometric symmetry breaking scenario for family-spin symmetry SP(5) as subgroup symmetry of hyperspin symmetry SP(1,18), where the chirality-correlation $Z_2$ discrete symmetry for the nineteenth dimension as bulk dimension has been proposed to provide a possible mechanism to understand why there appear only three families of chiral type leptons and quarks observed in current experiments and why there is maximum parity violation in nature. The fourth family is considered to be as vector-like lepton-quark states. 

In short, we have studied the foundation of the hyperunified field theory starting from the motional nature of fundamental building block, which brings on a consistent theoretical framework of unification theory to comprehend more fundamental questions which are unable to be answered in the standard models of particles physics and cosmology. Nevertheless, there are much more researches need to be carried out in order to reproduce the SM with quantitative predictions on the masses and mixings of leptons and quarks.


\centerline{{\bf Acknowledgement}}

This work was supported in part by the National Key Research and Development Program of China under Grant No.2020YFC2201501, the National Science Foundation of China (NSFC) under Grants  No.~11851302, No.~11851303, No.~11690022, No.~11747601 and special fund for theoretical physics, the Strategic Priority Research Program of the Chinese Academy of Sciences under Grant No. XDB23030100. It is grateful to the organizing committee of SUSY2021 for inviting me to deliver a plenary talk on "Foundation of Unification Theory and Space-based Gravitational Waves Detection".  I would like to dedicate this paper to Professor Zhou Guang-Zhao (K.C. Chou) and Professor Peng Huan-Wu for their great encouragements and stimulating discussions when I began to explore unified field theory.

\end{document}